%% file: MenssenThesis.tex
\documentclass[12pt]{ociamthesis}  %
\pdfoutput=1

\usepackage{amssymb}
\usepackage{setspace}
\usepackage[english]{babel}
\usepackage[utf8]{inputenc}
\usepackage{amsmath}
\usepackage[colorinlistoftodos]{todonotes}
\usepackage{graphicx}
\usepackage[colorinlistoftodos]{todonotes}
\usepackage{fancyhdr} 
\usepackage{lipsum}
\usepackage{mathrsfs}
\usepackage{youngtab}
\usepackage{bbold}
\usepackage{svg}
\usepackage{appendix}
\usepackage{afterpage}

\usepackage[colorlinks,
bookmarksnumbered,
citecolor=black,
urlcolor=black,
linkcolor=black
]{hyperref}
\usepackage[all]{hypcap} %
\hypersetup{
	unicode=false,          %
	pdftoolbar=true,        %
	pdfmenubar=true,        %
	pdffitwindow=false,     %
	pdfstartview={FitH},    %
	pdftitle={My Title},    %
	pdfauthor={My Name},     %
	pdfsubject={Subject of Document},   %
	pdfcreator={My Name},   %
	pdfproducer={My Name}, %
	pdfkeywords={Test} {Document}, %
	pdfnewwindow=true,      %
}

\newcommand\blankpage{%
	\null
	\thispagestyle{empty}%
	 \vspace{10cm}
	 {\centering Page intentionally left blank.\par}

	\addtocounter{page}{-1}%
	\newpage}

\newcommand\blankpageC{%
	\null
	\thispagestyle{empty}%
	\vspace{10cm}
	{\centering Page intentionally left blank.\par}

	\addtocounter{page}{-1}%
	\newpage}

\newcommand{\ket}[1]{|#1\rangle}
\newcommand{\braket}[2]{\langle{#1}|{#2}\rangle}
\newcommand{\ketbra}[1]{\ket{#1}\!\bra{#1}}
\newcommand{\bra}[1]{\langle#1|}

\def\eq{\begin{eqnarray}}
\def\en{\end{eqnarray}}

\def\beq{\begin{eqnarray}}
\def\een{\end{eqnarray}}
\def\bfig{\begin{figure}}
\def\efig{\end{figure}}

 \usepackage[normalem]{ulem}
\usepackage{bm}

 	\definecolor{forestgreen}{rgb}{0.13, 0.55, 0.13}

\title{Multi-photon interference phenomena}   %

\author{Adrian Johannes Menssen}             %
\college{Merton College}  %

\degree{Doctor of Philosophy}     %
\degreedate{Hilary 2019\\[0.5cm]Supervised by Prof. Ian A. Walmsley}         %

\begin{document}
\baselineskip=18pt plus1pt

\setcounter{secnumdepth}{3}
\setcounter{tocdepth}{3}

\maketitle                  %
\include{abstract}          %

\begin{romanpages}          %
\tableofcontents            %
\end{romanpages}            %

\begin{spacing}{2}
\pagestyle{fancy}
\fancyhf{}
\fancyhead[LO,RE]{\bfseries\footnotesize\thesection\hspace{1em}\rightmark\hspace{0cm}}
\fancyhead[RO,LE]{\thepage}
\renewcommand{\chaptermark}[1]{\markboth{\MakeUppercase{#1}}{}}
\renewcommand{\sectionmark}[1]{\markright{\MakeUppercase{#1}}{}}
\renewcommand{\headrulewidth}{.5pt}	
\pagestyle{plain}
\include{Dedication}
\include{Publications}
\include{Introduction}
\pagestyle{fancy}
\include{chapter2}
\include{chapter3}

\include{chapter4}
\include{chapter5}

\include{chapter6}
\include{chapter7}

\include{Conclusion}

\include{Appendix}
\afterpage{\blankpageC}
\end{spacing}

\addcontentsline{toc}{chapter}{Bibliography}
\bibliography{library}
\bibliographystyle{ieeetr}
\end{document}

%% file: abstract.tex
\begin{abstract}
In this work I demonstrate the interference of three photons, a generalisation of the famous Hong Ou Mandel (HOM) interference. I show that three-photon interference is governed by four parameters and measure three-photon interference independent of two-photon interference. Surprisingly, even when the states of the photons are highly distinguishable they can still exhibit strong quantum interference, challenging our intuition formed by the double slit and HOM interference. This will be followed by a demonstration of four-photon interference, where surprisingly we can still observe a fringe when involved particles are pairwise orthogonal. To explain these effects, I will be presenting a new framework to describe multi-photon interference in terms of a graph-theoretical approach, which illustrates the origin of different orders of multi-photon interference. My work leads to a more general definition of what we regard as an interference fringe in multi-photon scattering. 
This study of multi-photon interference is followed by an interdisciplinary work between photonics and solid state physics in the newly developing field of topological photonics. 
Interference phenomena are inextricably tied to exchange symmetries of the particle. I realise a simulation of the Jackiw-Rossi model as a localised topological mode in a photonic-crystal analogue of the 2D graphene lattice. These modes have previously been shown to obey non-abelian exchange statistics. I succeed in the experimental demonstration of a single such excitation and I am able to study the detailed mode structure for the first time.
The mode is a result of a topological defect and is, as such, protected against errors that do not change the topology of the system. Furthermore, I demonstrate adiabatic transport of the mode across the crystal lattice and show first attempts towards a demonstration of their non-abelian braiding statistics.
To realise these experiments, I developed a new method based on a spatial light modulator to excite large modes in photonic crystals. %
\end{abstract}

%% file: dedication.tex
\begin{dedication}
To my Mother
\end{dedication}
\afterpage{\blankpage}

%% file: Publications.tex
\chapter*{The author's publications}
\section*{Publications}
\begin{flushleft}
\begin{spacing}{1}
Adrian J. Menssen, Alex E. Jones, Benjamin J. Metcalf, Malte C. Tichy, Stefanie Barz, W. Steven Kolthammer, and Ian A. Walmsley. 2017. “Distinguishability and Many-Particle Interference.” Physical Review Letters 118 (15): 153603.\bigskip
	
Renema, J. J., Adrian J. Menssen, W. R. Clements, G. Triginer, W. S. Kolthammer, and I. A. Walmsley. 2018. “Efficient Classical Algorithm for Boson Sampling with Partially Distinguishable Photons.” Physical Review Letters 120 (22): 220502.\bigskip
	
Adrian J. Menssen, Jun Guan, David Felce, Martin J. Booth, and Ian A. Walmsley. 2019. “A Bulk Topological Photonic Zero Mode Bound to a Vortex Defect”, 2019 In preparation \bigskip
	
Guan, Jun, Xiang Liu, Adrian J. Menssen, and Martin J. Booth*. 2018. “Microscopic Characterisation of Laser-Written Phenomena for Component-Wise Testing of Photonic Integrated Circuits.” arXiv [physics.optics]. arXiv. http://arxiv.org/abs/1802.08016.\bigskip

Alexander E. Jones, Adrian. J. Menssen, Helen. M. Chrzanowski, Valery. S. Shchesnovich, and Ian A. Walmsley, ``Interfering photons in orthogonal states'' 2019 In preparation
	
\section*{Conference contributions}
\subsection*{Invited talks}
CLEO 2017, San Jose, US ``Multiparticle distinguishability: three photons are different in four ways''
\subsection*{Contributed talks}
CLEO 2018, San Jose, US ``Interfering Photons in Orthogonal States''\bigskip

Photon16, Leeds, UK ``Interference of three partially distinguishable photons''

\subsection*{Chairing}
CLEO 2018, ``Symposium on Emerging Quantum Sensing Techniques and Applications''
\subsection*{Posters}
QUTE Workshop 2015, Sweden

\end{spacing}	
\end{flushleft}

%% file: Introduction.tex
\chapter{Introduction}
Quantum interference lies at the heart of quantum theory. The famous double slit experiment gave a first glimpse of the exotic behaviour of quantum particles. Light hitting two slits and forming an interference pattern on a screen for the first time revealed its wave-like behaviour. Seemingly, this wave picture is in conflict with the corpuscular theory of light, which was demonstrated in a measurement of the photo effect \cite{einstein1905erzeugung}. Another landmark experiment came in 1987 when Hong, Ou and Mandel (HOM) \cite{hong1987measurement} demonstrated the interference of two photons. They showed that when two identical photons enter the two arms of a 50:50 beamsplitter, they interfere in such a way that they will always emerge as a pair from one of the output arms. More accurately: the probability amplitudes for the different paths the photons take through a beamsplitter interfere \cite{pittman1996can}. Two-photon interference has been explored extensively \cite{rarity1990two,kwiat1990correlated,bennett2009interference,patel2010two}. Their scattering behaviour is linked to the bosonic nature of the particles. Indeed, for fermions the behaviour is the opposite: they always emerge separately \cite{matthews2013observing}. The strength of the interference is linked to how distinguishable \cite{mandel1991coherence,brod2019witnessing} the participating particles are. If the photons are prepared in completely distinguishable states, the interference disappears and the photons behave as classical particles. If we ``mark'' a particle by preparing it in a distinguishable state, we can retain information about which path it took through the interferometer. The availability of ``which-way'' information removes interference capability \cite{kocsis2011observing,schmidt2013momentum}. Surprisingly, which-way information also destroys the interference pattern when obtained after the particle is detected \cite{scully1982quantum}. In classical single-photon interference at a double slit, as well as in the two-photon HOM effect, interference disappears gradually as the particles are made more distinguishable. Our intuition is largely shaped by these two hallmark experiments.
In this work I challenge this intuition by showing that in the case of three-photon interference particles can be made very distinguishable and still exhibit strong interference \cite{menssen2017distinguishability,Tichy2015,shchesnovich2015,ra2013nonmonotonic}, and that in the case of four-photon scattering the participating photons can even be made pair-wise completely distinguishable while still interfering collectively  \cite{tichy2011four,shchesnovich2017interference,jones2018interfering}.
Various extensions and generalisations of HOM interference to more photons have been suggested \cite{campos2000three,lim2005generalized,mahrlein2017hong}. In this work I present evidence for the observation of three-photon interference by showing that it occurs independently of lower order interference \cite{agne2017observation,menssen2017distinguishability,brod2019witnessing}. Further, I present theoretical motivation and experimental evidence that four-photon interference exists independently of two- and three-photon interference \cite{jones2018interfering}.

Another important factor in governing the scattering behaviour is the exchange symmetry of the participating particles: for fermions and bosons this behaviour is quite different. How can we probe such behaviour using the tools of photonics? Photons are bosons, at a first glance it seems impossible to study the influence of other exchange statistics. However, there are a few ways around this limitation. The first approach uses independent photons in partially distinguishable states. It was noted in \cite{tillmann2015generalized,stanisic2018discriminating} that when a measurement is performed on a system of photons, where certain degrees of freedom are not resolved\footnote{As in the HOM experiment, where the measurement is strictly performed on the two spatial modes of the interferometer and temporal, spectral, polarisation modes are not resolved.}, the reduced state of the system obtained by tracing over the unresolved modes is no longer completely symmetric. This can give access to fermionic (corresponding to the anti-symmetric subspace) and in principle exotic non-abelian statistics (higher-dimensional invariant sub-spaces). We can also access different exchange symmetries by using entangled states \cite{tichy2017extending}. A pair of photons can be prepared in an anti-symmetric superposition of states and exhibit purely fermionic behaviour \cite{green1953generalized,branning2000interferometric,tichy2012many,tichy2017extending,matthews2013observing}. To probe other more exotic statistics we would require large, well controlled entangled states \cite{tichy2017extending}. 

My interest in particles with non-abelian statistics motivated me to pursue yet another avenue: Particles of exotic exchange statistics are known to exist as quasi-particle excitations in two-dimensional solid state systems. These ``anyons'' can possess exotic non-abelian exchange statistics \cite{stern2004geometric}, where exchanging two particles changes the state of the wavefunction not by just a global phase, as in the bosonic and fermionic case. These exotic particles are linked to topological defects in the band structure of solid state systems \cite{teo2010topological,asboth2016short}. By virtue of an analogy between the Schr\"odinger equation of the tight binding model governing a solid state system and the par-axial equations of light travelling in a photonic crystal, we can simulate solid state effects in a  photonic system \cite{ozawa2018topological}. Indeed this analogy has spawned an entirely new research field of ``topological photonics'', which has led to remarkable results simulating topological features such as topological insulators and protected ``edge states'' \cite{rechtsman2013photonic,rechtsman2013strain,rechtsman2016topological}. 

The topological modes we want to investigate occur in a non-interacting tight binding model. They possess a remarkable property: because they are ``tied'' to a topological feature, they are robust to perturbations that leave the topology of the system invariant \cite{noh2018topological}. This could make them suitable to protect photonic quantum information \cite{blanco2018topological}.

I demonstrate experimentally a single localised photonic topological mode \cite{iadecola2016non,jackiw1981zero} in the bulk of a 2D+1 material (photonic graphene), show its topological protection and demonstrate that it can be adiabatically translated across the graphene lattice. Previous work has mostly studied topological modes at the edge of a the crystal material, so-called ``edge states'' \cite{rechtsman2013photonic,rechtsman2013strain,rechtsman2016topological,szameit2010discrete}. Here I study an excitation that is localised within the bulk. Enabled by major advances in femtosecond-laser waveguide-fabrication technology \cite{Huang2016} and by implementing a new experimental method to excite modes that are spread over many lattice sites, I am able for the first time to study the spatial features of such bulk topological states.

This is a first experimental implementation of the Jackiw Rossi model \cite{jackiw1981zero}, which describes a solution to the Dirac equation in the presence of coupling to a scalar field containing a vortex, with a localised excitation at the centre of the vortex. %
Further, these modes are similar to Majorana bound states \cite{teo2010topological,chamon2010quantizing,milovanovic2008fractionalization} in solid state systems and have been shown to obey non-abelian exchange statistics \cite{iadecola2016non} (Appendix). As two of these modes are braided around each-other, one accumulates a relative phase with respect to the other mode \cite{stern2004geometric}. This is different from for example the fermionic case, where the whole wavefunction gains a phase.

\section{Thesis structure}
In the first part of this thesis, I will probe the exchange symmetry of photons, which obey bosonic statistics, in multi-photon scattering experiments. In Chapter \ref{chap:TheoSinglePhot} I am going to give an introduction to the theory of multi-photon scattering. I motivate how their exchange symmetry gives rise to non-classical quantum interference and develop new concepts to describe multipartite correlations in single-photon measurements in terms of a graph theoretical approach \cite{jones2018interfering,shchesnovich2018collective}. In Chapter \ref{chap:ThreePhot} I demonstrate three-photon interference. I first show how three-photon interference \cite{menssen2017distinguishability} is governed by a geometric phase ``triad phase''. Surprisingly, we can achieve a strong interference fringe even for very distinguishable photons by tuning this phase. Next, I demonstrate how the three-photon interference contribution is isolated from two-photon interference. In Chapter \ref{chap:FourPhot} I illustrate how four-photon interference can be understood in terms of a graph theoretical picture and develop a state configuration which allows to isolate four-photon interference. At the end of the chapter I present experimental evidence for the observation of four-photon interference in independent photons. In Chapter \ref{chap:TheoryTopModes} I first give a brief introduction to basic concepts in topological solid state physics, I then illustrate an analogy between crystals in solid state physics and photonic crystals comprised of waveguides. I proceed to discuss the theory of ``topological zero modes'' of the Jackiw-Rebbi \cite{Jackiw1976} and Jackiw-Rossi model \cite{jackiw1981zero,Jackiw2007}, the latter of which I want to realise in a photonic crystal.
In Chapter \ref{chap:SLMControl} I discuss how photonic crystals are manufactured from individual femtosecond laser written waveguides and how optical modes that spread over many lattice sites can be excited. I present an experimental tool which allows to excite large modes in photonic crystals by individually exciting waveguides with directed, phase stable beams projected by a Spatial Light Modulator (SLM). I show how this method is used to excite topological modes.
After establishing experimental methods and theoretical concepts in Chapters \ref{chap:TheoryTopModes} and \ref{chap:SLMControl}, in Chapter \ref{chap:ExpTopModes} I present an experimental implementation of a photonic simulation of the Jackiw-Rossi model \cite{jackiw1981zero,Jackiw2007}. The realisation in a photonic system was proposed in \cite{iadecola2016non}. I succeed in the measurement of a single topological mode, demonstrate that it is topologically protected, and that it can be adiabatically translated across the lattice. These modes have in a recent proposal \cite{iadecola2016non} been shown to exhibit non-abelian exchange behaviour; I show a first attempt towards measuring this exotic exchange statistics that is exhibited when ``braiding'' two of these modes around each other. 

%% file: chapter2.tex
\chapter{Theory of multi-photon scattering}
\label{chap:TheoSinglePhot}
\section{Introduction}
In this chapter we are going to examine the fundamentals of photon scattering. Richard P. Feynman once remarked in a public lecture that it is a hallmark of any good physicist to always keep a set of several distinct theoretical models explaining the same physical phenomena in their mind. Following this philosophy I will demonstrate a few different approaches to the problem of multi-photon scattering. %
\section{Exchange symmetry}
The influence of particle exchange statistics in single-photon interference experiments has been studied in previous experimental works \cite{sansoni2012two,matthews2013observing} and theoretical studies \cite{tichy2012many,stanisic2018discriminating,tillmann2015generalized}.
In 3+1 dimensional space-time there exist two species of particles, bosons and fermions, which are distinct in their behaviour under particle exchange.
When interchanging the position of two identical fermions/bosons, the two-particle wavefunction receives an additional phase factor of -1 in the case of fermions and is left unchanged in the case of bosons.
\subsection*{Mathematical preliminaries} 
Let us consider a state of individual quantum particles described by a vector $\ket{\alpha_{1},...,\alpha_{N}}$ in a Hilbert space that is a product of single-particle Hilbert spaces: $\mathcal{H}^{\otimes N}=\mathcal{H}_1\otimes\mathcal{H}_2\otimes...\otimes\mathcal{H}_N$, where $N$ is the number of particles. The $N$ quantum numbers $\alpha_i$ uniquely define the vector. 
To formally describe the exchange of single-particle Hilbert-spaces, I introduce the symmetric group $S_N$, which contains all permutations $\sigma$ of a set of $N$ elements. For a thorough introduction to the symmetric group see for example: \cite{Barcy1977}. We write elements of the symmetric group in terms of ``cycles''. A cycle corresponds to a cyclic permutation of elements of a list. Let us take $S_2$ as a simple example. It contains two members:
the swap and identity. The swap is written as the cycle: $\sigma_2=(1,2):\rightarrow(1\rightarrow2,2\rightarrow1)$ and the identity is written as the product of two cycles of one element.
$\sigma_1=(1)(2):\rightarrow(1\rightarrow1)(2\rightarrow2)$. The action on a list of numbers is then: $(1,2)\{1,2\}\rightarrow\{2,1\}$ and $(1)(2)\{1,2\}=\{1,2\}$.

The representation of permutation operators on a tensor product space is defined by the following action $\sigma\in S_N$\footnote{In my notation I do not differentiate between the abstract group element and its representation.}:
$\sigma\ket{\alpha_{1}}\otimes\ket{\alpha_{2}}\otimes...\otimes\ket{\alpha_{N}}= \ket{\alpha_{\sigma(1)}}\otimes\ket{\alpha_{\sigma(2)}}\otimes...\otimes\ket{\alpha_{\sigma(N)}}$, 
where $\sigma\in S_N$ and $\sigma(i)=\sigma\{1,...,N\}(i).$
\subsection*{Particle exchange}
\begin{figure}[h]
	\centering
	\includegraphics[width=0.5\textwidth]{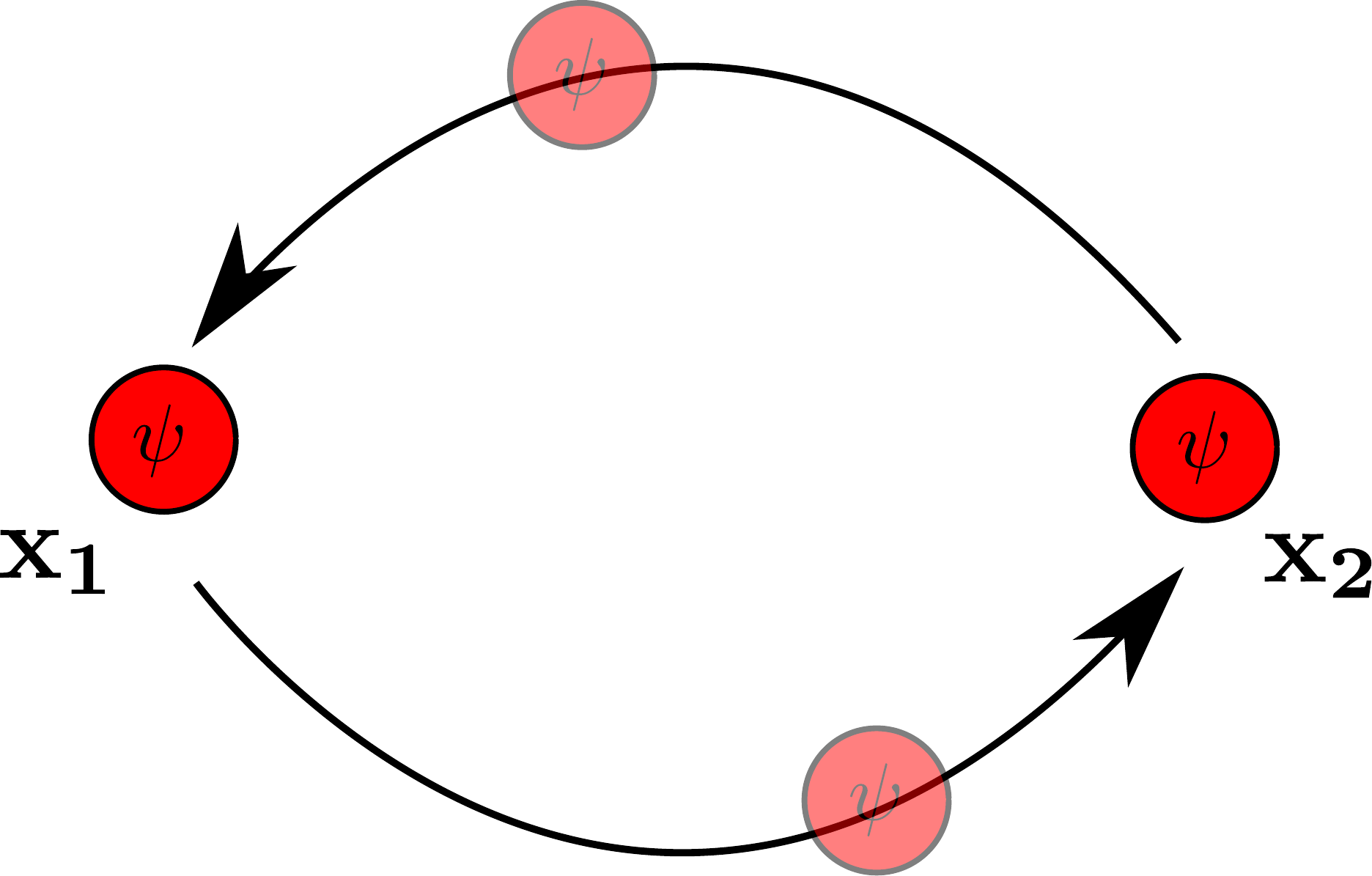}
	\label{fig:swap}
	\caption{Swapping two identical particles at different positions ${\bf x_1}$ and ${\bf x_2}$.}
\end{figure}

We can write a two-particle state as: $\ket{{\bf x_1},{\bf x_2}}$, where ${\bf x_1},{\bf x_2}$ are the quantum numbers of the first and second particle respectively. Swapping two identical particles should be indistinguishable by measurement, we can therefore at most gain a global phase \cite{leinaas1977theory}:
$\sigma_{2}\ket{{\bf x_1},{\bf x_2}}= e^{i\phi}\ket{{\bf x_1},{\bf x_2}}$, where $\sigma_{2} \in S_2$. Furthermore,
executing the swap operation twice should yield the initial state $\sigma_{2}^2\ket{{\bf x_1},{\bf x_2}}=e^{i2\phi}\ket{{\bf x_1},{\bf x_2}}$. It follows that $\phi=\{0,\pi\}$. 
Thus we have $\sigma_{2}\ket{{\bf x_1},{\bf x_2}}=\ket{{\bf x_1},{\bf x_2}}$ for bosons and $\sigma_{12}\ket{{\bf x_1},{\bf x_2}}=-\ket{{\bf x_1},{\bf x_2}}$ for fermions.
A general two-particle wavefunction that ensures this property takes the form 
\begin{equation}
	\ket{\Psi}=\ket{\alpha_1,\alpha_2}=\frac{1}{\sqrt{2}}(\ket{\alpha_1}\otimes\ket{\alpha_2} \pm \ket{\alpha_2}\otimes\ket{\alpha_1}).
\end{equation} 
$\alpha_{1/2}$ signifies a complete set of eigenvalues for particle 1 and 2 respectively, which will encompass spatial modes $\textbf{x}$.
In general
\begin{eqnarray}
	\ket{\Psi}_{Boson}&\in&\mathscr{S}(\mathcal{H}_1\otimes\mathcal{H}_2\otimes ... \mathcal{H}_N) \\ 
	\ket{\Psi}_{Fermion}&\in&\mathscr{A}(\mathcal{H}_1\otimes\mathcal{H}_2\otimes ... \mathcal{H}_N),
\end{eqnarray}
where $\mathscr{S}$ and $\mathscr{A}$ are the symmetriser and anti-symmetriser of the tensor product. They are projectors
on the symmetric/ anti-symmetric sub-spaces of the full tensor product space \cite{Barcy1977}.
The swapping operation can be regarded as a formal action on Hilbert space. However, we can also physically realise a particle swap:
we swap the position of a particle in Hilbert space $\mathcal{H}_1$ with the position of a particle in $\mathcal{H}_2$, where the particles are in the same state $\psi$ except for their position ${\bf x}\in R^3$. $\ket{{\bf x_1}}\in\mathcal{H}_1$ and $\ket{{\bf x_2}}\in\mathcal{H}_2$.  A particle swap is then realised by changing the state $\ket{{\bf x_1}}\rightarrow \ket{{\bf x_2}}$ and simultaneously moving the second particle $\ket{\bf x_2}\rightarrow \ket{\bf x_1}$. This can be realised by moving the particle adiabatically from state ${\bf x_1}$ to state ${\bf x_2}$ as illustrated in Figure \ref{fig:swap}. An experimental realisation of this could be two fermionic/bosonic atoms in optical traps which are slowly translated. This was suggested recently in \cite{roos2017revealing}.

A generalisation of this concept are (abelian) anyons
where under such an exchange the entire wavefunction gains an arbitrary phase
\begin{equation}
	\ket{\Psi}=\frac{1}{\sqrt{2}}(\ket{\alpha_1}\otimes\ket{\alpha_2} + e^{i\phi}\ket{\alpha_2}\otimes\ket{\alpha_1}).
\end{equation} 
Scattering statistics of abelian anyons has been simulated using entangled states \cite{matthews2013observing}.
In principle we could also imagine a type of particle that when exchanged in this way not only imparts a global phase on the multi particle wavefunction but can also change the state.
To capture this case we define a set of orthogonal states $\psi_{\alpha}$, $\alpha=1...$d spanning subspace of dimension d of the multi particle Hilbert space. $N$ Particles are localised at positions $r_1...r_N$. Upon particle exchange each component of the many-particle wavefunction is transformed in the following manner:
\begin{equation}
	\psi_\alpha\rightarrow U(\tau)_{\alpha}^\beta\psi_\beta ,
\end{equation}
where $U(\tau)$ can be a matrix representation of the element $\tau$ of the symmetric group of N particles $S_N$ \cite{Barcy1977,green1953generalized,carpenter1970green} or a matrix representation of an element of the braid group $B_N$ \cite{artin1947theory,nayak2008non}. This action on the particles' wavefunction under exchange can be used to realise quantum gates in a ``topological quantum computer'' \cite{nayak2008non}.
In both cases particles exhibiting this behaviour are called non-abelian, since the wavefunction of the particles exists in a degenerate irreducible subspace of the Hilbert space.  
It turns out that in 3 + 1 dimensions (3 spatial dimensions + time) particles are only either bosonic or fermionic \cite{leinaas1977theory}. However, in certain solid state systems quasi-particles live on lower-dimensional surfaces, such as a 2-D crystal lattice, these quasi-particles may exhibit non-abelian behaviour.
\section{The photon: state description and evolution}
\label{sect:singlephot}
Photons are excitations of the electromagnetic field. A  photon in a single temporal or spectral mode $\tau$ with polarisation $\sigma=$H/V is excited from the vacuum through the action of the creation operator in the following manner:
\begin{equation}
	a^{\dagger}(\tau,\sigma)\ket{0}=\ket{1(\tau,\sigma)}.
\end{equation}
This amounts to a single (delta-function normalised) excitation. For a photon at a single frequency this corresponds to a plane-wave.
The operators obey the usual commutation relations
\begin{equation}
	[a(\tau,\sigma),a^{\dagger}(\tau',\sigma')]=\delta(\tau,\tau')\delta(\sigma,\sigma').
\end{equation}
In quantum optics literature the validity of the concept of a photon wavefunction is discussed extensively, see for example \cite{smith2007photon,Fedorov2005}. In principle, the wavefunction of a photon cannot be defined in the context of a relativistic field theory. This is because it does not yield a valid probability density to calculate certain average quantities such as the photon's momentum. However, for photons of sufficiently narrow-band spectrum, when their spectral width $\Delta\omega$ is much smaller than their average frequency $\bar{\omega}$, $\Delta\omega \lll \bar{\omega}$, a single-photon wavefunction can be defined. This means that to a good approximation the wavefunction yields the correct probability density for calculating all average quantities of the photon. In the experimental work discussed in the following chapters this approximation is always well met.

The wavefunction of the photon can then be written as a superposition of the different amplitudes for the time/frequency and polarisation modes %
\begin{equation}
	\sum_\sigma\int  \text{d}\tau\psi(\tau,\sigma) a^{\dagger}(\tau,\sigma)\ket{0}=\psi^{\dagger}\ket{0}=\ket{\psi}.
\end{equation}

I will now discuss a specific example for a single-photon mode, a time delayed Gaussian wavepacket. Photons with a Gaussian shape are of particular relevance as they approximate well the state of a photon generated in the experiments presented in this thesis. I will also demonstrate how to calculate the overlap integrals of wavepackets at different times. These overlap integrals are of particular relevance in calculating event probabilities in multi-photon interference experiments, as will be discussed in the next section.
Taken from \cite{menssen2017distinguishability} (Appendix):
The state of a single-photon in the time-frequency modes $(\tau,\omega)$, delayed by time $t$ is given by
\begin{equation}
	\ket{t}=\int d\tau\phi(\tau-t)a^\dagger(\tau-t)\ket{0}.
\end{equation}
For a Gaussian wave-packet delayed by time $t$, central frequency $\Omega$ and variance in time $\sigma$,
$\phi(\tau;t)$  takes the form

\begin{equation}
	\phi(\tau-t)=\Big( \frac{1}{\pi\sigma^2}\Big)^{1/4}e^{-\frac{(\tau-t)^2}{2\sigma^2}+i\Omega(t-\tau)}.
\end{equation}

We can express the overlap of the temporal modes of two photons with identical Gaussian spectra at times $t_1$ and $t_2$ as
\begin{equation}
	\braket{t_1}{t_2}=\int_{-\infty}^{\infty}\phi^*(\tau-t_1)\phi(\tau-t_2)d\tau=e^{-\frac{(t_1-t_2)^2}{4\sigma^2}-i\Omega(t_1-t_2)}.
\end{equation}
\subsection*{Photon evolution in multi-mode interferometers}
We now introduce a mode index to label a discrete set of spatial modes the photon can occupy, e.g. the different modes of an interferometer. We treat this index differently since we are going to assume that the interferometer only acts on the spatial modes and does not change the polarisation or the time-frequency degrees of freedom of the photon. We choose our notation such that the order of the entries in the ``ket'' label the spatial modes, while the state the individual photons are in is indicated by their wavefunction symbol (e.g. $\psi/\phi$)
\begin{eqnarray}
	a^{\dagger}_k(\tau,\sigma)\ket{0}=\ket{0_1,..,1_k(\tau,\sigma),..0_m}\\
	\psi_k^{\dagger}\ket{0}=\ket{0_1,..,\psi_k,..0_m}.
\end{eqnarray}
Modes can also be occupied by more than one photon
\begin{equation}
	\psi_k^{\dagger}\phi_k^{\dagger}\ket{0}=\ket{0_1,..,\psi_k\phi_k,..0_m}.
\end{equation}
We can now establish commutation relations for the operators $\psi_k^{\dagger}$ and $\phi_j^{\dagger}$
\begin{equation}
	\bra{0}[\psi_k,\phi_j^{\dagger}]\ket{0}=\delta_{k,j}\braket{\psi}{\phi}.
\end{equation}
This is an important insight as it tells us that the properties of the photons under exchange are related to the scalar product of their wavefunctions. We now examine what happens if we send the photons through a linear optical interferometer. We can describe an interferometer with m inputs and m outputs as a unitary transformation U$\in$ SU(m). We evolve the operators in the Heisenberg picture, under the action of the unitary, in the following way \cite{schwinger1960unitary}:
\begin{equation}
	\psi^{\dagger}_k\rightarrow \sum_l\text{U}_{kl}\psi^{\dagger}_l.
\end{equation}
For two photons in states $\psi$ and $\phi$ sent into port 1 and 2 the output state of an interferometer would then be
\begin{equation}
	\label{eqn:heisenberg}
	\left(\sum_l\text{U}_{1l}\psi^{\dagger}_l\cdot\sum_m\text{U}_{2m}\phi^{\dagger}_m\right)\ket{0}.
\end{equation}
\section{Multi-photon interference of partially distinguishable photons}
In this section we introduce a formalism to describe the detection event statistics of multiple partially distinguishable photons in pure, separable states in multi-mode optical interferometers.
\subsection{Interference of two partially distinguishable photons}
In the Schr\"odinger picture we let the unitary transformation U act directly on the states. Consider a single-photon in state $\psi$ being inserted into the first port of a two-port interferometer. In a simplified approach we again regard ``internal'' states of the photon, such as polarisation and time-frequency modes, separate from a set of spatial modes which the interferometer acts on. The total Hilbert space then decomposes into a product: $\mathcal{H}_{\text{int}}\otimes\mathcal{H}_{\text{spatial}}$. When the unitary transformation of the interferometer acts on the spatial degree of freedom of the single-photon states, the output-state becomes a superposition across the two output modes, where the weights are given by the components of the unitary transformation
\begin{equation}
	\ket{\psi,1}\rightarrow \textcolor{red}{\text{U}_{11}}\ket{\psi,1}+\textcolor{violet}{\text{U}_{12}}\ket{\psi,2}.
\end{equation}
The first position in the ket contains the ``internal'' state, while the second labels the spatial modes.
For a second photon in state $\ket{\phi}$ inserted into the second port, we obtain analogously
\begin{equation}
	\ket{\phi,2}\rightarrow \textcolor{green}{\text{U}_{21}}\ket{\phi,1}+\textcolor{blue}{\text{U}_{22}}\ket{\phi,2}.
\end{equation}

\begin{figure}
	\centering
	\includegraphics[width=0.6\textwidth]{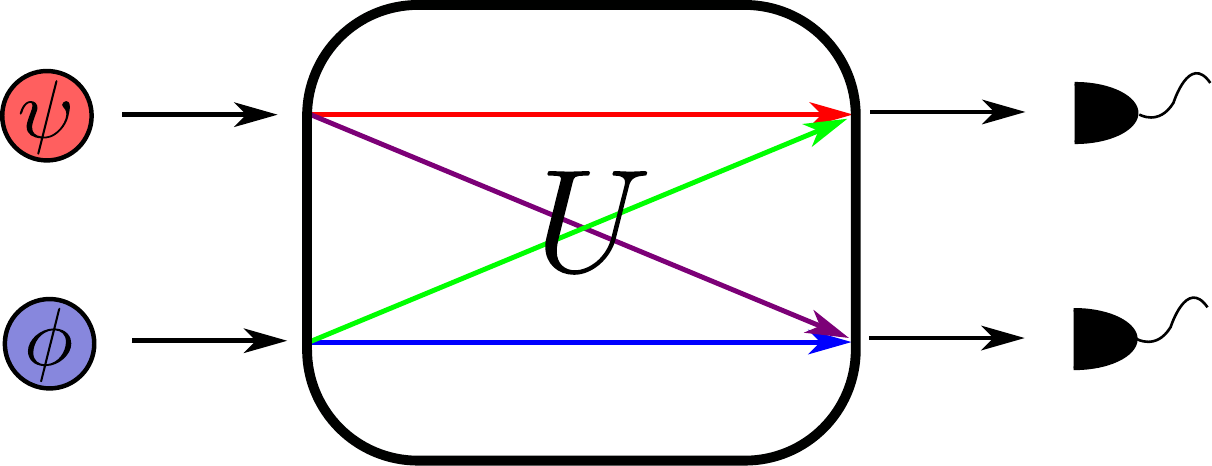}\label{fig:2x2interf}
	\caption{Single photons in states $\ket{\psi}$ and $\ket{\phi}$ entering a two-port interferometer. Detectors record events at the output.}
\end{figure}
The two-photon state is then in first quantisation
\begin{equation}
	(\text{U}_{11}\ket{\psi,1}+\text{U}_{12}\ket{\psi,2})\vee(\text{U}_{21}\ket{\phi,1}+\text{U}_{22}\ket{\phi,2})\label{eqn:1stqHOM},
\end{equation}
where $\vee$ is the symmetrised tensor product
\begin{equation}
	\ket{a}\vee\ket{b}=\frac{1}{\sqrt{2}}(\ket{a}\otimes \ket{b} + \ket{b}\otimes \ket{a}).
	\label{eqn:symprod}
\end{equation}  We now use the notation
\begin{eqnarray}
	\ket{\psi,i}\vee \ket{\phi,j}&=&\ket{0_1,...,0_{i-1},\psi_i,...,0_{j-1},\phi_j,...,0_m}\\
	\ket{\psi,i}\vee \ket{\phi,i}& =&\ket{0_1,...,0_{i-1},\psi_i\phi_i,...,0_m}.
\end{eqnarray}
This is the wavefunction expressed in terms of mode occupations (second quantisation). For example, a two-photon state where an H-polarised photon occupies spatial mode 1 and a V-polarised photon occupies spatial mode 2 is then written using this notation:
$\ket{H_1,V_2}$.
The state \ref{eqn:1stqHOM} in this notation then becomes, arriving at the same result as in the previous section (equation \ref{eqn:heisenberg}) (dropping position indices on the states for clarity)
\begin{equation}
	\text{U}_{11}\text{U}_{21}\ket{\psi\phi,0}+\text{U}_{11}\text{U}_{22}\ket{\psi,\phi}+\text{U}_{12}\text{U}_{21}\ket{\phi,\psi}+\text{U}_{12}\text{U}_{22}\ket{0,\psi\phi} \label{eqn:fockSt}.
\end{equation}
We aim to find the probabilities of all the possible detection outcomes. The first and last term in equation \ref{eqn:fockSt} correspond to events where two photons end up in the same mode (bunched events). In the second and third term the photons occupy different modes at the output of the interferometer. We can collect the terms for all possible events
\begin{eqnarray}
	\ket{1;1}_u&=&\text{U}_{11}\text{U}_{22}\ket{\psi,\phi}+\text{U}_{12}\text{U}_{21}\ket{\phi,\psi}\\
	\ket{2;0}_u&=&\text{U}_{11}\text{U}_{21}\ket{\psi\phi,0}\\
	\ket{0;2}_u&=&\text{U}_{12}\text{U}_{22}\ket{0,\psi\phi}.
\end{eqnarray}
The kets are named with respect to the detection outcome, e.g. $\ket{1;1}_u$ collects parts of the two-photon wavefunction where both output spatial modes are occupied.
Note that the kets in the previous expressions are not normalised, hence the subscript $u$.
The total state at the output of the interferometer is
\begin{equation}
	\ket{\psi}_{\text{out}}=\ket{1;1}_u+\ket{2;0}_u+\ket{0;2}_u.
\end{equation}

The corresponding output probabilities become
\begin{eqnarray}
	P_{11}=\braket{1;1}{1;1}_u\\
	P_{20}=\braket{2;0}{2;0}_u\\
	P_{02}=\braket{0;2}{0;2}_u.
\end{eqnarray}
We could also have written the state as
\begin{equation}
	\ket{\psi}_{\text{out}}=\sqrt{P_{11}}\ket{1;1}+\sqrt{P_{20}}\ket{2;0}+\sqrt{P_{02}}\ket{0;2},
\end{equation}
where the kets are normalised to 1.
Here, we can apply the familiar Born-Rule to obtain the probabilities for different detection outcomes:
\begin{equation}
	P_{nm}=\bra{\psi}\Pi_{nm}\ket{\psi}.
\end{equation}
The projectors onto different detection events are: 
\begin{eqnarray}
	\Pi_{11}=\ket{1;1}\bra{1;1}\\\nonumber
	\Pi_{20}=\ket{2;0}\bra{2;0}\\ \nonumber
	\Pi_{02}=\ket{0;2}\bra{0;2}. \nonumber
\end{eqnarray}
We now explicitly calculate the probabilities
\begin{eqnarray}
	P_{11}&=&|\text{U}_{11}\text{U}_{22}|^2+|\text{U}_{12}\text{U}_{21}|^2+\text{U}_{11}\text{U}_{22}\text{U}^*_{12}\text{U}^*_{21}\braket{\psi,\phi}{\phi,\psi}+\text{c.c.}\label{eqn:p11}\\
	P_{20}&=&|\text{U}_{11}\text{U}_{21}|^2\braket{\psi\phi,0}{\psi\phi,0}\\
	P_{02}&=&|\text{U}_{12}\text{U}_{22}|^2\braket{0,\psi\phi}{0,\psi\phi}.
\end{eqnarray}
To write the equations in a more compact form we introduce the permanent of a matrix
\begin{equation}
	\text{perm}(\text{M})=\sum_{\sigma\in\mathcal{S}_N}\prod_i{\text{M}_{i,\sigma(i)}}.
\end{equation}
$S_N$ is the symmetric group. $S_N$ contains as elements all possible permutations $\sigma$ of a list of $N$ objects. $S_2$ contains two elements, which can act on a list of integers: The identity $\sigma_1\{1,2\}\rightarrow\{1,2\}$ and the swap $\sigma_2\{1,2\}\rightarrow\{2,1\}$. $\sigma(i)$ then refers to the $i^{th}$ element of the list of integers permuted by the action of $\sigma$. The permanent of a 2x2 matrix M is then:
\begin{equation}
	\text{perm}(\text{M})=\text{M}_{11}\text{M}_{22}+\text{M}_{12}\text{M}_{21}.
\end{equation}

We make the observation that in equation \ref{eqn:p11} the first two terms correspond to $\text{perm}(|\text{U}|^2)$. The remaining terms are a bit more difficult to interpret but we can read off: $\text{U}_{11}\text{U}_{22}\text{U}^*_{12}\text{U}^*_{21}+\text{U}^*_{11}\text{U}^*_{22}\text{U}_{12}\text{U}_{21}=\text{perm}(\text{U}\star \text{U}_{\sigma_2,\mathbb{1}}^*)$, where the $\star$ product between the two matrices is the entry-wise matrix, or Hadamard product, and $\text{U}^*_{\sigma_2,\mathbb{1} }$ is the matrix 
$\text{U}^*$ with rows permuted by $\sigma_2$.
We can now proceed to write equation \ref{eqn:p11} in terms of matrix permanents
\begin{equation}
	P_{11}=\text{perm}(|\text{U}|^2)+\text{perm}(\text{U}\star \text{U}_{\sigma_2,\mathbb{1}}^*)\braket{\psi,\phi}{\phi,\psi}.
\end{equation}
Lastly, we need to evaluate the overlap $\braket{\psi,\phi}{\phi,\psi}$.
We have previously introduced the symmetrised tensor-product $\vee$ and used the notation: $\ket{\psi,\phi}=\ket{\psi,1}\vee\ket{\phi,2}=\frac{1}{\sqrt{2}}(\ket{\psi,1}\otimes\ket{\phi,2}+\ket{\phi,2}\otimes\ket{\psi,1})$. The overlap $\braket{\psi,\phi}{\phi,\psi}$ is therefore\\
\begin{eqnarray}\nonumber
	&1/2&(\bra{\psi,1}\otimes\bra{\phi,2}+\bra{\phi,2}\otimes\bra{\psi,1})\cdot(\ket{\phi,1}\otimes\ket{\psi,2}+\ket{\psi,2}\otimes\ket{\phi,1})\\
	&=&|\braket{\psi}{\phi}|^2.
\end{eqnarray}
Finally, we obtain
\begin{equation}
	P_{11}=\text{perm}(|\text{U}|^2)+\text{perm}(\text{U}\star \text{U}_{\sigma_2,\mathbb{1}}^*)|\braket{\psi}{\phi}|^2.\label{eqn:P11_fin}
\end{equation}
The probabilities for $P_{20}$ and $P_{02}$ events are calculated analogously.
We have now a complete description of the click statistics of two independent, non interacting, partially distinguishable bosons in states $\ket{\psi}$ and $\ket{\phi}$ scattering in a unitary interferometer with matrix U. At this point we may ask how this generalises to particles of different exchange symmetry. To investigate the scattering statistics of fermions we have to replace the symmetrised tensor product \ref{eqn:symprod} with an anti-symmetric tensor-product $\wedge$:
\begin{equation}
	\ket{a}\wedge\ket{b}=\frac{1}{\sqrt{2}}(\ket{a}\otimes \ket{b} - \ket{b}\otimes \ket{a}).
	\label{eqn:asymsymprod}
\end{equation} 
This results in a sign change in the sum
\begin{equation}
	P_{11}^{Fermion}=\text{perm}(|\text{U}|^2)-\text{perm}(\text{U}\star \text{U}_{\sigma_2,\mathbb{1}}^*)|\braket{\psi}{\phi}|^2.
\end{equation}
Another interesting question we might ask is: what happens if we had decided to not symmetrise the Hilbert-spaces at all. In other words abandoning the postulate of identical particles. The two-photon overlap then vanishes
\begin{eqnarray}
	\braket{\psi,\phi}{\psi,\phi}&=&(\bra{\psi,1}\otimes\bra{\phi,2})\cdot(\ket{\psi,2}\otimes\ket{\phi,1})=0\\
	P_{11}^{Classical}&=&\text{perm}(|\text{U}|^2).
\end{eqnarray}
We can see that two-photon interference arises as a direct consequence of the exchange symmetry of the particles involved. 

\subsection{HOM-Interference}
A special case of equation \ref{eqn:P11_fin} is where the unitary matrix is that of a 50:50 beam splitter \ref{eqn:bsmat}. This of course describes the classic Hong Ou Mandel experiment \cite{hong1987measurement}.
\begin{spacing}{1.5}
	\begin{equation}\label{eqn:bsmat}
		U=\frac{1}{\sqrt{2}}\begin{pmatrix}
			1 &1 \\ 
			1 &-1 
		\end{pmatrix}
	\end{equation}
\end{spacing}
The probability of detecting a coincidence is then:
\begin{equation}
	P_{11}=\frac{1}{2}(1-|\braket{\psi}{\phi}|^2).
\end{equation}
If we choose a Gaussian wave-packet with 
\begin{equation}
	|\braket{\psi}{\phi}|^2= e^{-\tau^2},
\end{equation}
\begin{figure}[h]
	\centering
	\includegraphics[scale=2]{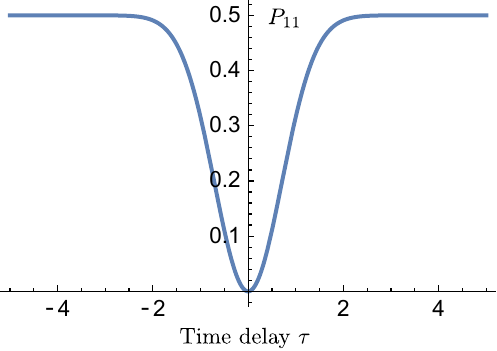}
	\caption{Hong Ou Mandel dip between two Gaussian shaped photon wave-packets.}
	\label{fig:hom_sim}
\end{figure}
where $\tau$ is the time-difference (or difference in mean frequency for the spectral domain) between two Gaussian shaped photon wave-packets, we obtain the Hong Ou Mandel dip in Figure \ref{fig:hom_sim}. When the overlap integral is small, the function asymptotically approaches a constant value of 0.5, which is the ``classical'', uncorrelated, per-shot probability of finding the two photons in different output ports. As the overlap increases, the coincidence probability decreases and approaches 0 when the photons are completely indistinguishable (the wavepackets completely overlap in time).
\subsection{Interference of more than two photons}
Equation \ref{eqn:P11_fin} can be generalised to an arbitrary number of photons impinging on an arbitrary interferometer. This was done in \cite{Tichy2015}. Tichy defines a distinguishability matrix $\mathcal{S}$, \ref{eqn:smatrix} containing all mutual overlaps between the $N$ photons sent into the interferometer
\begin{equation}\label{eqn:smatrix}
	\mathcal{S}=\begin{pmatrix}
		1   & \braket{\phi_1}{\phi_2} & \dots & \braket{\phi_1}{\phi_N} \\
		\braket{\phi_2}{\phi_1}       & 1 & \dots & \braket{\phi_2}{\phi_N}  \\
		\dots&\dots&\dots&\dots \\
		\braket{\phi_N}{\phi_1}       &\braket{\phi_N}{\phi_2}  &  \dots & 1
	\end{pmatrix}.
\end{equation}
Furthermore, he defines a scattering matrix M, which is constructed from the unitary interferometer matrix U. For an input state configuration with photon number occupations $n_i$, $r=(n_1,n_2,..,n_i,...n_m)$ and a measured output state configuration $s=(l_1,l_2,..,l_i,...l_m)$, mode assignment lists d(s/r) are defined\\ $d(r)=(\overbrace{1,..,1}^{n_1-\text{times}},\overbrace{2,..,2}^{n_2-\text{times}},...,\overbrace{m,...,m}^{n_m-\text{times}})$, which contain the mode indices for each photon as many times as the number of photons occupying that mode. The scattering matrix M is then $\text{M}=\text{U}_{d(r),d(s)}$.
The distinguishability matrix is also modified by the input state configuration:  $\mathcal{G}=\mathcal{S}_{d(r),d(r)}$.
The probability of detecting a click event $s$ with an input configuration $r$ of N photons with distinguishability matrix $\mathcal{S}$ is then given by \cite{Tichy2015}:
\begin{equation}
	P(s,r,\text{U},\mathcal{S})=\mathcal{N}\sum_{\sigma\in S_N}\bigg[\prod_{j=1}^{N}\mathcal{G}_{j,\sigma(j)}\bigg]\text{perm}(\text{M}\star \text{M}^*_{\sigma,\mathbb{1}}),\label{eqn:malte}
\end{equation}
where $\mathcal{N}=1/(\prod_{j}s_j!r_j!)$ is a constant normalising factor.
To calculate the probability, the first step is to construct the Matrices $\mathcal{G}$ and $\text{M}$ from the distinguishability matrix $\mathcal{S}$ and the unitary matrix of the interferometer $\text{U}$. For the earlier example of two photons entering in separate modes into a 2x2 interferometer and detecting coincidences, we have: $r=(1,1)$, $s=(1,1)$,
$d(r)=(1,2)$, $d(s)=(1,2)$, $\mathcal{N}=1$, $\mathcal{G}=\mathcal{S}$ and $\text{M}=\text{U}$.
Inserting these into equation \ref{eqn:malte}, we can easily retrieve equation \ref{eqn:P11_fin}.
Let us consider the case of sending three photons into the same 2x2 interferometer. For the input state occupation numbers we take: $r=(2,1)$ and measure the outputs $s=(1,2)$, i.e. sending two photons into the first mode and one into the second, while detecting one photon in the first output mode and two in the second. The mode-assignment lists are $d(r)=(1,1,2)$ and $d(s)=(1,2,2)$ respectively. As described previously, to construct these lists the mode index $i$ is repeated $n_i$ times, where $n_i$ is the number of photons occupying mode $i$. The matrix $\mathcal{G}$ then becomes $\mathcal{G}_{kl}=\mathcal{S}_{(1,1,2)_k,(1,1,2)_l}$, where $k$ and $l$ run from $1..3$ and $(1,1,2)_k$ denotes the $kth$ element of the mode assignment list $(1,1,2)$. We can also visualise how the matrix $\mathcal G$ is constructed in terms of block matrices.
\begin{equation}
	\includegraphics[scale=1.2]{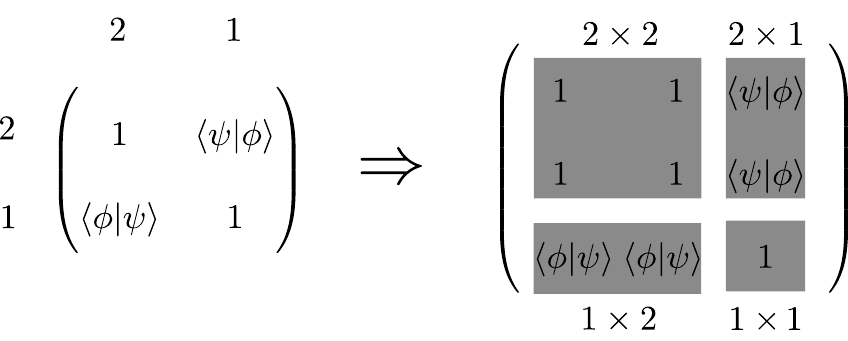}
\end{equation}
For every entry of the distinguishability matrix $\mathcal S_{kl}$ we construct a block of size $n_k\times n_l$, where $n_{k/l}$ is the number occupation of input port $k/l$. 
In an analogous manner the M matrix is constructed from the interferometer matrix U. Here, we use both, input and output mode assignment lists $d(r)$ and $d(s)$. $\text{M}_{k,l}=\text{U}_{(1,1,2)_k,(1,2,2)_l}$.
\begin{equation}
	\includegraphics[scale=1.2]{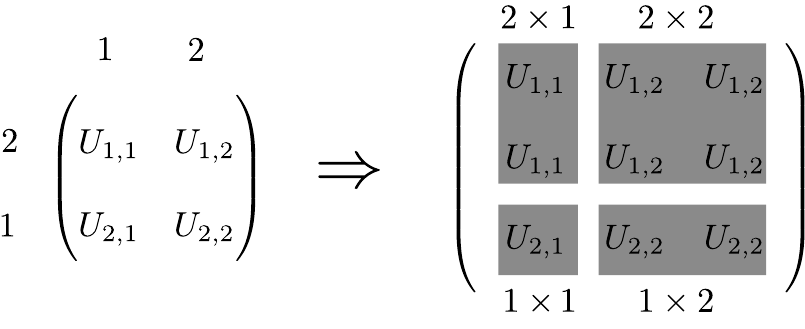}
\end{equation}

The outer sum of equation \ref{eqn:malte} is executed over all elements of the symmetric group $\sigma \in S_N$. I will examine more closely what each term of this sum for a given permutation $\sigma$ looks like. First, we have a product over the entries of the matrix $\mathcal{G}$, which we constructed from the distinguishability matrix: $\prod_{j=1}^{N}\mathcal{G}_{j,\sigma(j)}$. Let us take $\sigma=(1)(2,3)$ (swap of the last two elements) as an example. We let $\sigma$ act on a list of mode indices: $(1)(2,3)\{1,2,3\}=\{1,3,2\}$ and denote as $\sigma(j)$ the $jth$ entry of the permuted list and obtain: $\prod_{j=1}^{N}\mathcal{G}_{j,\{1,3,2\}(j)}=1*\braket{\phi}{\psi}\braket{\psi}{\phi}=|\braket{\phi}{\psi}|^2$. This corresponds to the product of diagonal elements of the matrix $\mathcal{G}$ with the order of columns permuted by $\sigma$. The next term contains the permanent of the entry wise product between the matrix M and its complex conjugate, where the order of rows has been permuted by $\sigma$:
$\text{perm}(\text{M}\star \text{M}^*_{\sigma,\mathbb{1}})$. For our example of $\sigma=(1)(2,3)$ this amounts to: $\text{perm}(\text{M}_{i,j}\text{M}^*_{{\{1,3,2\}(i),j}})$.  Collecting the contributions from the distinguishability matrix and the interferometer, we obtain for a single term in the sum over permutations in equation \ref{eqn:malte}: $1/4\cdot\text{perm}(\text{M}_{i,j}\text{M}^*_{{\{1,3,2\}(i),j}})|\braket{\phi}{\psi}|^2$,
where we have also included the normalisation: $\mathcal{N}=1/(\prod_{j}s_j!r_j!)=1/4$.

\subsection{Classifying interference terms}
We now examine more closely the form of equation \ref{eqn:malte}. The outer sum is executed over all elements of the symmetric group $S_N$. Each of the terms in the sum contain a product of entries of the distinguishability matrix, i.e., a product of overlaps between the wavefunctions of the photons involved. The order of the overlaps in this product is determined by the specific permutation $\sigma$. To illustrate, I will more closely study the example of three photons. If $\sigma$ is a simple swap $\sigma\{1,2,3\}\rightarrow\{2,1,3\}$, we obtain a term which only depends on the modulus of the overlap between two photons, i.e., something that has the same dependence on the state of the photons as HOM two-photon interference: $\braket{\phi_1}{\phi_2}\braket{\phi_2}{\phi_1}=|\braket{\phi_1}{\phi_2}|^2$. The term which corresponds to a full permutation, i.e., $\sigma\{1,2,3\}\rightarrow\{2,3,1\}$ is proportional to: $\braket{\phi_1}{\phi_2}\braket{\phi_2}{\phi_3}\braket{\phi_3}{\phi_1}$. The structure of the product of overlap integrals in these terms depends on the particular permutation $\sigma$ that is applied. We write the permutations in cycle form: the two-cycle, or swap, acts like: $(1,2)(3)\{1,2,3\}\rightarrow\{1\rightarrow 2,2 \rightarrow 1,3\rightarrow3\}=\{2,1,3\}\ $ and a three cycle has the following action on a list of numbers: $(1,2,3)\{1,2,3\}\rightarrow\{1\rightarrow 2,2\rightarrow 3,3\rightarrow 1\}=\{2,3,1\}$. 
Cycles of the same structure belong to the same ``conjugacy class''. The $3!=6$ elements of $S_3$ fall into 3 conjugacy classes
\begin{eqnarray}
	(1)(2)(3)&\rightarrow&\text{ Identity}\\
	(1,2)(3);(2,3)(1);(1,3)(2)&\rightarrow&\text{ Swaps}\\
	(1,2,3);(1,3,2)&\rightarrow& \text{ Full permutation}.
\end{eqnarray}
It is immediately clear that the form of the terms in the sum of equation \ref{eqn:malte} also fall into the same groups as the conjugacy classes. 
We can classify the structure of the terms appearing in the formula for the coincidence probability in terms of the conjugacy class $\sigma$ belongs to. The first conjugacy class consists of one element, the identity, that contributes a constant probability. There is no dependence on the state of the photons - as if each particle acts completely independent of the other particles in the interferometer -. There is still interference of the photon with itself (cf. double slit interference). In quantum optics this type of interference is often referred to as ``classical''. The second conjugacy class contains three members, all possible swaps between three elements. The terms in the scattering probability that correspond to this class are the HOM-like two-photon interference terms, which depend on the modulus of the overlap between two photons. The third class is that of full permutations of three elements. The corresponding terms in the scattering probability depend on the product of three overlaps, we refer to this as the three-photon interference contribution.
For some arbitrary interferometer the probability $P_{111}$ thus has the following form:
\begin{equation}
	P_{111}=\overbrace{c_1}^{\text{Self interf.}}+\overbrace{c_2|\braket{\phi_1}{\phi_2}|^2+...}^{\text{Two-photon interf.}}+\overbrace{c_5\braket{\phi_1}{\phi_2}\braket{\phi_2}{\phi_3}\braket{\phi_3}{\phi_1}+...}^{\text{Three-photon interf.}}.
	\label{eqn:p111_class}
\end{equation}

\section{Graph-theory of multi-photon scattering}
\label{sect:graphTh} 
\begin{figure}[h]
	\centering
	\includegraphics[width=0.7\textwidth]{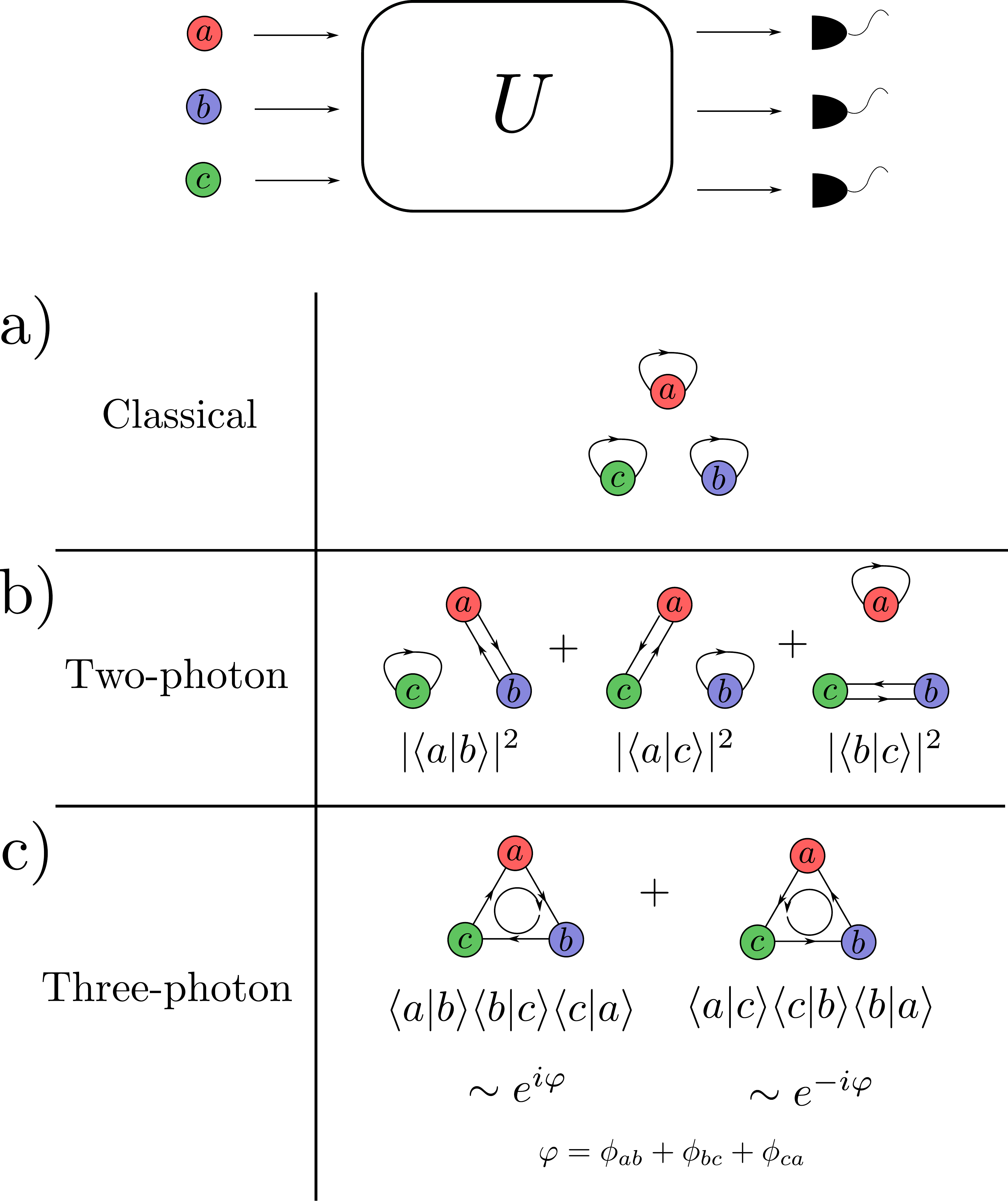}
	\caption{Graph theoretical description of three-photon interference. All possible closed paths on the three vertex graph are illustrated. a) Constant ``classical'' contribution. b) Two-photon - Hong ou Mandel like - interference terms. c) three-photon interference terms, with clockwise and counter clockwise paths.}
	\label{fig:graph}
\end{figure}

In this section I will discuss an approach that I developed to describe the terms appearing in equation \ref{eqn:malte}. This method was also independently developed by V. Shchesnovich et. al. around the same time \cite{shchesnovich2017interference,jones2018interfering}. There have been other approaches to describe multi-photon interference in terms of graph-theoretical approaches \cite{gu2019quantum} and even to solve graph theoretical problems \cite{bradler2018gaussian}. I need to emphasise here that this formalism is unrelated to graph or cluster states which are used to describe entangled states that are resources in a measurement based quantum computation scheme \cite{benjamin2006brokered,lu2007experimental}.
We can interpret equation \ref{eqn:malte} in terms of a graph. Each state is assigned a vertex on a 2D graph, a directed edge between vertices is identified with the overlap between the two states. If the overlap is zero, the edge is not drawn.
Two states $\ket{a}$ and $\ket{b}$ are connected by a single directed edge $\braket{a}{b}$, where the direction of the edge is given from bra to ket. In this case the edge points from $\bra{a}$ to $\ket{b}$. 
\begin{figure}[h]
	\centering
	\includegraphics[width=0.4\textwidth]{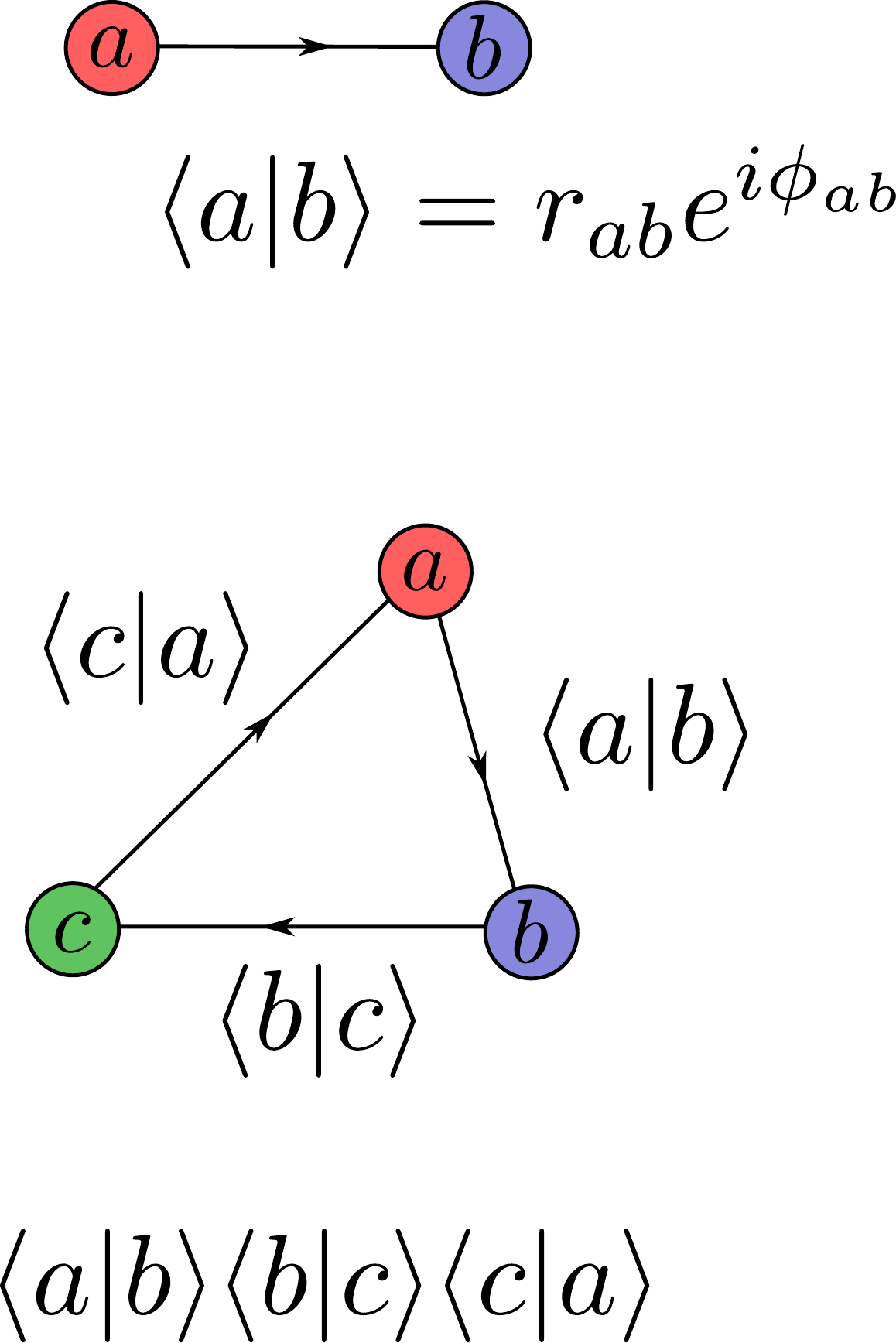}
	\caption{Single directed edge and a three vertex path over closed loop.}
	\label{fig:vertgraph}
\end{figure}
The overlaps appearing in equation \ref{eqn:malte} correspond to all possible coverings of the a graph with closed, directed paths. Each path is allowed to cross each vertex only once. The analogy to graphs stems from the fact that any permutation of N elements is in one to one correspondence to a covering of a graph of N vertices with closed, directed paths. We identify for example the cycle $(1,2,3)\{a,b,c\}\rightarrow \{a\rightarrow b,b\rightarrow c,c\rightarrow a\}$ with the loop along the three vertices a,b,c with the direction $a\rightarrow b\rightarrow c\rightarrow a$, as illustrated in Figure \ref{fig:vertgraph}. In Figure \ref{fig:graph} all possible coverings of the three-photon graph are depicted.
We assign to each overlap its complex phase $\phi_{ab}=\text{arg}(\braket{a}{b})$ and modulus $r_{ab}=|\braket{a}{b}|$. The phase cancels in case of a two-photon path, leaving only a dependence on the modulus squared of the overlap. For a path connecting three vertices however, we get a contribution from the sum of the complex arguments along the three edges. The path with reversed handedness also contributes a term with an argument of opposite sign. We have coined the term ``triad phase'' for this collective three-photon phase. In the experimental work of the next chapter this phase will play a prominent role in the demonstration of three-photon interference.\\ Now we are in a position to give a more precise notion of multi photon interference. M-photon interference arises if there exists a closed path traversing M vertices exactly once. For N photons entering an interferometer there will be N-photon interference if there exists a path traversing every vertex once. This path is also called Hamilton path in graph theory. The question if there exists a Hamilton path for a given graph is a prominent problem in graph theory. The answer to this question, for the kind of graphs we have introduced here, can therefore be answered experimentally by demonstrating N-photon interference in an ensemble of N photons which have been prepared to exhibit some interesting graph. If there is a quantum advantage to solving this problem using partially distinguishable photons is an interesting question, which could merit further investigation.
\begin{figure}[h]
	\centering	\includegraphics[width=0.6\textwidth]{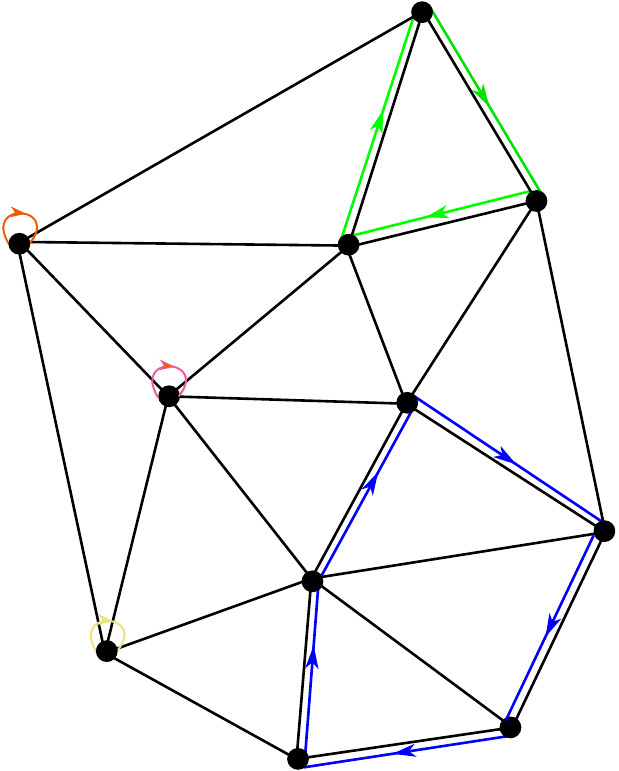}
	\caption{A possible complete covering of a graph with closed paths.}
	\label{fig:graph2}
\end{figure}
\clearpage
\subsection{Mixed states}
A mixed state is given by an ensemble of pure states $\ket{\psi_k}$ that are prepared with probability $p_k$. The density matrix is defined as:
\begin{eqnarray}
	\rho_j=\sum_{k=1}^Rp_{j,k}\ket{\psi_{j,k}}\bra{\psi_{j,k}},
\end{eqnarray} 
where $R$ is the dimension of the density matrix.
We can generalise our previous approach to calculate click probabilities to incorporate mixed states. For simplicity I will restrict myself to the case where each input $i$ of the interferometer is occupied by exactly one photon with density matrix $\rho_i$.
I will not give a rigorous proof but rather motivate the formalism through basic observations. A more rigorous description can be found in \cite{shchesnovich2015}. We have already discussed that the form of the overlaps follows the cycle structure of the symmetric group (conjugacy classes). Furthermore, we know that a cycle is cyclically invariant e.g., $(1,2,3)=(3,1,2)=(2,3,1)$. We are looking for a way to combine the density matrices describing some mixed input states $\rho_i$ in such a way that their order is also cyclically invariant. We know that the trace of a product of matrices is cyclically invariant $\text{Tr}(M_1M_2M_3)=\text{Tr}(M_3M_1M_2)=\text{Tr}(M_2M_3M_1)$. The most straightforward approach therefore seems to be to make the replacement
\begin{equation}
	\braket{\phi_1}{\phi_2}\braket{\phi_2}{\phi_3}\braket{\phi_3}{\phi_1}\rightarrow \text{Tr}(\rho_1\rho_2\rho_3).
\end{equation}
We can further strengthen our position by noting that for pure states our rule also holds true.
With $\rho_i=\ket{\phi_i}\bra{\phi_i}$:
\begin{equation}
	\text{Tr}(\ket{\phi_1}\bra{\phi_1}\ket{\phi_2}\bra{\phi_2}\ket{\phi_3}\bra{\phi_3})=\braket{\phi_1}{\phi_2}\braket{\phi_2}{\phi_3}\braket{\phi_3}{\phi_1},
\end{equation}
we take the product over disjoint cycles.
$(1,2)(3)\rightarrow \text{Tr}(\rho_1\rho_2)\overbrace{\text{Tr}(\rho_3)}^{1}$, $(1,2)(3,4)\rightarrow \text{Tr}(\rho_1\rho_2)\text{Tr}(\rho_3\rho_4)$.
We are now in a position to reformulate equation \ref{eqn:malte}
\begin{equation}
	P_{\rho_1,...,\rho_N}=\mathcal{N}\sum_{\sigma\in S_N}\bigg[\prod_{j=1}^{a}\text{Tr}(\rho_{\alpha_{j1}}\cdot...\cdot\rho_{\alpha_{jn}})\bigg]\text{perm}(\text{M}\star \text{M}^*_{\sigma,\mathbb{1}}),
\end{equation}
where $(\alpha_{j1},..,\alpha_{jn})$ is the structure of the $j^{th}$ disjoint cycle of $\sigma$. $a$ is the number of the disjoint cycles in $\sigma$ and n is the length of the $j^{th}$ cycle. 

Tichy also derives a formula for general mixed states \cite{Tichy2015} (Appendix F)
\begin{equation}
	P(s,r,U,\mathcal{S})=\sum_{k_1...k_d}^R\left(\prod_{j=1}^dp_{j,k_j}\right)P(s,r,U,\mathcal{S}(k)),
\end{equation}
which can be understood as an ensemble averaged probability. We prepare a photon entering port $j$ in state $\ket{\psi_{j,k}}$ with a probability of $p_{j,k}$. $\rho_j=\sum_{k=1}^Rp_{j,k}\ket{\psi_{j,k}}\bra{\psi_{j,k}}$. 
Then a sum over all possible input state realisations $\{k_1,...,k_n\}$ is performed, which occur with probability $\prod_jp_{j,k_j}$. The distinguishablility matrix for a specific realisation is given by $\mathcal{S}(k)_{j,l}=\braket{\psi_{j,k_j}}{\psi_{l,k_l}}$.
\clearpage
\section{Independent parameters in multi-photon scattering}
\label{sect:indepPar}
When performing multi-photon interference experiments on independent photons
one important question that arises is: what are the experimental degrees of freedom that need to be controlled in order to fully characterise the scattering dynamics of the process? For example, in two-photon scattering we know that there is one parameter that we need to adjust. This parameter may be a time delay between two-photon wavepackets. As a function of the time delay we measure a change in coincidence counts, observing two-photon bunching in a HOM dip, for example. Note that we could have also adjusted the polarisation of the photons; preparing one photon in horizontal polarisation and scanning the polarisation of the second from vertical to horizontal polarisation. It is obvious from the previous discussion that the only thing that matters is the change in the overlap integral between the photons. This is for two photons the single independent parameter that we need to adjust. We may ask: how many free parameters do we need to adjust for more than two photons? For three photons we might guess that the number of parameters is the number of all mutual overlaps between the three photons, i.e., three. However, it turns out this intuition is incorrect. To find the correct answer, we investigate the scattering probabilities for multi-photon coincidences. The scattering probability for some coincidence event and a given unitary is completely determined by the $\mathcal{S}$ matrix (equation \ref{eqn:smatrix}) $\mathcal{S}(x_1,...x_L)$, which is in general a function of $L$  parameters $x_i$.
We are going to need an explicit form of the distinguishability matrix $\mathcal{S}$. To do so we need to find a representation for the states $\ket{\phi_i}$. We define a set of orthogonal basis vectors $\ket{\alpha_i}$ and perform a Gram-Schmidt decomposition to construct a set of states in terms of this basis
\begin{equation}
	\ket{\phi_i}=\sqrt{1-\sum_{k=2}^i s_{i,k}^2}\ket{\alpha_1}+\sum_{k=2}^i s_{i,k}e^{i\gamma_{i,k}}\ket{\alpha_k}.
\end{equation}
We can easily see that these vectors are normalised. 
For three photons the distinguishability matrix takes the form: 
\begin{equation}\label{eqn:smatred}
	\left(
	\begin{array}{ccc}
		1 & \text{cc.} & \text{cc.} \\
		\sqrt{1-s_{2,2}^2} & 1 & \text{cc.} \\
		\sqrt{1-s_{3,2}^2-s_{3,3}^2} & \sqrt{\left(1-s_{2,2}^2\right) \left(1-s_{3,2}^2-s_{3,3}^2\right)}+e^{-i (\gamma _{2,2}-\gamma _{3,2})} s_{2,2} s_{3,2} & 1 \\
	\end{array}
	\right).
\end{equation}
This form contains the constraints for the choice of parameters. We can immediately read off the ranges for $s_{2,2}^2\leq 1$ and 
$s_{3,2}^2+s_{3,3}^2\leq 1$ (2D disk).
The only redundant parameter is going to be one of the phases $\gamma$, leaving us with four independent parameters. In a more rigorous approach to determine the number of free parameters, we can in a first step we write the $\mathcal{S}$ matrix as a continuous vector $\vec{\mathcal{S}}$= $(\braket{\phi_1}{\phi_2}, \braket{\phi_2}{\phi_3},..,\braket{\phi_i}{\phi_j})$.
We may imagine this vector as pointing along a surface in some high dimensional space. We want to ask what is the dimension of this surface? It is sufficient to determine the dimension of the tangent vector space for each point. The tangent vectors are defined as $\frac{\partial}{\partial x_i}$, where the $x_i$ encompass all degrees of freedom of Matrix \ref{eqn:smatred}: $s_{i,j}$ and $\gamma_{i,j}$. The dimension of the space spanned by the tangent vectors is then  rank$(\frac{\partial}{\partial x_i}\vec{\mathcal{S}}(x_1,...x_L)|_{x_0})$.
Generally the number of independent parameters is found to be at most $(N-1)^2$, where $N$ is the number of photons.
\subsection{Overlap integrals with time-delays}
\label{sect:Gaussian}
Next, we show that for time-delays the products of overlaps of Gaussian wavepackets, that appear in the expressions for the multi-photon coincidence probabilities, are always real and positive and hence do not give rise to a collective multi-photon phase, such as the three-photon ``triad phase''.
In the case of the two-photon interference terms, which contain expressions of the form:
\begin{equation}
	\braket{t_1}{t_2}\braket{t_2}{t_1}=|\braket{t_1}{t_2}|^2,
\end{equation}
this is easy to see, as the expression is purely real. For the three-photon interference term we obtain:
\begin{equation}
	\begin{split}
		\braket{t_1}{t_2}\braket{t_2}{t_3}\braket{t_3}{t_1}&=e^{-((t_1-t_2)^2+(t_2-t_3)^2+(t_3-t_1)^2)/(4\sigma^2)-i\Omega(t_1-t_2+t_2-t_3+t_3-t_1)}\\&=e^{-((t_1-t_2)^2+(t_2-t_3)^2+(t_3-t_1)^2)/(4\sigma^2)}.
	\end{split}
\end{equation}
As we can see this expression is also real. This holds for any number of photons. This illustrates that in order to access all independent parameters of multiphoton interference it is necessary to exploit different degrees of freedom other than time delays between Gaussian wavepackets. As we shall see in the next chapter, we can conveniently access multi-photon phases by preparing the interfering particles in different polarisation states.  
\section{Conclusion}
If have presented the reader with the necessary tools to understand the interference effects that occur in multiple, independent single photons \cite{Tichy2015,shchesnovich2015}. I motivated how the exchange symmetry of particles gives rise to non-classical interference. Furthermore, I described a new theoretical tool to characterise interference terms that appear in the multi-photon coincidence probability: each term can be associated with a closed path traversing a graph where the vertices represent states of interfering photons \cite{shchesnovich2018collective,jones2018interfering}. Using the graph theoretical picture allows us to clearly identify contributions arising from multi-photon interference. I discussed the parameters which determine multi photon interference and found that for $N$ photons at most $(N-1)^2$ parameters are independent. Furthermore, I have shown that time delays between Gaussian shaped wavepackets do not give access to all parameters governing multiphoton interference.

%% file: chapter3.tex
\chapter{Three-photon interference}
\label{chap:ThreePhot}
This chapter is in parts taken from a recently published paper \cite{menssen2017distinguishability}.  I conceived the state configuration (illustrated in Figure \ref{fig:SweepPol}) allowing to isolate three-photon interference with help from A.E. Jones. A.E. Jones and I conducted the experiment, with myself in the experimental lead. A.E. Jones and I modelled and analysed the experimental data. The paper \cite{menssen2017distinguishability} was written with contributions from all authors.
\section{Introduction}

two-photon interference has extensively been studied after first discovered by Hong Ou and Mandel \cite{hong1987measurement}. A natural question that arises is, how multi-photon scattering phenomena extend beyond the interference of two photons. Several previous experimental \cite{Spagnolo2013,tillmann2015generalized,spring2017chip} and theoretical studies
\cite{tichy2011four, tan2013, ra2013nonmonotonic, deGuise2014, shchesnovich2014, tamma2016multi, tichy2014interference, shchesnovich2015tight, shchesnovich2015, tamma2016multi} have investigated this subject. Spagnolo et. al. \cite{Spagnolo2013} and later Tillman et. al. \cite{tillmann2015generalized} probed the three-photon coincidence landscape with time delays between otherwise indistinguishable photons. Spring et. al. \cite{spring2017chip} mapped the entire coincidence landscape of three time-delayed photons. In \cite{ra2013nonmonotonic} it was already noticed that in the interference of three or more photons, the interference fringe is non-monotonic as a function of distinguishability. The theoretical framework developed by Tichy \cite{Tichy2015} especially illustrates the inadequacy of bi-partite distinguishability, as measured by the overlap integral of the wavefunctions, for characterising multi-photon interference. For example, we can increase multi-partite correlations even as we increase (mean) distinguishability. This goes against our intuition for interference phenomena which is shaped by the double-slit and HOM experiments. The origin of this is multi-photon interference. The conventional perception in preceding experiments has been that in observing multi-photon interference it does not matter how distinguishability is realised. For example, in HOM interference it does not in principle matter in what degree of freedom of the electrical field you make the two photons distinguishable as you observe the two-photon interference fringe. You may vary polarisation, time or spectral modes of the photons. The value range for the multi-partite correlation (coincidence probability) will be the identical. As we have shown previously in section \ref{sect:Gaussian}, restricting a multi-photon interference experiment to varying time delays between Gaussian wavepackets only will not enable one to explore the entire range of possible values for the multi-photon coincidence probability. The underlying reason is that the interference of three independent photons is governed by a collective geometric phase between the single photon states which can not be accessed through time delays. In this work I aim to access the previously unexplored range of multipartite interference using polarisation as an additional degree of freedom besides time delays. 
In a first experiment I will demonstrate how three-photon interference is determined by a multi-photon collective phase.
The second experiment will isolate three-photon interference: we can observe a variation in probability of the coincident measurement of three photons, while at the same time all two-photon interference is kept constant.
\section{Theory}

\subsection{A balanced unitary 3x3 interferometer}
Firstly, I will show how the matrix of the balanced 3x3 unitary interferometer or tritter, which will serve as the unitary interferometer for our experiments, can be constructed. We first require that all the amplitudes are identical to $\frac{1}{\sqrt{3}}$ to ensure an equal $\frac{1}{3}$ splitting across all ports.
Input phases of the interferometer are irrelevant for the scattering scenario we are examining, since the Fock state we send through the interferometer is only defined up to a global phase (the sum of all input phases of the interferometer). Likewise, output phases do not change the scattering statistics. This allows us to reduce the scattering matrix to a unit-bordered form. There are four free phases remaining
\begin{spacing}{1}
	\begin{equation}\label{eqn:TritterU1}
	U_{tritter}=\frac{1}{\sqrt{3}}\left(\begin{matrix}
	1&1&1\\
	1&e^{i\phi_1}&e^{i\phi_2}\\
	1&e^{i\phi_3}&e^{i\phi_4}
	\end{matrix}\right).
	\end{equation}
\end{spacing}
\vspace{1cm}
\noindent
From the unitarity condition $U_{tritter}U_{tritter}^\dagger=\mathbb{1}$ we obtain a set of three equations:

\begin{eqnarray}
\left(e^{i \phi _1}+e^{i \phi _2}+1\right)=0\label{eqn:cru1}\\ 
\left(e^{i \phi _3}+e^{i \phi _4}+1\right)=0\label{eqn:cru2}\\
\left(e^{i \left(\phi _1-\phi _3\right)}+e^{i \left(\phi _2-\phi _4\right)}+1\right)=0.\label{eqn:cru3}
\end{eqnarray}
To find a combination of phases which satisfy the above equations consider the cubic roots of unity: $z_j=e^{i2\pi/3(j-1)}$, which satisfy
\begin{eqnarray}
\sum_{j=1}^3 z_j=0.\label{eqn:cruRel1}
\end{eqnarray} We substitute into equations \ref{eqn:cru1} and \ref{eqn:cru2}: $e^{i \phi _1} \rightarrow z_3$, $e^{i \phi _2} \rightarrow z_2$, $e^{i \phi _3} \rightarrow z_2$, $e^{i \phi _4} \rightarrow z_3$. Equations \ref{eqn:cru1}, \ref{eqn:cru2}, and \ref{eqn:cru3} then satisfy equation \ref{eqn:cruRel1}.
The unitary matrix describing the interferometer is then, using $z_3=z_2^2$:
\begin{spacing}{1}
	\begin{equation}\label{eqn:TritterU}
	U_{tritter}=\frac{1}{\sqrt{3}}\left(\begin{matrix}
	1&1&1\\
	1&z_2^2&z_2\\
	1&z_2&z_2^2
	\end{matrix}\right),
	\end{equation}
\end{spacing}
\vspace{1cm}
with $z_2=e^{i2\pi/3}$. It is straightforward to see that no other set of phases yields a balanced unitary. The unitary itself comes in two possible configurations \ref{eqn:TritterU} and its complex conjugate \cite{zeilinger1993einstein}. 

\subsection{Detection event probabilities}
\begin{figure}
	\centering
	\includegraphics[width=0.8\textwidth]{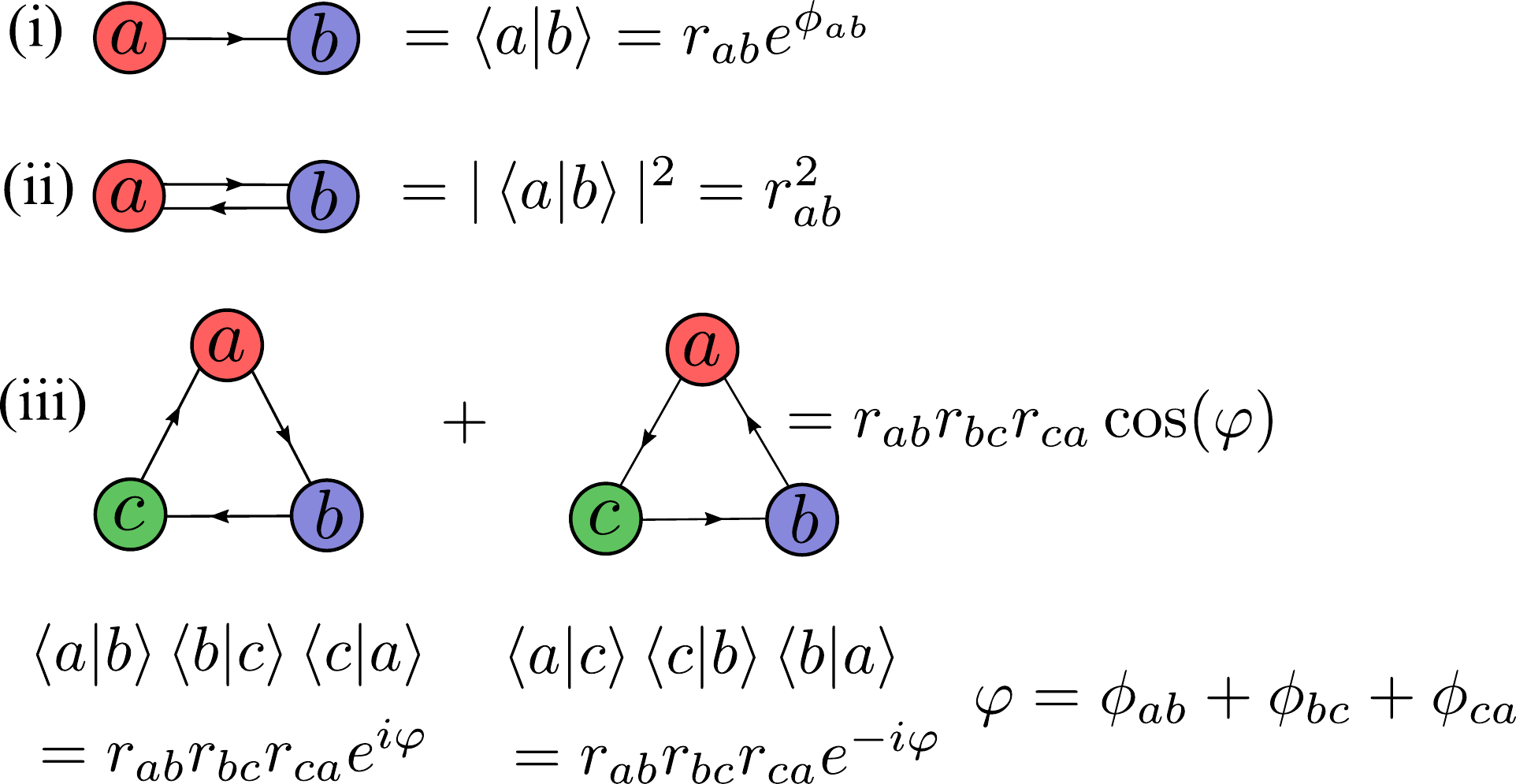}
	\caption{Graph picture of three-photon interference. i) Two-photon directed edge. ii) Closed loop path with modulus squared overlap contributions. iii) Three-vertex paths in clockwise and anti-clockwise directions which together contribute a $\cos(\varphi)$ dependence.}
	\label{fig:3PhotGraph}
\end{figure}

We can write the state overlap integrals in terms of the real moduli and a complex phase: $\braket{a}{b}=r_{ab}e^{i\phi_{ab}}$.
The three-photon coincidence probability for a balanced 3x3 unitary interferometer (equation \ref{eqn:TritterU})
can be obtained from equation \ref{eqn:malte}
\begin{align}\label{eqn:P1112}
P_{111} = \frac{1}{9}\left[2+ 4\:r_{ab} r_{bc} r_{ca}\cos(\varphi)
-r_{ab}^2-r_{bc}^2-r_{ca}^2\right].
\end{align} 
We define the collective ``triad'' phase as: $\varphi=\phi_{ab}+\phi_{bc}+\phi_{ca}$. Note that as discussed in section \ref{sect:indepPar}, three-photon interference depends on $(3-1)^2=4$ free parameters. Three of these are all possible modulus-squared overlap integrals $|\braket{\phi_i}{\phi_j}|^2=r_{i,j}^2$ and $i,j\in (a,b,c);i\neq j$. The fourth parameter is the ``triad-phase'' $\varphi$.
In terms of the graph theoretical approach discussed in section \ref{sect:graphTh}, the individual contributions can be easily understood and interpreted. In Figure \ref{fig:3PhotGraph} the paths on the three-vertex graph which contribute to the $P_{111}$ detection probability are illustrated. Note that there are two distinct three-vertex paths (clock- and anti-clockwise), which have a $e^{i\varphi}$ and $e^{-i\varphi}$ dependence. The sum of these two paths results in the overall $\cos(\varphi)$ dependence in equation \ref{eqn:P1112}.
The output event probabilities for all detection events are: 
\begin{eqnarray}\label{eqn:probs}
P_{111} &=& \frac{1}{9}\left[2+ 4\:r_{ab} r_{bc} r_{ba}\cos(\varphi)-r_{ab}^2-r_{bc}^2-r_{ca}^2\right] \\
P_{300} &=& P_{030} =  P_{003}=  \frac{1}{27} \left(1+r_{ab}^2+r_{bc}^2+r_{ca}^2 + 2 r_{ab} r_{bc} r_{ca} \cos (\varphi) \right)  \\\label{eqn:probs6}
P_{120} &=& P_{012} = P_{201} = \frac{1}{9} \left(1-2 r_{ab} r_{bc} r_{ca} \cos (\varphi+\pi/3)\right) \\\label{eqn:probs7}
P_{021} &=& P_{210} = P_{102} = \frac{1}{9} \left(1-2 r_{ab} r_{bc} r_{ca}   \cos (\varphi-\pi/3) \right).
\end{eqnarray}
The coincidence probability for the detection of two photons only depends on the modulus squared overlap of the inserted photons
\begin{equation}\label{eqn:P0112}
P_{011}=P_{101}=P_{110}=\frac{1}{9}\left(2-r_{ij}^2\right),\mbox{ $i,j\in(a,b,c)$; $i\neq j$}.
\end{equation}

\section{Constant triad phase}
\begin{figure}[h!]
	\centering
	\includegraphics[width=0.7\textwidth]{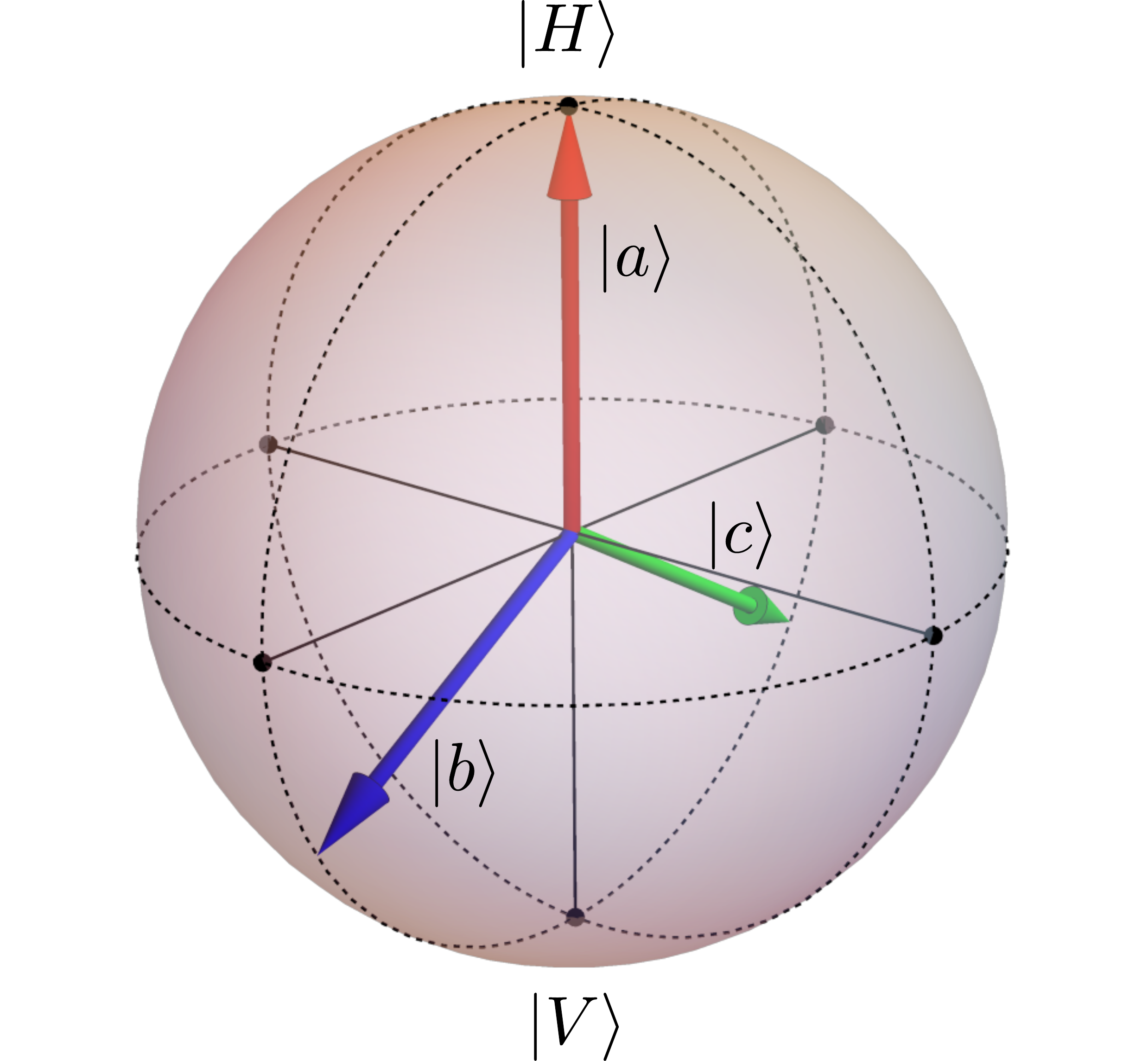}
	\caption{\label{fig:MercBloch} ``Mercedes Star'' configuration illustrated on the Bloch-sphere of polarisation.}
\end{figure}

In a first experiment we aim to show the effect of the triad phase and demonstrate how non-zero values for $\varphi$ can be accessed. When preparing states in identical polarisations and only varying time delays, the triad phase will be zero $\varphi=0$. As was shown in section \ref{sect:Gaussian}, all products over overlap integrals in this case will be purely real, leading to an overall real and positive three-photon interference term: $\braket{a}{b}\braket{b}{c}\braket{c}{a}=\braket{t_a}{t_b}\braket{t_b}{t_c}\braket{t_c}{t_a}$. For example, we can prepare the photons in different temporal states and identical polarisations
\begin{eqnarray}\label{eqn:staticpol1}
\ket{a} &=&\ket{ t_a}\otimes\ket{H} \\\nonumber
\ket{b}&=&\ket{ t_b}\otimes\ket{H} \\\nonumber
\ket{c} &=&\ket{ t_c}\otimes\ket{H}. 
\end{eqnarray}
If, in contrast, we prepare the polarisations in a ``Mercedes-Star'' configuration, as shown in Figure \ref{fig:MercBloch}, we obtain a triad phase of $\varphi=\pi$

\begin{eqnarray}\label{eqn:staticpol2}
\ket{a} &=&\ket{ t_a}\otimes\ket{H} \\\nonumber
\ket{b}&=&\ket{ t_b}\otimes\frac{1}{2}(\ket{H} +\sqrt{3}\ket{V})\\\nonumber
\ket{c} &=&\ket{ t_c}\otimes\frac{1}{2}(\ket{H} -\sqrt{3}\ket{V}).
\end{eqnarray}
This can be seen straightforwardly by considering the values of the overlap integrals of the states defined in equation \ref{eqn:staticpol2}:
$\braket{a}{b}=\frac{1}{2} \braket{t_a}{t_b}$ and $\braket{c}{a}=\frac{1}{2} \braket{t_c}{t_a}$, but $\braket{b}{c}=-\frac{1}{2} \braket{t_b}{t_c}$.
This leads to a real, negative three-photon term:
$\braket{a}{b}\braket{b}{c}\braket{c}{a}=-\frac{1}{8}\braket{t_a}{t_b}\braket{t_b}{t_c}\braket{t_c}{t_a}$, which implies: $\cos(\varphi)=-1$ or $\varphi=\pi$.
 This sign flip in the three-photon interference term that occurs when we set $\varphi=\pi$ has a profound consequence: since the two-photon and three-photon terms now both contribute negatively to the coincidence probability (equation \ref{eqn:P1112}), their effects compound, enhancing the interference. 
\section{Isolating three-photon interference}
\begin{figure}[h!]
	\centering
	\includegraphics[width=0.7\textwidth]{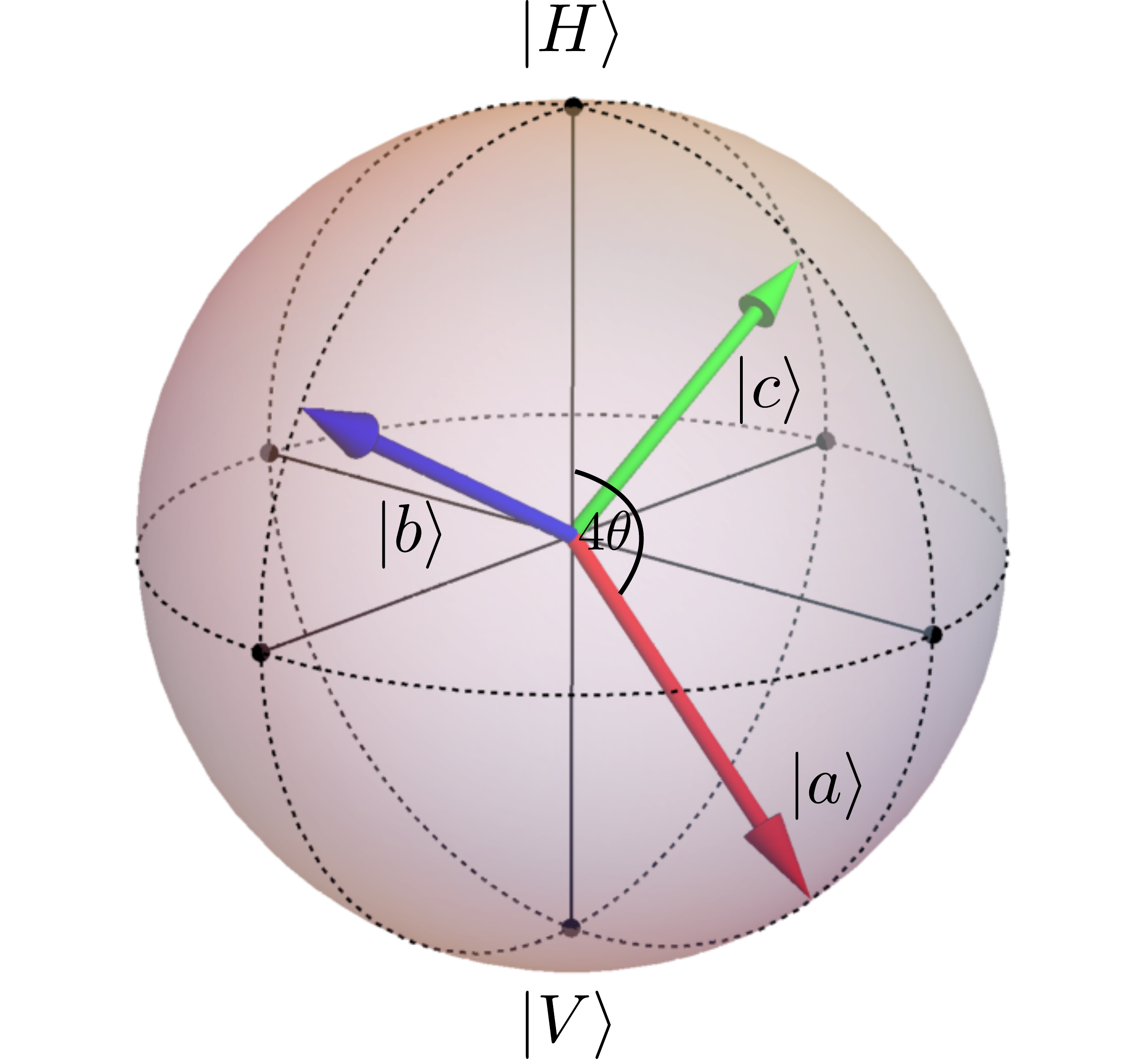}
	\caption{\label{fig:SweepPol} Polarisation configuration on the Bloch-sphere enabling to access triad phases between $0$ and $2\pi$ by rotating the polarisation vector $\ket{a}$. }
\end{figure}

We want to investigate the question whether we can observe three-photon interference independently of lower order interference contributions. In other words: can we observe a variation in the three-fold coincidence probability $P_{111}$ as we vary some parameter in the preparation of the three-photon input state, while all lower order interference terms are kept constant? The two-photon coincidence probabilities (equations \ref{eqn:P0112}) are only functions of the moduli squared overlaps $|\braket{a}{b}|^2$. The three-photon coincidence probability (equation \ref{eqn:P1112}) depends on an additional independent parameter, the triad phase. Therefore, we should always be able to keep second order correlations constant by keeping the moduli squared overlaps constant. We can then simultaneously vary the three-fold coincidence probability by changing the triad phase. We have noted previously that the triad phase can be accessed through polarisation and identified $\varphi=\pi$ with the ``Mercedes-star'' configuration of polarisations on the Bloch-Sphere. It is possible to vary the triad phase continuously from 0 to $2\pi$ by rotating the polarisation of state $\ket{a}$ in the plane orthogonal to the plane spanned by vectors $\ket{b}$ and $\ket{c}$. This particular path is also favourable since at all times the moduli overlap integrals of the two stationary states with the rotating one are identical: $|\braket{a}{b}|^2 =|\braket{a}{c}|^2$. This means we only have to keep one remaining parameter constant.
The polarisation configuration is illustrated in Figure \ref{fig:SweepPol}.
\begin{align}\label{eqn:dynamicpol2}
\ket{a}&=\ket{t_a}\otimes \left[ \cos{\left(2\theta\right)}\ket{H}+i\sin{\left(2\theta\right)}\ket{V}\right]\\\nonumber
\ket{b}&=\ket{t_b} \otimes \left[\frac{1}{2}(\sqrt{3}\ket{H}+\ket{V})\right]\\\nonumber
\ket{c}&=\ket{t_c} \otimes \left[\frac{1}{2}(\sqrt{3}\ket{H}-\ket{V})\right].
\end{align}
How do we keep the overlap integrals $|\braket{a}{b}|^2 =|\braket{a}{c}|^2$ constant? We have one more degree of freedom in the preparation of our state which we have not used so far: time. The idea is relatively simple. First assume the photons fully overlap in time: As the polarisation vector of state $\ket{a}$ is rotated the overlap integral will vary, when $\theta=0$ it will be at a maximum value of $|\braket{a}{b}|^2 =|\braket{a}{c}|^2=0.75$ and when $4\theta=\pi$ the minimum of $0.25$ is reached. Consider starting the path on the polarisation Bloch sphere at $4\theta=\pi$; as we rotate the polarisation vector by a small amount, the overlap starts to increase. In order to maintain the previous value of $0.25$, we can reduce the overlap in the temporal degree of freedom. At $4\theta=\pi$ all photons will be completely indistinguishable in time. As the polarisation angle changes, we walk off $\ket{a}$ in time and reduce the temporal overlap $|\braket{t_a}{t_b}|=|\braket{t_a}{t_c}|$ by the same amount that the overlap increases in the polarisation degree of freedom.

The moduli overlap between the states are (with the polarisation configuration of equations \ref{eqn:dynamicpol2}):
\begin{align}
r_{ab}&=\frac{1}{2}|\braket{t_a}{t_b}|\sqrt{2+\cos(4\theta)}\\
r_{ca}&=\frac{1}{2}|\braket{t_c}{t_a}|\sqrt{2+\cos(4\theta)}\\
r_{bc}&=\frac{1}{2} |\braket{t_b}{t_c}|.
\end{align}
We assume for the temporal shape of the photons a Gaussian wavepacket with standard deviation $\sigma$. The overlap integral then becomes: $|\braket{t_a}{t_b}|=\exp(-\frac{|t_a-t_b|^2}{4\sigma^2})$ (cf. section \ref{sect:singlephot} and \ref{sect:Gaussian}). To find the time delays that maintain a constant overlap of $0.5$, we need to solve
\begin{equation}
\frac{1}{2}\exp(-\frac{|t_a-t_b|^2}{4\sigma^2})\sqrt{2+\cos(4\theta)}=0.5.
\end{equation}
The solution is given by:
\begin{equation}\label{eqn:timediff}
|t_a-t_b|=|t_a-t_c|=\sigma\sqrt{2 \ln[2+\cos(4\theta)]}.
\end{equation}

The triad phase is straightforwardly obtained from equations \ref{eqn:dynamicpol2}:
\begin{align}
\varphi&=\arg(\braket{a}{b}\braket{b}{c}\braket{c}{a})=2\,\arg\left(\sqrt{3}\cos(2\theta)-i\sin(2\theta)\right).
\end{align}
\section{Experimental setup}
\subsection{Interferometer}\label{sect:tritterInterf}
\begin{figure}[h!]
	\centering
	\includegraphics[width=0.7\textwidth]{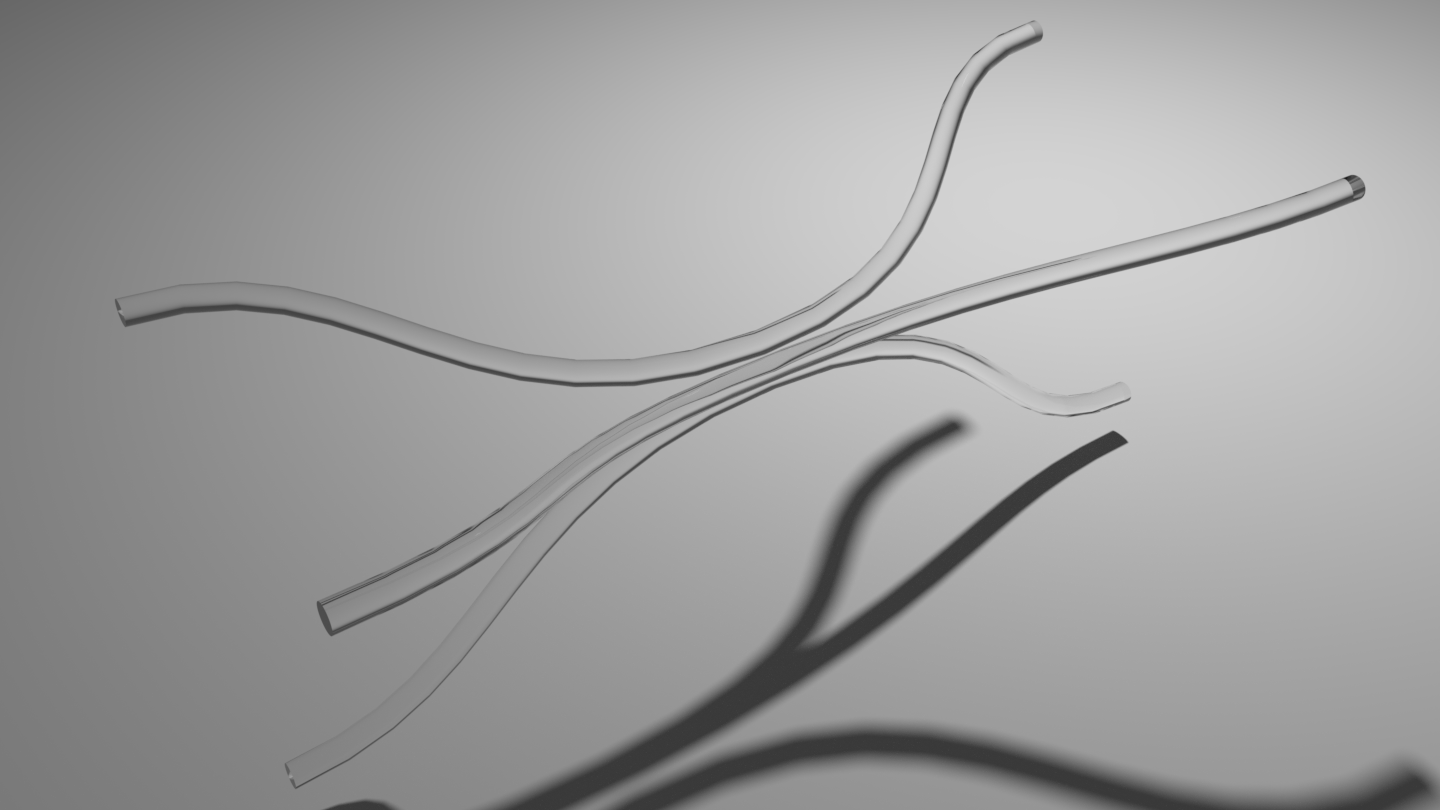}
	\caption{\label{fig:Tritter} Three-mode coupler ``Tritter''.
	}
\end{figure}
\vspace{0.5cm}
A balanced three mode interferometer can be realised in several ways. Here, we use a fibre integrated tritter. Three fibres are fused together in such a way that their cores are in close proximity such that light can couple between the fibres. In Appendix \ref{app:FSwriting} and section \ref{sect:modeCoup} couplers in integrated photonic circuits, which also consist of evanescently coupled waveguides are discussed in more detail. In Figure \ref{fig:Tritter} the basic layout of the coupler is illustrated.
The matrix of the fibre tritter was determined by inserting single photons into each input port and measuring the detection probability at each of the outputs. The phases were obtained by requiring unitarity of the matrix for the measured set of amplitudes.
\begin{equation}\label{eqn:tritterexp}
U_{\text{real}}= \left(\begin{matrix}0.6&0.6&0.53\\0.6&-0.28+0.48i&-0.27-0.48i\\0.6&-0.28-0.5i&-0.27+0.48i\end{matrix}\right) 
\end{equation}
The measured unitary interferometer matrix \ref{eqn:tritterexp} has a high fidelity of 0.99 with respect to the ideal tritter matrix \ref{eqn:TritterU}. The influence of slight variations in coupling strength between different ports is negligible. 
\subsection{Single photon creation, preparation, and detection}
\begin{figure}[h!]
	\centering
	\includegraphics[width=1\textwidth]{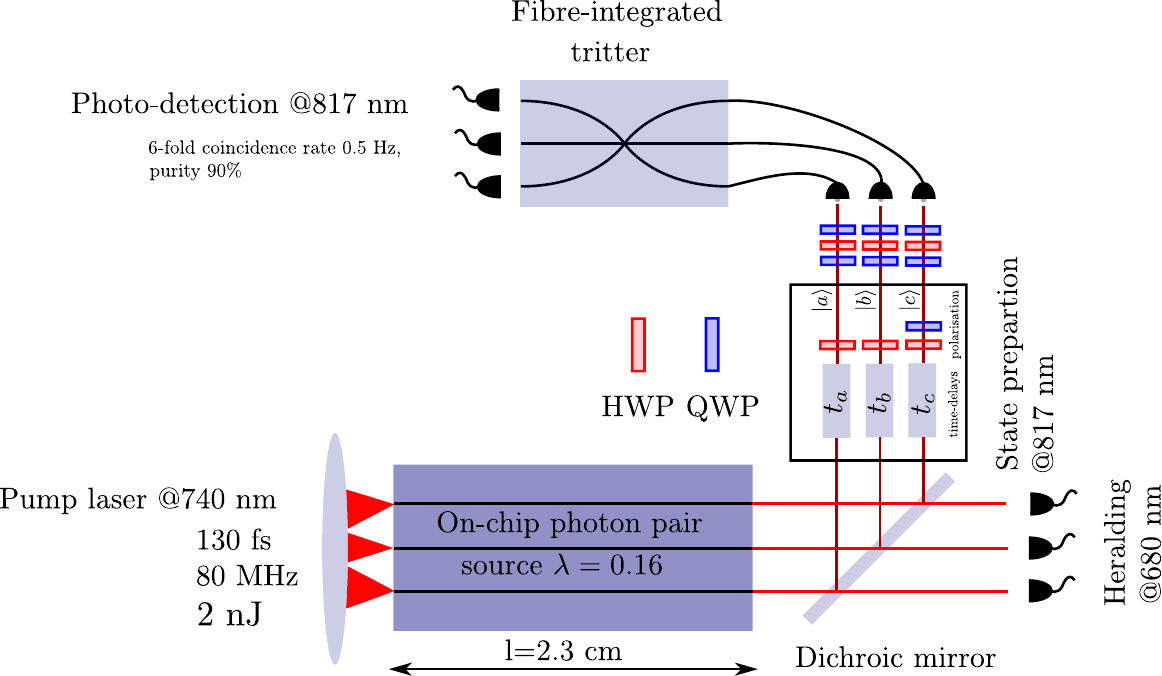}
	\caption{\label{fig:ExpSetup} Scheme of the experimental setup.
		The relative temporal delays of the three photons are adjusted using delay stages.
		We use sets of quarter-wave plates (QWPs) and half-wave plates (HWPs) to prepare the polarisation state of each photon and to compensate for polarisation rotations in the fibres.  The outputs of the fibre tritter are monitored using multiplexed commercial avalanche photodiodes.}
\end{figure}
The experimental setup is depicted in Figure \ref{fig:ExpSetup}. A detailed description of the heralded three-photon source can be found in \cite{spring2013chip}. We use a mode-locked pulsed laser at 740 nm, 80 MHz, with 130 fs pulse-width. The pump beam is first split into three separate beams using bulk-optic components. The three beams pass through two delay stages which apply relative time-delays between the pump-pulses. The beams are then focussed and coupled into three separate silica waveguides within a silica on silicon chip. The on-chip waveguides were fabricated in a uv-writing process described in \cite{lepert2011demonstration}. The third order non-linearity of glass gives rise to a four wave mixing process (described in more detail in Appendix \ref{app:FWM}), creating spectrally separate signal and idler photons at 680 and 817 nm. The squeezing paramater $\lambda$ is $\sim 0.16$ cf. Appendix \ref{app:higherord}. Signal and idler photons are orthogonally polarised to the pump beam. After the silica on silicon source chip, a polarising beam splitter is used to separate the pump light at 740 nm from the signal and idler photons. Subsequently a pair of angle tuned dielectric filters is used to filter the bandwidth of signal and idler photons.

\begin{figure}
	\centering
	\includegraphics[width=1\textwidth]{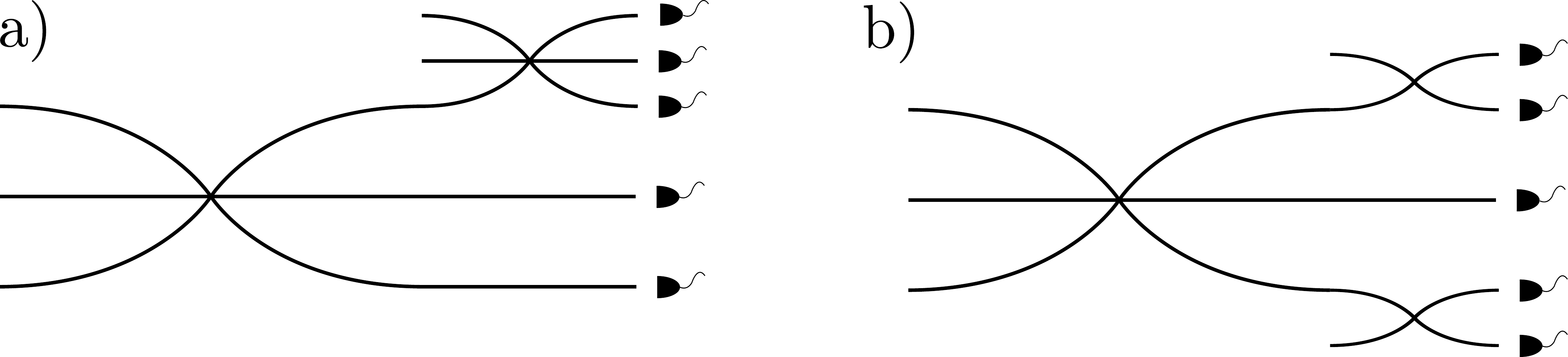}
	\caption{Detector configurations for: a) constant triad phase, b) isolating three-photon interference.}
	\label{fig:DetConf}
\end{figure}

Next, the idler photons at $817$ nm are prepared in polarisation using sets of half and quarter waveplates.
After preparing the polarisations the photons are injected into the tritter.
At the end the photons are detected using a series of avalanche photo-diodes. The detector configuration is shown in Figure \ref{fig:DetConf}. Placing beam-splitters or additional tritters at the output of the interferometer affords pseudo number resolution: with some probability, depending on the interferometer used, bunched outputs of two or three photons separate in the additional beam-splitter or tritter and can be resolved. Configuration a) in Figure \ref{fig:DetConf} allows for the detection of $(1,1,1),(3,0,0),(2,1,0),(2,0,1)$ events. Configuration b) allows us to detect $(1,1,1),(2,1,0),(2,0,1),(1,0,2),(0,1,2)$ events (see Appendix \ref{app:additionalEv}).  

\section{Experimental data}
\subsection{Constant triad phase}
\begin{figure}[h!]
	\centering
	\includegraphics[width=1\textwidth]{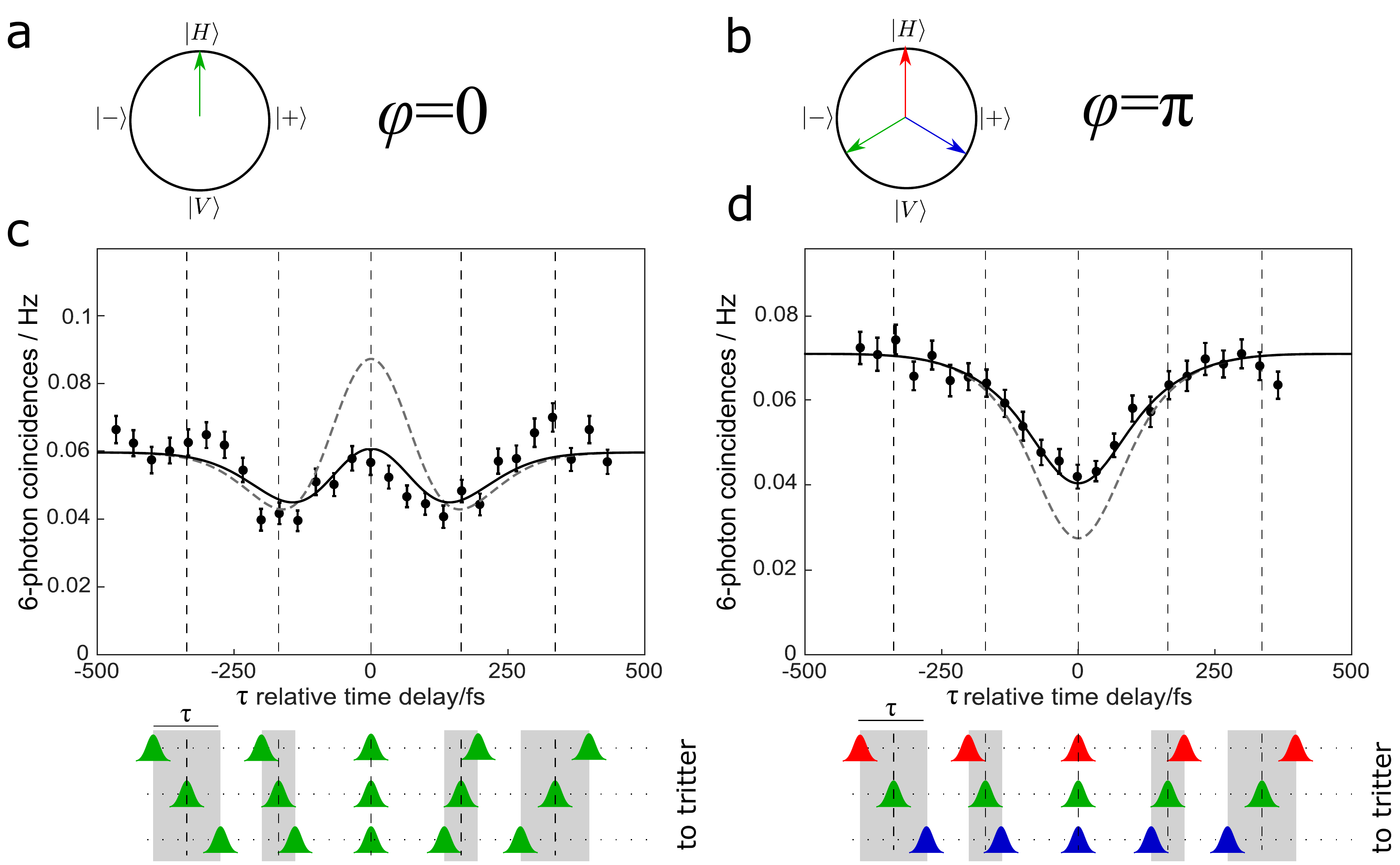}
	\caption{\label{fig:StaticPol} 
		Measured three fold coincidence event rates for two polarisation configurations a) and b), corresponding to values of 0 and $\pi$ for the triad phase. The time delays of the three photons are varied as illustrated in the lower part of the figure. We plot the three-fold coincidence events as a function of the time delay $\tau$:  $t_1=t_2-\tau/2, t_3=t_2+\tau/2$. The dotted lines indicate the ideal, simulated curve. The solid line is a fit to the experimental data using a model discussed in Appendix \ref{sect:simExp}.  The total number of heralded three-photon counts is between 200 and 350 for a) and between 250 and 450 for b). Figure from \cite{menssen2017distinguishability}.
	}
\end{figure}

In this section I present the experimental data collected. The first experiment explores three-photon interference by choosing two distinct values of the triad phase, $\varphi=0$, $\pi$. In Figure \ref{fig:StaticPol} the results are shown. In each case we vary the relative time delays between two of the photons symmetrically around the third, which is kept  stationary. This is illustrated beneath the experimental data in Figure $\ref{fig:StaticPol}$. 
For all identical polarisations, as a function of the relative time delay $\tau$, we observe a ``W'' shape. The origin of this feature can be explained by the effect of contributions from different terms in the three-fold detection probability in equation \ref{eqn:P1112}. For large $\tau$ there is no overlap between any of the states and the detection probability asymptotically approaches a constant value. As $\tau$ decreases, the negative two-photon interference terms, which scale quadratically in the overlap-integral, contribute and we start to see a dip. As $\tau$ approaches zero, the positive three-photon term, which scales as the cube of the overlap, dominates over the negative two-photon terms and a peak forms.
In contrast, when we choose a triad phase of $\varphi=\pi$ by preparing polarisations in the ``Mercedes-star'' configuration on the Bloch sphere, the three-photon term will contribute negatively, adding to the effect of the two-photon interference terms. The result is an enhanced dip in the third order correlations, shown in Figure \ref{fig:StaticPol} on the right. %
This is rather surprising since the participating photons have a maximal modulus squared overlap of only $0.25$, indeed they are in a ``maximally'' distinguishable configuration in polarisation space (excluding two being orthogonal). This highlights especially the importance of the triad phase in multi-photon interference and challenges our perception on the relationship between distinguishability and strength of interference. 

\subsection{Isolating three-photon interference}
\begin{figure}[h!]
	\centering
	\includegraphics[width=1\textwidth]{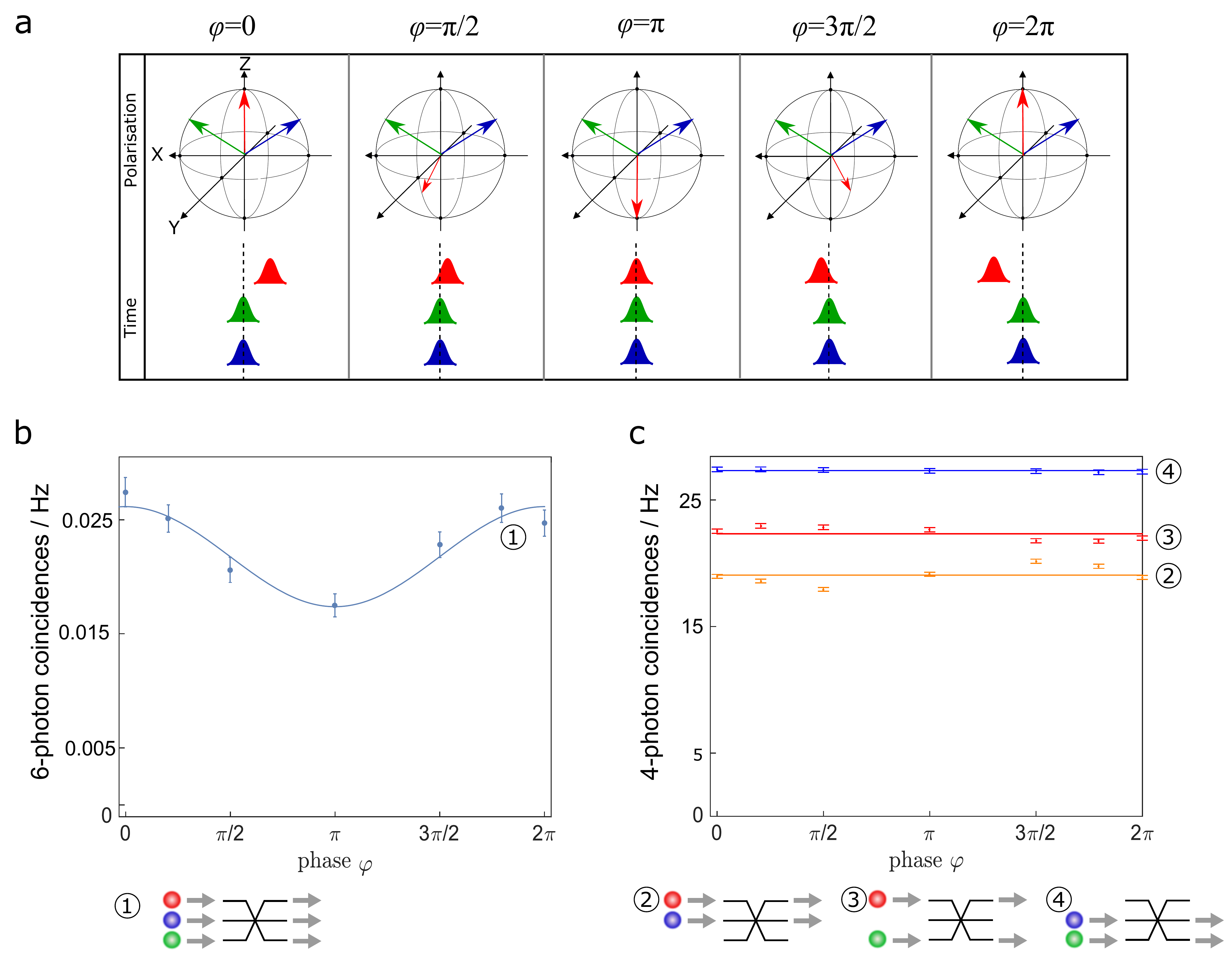}
	\caption{\label{fig:SweepP} Experimental data for three and two-fold detection events. a) Each polarisation configuration corresponding to a specific triad phase is illustrated. Two photons remain at a constant polarisation while the third polarisation is rotated in the plane of mirror symmetry. As the polarisation of the photon (marked in red) is rotated it is simultaneously delayed in time. b) Rate of three-fold coincidences as a function of the triad phase. We clearly observe a dip at $\varphi=\pi$. c) Rate of two-fold detection events. A representative sub-set, corresponding to the input/output configurations is shown.  The total number of heralded three-photon counts is between 330 and 550 per point. Figure from \cite{menssen2017distinguishability}.  
	}
\end{figure}

Here I present experimental results demonstrating that we can isolate three-photon interference from lower order interference. As previously discussed, we prepare polarisation and time in such a way that we keep all moduli squared overlap integrals and hence all second order correlations constant. The triad phase is adjusted from 0 to $2\pi$ by rotating the polarisation around the Bloch sphere. In Figure \ref{fig:SweepP} b) we can clearly observe a dip at a $\varphi=\pi$. 
The three-photon coincidence probability rate has the expected cosine dependence on the triad phase.
At the same time two-photon coincidence measurements are constant (Figure \ref{fig:SweepP} c)). Their average variation is  $6\%$. For single counts (not shown), the variation is $3\%$. 
I have chosen to display three representative two-fold coincidence measurements. In contrast, the visibility of the three-fold coincidence dip is $43\%.$

\section{Conclusion and outlook}
In this chapter I have demonstrated three-photon interference in independent photons and presented two separate experiments.
I have shown how three-photon interference is governed by a collective phase, the ``triad'' phase, which can not be accessed through time delays. A surprising consequence of three-photon interference is that three photons can interfere strongly, even when they have a very low state overlap. Further, I have demonstrated that I can isolate three-photon interference from contributions of lower order correlations. In a simultaneous work this was also demonstrated by Agne et. al. \cite{agne2017observation}, however using three entangled photons.
This experiment shows the richness of multi-photon interference and opens up new avenues in studying fundamental effects in multi-photon interference. Beyond the study of foundational physics, three-photon interference could have implications in quantum technologies. The HOM effect is a widely used in linear optical quantum computation to realise quantum gates between photonic qubits \cite{knill2001scheme} and is an essential ingredient to achieving entanglement of atomic qubits in remote locations \cite{hofmann2012heralded}. Extending these protocols to utilise the kind of multi-photon interference we demonstrate here could be an exciting perspective. Further, I note that the triad phase bears similarity to a geometric phase. This will be more closely explored in Chapter \ref{chap:TheoryTopModes}.
In the next chapter I will explore an extension of the triad phase to four photons and show how four-photon interference can likewise be demonstrated.

%% file: chapter4.tex
\chapter{Four-photon interference}
\label{chap:FourPhot}
In this chapter I will examine the interference of four independent single photons \cite{tichy2011four} in an eight mode interferometer or ``quitter''. This work was done together with A.E. Jones. A.E. Jones lead the experiment.  I worked on the theory of four-photon interference. Specifically, I worked on the development of the state configuration which exhibits four-photon interference (shown in Figure \ref{fig:4photState}). I also worked on a graph theoretical approach to multi-photon interference, which was independently developed by me and V. Shchesnovich \cite{shchesnovich2017interference,I4phot}. %
The data presented here was collected by A.E. Jones. I assisted with the data analysis; we both worked on modelling the experimental results. The results discussed in this section are currently in preparation for publication \cite{I4phot}.

\section{Introduction}
Having examined three-photon interference, the question arises whether adding more photons will still reveal new physics that was not captured in three (and two) photon interference. We have already seen that three-photon interference challenges our intuition about interference phenomena. Very different photons can interfere strongly in collective three-photon interference. For four-photon interference \cite{tichy2011four,shchesnovich2018collective} this is taken to an extreme. I will demonstrate how we can see collective four-photon interference even if some of the photons involved are in orthogonal states. The graph theoretical picture, illustrated in Chapter \ref{sect:graphTh}, is particularly suitable to describe this type of interference. four-photon interference phenomena have been studied in previous works \cite{shih1993four,ou1999observation,de2003quantum,liu2007four}. We succeed in demonstrating, for the first time, four-photon interference that is independent of two- and three-photon interference contributions.

\begin{figure}[h]
	\centering
	\includegraphics[width=1\textwidth]{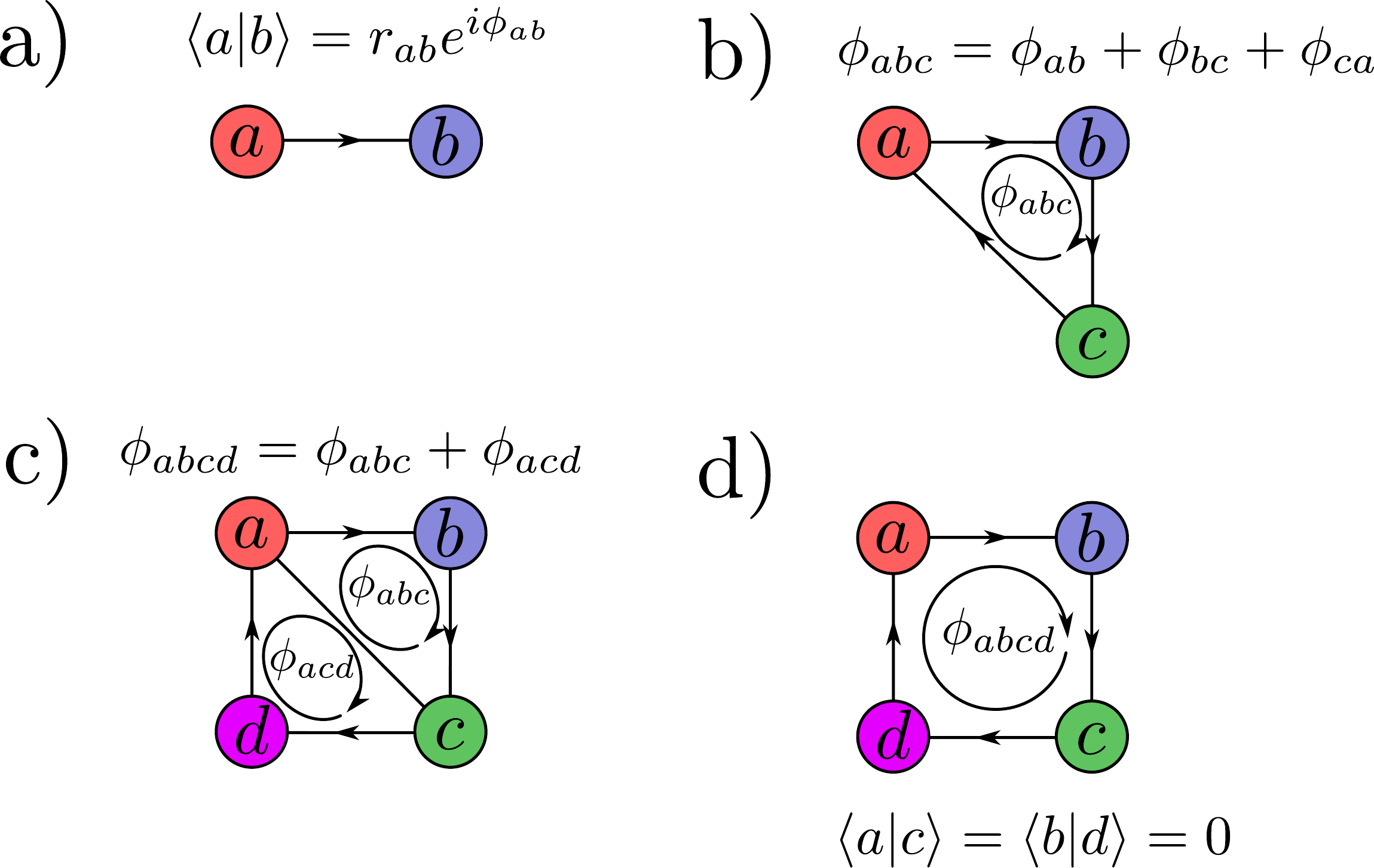}
	\caption{Graphs a) Single path connecting two states.  b) three-photon closed loop. c) Fully connected graph of four photons (diag. connection omitted). d) Partially disconnected circular graph.}
	
	\label{fig:4photGraph}
\end{figure}
\begin{figure}[h]
	\centering
	\includegraphics[width=1\textwidth]{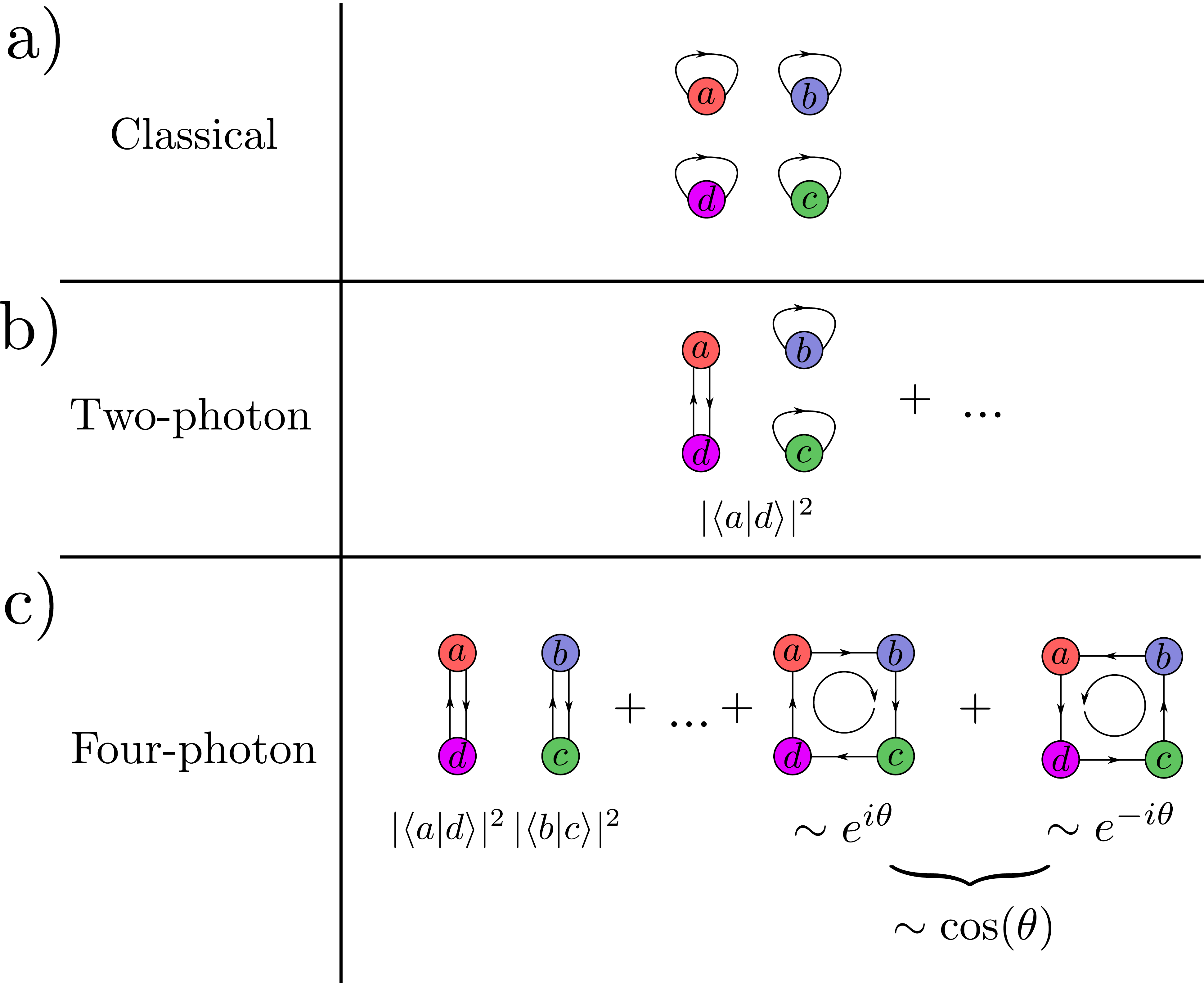}
\caption{Four-photon graph with possible paths illustrated, sorted by total number of photons exchanged. a) The contribution from single photon (self-interference) is illustrated. b) Closed paths contributing two-photon interference are shown. c) Paths containing four vertices are depicted. The last two paths are dependent on the four-photon phase $\theta=\phi_{abcd}$.}
	\label{fig:4photGraph2}
\end{figure}

In Figure \ref{fig:4photGraph2} all possible closed paths on the four vertex graph contributing to four-photon interference are illustrated.
When the graph is fully connected, i.e. all photons have some finite overlap, as illustrated in Figure \ref{fig:4photGraph} c), the four-fold coincidence probability function will contain contributions from two, three, and four-photon interference. However, the astute observer will notice that the four-photon phase, which is associated to the path $a\rightarrow b \rightarrow c \rightarrow d \rightarrow a$, is redundant. Namely, it can be expressed as a sum over the two triad phases, as shown in Figure \ref{fig:4photGraph} c). As a result it is not an independent parameter of the interference. Performing only two and three-photon experiments, with photons prepared in all combinations of the four states appearing, we could infer the four-photon phase.
However, as was first pointed out in \cite{shchesnovich2017interference},
if the photons on the four-photon graph are made pair-wise orthogonal \cite{jones2018interfering}, i.e. severing diagonal connections between the vertices, the triad phase becomes undefined. As a result we obtain another interference phenomenon, for which V. Shchesnovich coined the term ``Circle-dance-interference'' \cite{shchesnovich2017interference}: Photons interfere on a circular graph.
\begin{figure}[h]
	\centering
	\includegraphics[width=0.7\textwidth]{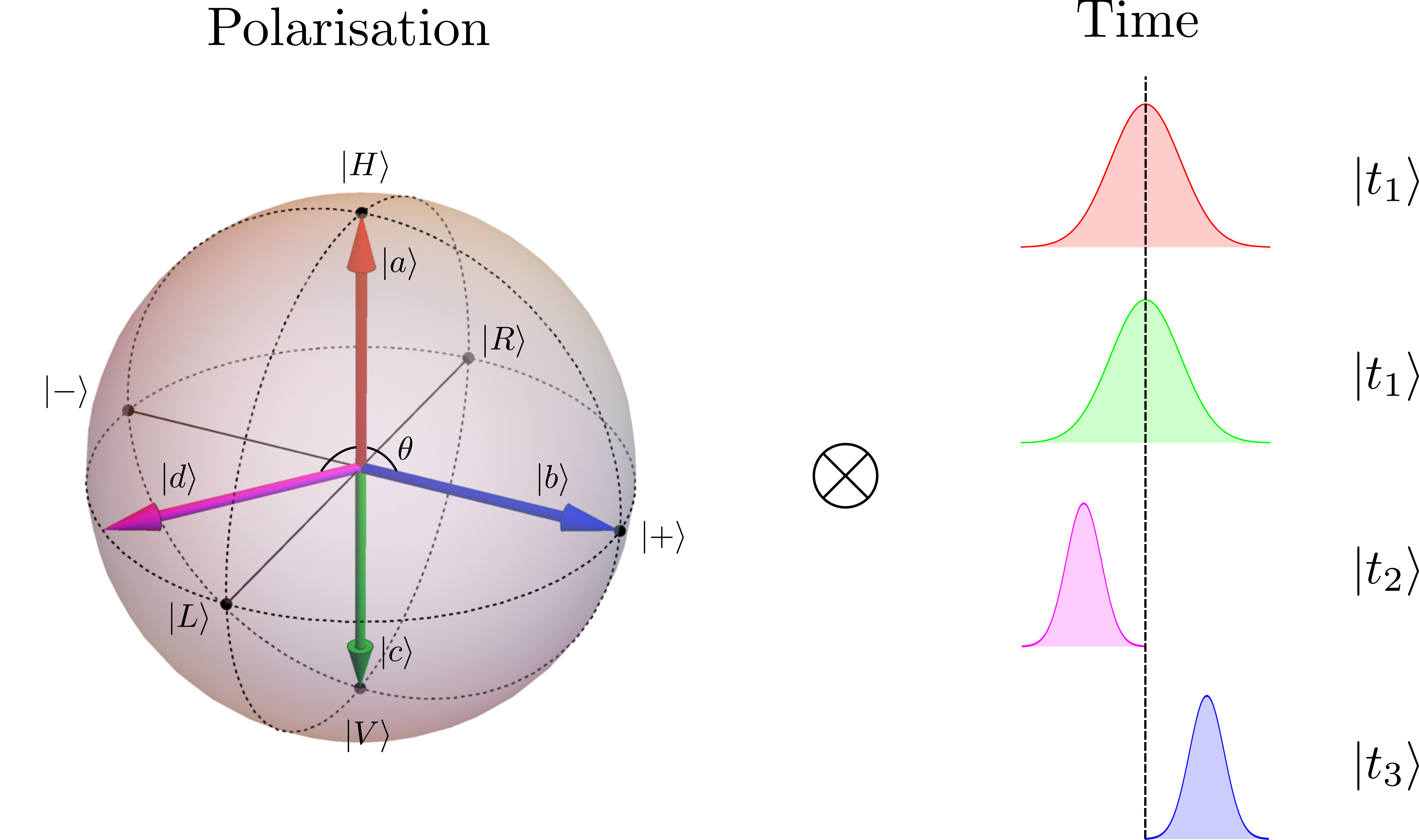}
	\caption{Illustration of the four-photon state-configuration in polarisation and temporal degrees of freedom.}
	\label{fig:4photState}
\end{figure}

In order to observe four-photon interference, we are faced with a similar challenge as in three-photon interference. We need to keep all coincidence detection probabilities for less than four photons constant as we vary the four-fold coincidence probability. In principle this is only possible when the four-photon phase appears as an independent parameter in the scattering probability, which we accomplish by preparing the states in a circular graph, as illustrated in Figure \ref{fig:4photGraph} d). In four photons this state configuration can be achieved with polarisation and temporal degrees of freedom. In Figure \ref{fig:4photState} it is illustrated how the state is prepared. Photons $a$, $b$ and $c$ are kept static at 90 deg. angles in the Bloch-sphere, in polarisation states: $\ket{H},\ket{+},\ket{V}$. The fourth photon ($d$) is rotated in the plane orthogonal to the $\ket{H}$ polarisation Bloch vector. In time, photon $a$ and $c$ are in an identical broad temporal profile, while photon $b$ and $d$ are narrow in time and non-overlapping with each other, but with some overlap with photons $a$ and $c$. This configuration realises the path connections in the graph illustrated in Figure \ref{fig:4photGraph} d). 
We can write the state configuration as:
\begin{eqnarray}
\ket{a}&=&\ket{H,t_1}\\
\ket{b}&=&\ket{+,t_2}\\
\ket{c}&=&\ket{V,t_1}\\
\ket{d}&=&\frac{1}{\sqrt{2}}(\ket{H,t_3}+e^{i\theta}\ket{V,t_3}).
\end{eqnarray} 

The different paths on the four-photon graph that yield contributions to the four-fold coincidence probability are illustrated in Figure \ref{fig:4photGraph2}. 
The contribution to four-photon interference comes from the paths which are illustrated in Figure \ref{fig:4photGraph2} c). The clockwise and anti-clockwise paths contribute a cosine dependence on the multi-photon phase, as was the case for three-photon interference. All other contributions are either constant (two-photon interference) or vanish (three-photon interference).
The non-vanishing overlap integrals of the photons are:
\begin{eqnarray}
\braket{a}{b}&=&\frac{1}{\sqrt{2}}\braket{t_1}{t_2}\\
\braket{b}{c}&=&\frac{1}{\sqrt{2}}\braket{t_2}{t_1}\\
\braket{c}{d}&=&\frac{1}{\sqrt{2}}e^{i\theta}\braket{t_1}{t_3}\\
\braket{d}{a}&=&\frac{1}{\sqrt{2}}\braket{t_3}{t_1}.
\end{eqnarray}
As we can see, the $\braket{c}{d}$ overlap is responsible for contributing the four-photon phase.
There is a further important aspect. We note that the phase $\theta$ can be measured by observing four-photon coincidence events. However, if we examine the state configuration as depicted in Figure \ref{fig:4photState} this seems rather surprising since the ``reference'' state on the Bloch-sphere $\ket{+}/\ket{b}$ is orthogonal to $\ket{d}$ (in time). Further, the states $\ket{H}$ and $\ket{V}$ alone can not provide a phase reference, since they are orthogonal to the plane in which $\ket{d}$ rotates. Indeed, in this configuration we measure a phase between orthogonal states. The $\ket{H}$ and $\ket{V}$ merely act as a ``mediator''. This situation has been explored in \cite{wong2005quantum}.

We realise the interference by sending the photons through a balanced four-port interferometer or ``quitter''.
The unitary matrix of the quitter is:
\begin{spacing}{1}
	\begin{equation}
	\label{eqn:appQuitterUnitary}
	U_{quit}=\frac{1}{2}\left(\begin{matrix}
	1&1&1&1\\
	1&1&-1&-1\\
	1&-1&e^{i\chi}&-e^{i\chi}\\
	1&-1&-e^{i\chi}&e^{i\chi}
	\end{matrix}\right).
	\end{equation}
\end{spacing}
\vspace{1cm}
The balanced quitter, unlike the tritter, has a free phase $\chi$ \cite{bernstein1974must}.
Other representations of the quitter matrix shown here can be obtained by permuting rows and columns. The form of this matrix can be obtained in a similar fashion to the tritter matrix discussed in the previous chapter. First, we set the moduli of all entries to $1/2$ ensuring a balanced output. We can again remove phases at the borders of the matrix since we are free to choose input and output phases of the interferometer. From the unitarity condition, $U_{quit}U_{quit}^{\dagger}=\mathbb{1}$, a set of equations can be obtained. We examine one of these equations as an example:
\begin{equation}
1+e^{i\phi_1}+e^{i\phi_2}+e^{i\phi_3}=0.
\end{equation}
Possible solutions are:
$\phi_1=0$, $\phi_2=\pi$, $\phi_3=\pi$,
corresponding to row 2 of the quitter matrix \ref{eqn:appQuitterUnitary}.
Another possible solution is:
$\phi_1=\pi$, $\phi_2=\chi$, $\phi_3=\chi+\pi$,
corresponding to row 3 of the quitter matrix.

\begin{figure}[h!]
	\centering
	\includegraphics[width=1\textwidth]{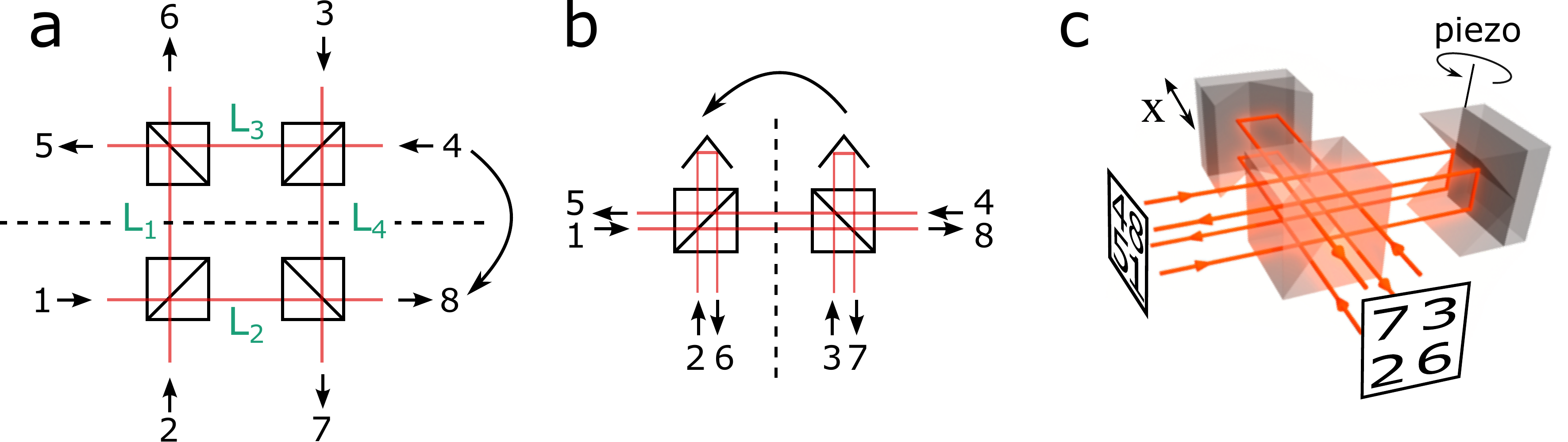}
	\caption{Illustration of the folded bulk-quitter design. A piezo-stage is used to adjust the internal phase of the quitter. Figure from \cite{I4phot}. Credit: A.E. Jones.}
	\label{fig:Quitter}
\end{figure}

The visibility of four-photon interference crucially depends on the internal phase $\chi$. The fourfold coincidence probability as a function of the overlap integrals between the single photons $r$, the internal phase $\chi$ of the quitter, and the four-photon phase $\theta$ (cf. Figure \ref{fig:4photState}) is given by:
\begin{eqnarray}
P_{5678}&=&\frac{1}{32}[3-r_{ab}^{2}-r_{bc}^{2}-r_{cd}^{2}-r_{ad}^{2} +(2+\cos (2\chi))(r_{ab}^{2}r_{cd}^{2}+r_{ad}^{2}r_{bc}^{2})\\ \nonumber
&+&2(\cos 2\chi-2)r_{ab}r_{bc}r_{cd}r_{ad}\cos(\theta)].
\end{eqnarray}
Input ports are labelled 1234 and output ports 5678. The port configuration in the bulk-optic quitter is illustrated in Figure \ref{fig:Quitter}.
The overlap integral is written as: $\braket{a}{b}=r_{ab}e^{i\phi_{ab}}$. The injection order is given by expression \ref{eqn:stateOrder}.
Each of the terms appearing corresponds to a covering of the four-photon graph in closed loops as illustrated in Figure \ref{fig:4photGraph2}.

\section{Experimental methods}
\begin{figure}[h!]
	\centering
	\includegraphics[width=0.8\textwidth]{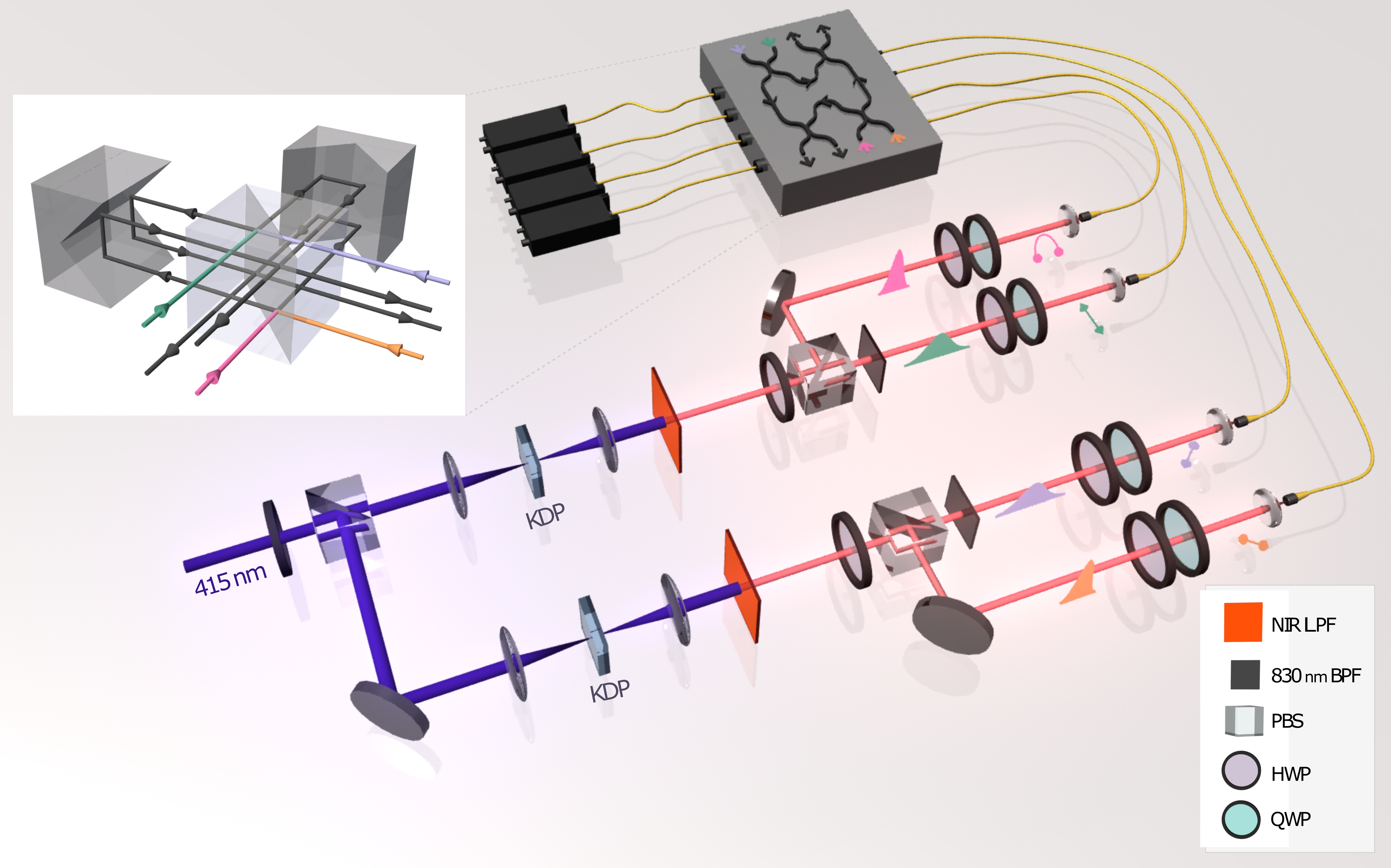}
	\caption{Experimental setup: i) Pump light at 415 nm is sent into two KDP crystals. ii) Filters remove the pump light. iii) Signal and idler photons are split by polarising beamsplitters. iv) Polarisation and spectral states are prepared through dielectric filters and waveplates. v) Single photons are sent through the interferometer (quitter) and detected using avalanche photo diodes (APDs). Figure from \cite{I4phot}.}
	\label{fig:setupFigv}
\end{figure}

To generate a four-photon state we use two KDP bulk crystal down-conversion sources \cite{mosley2008heralded}. Each of the crystals is pumped with light at 415 nm. A pump photon in a type II spontaneous parametric down conversion process (SPDC) is converted into two orthogonally polarised signal and idler photons with a central wavelength at 830 nm ($\omega_p$=$\omega_s+\omega_i$). Due to the elliptical shape of the joint spectral amplitude, the idler photon is spectrally broader than the signal. The difference in spectral width means that their Fourier transformed temporal modes are narrow for the spectrally broad idler and broad for the spectrally narrower signal. This allows us to implement the temporal modes illustrated in Figure \ref{fig:4photState} with only small adjustments to the bandwidth of the photons. In deciding the experimental bandwidth of the photons, a balance between four-photon interference fringe visibility and maintaining orthogonality between the temporal modes needs to be found. In the experiment the bandwidth ratios for the photons are 2.2:1. They are temporally walked off, as illustrated in Figure \ref{fig:4photState}, to realise orthogonality between the $\ket{t_2}$ and $\ket{t_3}$ temporal modes. The residual overlap is $\braket{t_2}{t_3}\leq0.1$. In Figure \ref{fig:setupFigv} the experimental setup is illustrated. 

In the experiment the order of injection is given by:
\begin{equation}
\begin{aligned}
\label{eqn:stateOrder}
\ket{a}&\rightarrow \mathrm{input}\enskip4\\
\ket{b}&\rightarrow \mathrm{input}\enskip2\\
\ket{c}&\rightarrow \mathrm{input}\enskip3\\
\ket{d}&\rightarrow \mathrm{input}\enskip1.
\end{aligned}
\end{equation}
Unlike in the three-photon interference experiment, where we had three heralded single photons available, the photons are only detected after passing through the interferometer. This means we cannot distinguish between events where each source creates a signal-idler pair and where a single source generates two pairs of signal and idlers \cite{zou1991induced}. These events are equally likely and scale with $\lambda^4$, where lambda is the squeezing parameter. By distributing the preparation of the single photon states across the two sources, where
each PDC source generates the pairs $\ket{a}/ \ket{d}$ and $\ket{b}/ \ket{c}$, any interference that might occur in the double emission event will never depend on the four-photon phase. Thus, these events will contribute a constant background which reduces the overall visibility significantly.
\subsection{Locking on quantum signal}
\begin{figure}[h!]
	\centering
	\includegraphics[width=0.7\textwidth]{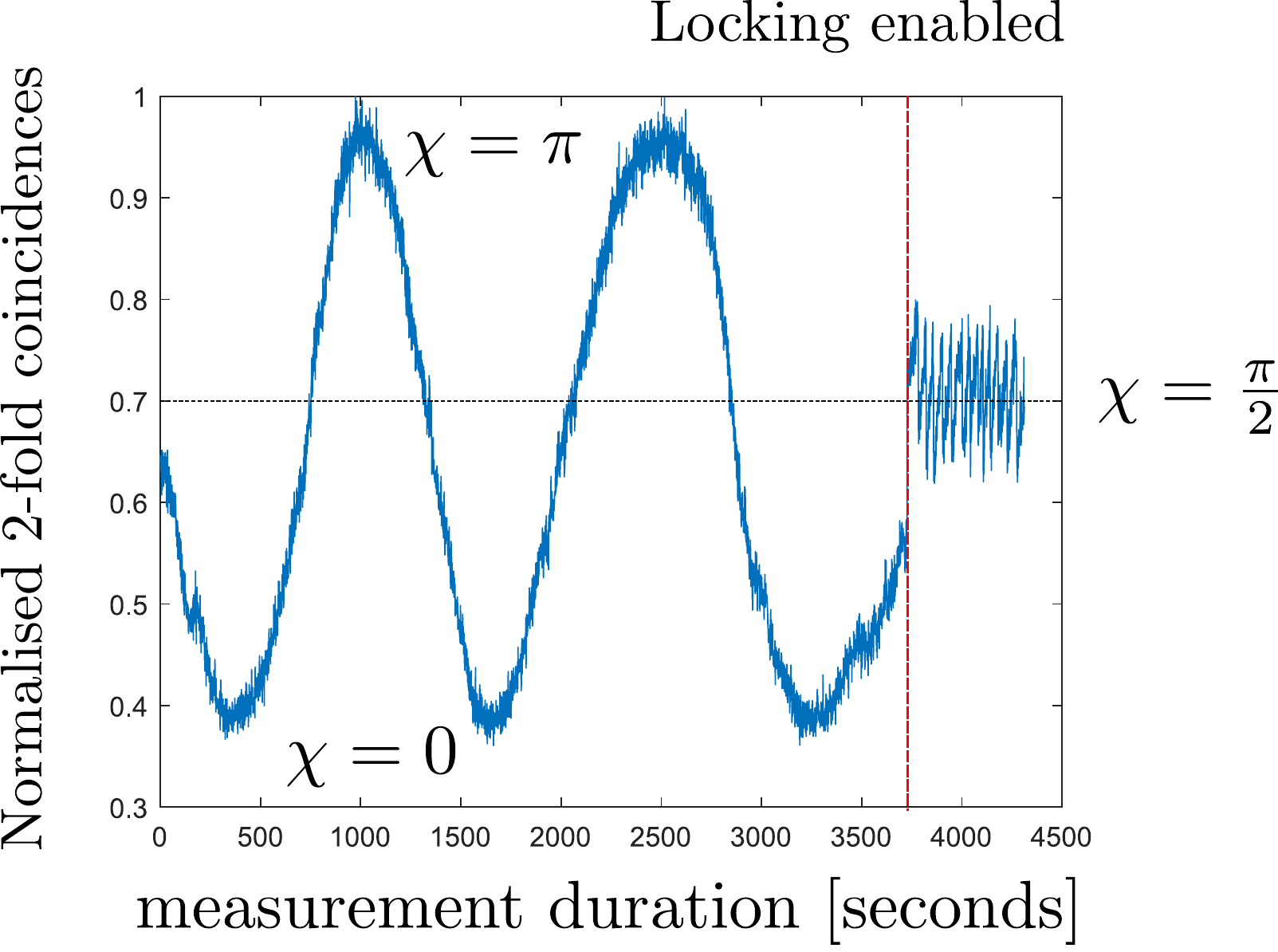}
	\caption{Normalised locking signal. A single two-fold channel, sensitive to the interferometer phase, is observed as a function of time. As the interferometer phase $\chi$ drifts, the coincidences vary. As soon as the locking is turned on, the phase is kept at a constant value of $\frac{\pi}{2}$.}
	\label{fig:Locking}
\end{figure}

We are presented with the challenge of stabilising the bulk interferometer, i.e. fixing the phase $\chi$ at a value that maximises the visibility of the four-photon interference fringe ($\pi/2$).
It can be easily verified that some of the two-fold HOM dips depend on the internal phase and can thus be used as a monitor. The two-fold coincidence probability for photons inserted into port 1,3 and detected in ports 5,7 is:
\begin{equation}
\label{eqn:phaseDepHOM}
P_{13}^{57}=\frac{1}{8}(1-r_{\alpha\beta}^2\cos(\chi)).
\end{equation}
We can change $\chi$ by adjusting the internal path-length inside of the folded interferometer by tilting one of the retro-reflectors using a piezo-motor as illustrated in Figure \ref{fig:Quitter} c). 
We then plot the counts in the $P_{13}^{57}$ channel as a function of time.
A locking algorithm can then be used to lock the value of the two-photon interference signal to the required phase of $\chi=\pi/2$, as illustrated in Figure \ref{fig:Locking}.

\section{Experimental data}
\begin{figure}[h!]
	\centering
	\includegraphics[width=1\textwidth]{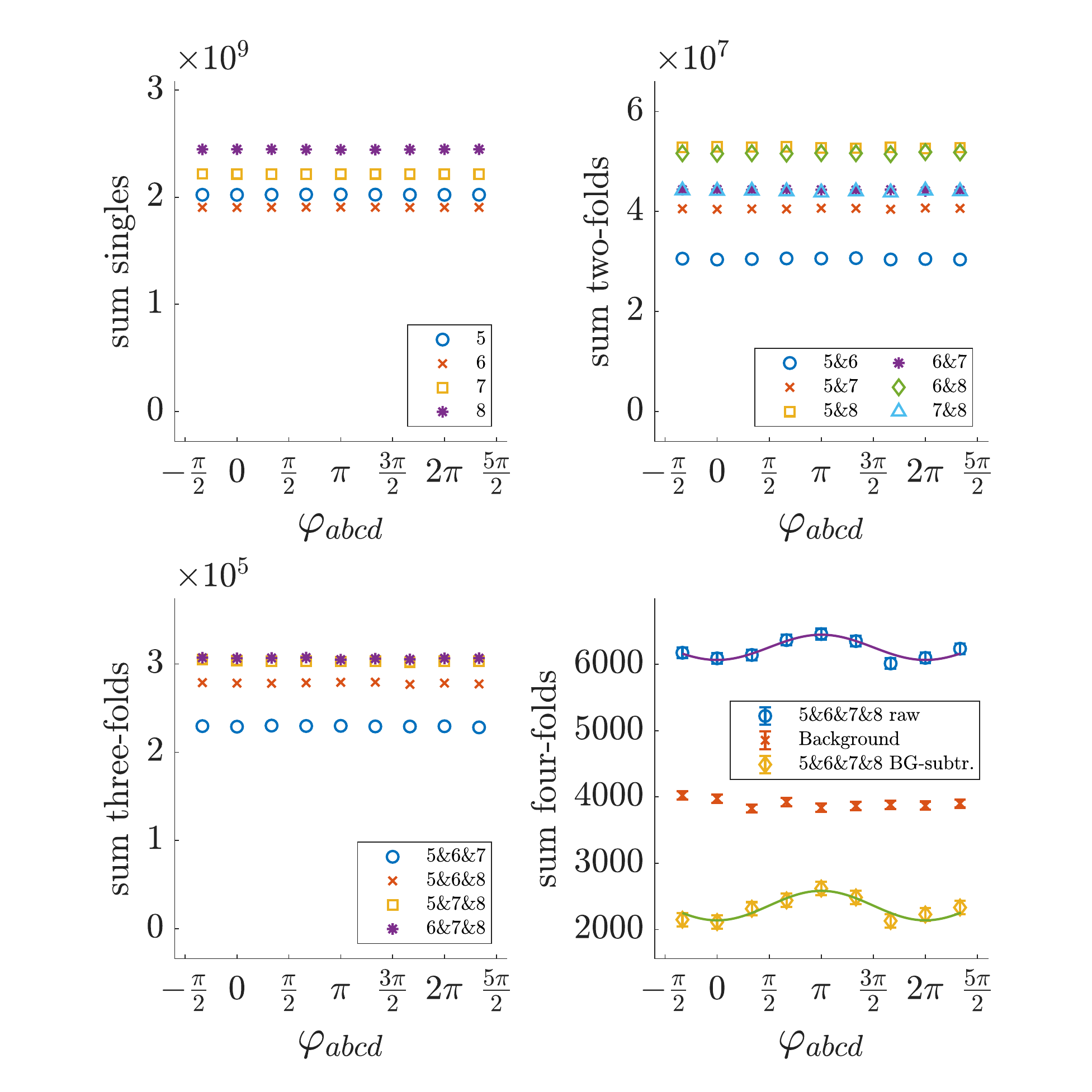}
	\caption{In this figure measurements of singles, two-, three-, and four-fold coincidences are shown. As expected, singles, two-fold, and three-fold coincidences are flat, while we observe a significant peak in the four-fold coincidences. The visibility of four-photon interference is $5.9\% \pm 1\%$ for the raw counts and $17.2\% \pm 2.8\%$ after background subtraction.}
	\label{fig:CombinedCounts}
\end{figure}

In Figure \ref{fig:CombinedCounts} the experimental data is shown. The theoretical maximum visibility of the four-photon interference fringe is $30\%$. The lower observed visibility results mainly from the flat double-emission background as well as imperfect preparation of the interferometer phase. The total counts for the entire duration of the measurement are shown. As expected, the state configuration of Figure \ref{fig:4photState} leads to a flat signal in the case of single, two and three-fold counts, while a peak with the expected cosine shape can be observed in the four-fold counts. Similarly to the three-photon interference, which was presented in the previous chapter, this demonstrates four-photon interference independent of three and two-photon interference.  
\section{Concluding remarks and outlook}
\subsection{Interference in independent particles}
 
In the last two chapters I have investigated the quantum interference of three and four photons in optical interferometers. I presented a generalised notion of interference. Our conventional intuition is shaped by the classic double slit experiment, where interference fringes are formed when two beams of coherent light, two indistinguishable paths, interfere and give rise to a varying pattern on a screen.
In this experiment the intensity of the electrical field is measured by a detector capturing light from a single point at some distance from the slit. As we vary some degree of freedom of the field, such as the relative phase between the two interfering paths (for example by adding a variable retarder at one of the slits), we observe an interference fringe. We can imagine that we trace a path $\gamma(\tau)$ in the space of all field parameters and record detector events as a function of $\tau$. In the case of the double slit, we use a single detector to measure intensity. Let us now generalise this notion of interference fringe to include coincident events measured by multiple detectors in different optical modes. For a two-photon interference fringe we measure two-photon coincidences with associated detection probability $P_{12}(\tau)$, where detectors are capturing light in modes 1 and 2. However, observing a fringe in the coincidence probability is not a sufficient condition for claiming to have observed a two-photon interference effect. Consider that we could have performed the classical Hong Ou Mandel experiment with two completely distinguishable photons (say in H and V polarisations), but varying the intensity of single photon events as a function of the time delay as well, turning it down to zero when the photons completely overlap in time. This would deliver the false impression of having observed two-photon interference. Therefore it becomes clear that in addition to requiring an interference fringe in the $P_{12}(\tau)$, we must also verify that we have kept all single photon detection events constant as a function of the parameter $\tau$. Extending this requirement to more than two photons, we note that when observing three-photon interference we must also demonstrate that we have kept both single and two-fold detection events constant. Since, in a similar way, we could have observed a fringe in the three-fold detection probability having resulted from two-particle interference. For asserting interference between $n$ photons, we therefore need to demonstrate that there are no fringes in all m-fold coincidence measurements with $m<n$. \\
We need to clarify a bit our intuition as to what we mean when we talk about n-particle interference. Let us construct a fully connected unitary interferometer $U_{nxn}$ with $n$ input and $n$ output ports. We can build an interferometer composed of two fully connected interferometers which are not inter-connected. Let us take four-photon interference as an example: there are three ways in which we can construct a unitary interferometer: $U_{4x4}$, $U_{2x2}\bigoplus, U_{2x2}$, $U_{3x3}\bigoplus U_{1x1}$. We now measure coincidence events at the output of all of these interferometers and vary some parameter of our input field (how we prepare the photons) in an attempt to observe a variation of the four fold coincidence probability, a four-photon interference fringe. In the case of the interferometer $U_{2x2}\bigoplus U_{2x2}$ we will find that indeed a fringe in the four-fold coincidence probability can be observed as we tune for example the time delay between the photons sent into the first/second independent 2x2 interferometer. The same applies to the $U_{3x3}\bigoplus U_{1x1}$ interferometer. Here we can vary both the intensity on one source (connected to the single optical mode $U_{1x1}$) or two-/three-fold coincidences by varying time delays and polarisations. These features in the four-fold coincidence probability evidently cannot be a signature of four-photon interference, since at no point do the paths of all four photons interfere in these partially disconnected interferometers. A variation of the four-fold coincidence probability can only be observed when varying either three-fold, two-fold coincidences or singles. In the fully connected $U_{4x4}$ in contrast, it is always possible to observe a fringe in the four fold coincidence events, independent of changes in lower order interference. This motivates and defines our notion of n-particle interference for independent photons.  
\subsection{Interference of entangled particles}
So far we have restricted ourselves to describing the interference of independent particles, where the input state can be written as a product of the single photon states. However, the requirement for n-photon interference: to observe a variation in the n-particle coincidences while all lower order coincidence counts are kept constant as a function of some field parameter, applies of course generally to any field. We may therefore ask ourselves if there are other states that satisfy the requirements of exhibiting n-photon interference if we include entangled states. Indeed, the answer was given in \cite{rice1994multiparticle}. Here it was noted that a fully n-partite entangled state can exhibit ``genuine'' n-particle interference. Subsequently this was demonstrated experimentally in \cite{agne2017observation} for the case of three photons. This work was published together with our work \cite{menssen2017distinguishability}.
\begin{figure}[h!]
	\centering
	\includegraphics[width=0.7\textwidth]{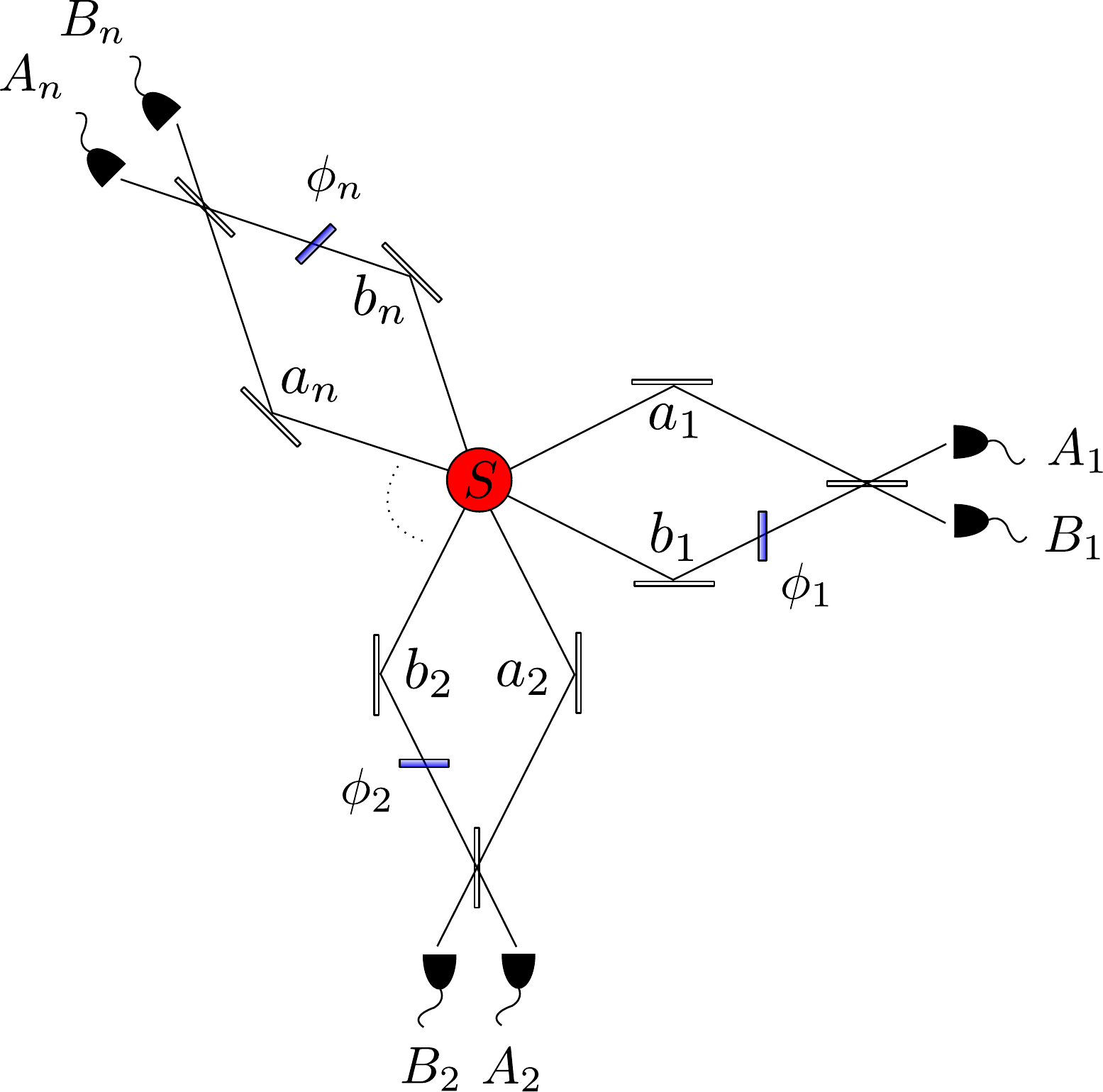}
	\caption{n-particle GHZ-Interferometer adapted from \cite{rice1994multiparticle}.}
	\label{fig:EntInterf}
\end{figure}
In Figure \ref{fig:EntInterf} an n-particle GHZ interferometer, introduced in \cite{greenberger1990bell,rice1994multiparticle}, is depicted.
The state produced by the source $S$ is:
\begin{equation}
\ket{\psi}=\frac{1}{\sqrt{2}}(\ket{a_1,a_2,..,a_n}+\ket{b_1,b_2,..,b_n}).
\end{equation}
When sending this GHZ-state into the interferometer, the
probability of detecting $n$ photons, where $k$ are detected in modes $B$, and $n-k$ are detected in modes $A$, is derived to be \cite{rice1994multiparticle} (for even n\footnote{We restrict ourselves to the case of even n for simplicity; all arguments also apply for an odd number of total photons.}):
\begin{equation}
P(\text{n-k A, k B events})=(1/2)^{2n-1}(1+(-1)^{k+n/2}\cos(\Phi)),
\end{equation}
with the total n-particle phase
$\Phi=\sum_{j=1}^{n}\phi_j$.
Further, the lower order correlations were also calculated: the probability of detecting $j$ photons in detectors $A$ and $k$ photons in detectors $B$, with the number of photons detected $m=j+k$ less than the total number of photons $m<n$, (and an even number of total photons) was derived to be:
\begin{equation}
P(\text{A, k B events})=(1/2)^n\sum_{p=0}^q\binom{q}{p}+(1/2)^n(-1)^{k+n/2}\cos(\Phi)\sum_{p=0}^{q}(-1)^p\binom{q}{p},
\end{equation}
with $q=n-j-k$,
and noting that by the binomial theorem
\begin{equation}
\sum_{p=0}^{q}(-1)^p\binom{q}{p}=0,
\end{equation}
the authors of \cite{rice1994multiparticle} find that the detection probability is constant as a function of the phase $\Phi$
\begin{equation}
(1/2)^n\sum_{p=0}^q\binom{q}{p}=(1/2)^{j+k}
\end{equation}
for all lower order correlations.
Thus, a fully entangled state observed through a GHZ interferometer, as illustrated in Figure \ref{fig:EntInterf}, also allows the observation of n-photon interference, independent of lower order correlations.

%% file: chapter5.tex
\chapter{Topological photonics}
\label{chap:TheoryTopModes}
The work on this project was done with myself in the lead, together with D. Felce. He contributed to writing routines in the SLM control-software, built the experimental setup with me and helped with data taking and analysis. The results of this project are currently in preparation for publication \cite{menssen2019photonic}.
\section{Introduction}

Topology is a field in mathematics that describes the properties of objects up to a deformation. If an object can, by smooth changes, be deformed into a different shape, they are topologically equivalent, e.g. a sphere can be deformed into an ellipsoid but not a torus, because the torus contains an additional hole. 
Phenomena associated with the topological characteristics of physical systems are of wide-reaching interest in many fields in physics with applications ranging from condensed matter physics to particle physics \cite{jackiw1981zero} and cosmology \cite{WITTEN1985557}. Most prominently topology has been applied in condensed matter physics where the importance of the topology of the band structure was first recognised in the discovery of the integer quantum Hall effect \cite{thouless1982}.  Recently, work by Haldane on the topological effects in the band structure of Graphene was recognised with a Nobel prize \cite{haldane2017nobel}. Subsequently many classes of topological insulators and superconductors have been discovered \cite{teo2010topological}.  In photonics we are able to investigate topological effects through an analogy between crystals in solid state-physics and photonic crystals \cite{ozawa2018topological}. Thus, topological effects also govern the way light propagates in photonic crystals. This realisation has spawned an entirely new research field of ``topological photonics''. \\
A key characteristic of a system with non-trivial topology is the presence of topological invariants, most commonly integer numbers that classify the topological structure. They are preserved under smooth deformations of the Hamiltonian. At the boundary between domains governed by different such invariants, where the topology abruptly changes, a topological defect occurs. Localised at these defects are states protected by the topology of the system: they are robust to errors in the underlying Hamiltonian. These ‘edge states’ have been investigated extensively in photonic platforms \cite{rechtsman2013topological,plotnik2014observation,stutzer2018photonic,noh2018topological,hafezi2013imaging}. Their study generated important insight into the physics of topological insulators and spawned technological advances such as the development of topological lasers \cite{bandres2018topological}, where lasing occurs in edge states, protected from imperfections.

In this chapter I will first introduce some basic notions of topology and its significance in describing crystal systems. I present the concept of topological invariants and discuss the ``Chern-number'' as an example. I will then draw a connection between this invariant and the multi-photon phases we have encountered in previous chapters.
Furthermore, I will illustrate the analogy between non-interacting electrons in a crystal lattice and
photons in a lattice of waveguides. The dynamics of the ``tight binding model'' which describes this situation in a solid state system will be shown to be analogous to the coupled-mode equations of light in a photonic crystal.
I will then proceed to investigate a simple tight binding model for a chain of sites with anisotropic hopping, the SSH model \cite{su1979solitons}. I then introduce the ``winding number'' as a topological invariant of this model and demonstrate how localised modes appear at interfaces between regions that are governed by different topological invariants. I motivate that these modes are protected by the topology of the system. Informed by this example I will then outline the basic band theory of graphene and motivate the existence of topologically protected modes in this two-dimensional system.

\section{Topology}
\begin{figure}[h]
	\centering
	\includegraphics[width=0.8\textwidth]{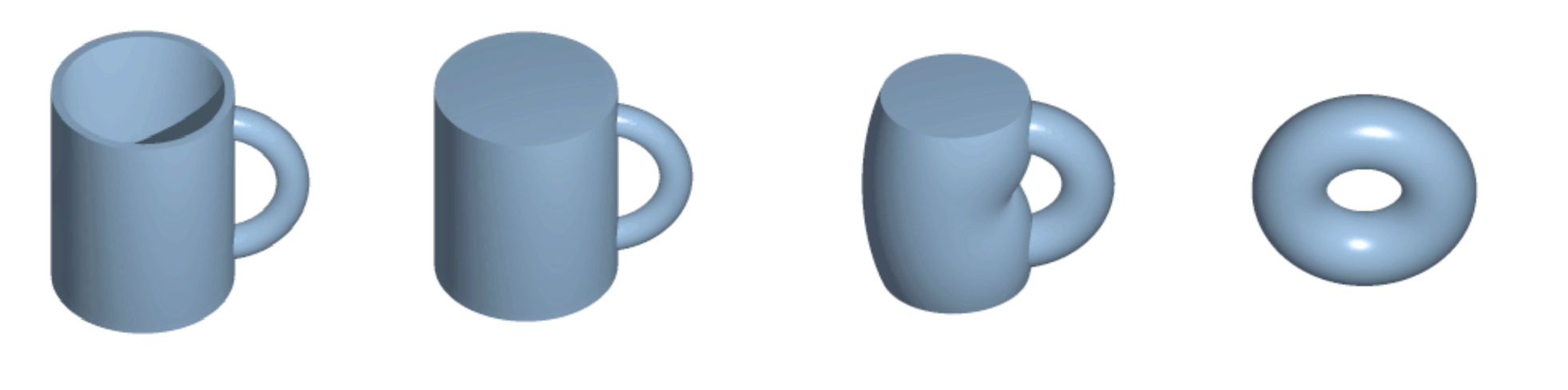}
	\caption{Continuous deformation of a cup into a torus. (Pictures taken from Wikipedia article on topology.)}
	\label{fig:TopDef}
\end{figure}

\begin{figure}[h]
	\centering
	\includegraphics[width=0.7\textwidth]{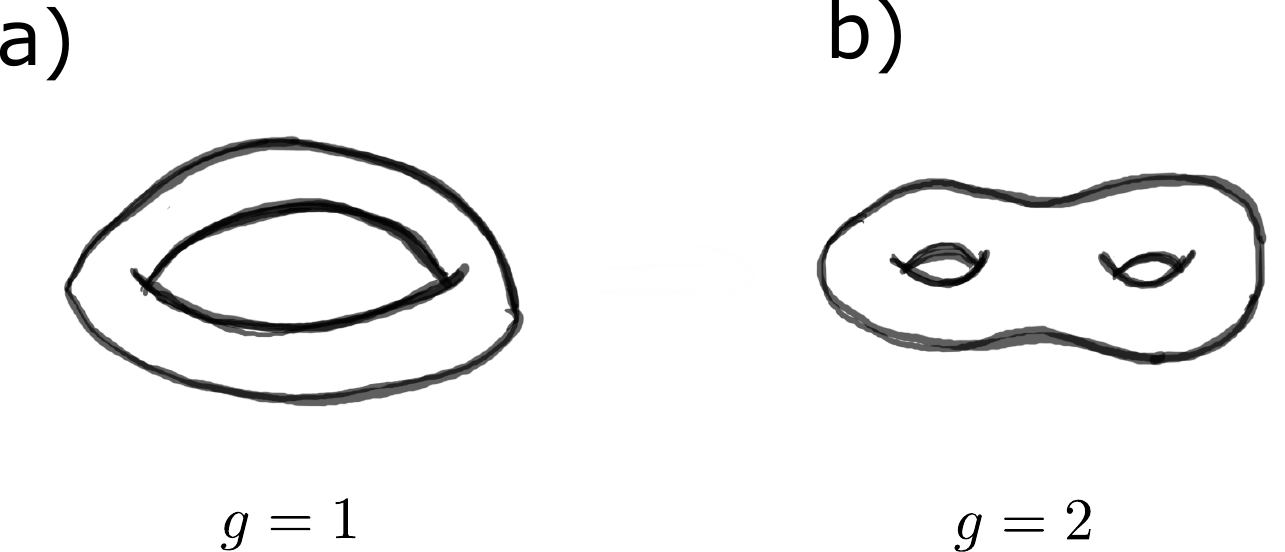}
	\caption{Example of two objects of genus 1 and 2 that are not homeomorphic.}
	\label{fig:ToriDef}
\end{figure}

In this section I will introduce the central concepts of topology. I will rely more on delivering intuitive notions instead of giving a stringent mathematical description. For a more rigorous treatment of the matter I recommend \cite{nakahara2003geometry}. Topology describes the properties of objects up to a continuous deformation. A deformation is continuous if it does not contain any cuts or pasting operations. An example of this is given in Figure \ref{fig:TopDef}: a teacup can be deformed continuously into a torus.
In mathematical terms we say that the teacup and the torus are homeomorphic.
As a counterexample consider the two objects in Figure \ref{fig:ToriDef}  which can not be continuously deformed into each other. One could construct the object in Figure \ref{fig:ToriDef} b) by attaching two tori at the sides, this however requires both a cutting and a pasting operation. 

A specific topology can be associated with a topological invariant. This is a quantity which is preserved under a continuous deformation of the object carrying the topological structure.  
A topological invariant for a torus and objects that are obtained by successively stitching tori together could for example be the number of ``holes'' in the geometry, which is one for a torus and two for the object shown in Figure \ref{fig:ToriDef} b). This invariant is called the `genus' of the manifold M, $g$.  
The genus of a closed, Riemannian manifold can be calculated with the Gauss-Bonnett theorem \cite{lu2014topological}
\begin{equation}
\int_{M}\Omega dA=2\pi(2-2g),
\end{equation}
where $\Omega$ is the Gaussian curvature of the surface.
In solid state physics we are interested in the topological structure of the energy bands. A major challenge is the search for suitable topological invariants. As we shall see in the following chapter, specific (topological) modes in solid-state systems can be associated with a topological invariant. Because the invariant is unaffected by a change of the Hamiltonian which is topology preserving, these modes are robust: even if large distortions are introduced to the Hamiltonian of the system, these modes persist. This can for example be used in the context of quantum information science to protect quantum states against gate imperfections \cite{raussendorf2007topological,blanco2018topological}.

\section{Berry curvature and Chern number}
\label{sect:Chern}
I will summarise the main points from \cite{ozawa2018topological} and \cite{nakahara2003geometry}.
A Hamiltonian ${H}(\bm{ {r},{p}})$ (with momentum and position operators $\bm{ {r},{p}})$ which is translation invariant  ${H}(\bm{ {r} +a_i,{p}})={H}(\bm{ {r},{p}})$ (with lattice vectors $\bm{a_i}$) has eigenfunctions which can be written
in terms of the Bloch basis
\begin{equation}
\psi_{n,\bm {k}}(\bm{ r})=e^{i\bm{ kr}}u_{n,\bm{ k}}(\bm{ r}).
\end{equation}
$n$ is the index of the band and $\bm{ k}$ is a momentum vector defined within the Brillouin zone (BZ). $u_{n,\bm{ k}}(\bm{ r})$ has the same translation invariance as the Hamiltonian. $u_{n,\bm{ k}}(\bm{ r})$ is an eigenstate of the Bloch Hamiltonian
\begin{equation}
{h}(\bm{ k})=  e^{-i\bm{ k{r}}}{H}(\bm{ {r},{p}})e^{i\bm{ k{r}}}.
\end{equation}
The equations for the eigenstates are:
\begin{equation}
{h}(\bm{ k})u_{n,\bm{ k}}(\bm{ r})= E_n(\bm{ k})u_{n,\bm{ k}}(\bm{ r}).
\end{equation}
In optics we are already familiar with the concept of the Berry phase $\gamma$ which is defined for a state that is a function of a continuous parameter as the closed line integral over Hilbert space
\begin{equation}
\gamma=\oint\mathcal{A}_n(\bm{k})d\bm{k}.
\end{equation}
The ``Berry connection'' is defined as:
\begin{equation}
\mathcal{A}_n(\bm{k})=i\bra{u_{n,\bm{k}}}\nabla_{\bm{k}} \ket{u_{n,\bm{k}}}.
\label{eqn:berry}
\end{equation}
An associated quantity is the ``Berry Curvature'', which in 3 dimensions is:

\begin{equation}
\Omega_n=\nabla_{\bm{k}}\times \mathcal{A}_{n}(\bm{ k}).
\end{equation}
The Berry curvature has the property that it is gauge invariant. In this case this implies that it is invariant under the $U(1)$ gauge transformation: $\ket{u_{n,\bm{ k}}}\rightarrow e^{i\Phi(\bm{k})}\ket{u_{n,\bm{ k}}}$ \cite{nakahara2003geometry}.
We can define the Chern number with the Berry curvature for each band $n$ as:
\begin{equation}
\mathcal{C}_n=\frac{1}{2\pi i}\int_{BZ}\Omega_n dV.
\label{eqn:chern}
\end{equation}
This is analogous to the Gauss-Bonnet theorem but applied to the band structure of a solid state system. The Manifold we execute the integration over is the Brillouin zone and the Gaussian curvature is replaced with the Berry curvature \cite{haldane2017nobel}.
It can similarly be shown that the Chern number is always an integer (see Appendix \ref{app:proofChern}). The Chern number is a topological invariant associated with an energy band of a solid state system. It is widely used to characterises systems such as topological insulators and superconductors.  
\subsection{The multi-photon phase as Berry phase}
We have already alluded to a the fact that the triad phase bears similarity to a Berry phase. We define the (discrete) Berry phase between two quantum states as \cite{asboth2016short}:
\begin{equation}
\gamma_{12}=\arg(\braket{\psi_1}{\psi_2})=\arg(r_{12}e^{i\gamma_{12}}),
\end{equation}
and the Berry phase for a set of states:
\begin{equation}
\gamma_{1..N}=\arg(\prod_{i=1}^{N}\braket{\psi_i}{\psi_{i+1}}).
\end{equation}
An important property of the Berry phase is that it is $U(1)$ gauge invariant when taken over a closed loop of states:
\begin{equation}
\gamma_{(1..N)}=\arg(\prod_{i=1}^{N-1}\braket{\psi_i}{\psi_{i+1}}\braket{\psi_N}{\psi_{1}}).
\label{eqn:BerryGauge}
\end{equation}
The gauge transformation on the states corresponds to applying a phase
\begin{equation}
\ket{\psi_{i}}\rightarrow e^{i\alpha_i}\ket{\psi_{i}}.
\end{equation}
For three states:
\begin{equation}
\text{arg}(\braket{\phi_1}{\phi_2}\braket{\phi_2}{\phi_3}\braket{\phi_3}{\phi_1})=\phi_{(123)}+\alpha_2-\alpha_1+\alpha_3-\alpha_2+\alpha_1-\alpha_3=\phi_{(123)}.
\end{equation}
The gauge invariance can even more readily be seen if we write the above expression as a trace over projectors, which are gauge invariant,
\begin{equation}
\phi_{123}=\arg(\text{Tr}(\ketbra{\psi_1}\ketbra{\psi_2}\ketbra{\psi_3})).
\end{equation} 
This should remind us of the type of overlap integrals we have seen to appear in multi-photon coincidence probabilities. Indeed, we have determined that the terms appearing in equation \ref{eqn:malte} are the product of overlaps over a closed path of states. This type of invariant has been previously investigated by Bargmann \cite{bargmann1964note,rabei1999bargmann}.
To explain the presence of these invariants let us make a general observation: consider a number of single photons in states $\ket{\psi_{1}},...,\ket{\psi_{N}}$, which describe a separable state of N photons inserted into an interferometer. The requirements on any measurement of that ensemble should necessarily be that it is independent of any single photon phase. In other words independent of a $U(1)$ gauge transformation: $\ket{\psi_{i}}\rightarrow e^{i\alpha_i}\ket{\psi_{i}}$. The presence of terms $\text{Tr}(\ketbra{\psi_1},...,\ketbra{\psi_m})$, where the indices $1...m$ are taken over any loop on the multi-photon graph, is thus simply a consequence of gauge invariance.
In solid state physics we are interested in the gauge invariant quantities that determine the topology of the band structure for a given material. We consider therefore a continuous analogue of the Berry phase, where the states are defined over a continuous parameter $\bm{R}$, such as a momentum vector in the Brillouin zone.
We can make the transition from the discrete case of the Berry phase to the continuous case \cite{asboth2016short}:
\begin{eqnarray}
e^{-i\Delta\phi}&=&\frac{\braket{\psi(\bm{R})}{\psi(\bm{R}+d\bm{R})}}{|\braket{\psi(\bm{R})}{\psi(\bm{R}+d\bm{R})}|}\\
\Delta\phi&=&i\bra{\psi(\bm{R})}\nabla_{\bm{R}}\ket{\psi(\bm{R})}d\bm{R}
\end{eqnarray}
and find the the Berry connection: $\mathcal{A}(\bm{R})=i\bra{\psi(\bm{R})}\nabla_{\bm{R}}\ket{\psi(\bm{R})}$.
Products of overlaps for different states then become line integrals.

\section{A photonic crystal as model for a solid state system} 
A photonic crystal in our applications is comprised of a large set of waveguides which are in close proximity such that light fields are allowed to couple between them. There is a straightforward analogy with a solid state system of non-interacting electrons on a lattice of atoms which tunnel from atom-site to atom site. In solid state physics this model is referred to as ``tight binding model'', because the electron wavefunctions are assumed to be tightly confined to each atom site. The electron can tunnel between atom sites A and B with a `hopping amplitude' that is given by a transition element which is proportional to the overlap of the initial state wavefunction at location A and the final location B. The analogous case is that of non-interacting photons which are tightly confined to waveguides. In section \ref{sect:modeCoup} the mode-coupling theory of light is described in more detail. Here, similarly, the probability of a photon hopping from site to site is given by the mode-function overlap of the waveguides. The time dimension in the Schrödinger equation takes the role of the propagation direction of the light through the crystal. 

A Hamiltonian describing such a system is given simply by:
\begin{equation}\label{eqn:parax2}
H=-\sum_{i,j}t_{\bm{r}_i,\bm{r}_j}a(\bm{r}_i)^{\dagger}a(\bm{r}_j) +\text{h.c.}.
\end{equation}
Here, $t_{\bm{r}_i,\bm{r}_j}$ is the hopping amplitude of a photon coupling to a waveguide/an electron at lattice site $\bm{r}_j$ tunnelling to another atom lattice site at location $\bm{r}_i$. The operator $a(\bm{r}_j)$ annihilates the particle at lattice site $\bm{r}_j$ and $a^{\dagger}(\bm{r}_i)$ creates it at site $\bm{r}_i$.

\begin{figure}[h!]
	\centering
	\includegraphics[width=0.5\textwidth]{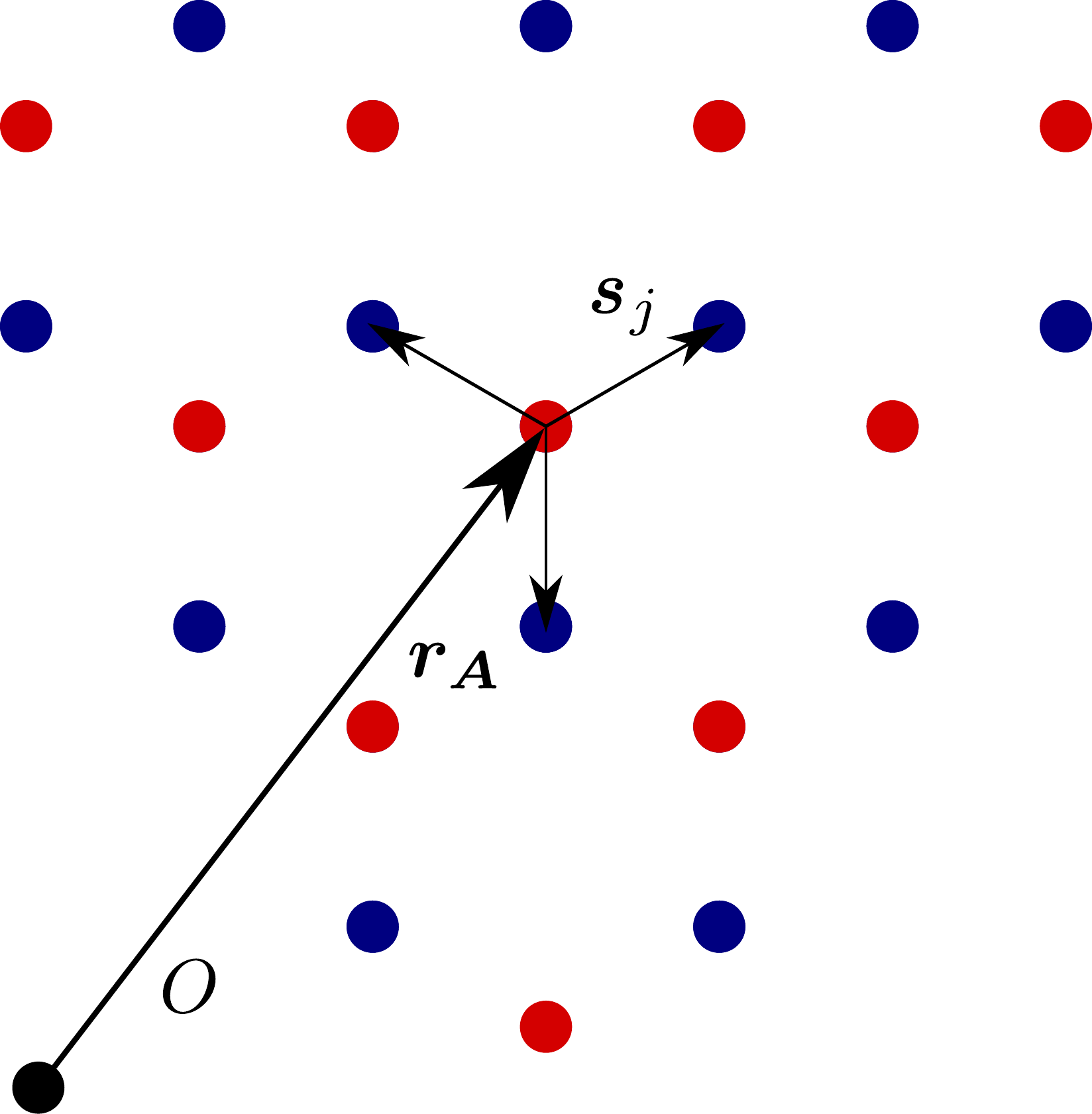}
	\caption{Graphene lattice with two triangular sublattices A/B in red and blue, lattice vector $\bm{r_A}$ and vectors between the (M=3) nearest neighbours $\bm{s}_j$ indicated.
		Hopping of electrons between lattice sites is analogous to photons hopping between the waveguides comprising the lattice. }
	\label{fig:sublat3}
\end{figure}

In the case of nearest-neighbour only coupling the Hamiltonian reduces to
\begin{equation}\label{eqn:parax3}
H=-\sum_{i}^N\sum_{j}^Mt_{\bm{r}_i,\bm{r}_i+\bm{s}_j}a(\bm{r}_i)^{\dagger}b({\bm{r}_i+\bm{s}_j}) +\text{h.c.},
\end{equation}
where N is the number of lattice sites, M is the number of nearest neighbours, $b({\bm{r}_i+\bm{s}_j})$ is an operator annihilating a particle on a nearest neighbour site at position $\bm{r}_i+\bm{s}_j$. In this work we frequently regard a single particle basis of a crystal system comprised of two lattices, which couple to each other and contain the sites that are the nearest neighbours of every site it the other lattice. In Figure \ref{fig:sublat3} this is illustrated for graphene.

We can interpret the scattering of classical light in a photonic crystal of waveguides, where only the nearest neighbour waveguides couple, as the single particle solution to the Hamiltonian \ref{eqn:parax3}.
The coupled mode equations of light propagating in a crystal lattice, including only nearest neighbour coupling, take the form

\begin{equation}
\label{eqn:parax}
i\partial_z\psi_{\bm{r}}(z)=\sum_{j=1}^{M}H_{\bm{r},\bm{r}+\bm{s}_j}\psi_{\bm{r}+\bm{s}_j}(z),
\end{equation}
where $\psi_{\bm{r}}(z)$ is the wavefunction at lattice site $\bm{r}$ and z is the length of the crystal in the direction of light propagation. 
This is a purely classical expression, describing a beam of light propagating in a photonic crystal (or equivalently a single photon).

To find the evolution of the wavefunction \ref{eqn:parax} we can exponentiate the coupling matrix/Hamiltonian \ref{eqn:parax2} (expressed in a single particle basis)
\begin{equation}
\bm{\psi}(z)=\bm{\psi}_0\exp(-i\hat{H}z),
\end{equation}
with $\hat{H}_{i,j}=t_{\bm{r}_i,\bm{r}_i+\bm{s}_j}$ and  $\bm{\psi}(z)=(\bm{\psi}_{\bm{r}_1},...,\bm{\psi}_{\bm{r}_N})$.
Here $t_{\bm{r}_i,\bm{r}_i+\bm{s}_j}$ is again the hopping amplitude of a photon coupling to a waveguide/an electron tunnelling to another atom lattice site. 

We conclude this section by noting that whenever a system is described by equations of the form \ref{eqn:parax2}, we can model its behaviour (for bosons) in a system of photonic waveguides. In the following sections, I will discuss Hamiltonians of the form \ref{eqn:parax2}. By the analogy detailed above, we can always find for these an implementation in a system of waveguides. 
\section{The SSH model}
Here, I present the Su-Schrieffer-Heeger (SSH) model \cite{su1979solitons} to illustrate the basic concepts of topological phases of matter and demonstrate how topological properties of the crystal structure give rise to localised, topologically protected modes. The SSH model is a simple 1D example of a tight binding model. It was originally introduced to describe an acetylene chain, where alternating single and double bonds between carbon atoms result in a difference of the hopping strengths between left and right hops. Here, I develop important concepts which transfer onto the more elaborate two-dimensional tight binding model of graphene.
\begin{figure}[h!]
	\centering
	\includegraphics[width=0.7\textwidth]{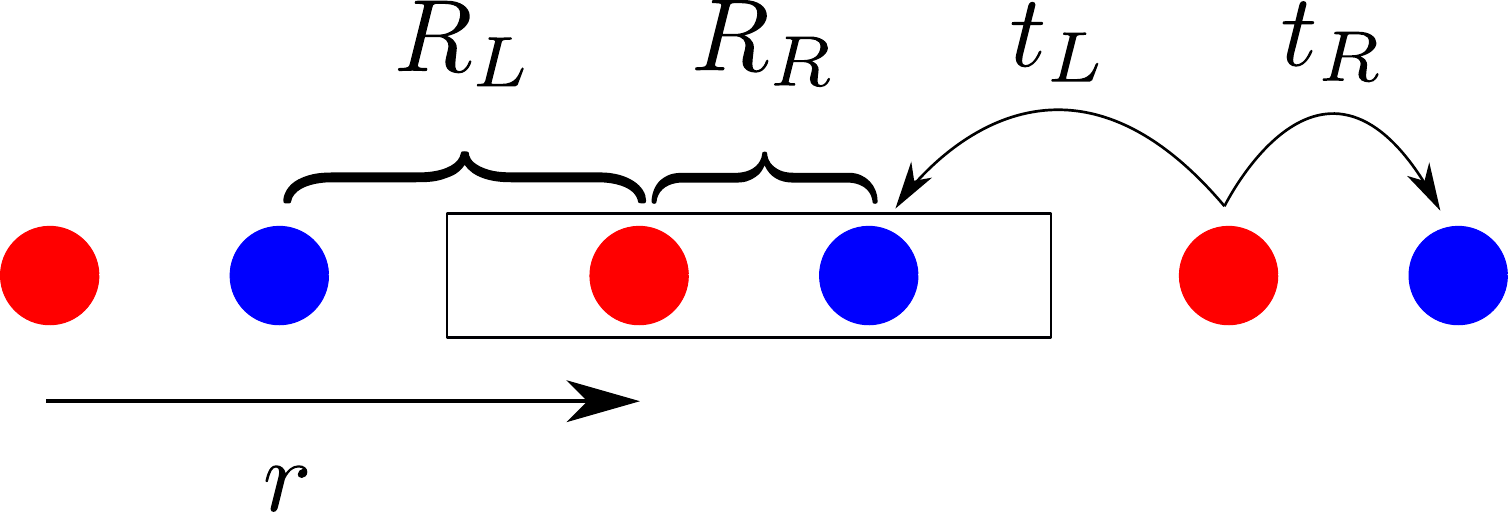}
	\caption{SSH lattice. Two sub-lattices A/B indicated in red/blue. Anisotropic hopping between sites $t_L$ and $t_R$. Unit cell containing two sites indicated by box.}
	\label{fig:SSH}
\end{figure}
In \cite{su1979solitons,asboth2016short} the Hamiltonian of a 1D chain of sites in the basis of two sub-lattices $A/B$ is expressed in the following way:
\begin{equation}
H=-\sum_{r\in A}t_La^\dagger(r)b(r-R_L)+t_Ra^\dagger(r)b(r+R_R)+\text{h.c.},
\end{equation}
where $a(r)$ and $b(r)$ annihilate particles on sublattice $A/B$ and with $t_L/t_R$ the transition elements for making a left/right hop\footnote{Left and right hops are defined as such from an atom in sub-lattice $A$. Viewed from sub-lattice $B$ left and right would be interchanged.}.
In \cite{asboth2016short} a Fourier transform to momentum space is performed
\begin{eqnarray}
a(r)&=&\frac{1}{\sqrt{N}}\sum_{n=1}^{N}e^{i\frac{2\pi}{N(R_R+R_L)}nr}a(k)\\
&=&\frac{1}{\sqrt{N}}\sum_{k\in BZ}e^{ikr}a(k), \nonumber
\end{eqnarray}
with: $k(n)=\frac{2\pi}{N(R_R+R_L)}n$.
$k$ is 1-dimensional and runs from $0$ to $2\pi$ in the 1D Brillouin zone (BZ) of this system. 
\begin{eqnarray}\nonumber\label{eqn:sshhamilt}
H&=&-\frac{1}{N}\sum_{r\in A}\sum_{k,q}t_La^\dagger(k)e^{-ikr}b(q)e^{iq(r-R_L)}+t_Ra^\dagger(k)e^{-ikr}b(q)e^{iq(r+R_R)}+\text{h.c.}\\ 
&=&-\sum_{k\in BZ}t_La^\dagger(k)b(k)e^{-ikR_L}+t_Ra^\dagger(k)b(k)e^{ikR_R}+\text{h.c.}\\ \nonumber
&=&-\sum_{k\in BZ}q(k)a^\dagger(k)b(k)+\text{h.c.} 
\end{eqnarray}
We substitute: $q(k)=t_Le^{-ikR_L}+t_Re^{ikR_R}$ and note that $\frac{1}{N}\sum_{ r\in A}e^{i{( q-k)}r}=\delta_{{q},{k}}$ is the Dirac delta function to eliminate one of the sums over momentum space.

The total Hamiltonian can be written in the following compact form:
\begin{spacing}{1}
	\begin{equation}
	H=-\sum_{k\in BZ}(a^\dagger(k),b^\dagger(k))
H(k)
	\binom{a(k)}{b(k)},
	\end{equation}
\end{spacing} with
\vspace{0.5cm}
\begin{spacing}{1}
	\begin{equation}
	H(k)=\begin{pmatrix}
	0&q(k)\\
	q^*(k)&0
	\end{pmatrix}.
	\label{eqn:HSSH}
	\end{equation}
	
\begin{figure}[h!]
	\centering
	\includegraphics[width=0.7\textwidth]{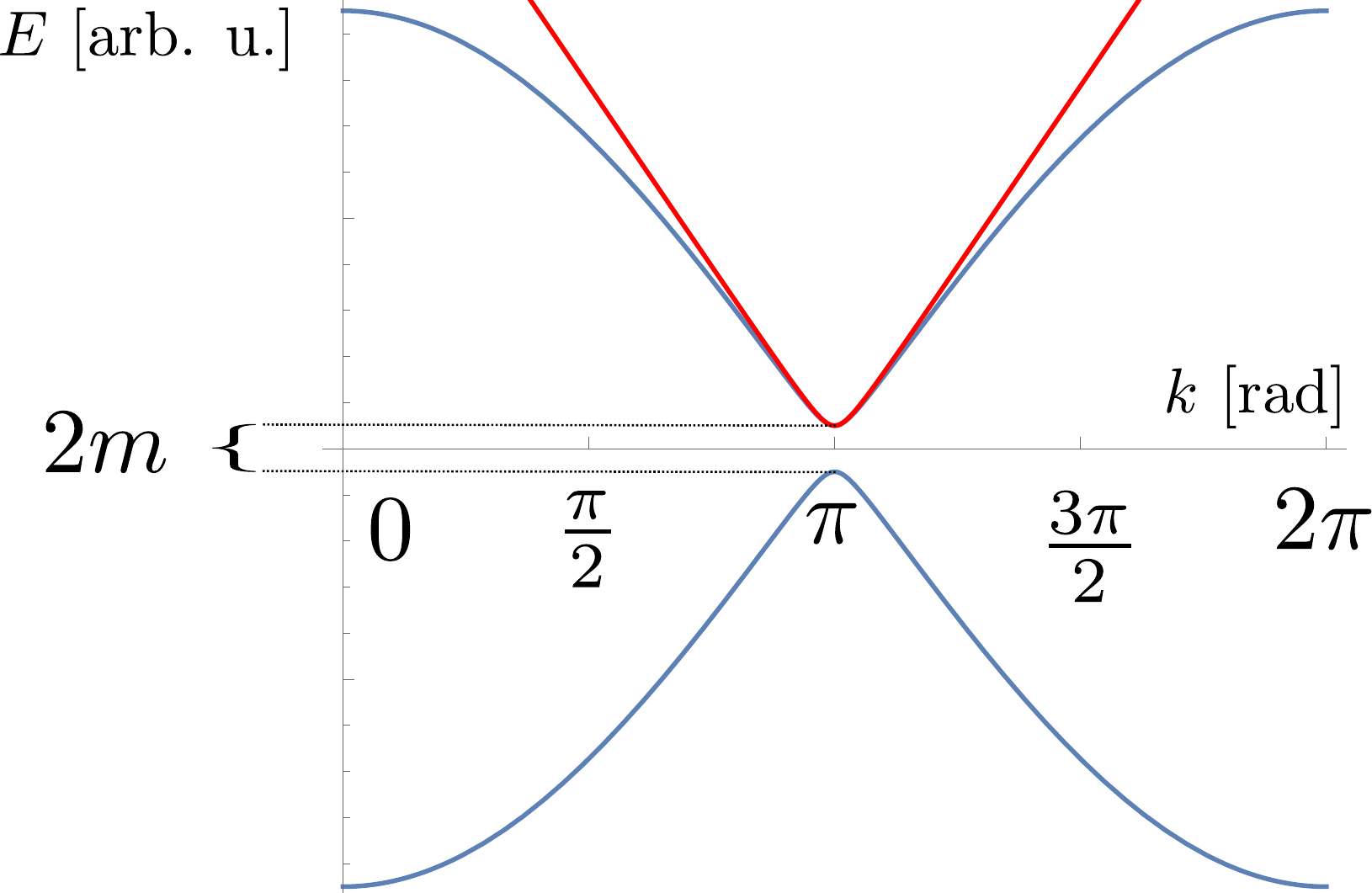}
	\caption{Band structure of the SSH model with linearisation around the band-gap in red.}
	\label{fig:SSHLin}
\end{figure}

\end{spacing}
\vspace{0.5cm}
We can make the simplification that sites are at the same distance (the hopping is still anisotropic) $R/2=R_L=R_R$. We choose $k\in BZ$ and $k=\frac{2\pi n}{RN}$. The distance is set to $R=1$ for convenience. We obtain: $q(k)=t_R+t_Le^{-ik}$ (up to a phase). The Hamiltonian \ref{eqn:HSSH} can be expressed in terms of Pauli-Matrices
\vspace{0.5cm}
\begin{spacing}{1}
	${\sigma}_{x} = \left( \begin{array}{cc} 0 & 1 \\ 1 & 0 \end{array} \right), \;\;\;
	{\sigma}_{y} = \left( \begin{array}{cc} 0 & -i \\ i & 0 \end{array} \right), \;\;\;
	{\sigma}_{z} = \left( \begin{array}{cc} 1 & 0 \\ 0 & -1 \end{array} \right)$
\end{spacing}
\vspace{0.5cm}

\begin{equation}
H(k)=(t_R+t_L\cos(k))\sigma_x+t_L\sin(k)\sigma_y.
\end{equation}
One can expand H to first order in a Taylor series expansion in $p$: $p=k\pm\pi$, to obtain a linearised expression \cite{asboth2016short}
\begin{equation}
\label{eqn:LinHamSSH}
H(p)=(t_R-t_L)\sigma_x-t_Lp\sigma_y.
\end{equation}
The dispersion can be found by determining the eigenvalues of the linearised Hamiltonian \ref{eqn:LinHamSSH}.
The two eigenvalues are:
\begin{equation}
E(k)=\pm\sqrt{(t_Lp)^2+(t_R-t_L)^2}.
\end{equation}
We can recognise this as the dispersion relation of a massive relativistic particle, interpreting $m=t_R-t_L$ as ``mass''.
The hopping anisotropy gives rise to a massive (relativistic) dispersion relation, a band gap opens around $k=\pi$, as shown in Figure \ref{fig:SSHLin}.
\section{The topological invariant of the SSH model}
My discussion of topological invariants will deliberately be kept brief. For more information I recommend \cite{asboth2016short} for a short introduction in the context of solid state physics and \cite{nakahara2003geometry} for a very detailed treatment and applications to many fields in physics.
For the present system we need to find an adequate topological invariant. 
This invariant plays an important role in explaining the origin of localised ``topological'' modes, which occur at boundary regions where the topological invariant changes. These modes are discussed in the next section. I will briefly explain the origin of the topological invariant which is different from the Chern number. It is clear that the topology of the band structure does not change under a pure deformation. However, the situation is different when two bands cross: there will be a discontinuity at the point of band degeneracy, where the gap closes. For our Hamiltonian this happens if $q(k)$ (defined in equation \ref{eqn:sshhamilt}) is zero. Because $q(k)$ is a complex function we can plot the path it traces in the complex plane as a function of $k$ \cite{asboth2016short}. Since $q(k)=q(k+2\pi)$, the path will be closed. The integer $\mathbb{Z}$ topological invariant we are looking for is the ``winding number''
\begin{equation}
Q(H)=\frac{1}{2\pi i}\int_{0}^{2\pi}dk\frac{d}{dk}\log (q(k)).
\label{eqn:winding}
\end{equation}
It counts the number of times that $q(k)$ winds around the origin of the complex plane \cite{asboth2016short}.
\begin{figure}[h]
	\centering
	\includegraphics[width=0.8\textwidth]{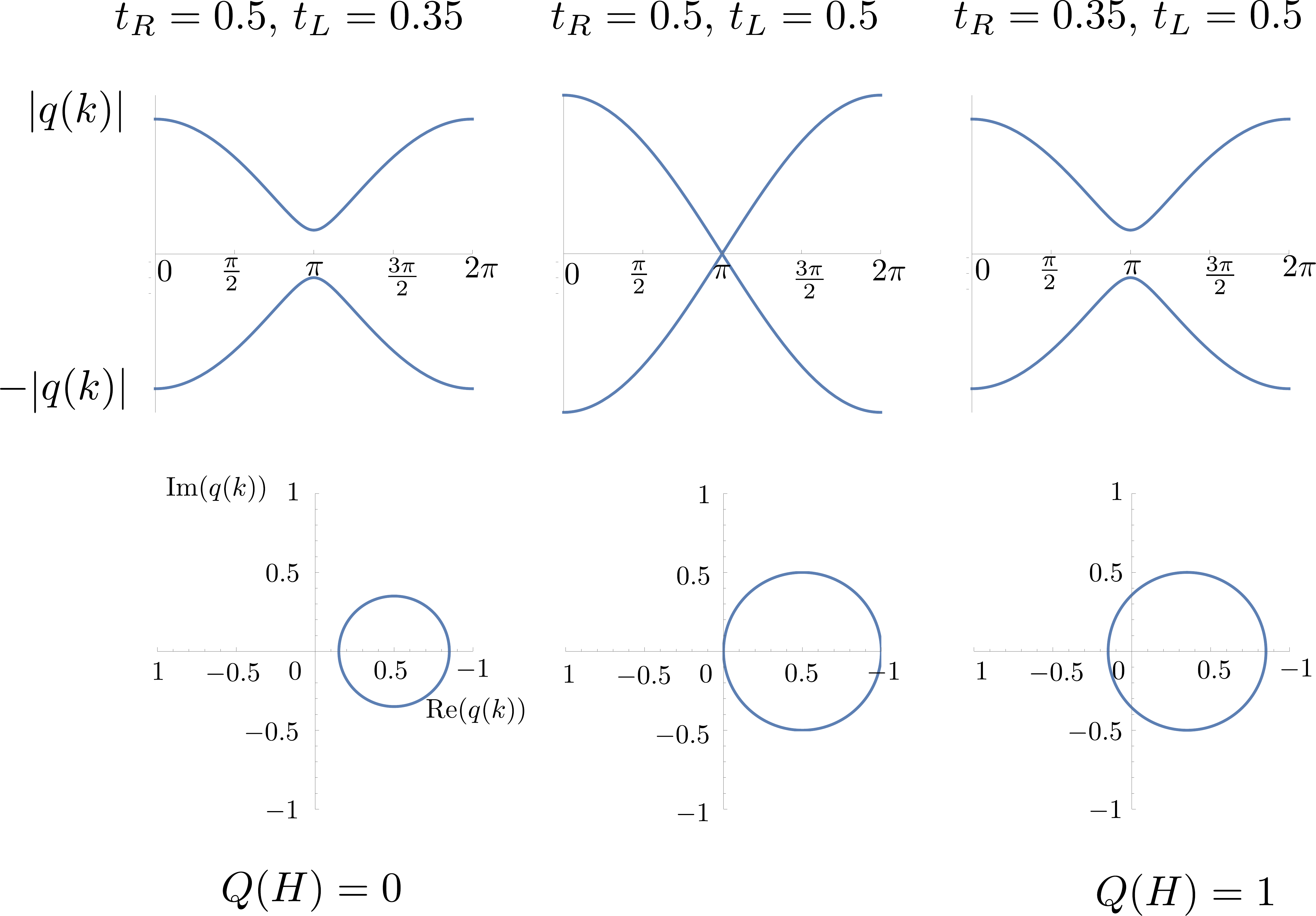}
	\caption{Band structure and winding number of the SSH model for different values of $t_R$ and $t_L$. Adapted for current example from \cite{asboth2016short}.}
	\label{fig:SSHWinding}
\end{figure}
In Figure \ref{fig:SSHWinding} the band structure of the SSH Hamiltonian and the path winding number for different value of the hopping strengths $t_R$ and $t_L$ are illustrated.
What the precise path traced in the complex plane is, does not matter. For the winding number it is only relevant if the origin of the complex plane is enclosed by the path or not. The topology of the band structure only changes when the path passes through the origin and so does our winding number, (the band gap closes), all other changes are homeomorphic. It is thus a suitable topological invariant associated with each band.
The meaning of this rather abstract quantity will become clear in the next section.
The key insight is that the invariant is a quantity that only changes when the bands cross. \\
The classification of topological defects is an active field of research.
Teo and Kane in their paper \cite{teo2010topological} discuss more generally different classes of invariants associated with topological defects in different dimensions. %
\section{Bulk-edge correspondence and the Jackiw-Rebbi model}
I now want to illustrate how a simple, localised topological defect can be introduced to a solid state system.
The ``Bulk-Edge correspondence'' \cite{hatsugai1993edge} states that whenever there are two materials governed by different topological invariants put into contact, there will be a localised state at their boundary. We can understand this by remembering that we can introduce a topological change to the band structure only if we produce a crossing between two bands. Any other shift or deformation will necessarily be homeomorphic. Imagine we vary the ``mass'' term slowly from a negative to a positive value as a function of lattice position. The band-gap will close at zero energy (cf. Figure \ref{fig:SSHWinding})  when $m=0$ and then reopen again. Since the energy is zero where the band-crossing occurs, there will be a localised mode at zero energy, a ``zero-mode''. %
A prominent example of this is the Jackiw-Rebbi model \cite{Jackiw1976}. We take the SSH Hamiltonian
and vary the mass term continuously as a function of the lattice coordinate $m(r)$, such that it changes sign. We have noted in the previous section that the winding number of the SSH chain changes by 1 when the sign of the mass term switches \cite{asboth2016short,ozawa2018topological}.
Our starting point is the linearised equation of the SSH model
\begin{equation}
H(p)=m(r)\sigma_x-t_Lp\sigma_y.
\end{equation}
In the continuum approximation we make the replacement $p\rightarrow-i\partial_r$
\begin{equation}
H(p\rightarrow-i\partial_r)=m(r)\sigma_x+it_L\partial_r\sigma_y.
\end{equation}
We then search for the expected zero mode by explicitly looking for solutions at 0 energy
\begin{equation}
H(-i\partial_r)\bm{\psi}=0.
\end{equation}
A normalisable solution to the first order system of differential equations is \cite{ozawa2018topological}:
\begin{spacing}{1}
	\begin{equation}
	\psi(r)\sim \exp(-\frac{1}{t_L}\int_{0}^{r}m(r')dr')\begin{pmatrix}
	0\\
	1
	\end{pmatrix}.
	\end{equation}
\end{spacing}
\clearpage
\begin{figure}[h]
	\centering
	\includegraphics[width=0.7\textwidth]{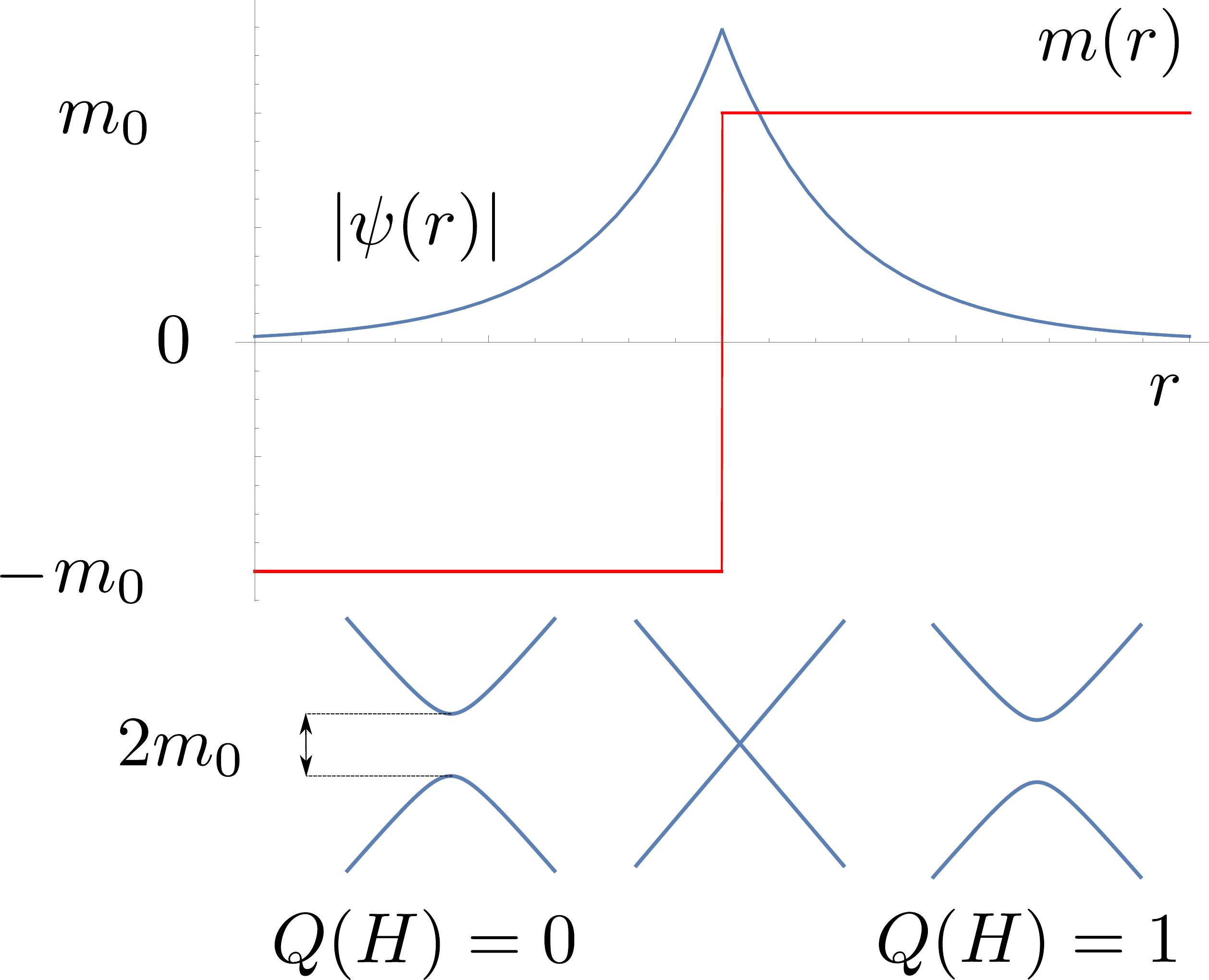}
	\caption{Illustration of the shape of the localised zero-mode and the corresponding mass-term.}
	\label{fig:JackiwRebbiComb2}
\end{figure}

As an example we can take a simple Heaviside function for the mass term
\begin{equation}
m(r)=2m_0(\theta(r)-1/2)
\end{equation}
\begin{spacing}{1}
	\begin{equation}
	\psi(r)\sim \exp(-\frac{1}{t_L}2m_0 r(\theta(r)-1/2))\begin{pmatrix}
	0\\
	1
	\end{pmatrix}.
	\end{equation}
\end{spacing}
\vspace{0.5cm}
This also serves as a good example for topological protection. The energy of the zero-mode and its position are unaffected by the exact shape of the mass term $m(r)$ in the Hamiltonian (or indeed any other perturbation to the system which preserves the symmetry of the Hamiltonian \cite{altland1997nonstandard,teo2010topological}). The only important factor is that the topological invariants left and right of the boundary retain the same value. In a photonic system the Jackiw Rebbi model has already been demonstrated \cite{angelakis2014probing}.

\section{Graphene tight binding model}
\begin{figure}[h!]
	\centering
	\includegraphics[width=0.55\textwidth]{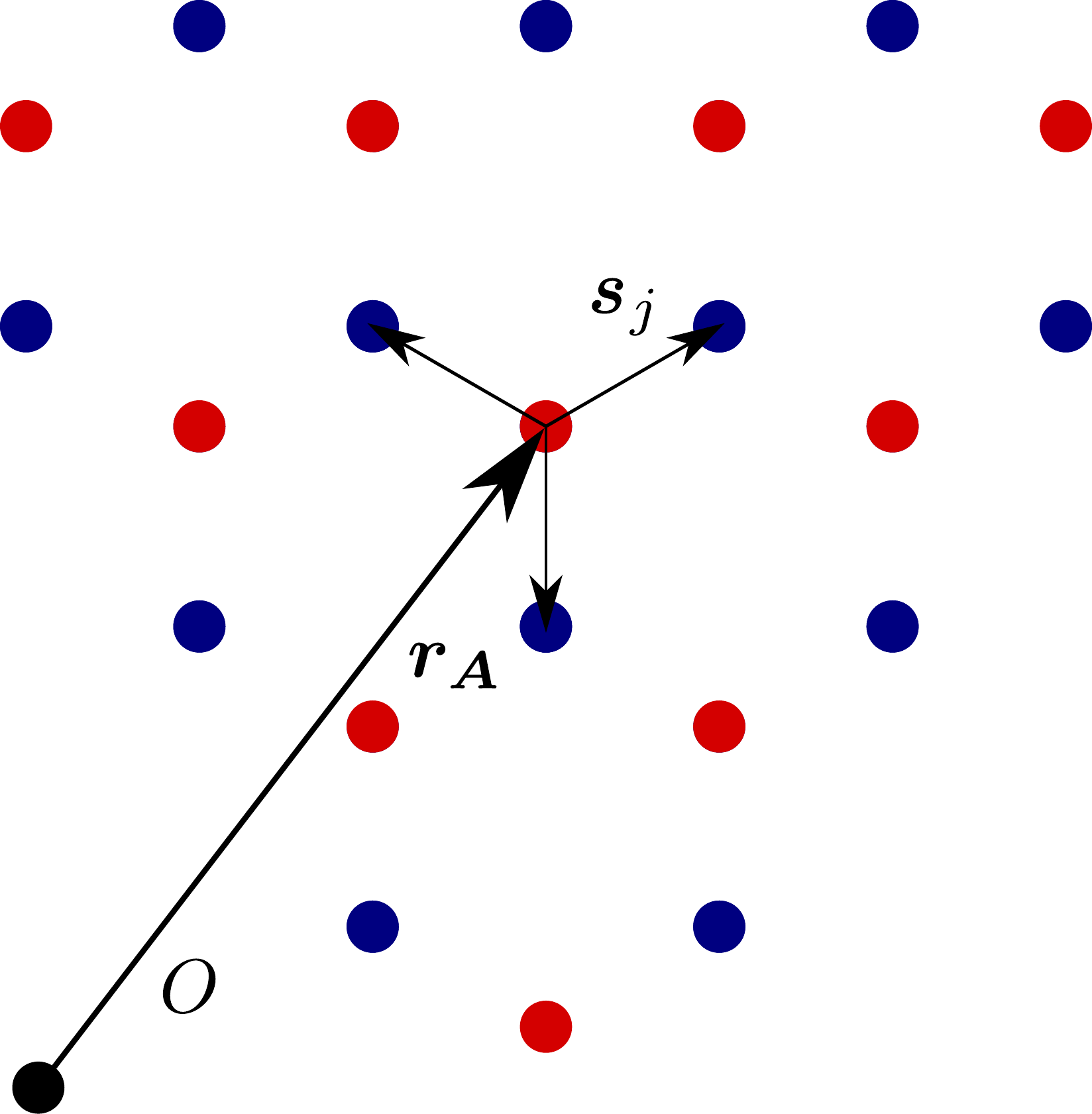}
	\caption{Graphene lattice with two triangular sub-lattices in red (A) and blue (B), lattice vector $\bm{r}$ and vectors between nearest neighbours $\bm{s_j}$ indicated.}
	\label{fig:sublatb}
\end{figure}

In this section I will briefly revise the electronic band structure of graphene. %
The discussion in this chapter will closely follow and summarise the results discussed in \cite{hou2007electron,Jackiw2007,iadecola2016non}. The discussion is kept analogous to the Jackiw Rebbi model.\\
In graphene carbon atoms are arranged in a hexagonal pattern.
We can decompose the lattice into two triangular sub-lattices A and B, marked in Figure \ref{fig:sublatb} in red and blue. The nearest neighbour lattice vectors of $\bm{r}$ are: $\bm{s_1}=a_0\big(0,\pm1\big),\bm{s_2}=a_0\bigg(\sqrt{3}/2,\mp1/2\bigg),\bm{s_3}=a_0\bigg(-\sqrt{3}/2,\mp1/2\bigg)$. The choice of the $\bm{s_1}$ sign depends on the sub-lattice the site $\bm{r}$ is on, $-$ for sublattice A $+$ for sublattice B.

\begin{figure}[h]
	\centering
	\includegraphics[width=1\textwidth]{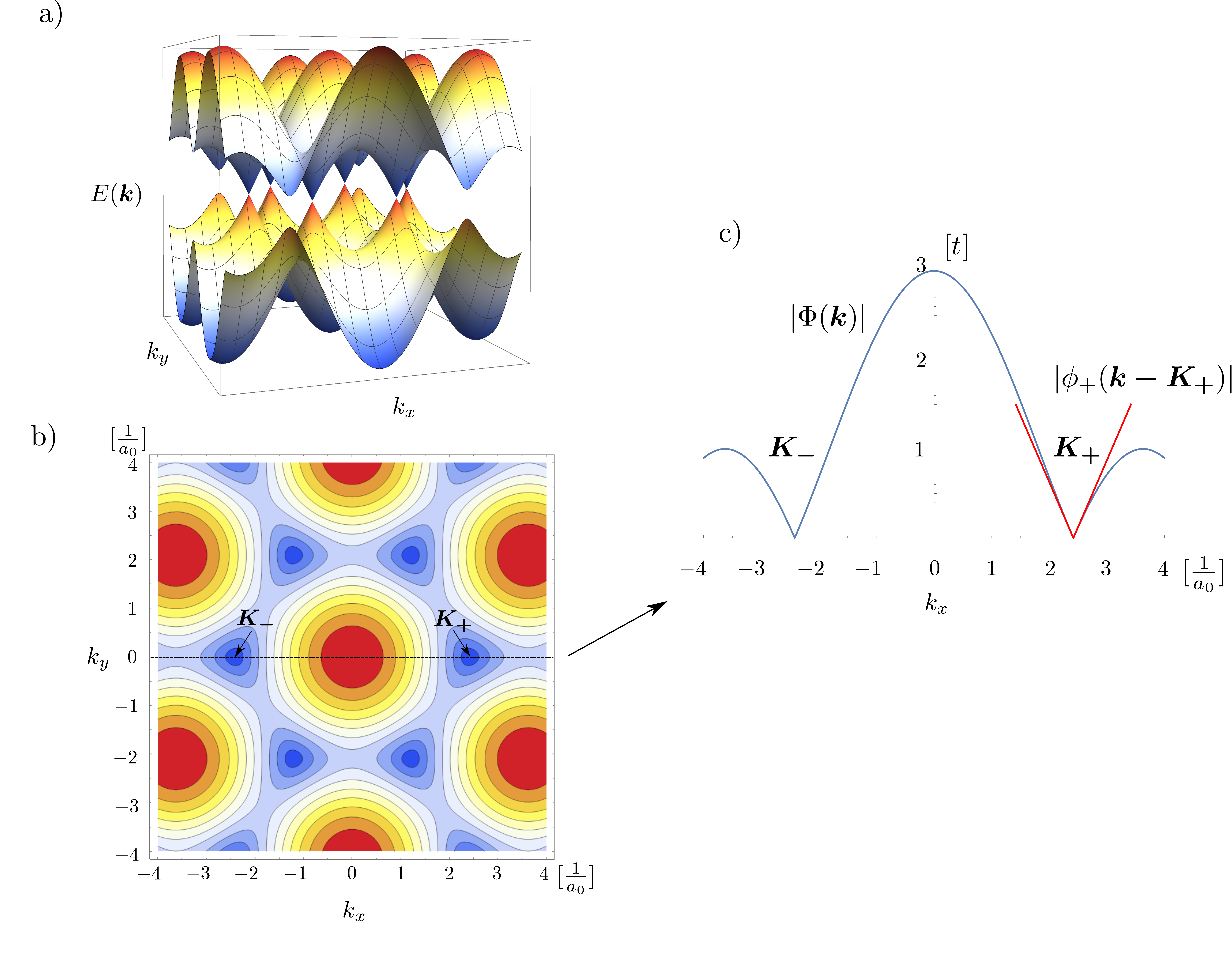}
	\caption{Band structure of graphene. a) The two bands of graphene. b) contour-plot of the positive energy band with the Dirac points indicated. c) Cut along $k_y=0$, as indicated by dotted line in sub-figure b). Blue curve: positive energy band $|\Phi(\bm{ k})|$ Red curve: linearisation $|\phi_+(\bm{ p})|$ of  $|\Phi(\bm{ k})|$ around $\bm{K_+}$. }
	\label{fig:BandStruct}
\end{figure}

In this part we follow \cite{Jackiw2007}.
The tight binding Hamiltonian takes the form:
\begin{equation}
H_0=-t\sum_{\bm{ r\in A}}\sum_{i=1}^{3}a(\bm{r})^\dagger b(\bm{r}+\bm{s}_i)+\text{h.c.},%
\label{eqn:H0}
\end{equation}
$t$ is the hopping strength. $a$ and  $b$ are annihilation/creation operators acting on sub-lattice A and B respectively. We introduce momentum space operators
\begin{spacing}{1.2}
	
	\begin{equation}
	\Bigg\{ \begin{matrix}
	\displaystyle a(\bm{r}) \\[0.0cm]
	\displaystyle b(\bm{r})
	\end{matrix} \Bigg\}=\frac{1}{\sqrt{N}}\sum_{\bm{ k\in BZ}}e^{i\bm{ k r}}
	\Bigg\{ \begin{matrix}
	\displaystyle a(\bm{k}) \\[0.0cm]
	\displaystyle b(\bm{k})
	\end{matrix} \Bigg\}.
	\label{eqn:opfourier}
	\end{equation}
\end{spacing}
\vspace{0.5cm}
The Hamiltonian is diagonal in momentum space representation. We insert equation \ref{eqn:opfourier}
into \ref{eqn:H0} to obtain: 
\begin{equation}
H_0=-t\frac{1}{N}\sum_{\bm{ r\in A}}\sum_{j=1}^{3}\sum_{\bm{k,q}\in BZ}e^{i{(\bm {q-k}) r}}e^{i\bm{q}\bm{s}_j}a(\bm{k})^\dagger b(\bm{k})+\text{h.c.}.
\end{equation}
We note that $\frac{1}{N}\sum_{\bm{ r\in A}}e^{i{(\bm {q-k})}\bm{ r}}=\delta_{\bm{q},\bm{k}}$ is the Dirac delta function, and eliminate one of the sums over momentum space:
\begin{equation}
H_0=-t\sum_{j=1}^{3}\sum_{\bm{k}\in BZ}e^{i\bm{k}\bm{s}_j}a(\bm{k})^\dagger b(\bm{k})+\text{h.c.}
\end{equation}
\begin{eqnarray}
H_0&=&\sum_{\bm{ k}}\Phi(\bm{k})a(\bm{ k})^\dagger b(\bm{ k})+\Phi(\bm{k})^*a(\bm{ k}) b(\bm{k})^\dagger\\
\Phi(\bm{k})&=&-t\sum_{j=1}^{3}e^{i\bm{ k} \bm{s}_j}.
\end{eqnarray} 
For a single particle the dispersion relation is then $E(\bm{ k})=\pm |\Phi(\bm{k})|$.
In Figure \ref{fig:BandStruct} $E(\bm{ k})$, the band structure of graphene in the Brillouin zone is plotted.
We consider the two Dirac points at opposing sides in the Brillouin zone $\bm{K_\pm}= \pm(\frac{4\pi}{3\sqrt{3}a_0},0)$ and linearise $\Phi(\bm{ k})$ around  $\bm{p_\pm}=\bm{k}-\bm{K_\pm}$. The linearised function $\phi_\pm(\bm{ p})$ is depicted in Figure \ref{fig:BandStruct} c)
\begin{equation}
\phi_\pm(\bm{ p})=\pm v_F(p_x\pm ip_y),
\end{equation}
with $v_F=3ta_0/2$. The linearised Hamiltonian $H_0$ then takes the form:
\clearpage
\begin{equation}
\begin{split}
H_0&=\sum_{\bm{ p}}\phi_+(\bm{p})a^\dagger_+(\bm{p}) b_+(\bm{p})+\phi_-(\bm{p})a^\dagger_-(\bm{ p}) b_-(\bm{ p})+\text{h.c.}\\
\end{split},
\label{eqn:Hamgraphenelin}
\end{equation}
\vspace{0.5cm}
where $a^\dagger_\pm(\bm{ p}),b^\dagger_\pm(\bm{ p}),a_\pm(\bm{ p}),b_\pm(\bm{ p})$ are operators which create/annihilate particles around the $\bm{K_\pm}$ Dirac points. 
In a single particle basis, we can express the Hamiltonian $H_0$ at one of the two Dirac points (in a basis of the two sub-lattices A/B) as:
\begin{equation}
H_0=\pm v_F (p_x\sigma_x\pm p_y\sigma_y)
\label{eqn:graphen0}.
\end{equation}
Hereafter we set $v_F=1$.
\section{Jackiw-Rossi model}
The Jackiw-Rossi model \cite{jackiw1981zero} that we examine in this work is similar in nature to the Jackiw-Rebbi  \cite{Jackiw1976} model. We study again a topological defect, this time however the ``mass-term'' henceforth labelled as $\Delta(\bm{ r})$ to differentiate it from the Jackiw-Rebbi case has a dependence on both $x$ and $y$ lattice coordinates and it introduces a coupling between the two Dirac points instead of the two sub-lattices.  
The authors of \cite{hou2007electron} have studied this situation by introducing an additional term $H_c$ to the Hamiltonian in equation \ref{eqn:Hamgraphenelin} which couples the two Dirac points $K_+$ and $K_-$ with a coupling strength $\Delta_0$
\begin{equation}
H_c=-\sum_{\bm{ p}}\Delta_0 a^\dagger_+(\bm{ p}) b_-(\bm{ p})+\Delta_0^* a^\dagger_-(\bm{ p}) b_+(\bm{ p})+\text{h.c.}.
\label{eqn:HC}
\end{equation}
The total Hamiltonian is then:
\begin{equation}
H=H_0+H_c.
\label{eqn:fullHamilt}
\end{equation}
C. Chamon first demonstrated in \cite{chamon2000solitons} how the coupling \ref{eqn:HC} could be introduced to the graphene Hamiltonian by applying small distortions to the position of the sites in the hexagonal lattice (Kekule distortion). This was expanded on in subsequent works: \cite{hou2007electron,iadecola2016non,Jackiw2007}.

\begin{figure}[h!]
	\centering
	\includegraphics[width=0.5\textwidth]{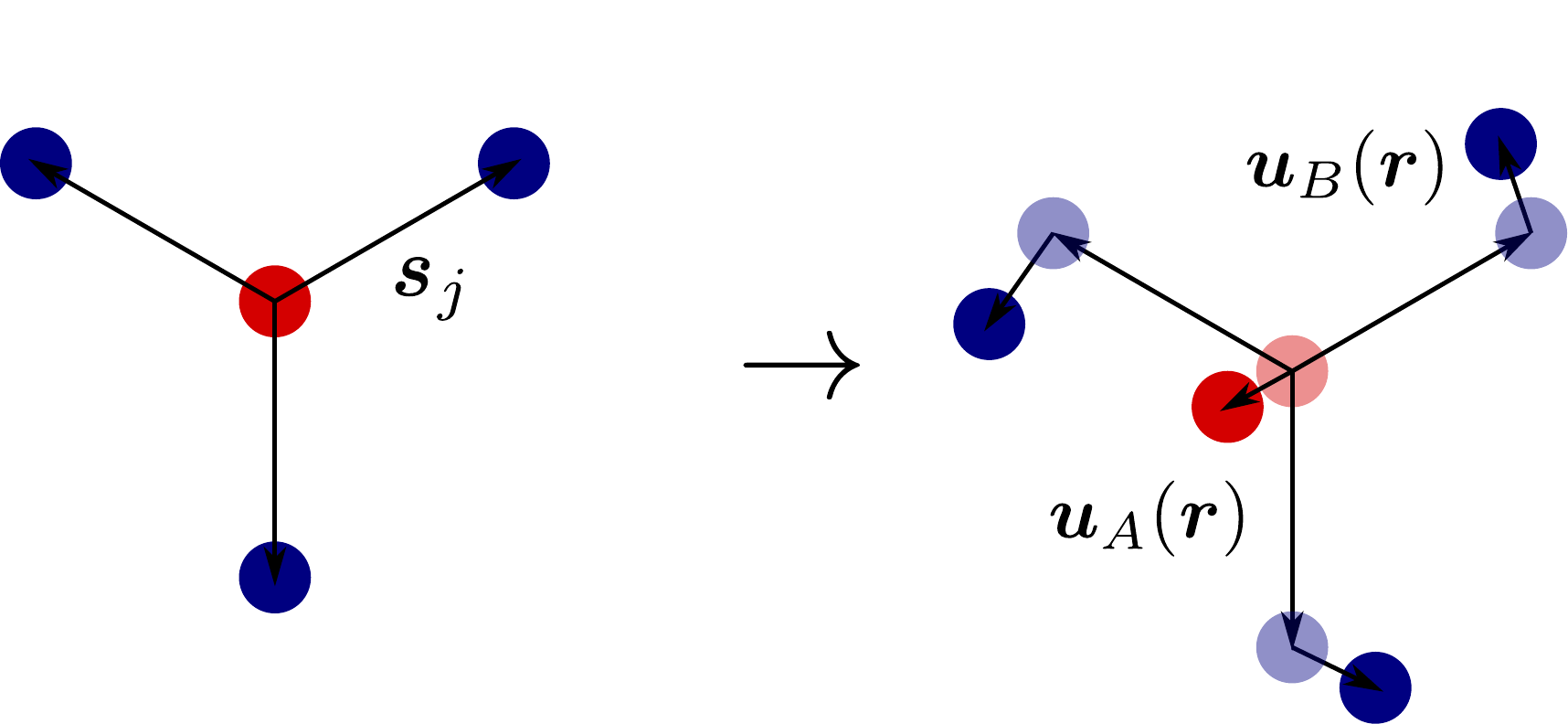}
	\caption{Graphene lattice displacement with two triangular sub-lattices in blue and red, vectors between nearest neighbours $\bm{s_j}$ and displacement vectors $\bm{u}_A(\bm{r})$ and $\bm{u}_B(\bm{r})$ indicated.}
	\label{fig:sublatDisp}
\end{figure}
In Figure \ref{fig:sublatDisp} the modification of the graphene lattice is illustrated. $\bm{u}_A(\bm{r})$ and $\bm{u}_B(\bm{r})$ are the displacement vectors on the respective sublattice A and B. The displacements in \cite{iadecola2016non} (Appendix) are taken of the form:
\begin{eqnarray}
\bm{u}_A(\bm{r})&=&\frac{i}{2}\xi\Delta_0e^{-i\bm{K}_+\bm{r}}\binom{1}{i}+\text{c.c.}\\
\bm{u}_B(\bm{r})&=&\frac{i}{2}\xi\Delta_0e^{-i\bm{K}_+\bm{r}}\binom{1}{-i}+\text{c.c.},
\end{eqnarray}
$\xi$ is a constant of units $[\text{length}^2]$. A detailed derivation of the waveguide distortion can be found in Appendix \ref{app:WGDist}.
The effect of the coupling is to introduce a band-gap at the Dirac points. The energy dispersion is again modified:
\begin{equation}
\epsilon(\bm{ p})=\pm\sqrt{|\bm{ p}|^2+|\Delta_0|^2}.
\end{equation}
The authors \cite{hou2007electron} now generalise the expression \ref{eqn:HC} by replacing the constant $\Delta_0$ with a  complex scalar field which varies as a function of the lattice coordinates $\bm{ r}$: $\Delta_0\rightarrow \Delta(\bm{ r})$. In a position representation we can write the Hamiltonian \ref{eqn:fullHamilt} in terms of the ``Bogoliubov-De-Gennes'' matrix $\mathcal{H}(\bm{ r,p})$ in a more compact form \cite{hou2007electron}:
\begin{eqnarray}
H=\int d\bm{ r}\Psi(\bm{ r})^\dagger\mathcal{H}(\bm{ r})\Psi(\bm{ r}),
\end{eqnarray}
where $\bm{ r}$ is the position coordinate on the 2D surface of the lattice $\bm{ r}=(x,y)$ and
$\bm{ p}$ is the momentum in the Brillouin zone $\bm{ p}=-i(\partial_x,\partial_y)$.

We introduce a four component ``spinor'': 
\begin{spacing}{1.2}
	\begin{equation}
	\Psi = \begin{pmatrix}
	\displaystyle \psi^a_+ \\[0.0cm]
	\displaystyle \psi^a_- \\[0.0cm]
	\displaystyle \psi^b_+ \\[0.0cm]
	\displaystyle \psi^b_-
	\end{pmatrix},
	\end{equation}
\end{spacing}
which in general has components:
\begin{eqnarray}
\psi^a_+=\int d^2r\; u_a^*(\bm{r}) a_+(\bm{r}) \\
\psi^a_-=\int d^2r\; v_a^*(\bm{r}) a_-(\bm{r}) \\
\psi^b_+=\int d^2r\; u_b^*(\bm{r}) b_+(\bm{r}) \\
\psi^b_-=\int d^2r\; v_b^*(\bm{r}) b_-(\bm{r}).
\end{eqnarray}
If we take the annihilation operators in position representation as spinor components:
\begin{eqnarray}
a_\pm(\bm{r})=\int d^2p\;e^{i\bm{ p r}}a_\pm(\bm{ p})\\
b_\pm(\bm{r})=\int d^2p\;e^{i\bm{ p r}}b_\pm(\bm{ p}),
\end{eqnarray}
the matrix $\mathcal{H}(\bm{ r,p})$, which is first introduced in  \cite{hou2007electron}, takes the form:
\begin{spacing}{1.2}
	\begin{eqnarray}
		\mathcal{H}(\bm{ r,p}) &=&
		\begin{pmatrix}
			0^{2x2} & q(\bm{ r},\bm{ p}) \\
			q^{\dagger}(\bm{ r},\bm{ p}) & 0^{2x2} 
		\end{pmatrix} \\	
		q(\bm{ r})&=& 
		\begin{pmatrix}
			-i2\partial_z^* & -\Delta(\bm{ r}) \\
			-\Delta(\bm{ r})^* & i2\partial_z 
		\end{pmatrix},
	\end{eqnarray}
	\label{eqn:BdG}
\end{spacing}

\vspace{0.5cm}
with $\bm{ p}=-i(\partial_x,\partial_y)$ and  $\partial_z=\frac{1}{2}(\partial_x-i\partial_y)$.

\subsection{Solutions of the Dirac kernel}
We now aim to find zero-energy solutions: As in the Jackiw-Rebbi case, we need to solve
\begin{equation}
\mathcal{H}(\bm{ r})\Psi=0.
\end{equation}
Upon transforming the partial derivatives $\partial_z,\partial_x,\partial_y$ which appear in $\mathcal{H}(\bm{ r})$ into a cylindrical coordinate frame $\partial_r,\partial_\theta$, this leads to two coupled equations for the spinor components
\begin{eqnarray}\label{eqn:zmode1}
(\partial_r+ir^{-1}\partial_\theta)u_b(\bm{r})-ie^{-i\theta}\Delta(\bm{r})v_b(\bm{r})&=&0\\
\label{eqn:zmode2}
(\partial_r-ir^{-1}\partial_\theta)v_b(\bm{r})+ie^{i\theta}\Delta^*(\bm{r})u_b(\bm{r})&=&0,
\end{eqnarray}
which are identical to the equations of the continuous Dirac theory explored in \cite{Jackiw1976}. The authors of \cite{hou2007electron} take the complex valued function $\Delta(\bm{r})$ to contain a vortex of the following form, where $n$ is a winding number
\begin{eqnarray}
\Delta(\bm{r})=\Delta_0(r)e^{i(\alpha+n\theta)}\\
\Delta_0(r)=\tanh{(r/l_0)}.
\end{eqnarray}
We will later motivate that this is the topological invariant of this system.
Equations \ref{eqn:zmode1} and \ref{eqn:zmode2} then have the solution \cite{hou2007electron} for a choice of $n=1$
\begin{equation}
u_b(\bm{r})=\frac{e^{i(\alpha/2-\pi/4)}}{2\sqrt{\pi}} \frac{e^{-\int_{0}^{r}dr'\Delta_0(r')}}{\sqrt{\int_0^\infty dr r e^{-2\int_{0}^{r}dr'\Delta_0(r')}}}.
\label{eqn:JackiwRossiMode}
\end{equation}
\begin{figure}[h]
	\centering
	\includegraphics[width=0.62\textwidth]{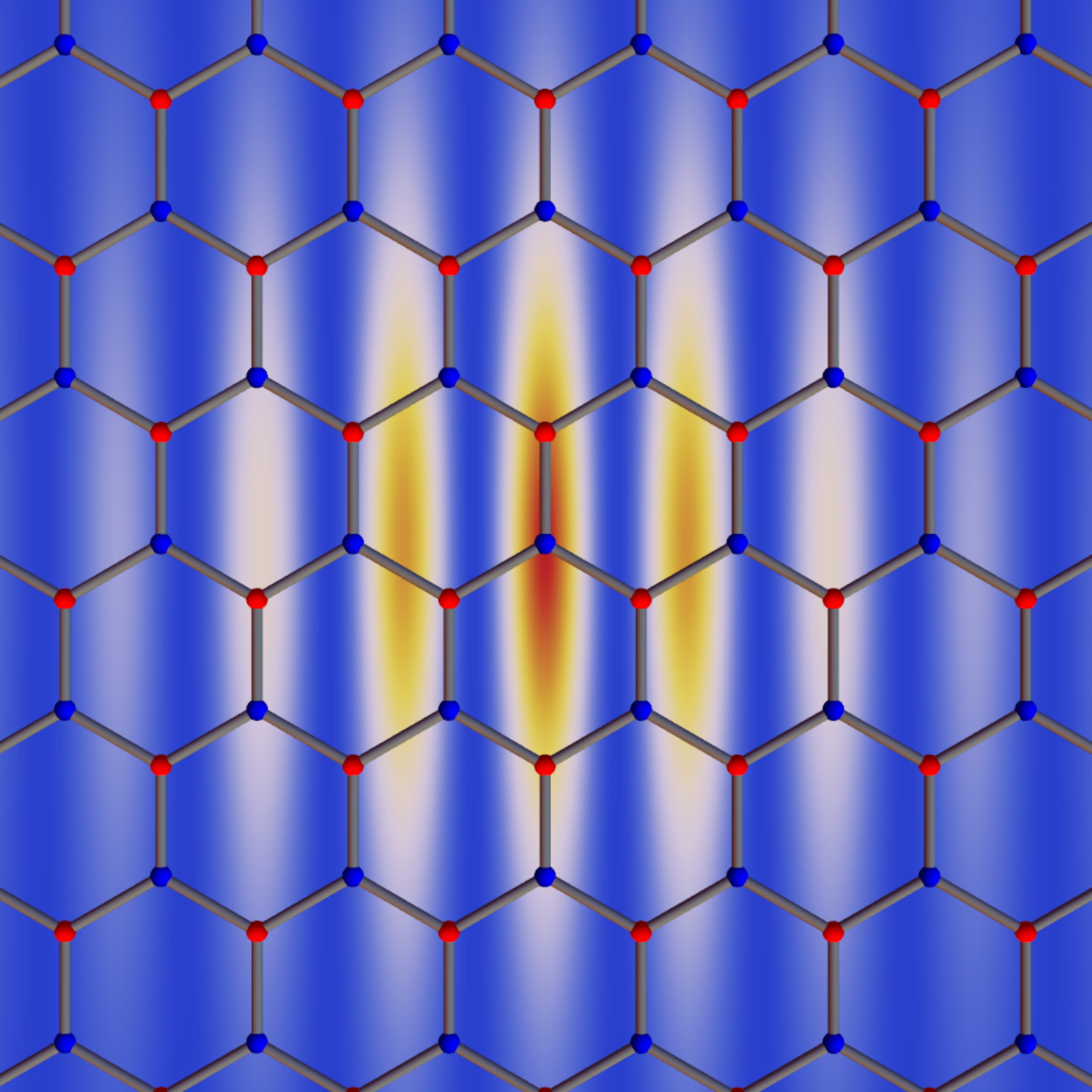}
	\caption{Hexagonal lattice with zero mode at the centre. Red and blue sites mark A/B sub-lattices. The absolute value of the zero-mode wavefunction is shown.}
	\label{fig:DiracLat}
\end{figure}
The zero-mode is tightly localised near the core of the vortex. The spatial dependence of the mode equation \ref{eqn:JackiwRossiMode} is reminiscent of the Jackiw-Rebbi case. Indeed the dependence on the mass term $\Delta_0$ is identical. In \ref{fig:DiracLat} the spatial dependence of the mode is depicted on the graphene lattice. The amplitude at each lattice site is given by the value of the function at that lattice site. The solution shown is localised only on sublattice B (blue sites in the figure). The additional modulation of the wavefunction originates from our initial momentum translation to the $\bm{K_+}$ Dirac point around which we linearised $\bm{p_\pm}=\bm{k}-\bm{K_\pm}$, $u_b(\bm{ r})\rightarrow u_b(\bm{ r})e^{i\bm{K_+r}}$. The operator creating the zero mode on the graphene lattice is then $\Psi=\int d^2r\hspace{1mm}  u_b(\bm{ r}) b^\dagger(\bm{ r})e^{i\bm{ K_+r}}+u_b^*(\bm{ r}) b^\dagger(\bm{ r})e^{-i\bm{ K_+r}}$. %
\subsection{The topological invariant of the Jackiw-Rossi model}

The band gap in the Jackiw-Rebbi case closes as the sign of the mass switches. This results in a defect in a 1-dimensional chain, at which a zero mode forms. The Jackiw-Rossi case can be regarded of an extension of this concept to 2D, where a vortex in the mass results in a defect at the centre of the vortex (with the mass vanishing and the band gap closing in the centre of the vortex), around which, again, a zero mode forms. The relation to the Jackiw Rebbi model is explored in more detail in \cite{teo2010topological}. 
Although the derivation of the winding number for the Jackiw-Rossi model \cite{teo2010topological} is beyond the scope of this work and will not be discussed here, the end result of how it can be calculated is relatively simple and is expressed, as before as a function of the underlying Hamiltonian.
In the present case the winding number is expressed as a function of the mass term in the Hamiltonian $\Delta(\bm{r})$, which contains the vortex
\begin{equation}
\Delta(\bm{r})=\Delta_0(r)e^{i\phi}.
\end{equation}
The winding number takes the form
\begin{equation}
Q(H)=\frac{1}{2\pi}\int_{S^1}d\phi.
\end{equation}
In our case with $\phi=(\alpha+n\theta)$ the winding number is simply $Q(H)=n$.
 
\section{Conclusion}
In this chapter I have introduced the reader to basic concepts in topology and their relation to photonics.
I introduced topological invariants as quantities that are left unchanged if the underlying space is subjected to ``homeomorphic'' changes. The Chern number as an example of such an invariant was discussed. I demonstrated that it can be identified with a gauge-invariant quantity: the Berry phase over a closed loop of states. I related this gauge invariant to the closed loop of overlap integrals which appear in the graph theory interpretation of multi-photon interference, as discussed previously. Next, I demonstrated how the tight binding model of solid state physics maps onto the coupled mode equations describing the propagation of light in a photonic crystal. A simple example for such a tight binding system was discussed: the SSH model. I then introduced the winding number as topological invariant for the SSH model. It is shown how localised bound states at zero energy appear at the interface between regions governed by different winding numbers. Finally, the Jackiw Rossi model for localised topological states in graphene is reviewed. I present a solution for a zero-mode that appears at the centre of a vortex in the ``mass'' $\Delta$, where similar to the Jackiw-Rebbi case the band-gap closes to produce a defect.

%% file: chapter6.tex
\chapter{Fabrication, excitation and characterisation of photonic crystals}
\label{chap:SLMControl}

In this chapter I will first illustrate how the graphene photonic crystal is fabricated through femtosecond laser writing. I will then show how to excite the localised photonic topological modes. To this end I developed a new method of exciting large modes localised over many sites of a photonic crystal lattice. I will in detail describe this tool, which relies on controlling the light field using a spatial light modulator (SLM). Further, I illustrate how the coupling strengths between individual waveguides can be measured and I show how this data can be used to construct a numerical simulation of the photonic crystal. 

\section{Femtosecond-laser fabrication of photonic crystals}
Femtosecond writing is a method to inscribe waveguide structures in a transparent substrate using pulsed laser light \cite{bhardwaj2005femtosecond,szameit2010discrete}. Strong laser pulses cause a change in the base substrate which manifests as an increase or decrease in the index of refraction in the vicinity of the laser focus. We use a regeneratively amplified Yb:KGW laser. The laser light at 1030 nm is frequency doubled in a second harmonic generation stage (SHG) to 515 nm. It is mode locked and emits pulses which are on the order of  170 fs. We use around 90 nJ per pulse to write the waveguides. As a substrate we use a Borosilicate glass (Eagle2000). This material was chosen since the required exposure time to cause a refractive index change is much lower compared to fused silica, resulting in a faster fabrication process \cite{eaton2005heat}. The laser is focussed by a microscope objective. The material change occurs a threshold area intensity which is reached at the focus. The focus spot is scanned along a programmed trajectory at a speed of 15 mm/s, creating a waveguide in its path.

\begin{figure}[h]
	\centering
	\includegraphics[width=0.4\textwidth]{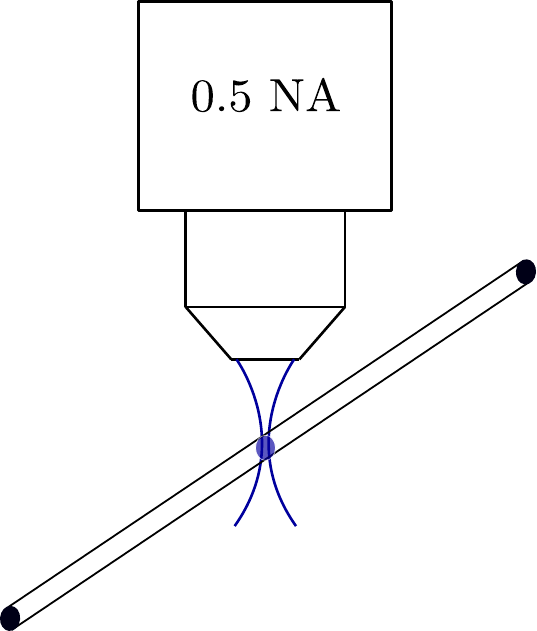}
	\caption{Waveguide written in the focus of an objective. Substrate omitted.}
	\label{fig:WGWriting}
\end{figure}
\begin{figure}[h!]
	\centering
	\includegraphics[width=0.8\textwidth]{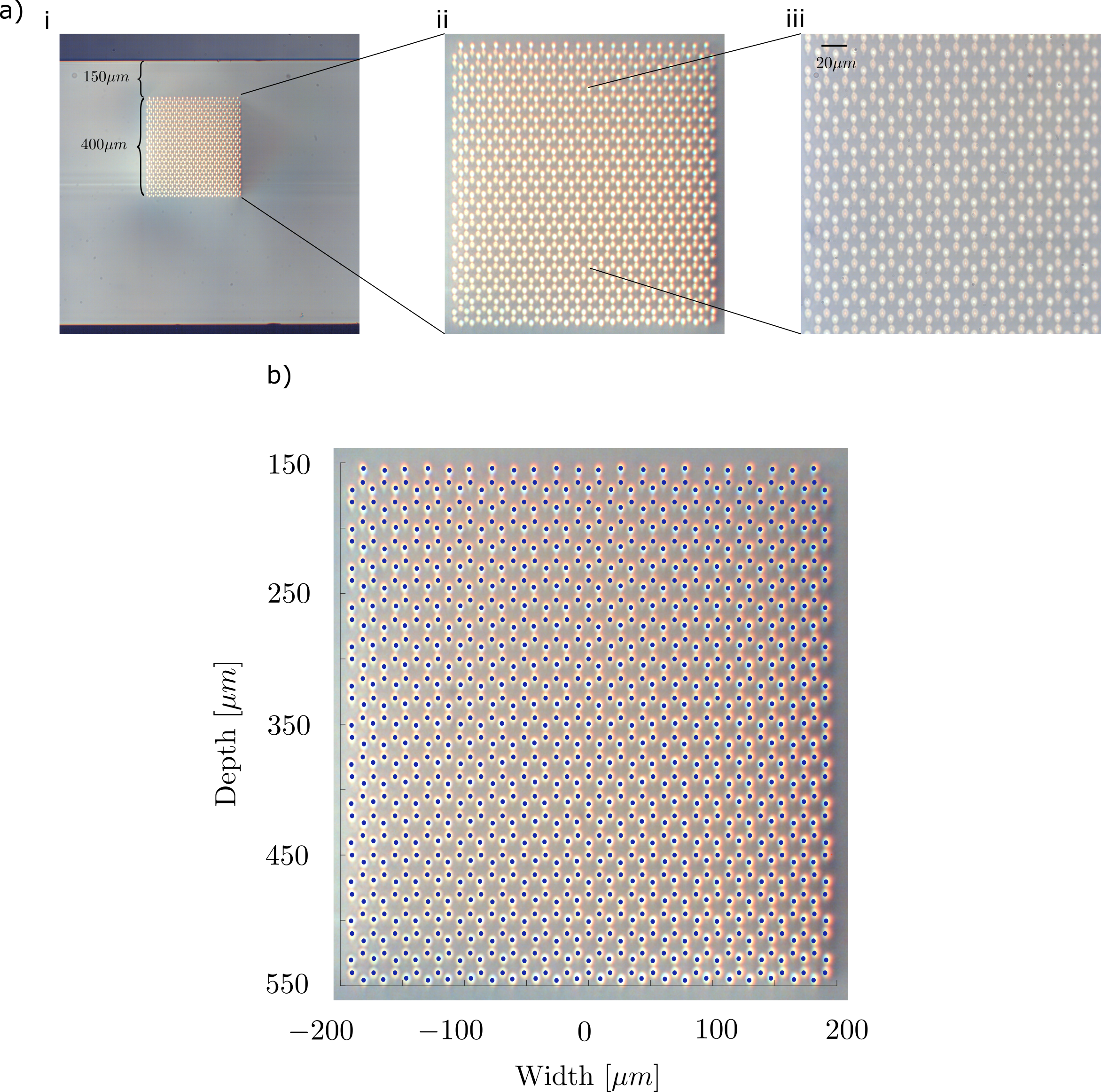}
	\caption{ a) i  wide field microscope image of the front face of the silica chip, the lattice is located at 150 $\mu m$ depth. The chip is 2.1 mm thick. ii zoomed in image, showing the entire lattice. iii Narrow field microscope image of the centre of the lattice showing the structure of individual waveguides.
		b) Comparison with desired placement of waveguides. Blue dots mark the positions of waveguides as designed.
	}
	\label{fig:FSLatComb}
\end{figure}

One of the main challenges in fabricating the large arrays of waveguides used in our experiments lies in making them homogeneous across a wide range of depths. To achieve this an SLM-based aberration correction \cite{Huang2016} is employed. To write the photonic crystal into the glass substrate a precision stage moves the sample Eagle2000 glass wafer along a pre-programmed set of trajectories.
\begin{figure}[h!]
	\centering
	\includegraphics[width=0.7\textwidth]{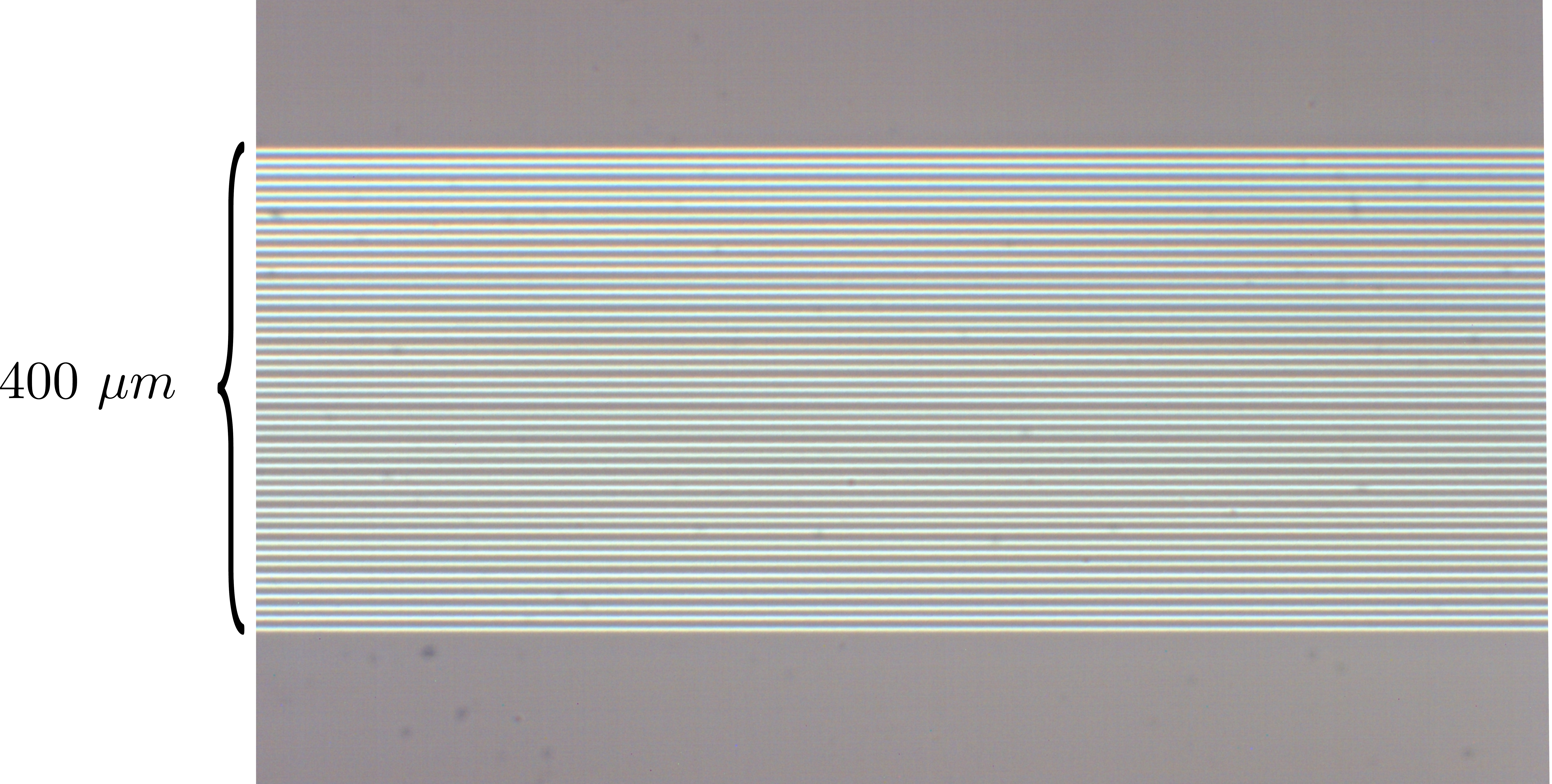}
	\caption{Top view of the waveguide lattice. The layer closest to the top is brought into focus.
	}
	\label{fig:WGstop}
\end{figure} 
In Figure \ref{fig:WGstop} a microscope image showing the top of the photonic crystal lattice is shown.
In Figure \ref{fig:FSLatComb} the front face of the photonic crystal lattice is depicted. The lattice is inscribed into the glass chip, 150 $\mu m$ beneath the surface. 
The photonic crystals were fabricated in collaboration with Martin J. Booth's group at the Oxford University Department of Engineering Science.
\section{Controlling the light field with a spatial light modulator}
\subsection{Preparing the light field}

To be able to excite eigen modes of the lattice Hamiltonian, such as the zero modes, we need to simultaneously illuminate multiple waveguides at specified amplitudes and phases. This can be achieved by means of a spatial light modulator. A spatial light modulator consists of an array of pixels that allow some control over the lights phase and amplitude a the pixels' location. This control is typically achieved by means of optically active liquid crystals. A voltage applied across the pixel induces a change in the orientation of the liquid crystal, changing phase or polarisation of the light interacting with the pixel. We use a reflective ferroelectric liquid crystal SLM. This SLM allows control over the optical axis orientation of each pixel.

\begin{figure}
	\centering
	\includegraphics[width=0.7\textwidth]{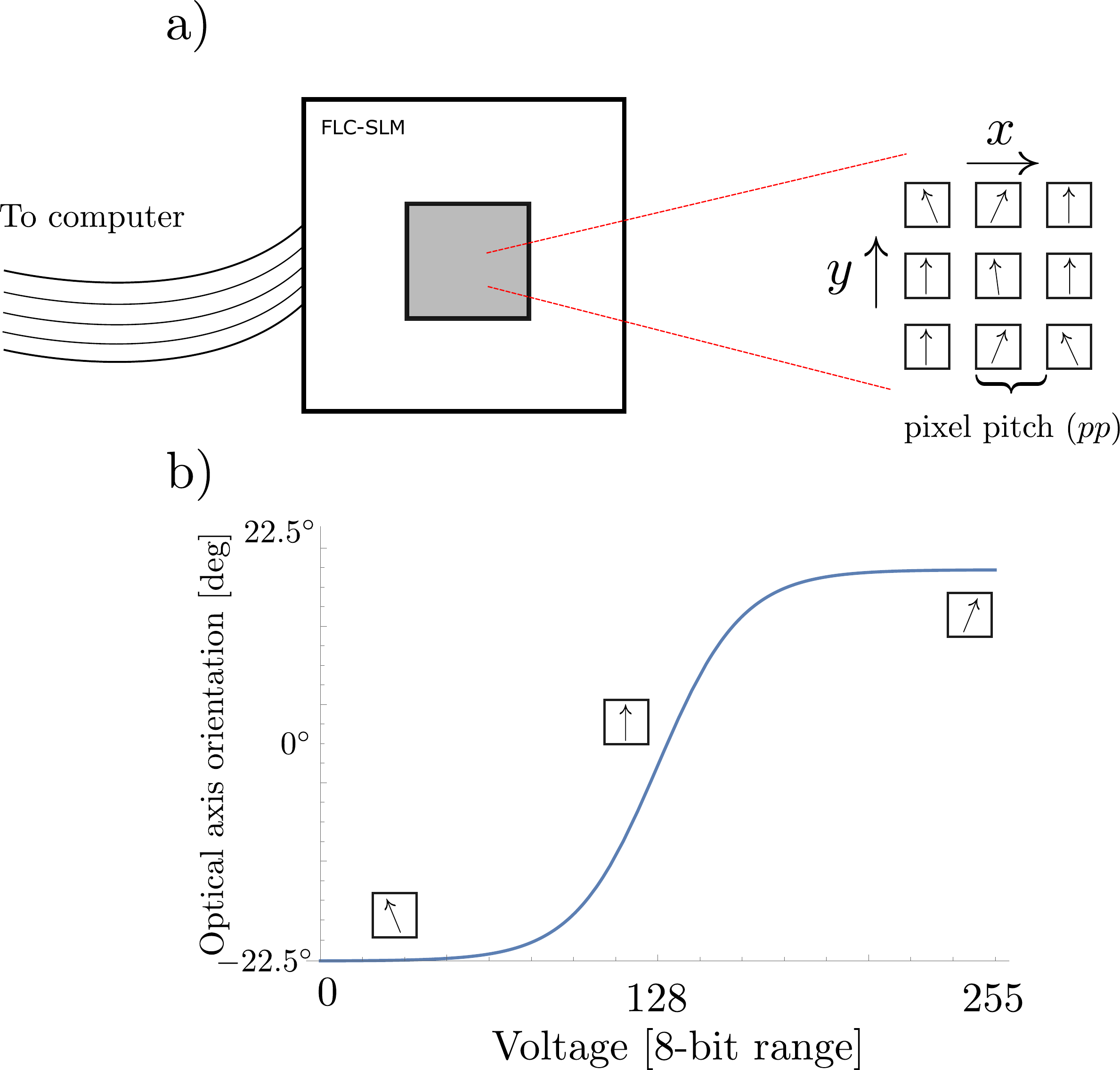} 
	\caption{a) Schematic illustration of the FLC-SLM. The SLM pixels are grouped in a 2D array at a pixel pitch of $pp=15 \mu m$. Optical axis orientation of each pixel indicated by arrow. b) The optical axis orientation is smoothly controlled by applying a voltage to the pixels. The voltage is specified by an 8-Bit value. The curve shows a typical pixel response, relating optical axis orientation to applied voltage value.  
	}
	\label{fig:SLMSchematic}
\end{figure}

Each pixel effectively acts as a half-wave plate. The optical axis can be rotated smoothly as a function of control voltage. In Figure \ref{fig:SLMSchematic} a schematic of the SLM and the pixel response function is depicted.

\subsubsection*{Amplitude control}
It is straightforward to see how we can can use this SLM to control the field amplitude for each pixel. We illuminate the SLM with linearly polarised light and position a polariser in the beam reflected from the SLM. The proportion of transmitted light then depends on the orientation of the optical axis in each pixel. In the experiment we prepare H-polarisation before the SLM. For an optical axis orientation of $22.5^\circ/-22.5^\circ$ the light will be rotated to diagonal/anti-diagonal polarisation upon reflection. After the SLM we position a fixed HWP at an optical axis orientation of $22.5^\circ$. If the optical axis of the pixels is $-22.5^\circ$, the combined action of the SLM and HWP takes the input H-polarised light to V polarisation: $H\rightarrow V$. Conversely, If the pixels' optical axis is at $22.5^\circ$ $H\rightarrow H$. For any other orientation of the pixels' optical axis between $-22.5^\circ$ and
$22.5^\circ$ we obtain a linear polarisation between $H$ and $V$. In our setup a PBS acts as a polariser, rejecting the $V$-polarised component of the light, enabling a smooth variation of light amplitude from completely dark to maximum transmission.
\subsubsection*{Phase control}
\begin{figure}
	\centering
	\includegraphics[width=1\textwidth]{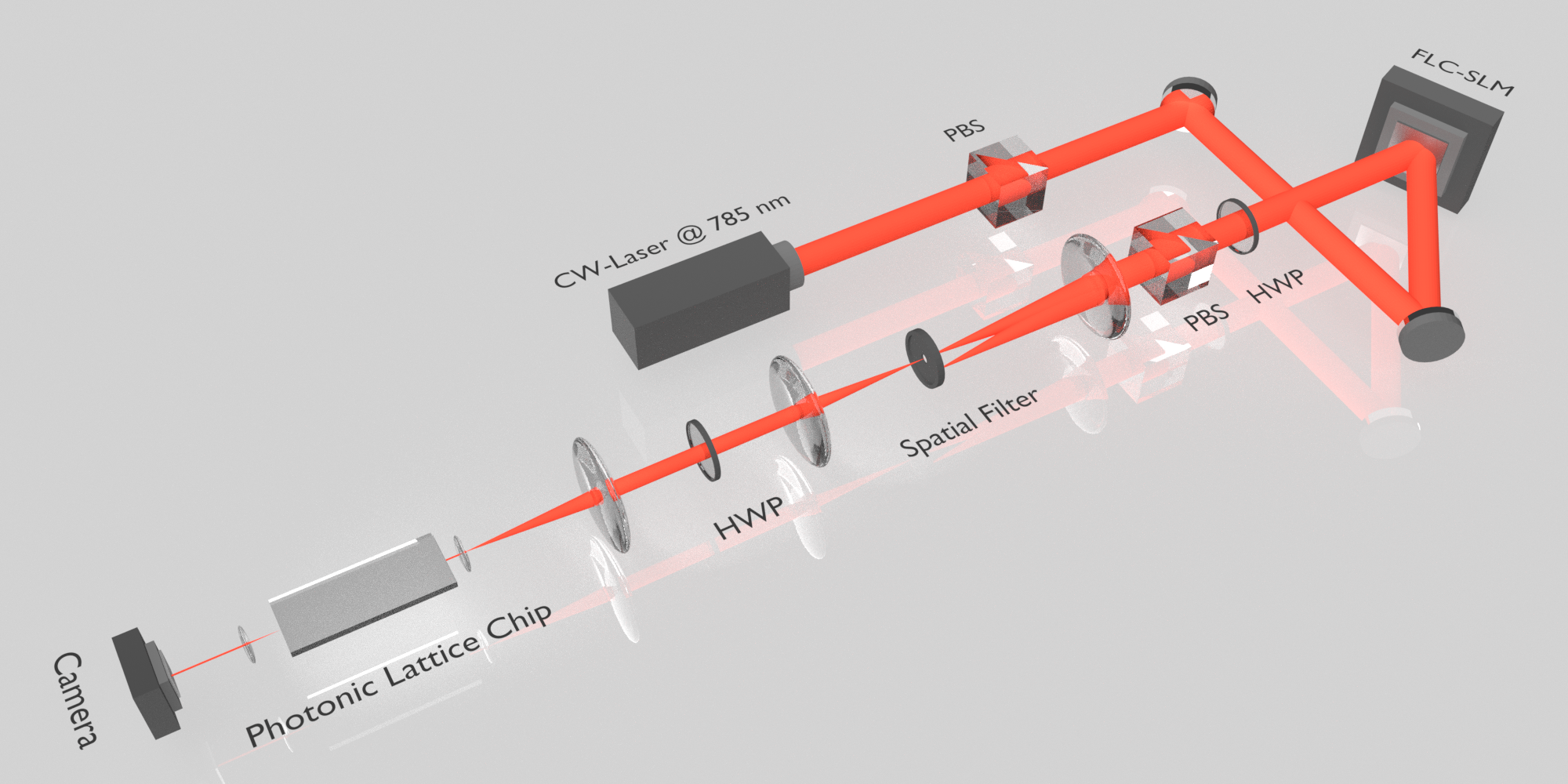} 
	\caption{Setup: As a coherent source of light we use a CW laser diode at a wavelength of 785 nm.  1. We send the light through a Polarising Beam Splitter (PBS) to to prepare a horizontal polarisation. 2. The beam is reflected from the surface of the Ferroelectric Liquid Crystal Spatial Light Modulator (FLC-SLM); each pixel rotates the light's polarisation according to its optical axis orientation. 3. After passing through a Half-Wave Plate (HWP), the vertically polarised component of the light is rejected, implementing the desired field amplitude. 4. The light passes through a 4-f system with an off-axis pinhole, setting the phase gradient and enabling phase control. 5. The polarisation of the light is prepared by a HWP. 6. The light beam is imaged onto the surface of the chip. The total (de-) magnification is $1/35$. 7. The light at the output-facet is imaged onto a camera. %
	}
	\label{fig:setupTopo}
\end{figure}

\begin{figure}
	\centering
	\includegraphics[width=0.7\textwidth]{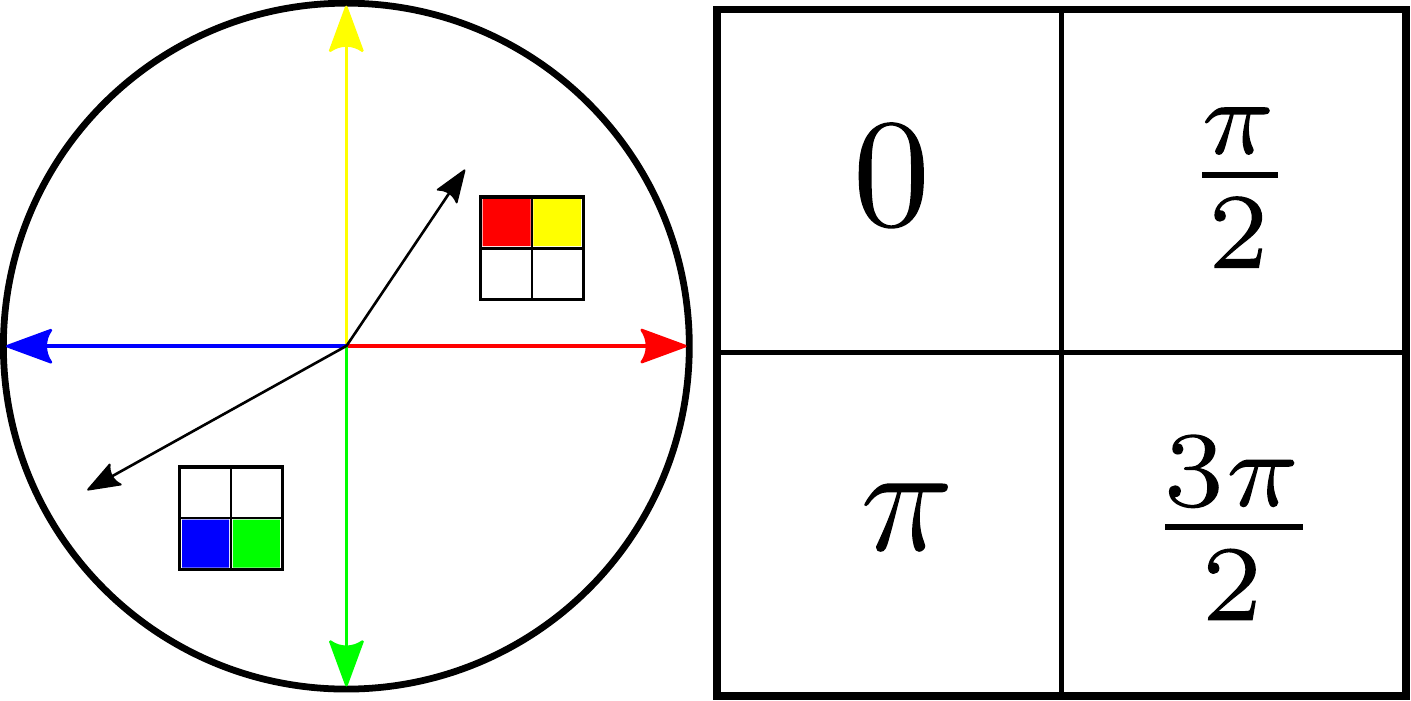} 
	\caption{Illustration of how the field amplitude and phase are realised: Field amplitude in the complex plane (left) with coloured arrows corresponding to individual pixels of different phase (red: 0, green=$\pi/2$, yellow=$3\pi/2$, blue=$\pi$). Spatial arrangement of a single 2x2 super pixel (right).}
	\label{fig:superpixel}
\end{figure}

We also need to be able to control the phase of the field. We can achieve this in a 4-f configuration illustrated in Figure \ref{fig:setupTopo}.
We recall that a lens performs a Fourier transform of the light in the focal plane \cite{Saleh2007}, $g(\nu_x,\nu_y)$ is the amplitude of the light field after the lens
\begin{equation}
g(\nu_x,\nu_y)=sF(\frac{\nu_x}{\lambda f},\frac{\nu_y}{\lambda f}),
\end{equation}
where $s=\frac{i}{\lambda f}e^{-2if2\pi/\lambda}$
and $F$ is the Fourier transform of the light amplitude $h$ before the lens
\begin{equation}
F(v_x,v_y)=\mathcal{F}(h)(v_x,v_y)\\
=\int\int h(x,y)\exp[2\pi i(v_xx+v_yy)]dxdy.
\end{equation}
Furthermore, a shift in position corresponds to a linear phase in the Fourier transform
\begin{eqnarray}
1/s\mathcal{F}^{-1}(g(\nu_x+d\nu_x,\nu_y+d\nu_y))(x,y)
&=&h(x,y)\exp[2\pi i(\frac{x}{\lambda f}d\nu_x+\frac{y}{\lambda f}d\nu_y)].
\end{eqnarray}
If we insert a pinhole in the Fourier plane of the first lens after the SLM (cf. Figure \ref{fig:setupTopo}) at an offset, we impart a linear phase onto the surface of the SLM, resulting in each pixel contributing light at a different phase. We have adjusted the phase gradient across the SLM such that for a group of four pixels each pixel has a phase as shown in Figure \ref{fig:superpixel}. 

The phase gradient imparted is $\pi/2/[\text{pixel}]$ in the x-direction and $\pi/[\text{pixel}]$ in the y-direction.
We can see (Figure \ref{fig:superpixel}) that this phase configuration is such that the field of each pixel in the complex plane can be represented by a vector along all four negative and positive complex coordinate directions. We aim to get a specific electric field by adding the field contributions from each pixel. To achieve this we need to filter the Fourier components of the beam such that individual pixels within a superpixel cannot be resolved and are in effect `blurred' together. This can be accomplished if we chose the radius of the aperture $r<\lambda f
/(2pp)$ (or $\pi$ [rad/pixel]), where $f$ is the focal length of the first lens placed after the SLM and $pp$ is the pixel pitch (distance between adjacent pixels), $pp=15$ $\mu m$. We can now set a specific electrical field, amplitude and phase, choosing the amplitudes contributed by each pixel which for a desired electrical field vector are given by its projection onto the pixel basis-vectors, as illustrated in Figure \ref{fig:superpixel}.

\begin{figure}
	\centering
	\includegraphics[width=0.5\textwidth]{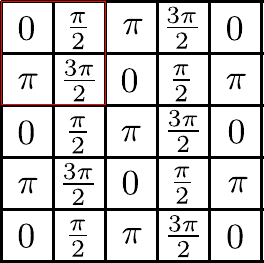} 
	\caption{Distribution of phases across the SLM. A single superpixel is marked in red.}
	\label{fig:superpixel2}
\end{figure}

 With an imaging lens $f=15$ cm, which images the plane of the SLM onto the Fourier plane, the Fourier filter has to be placed at an offset of $d\nu_x=\frac{\lambda f}{4pp}=1.9$ mm and $d\nu_y=3.9$ mm. To determine the position of the pinhole centre, we realise a beam with constant phase by switching on only the pixels corresponding to a phase of 0 in every 2x2 ``superpixel''. Effectively, this realises a tilted grating across the SLM with a maximum at the desired $d\nu_x$, $d\nu_y$. The position of maxima in the Fourier plane for a grating in 2D with a period of $\Delta x$ in x and $\Delta y$ in y is: $n_{x/y}\lambda f/\Delta (x/y)$ with integer $n_{(x/
y)}$. As we can see in Figure \ref{fig:superpixel2}, the grating period in x is 2 superpixels = $4pp$ and 1 superpixel in y = $2pp$. Thus, we expect the first order $(n_x=1,n_y=1)$ maximum of this grating at $\lambda f\binom{\frac{1}{4pp}}{\frac{1}{2pp}}$, which is what we previously determined for $\binom{d\nu_x}{d\nu_y}$.

 Our work draws on that by Goorden et.al. \cite{Goorden2014}, who introduced the ``superpixel'' method we use here. However, we adapted their method to work with continuous amplitude control, this allows us to operate with smaller superpixel sizes and retaining the same or better control over the light field at a higher resolution.
\subsubsection*{Generating individual beams}
We want to be able to control individual beams with a Gaussian intensity profile.
There are eight parameters that can be set for each beam:
\capstartfalse

\begin{table}[h]
	\centering
	\begin{tabular}{|l|l|l|l|l|l|l|l|}
		\hline
		$e2_x$ & $e2_y$ & Amplitude & Phase $\phi$ & x & y & $k_x$ & $k_y$ \\ 
		\hline
	\end{tabular}
\end{table}
\capstarttrue

\begin{figure}[h!]
	\centering
	\includegraphics[width=0.8\textwidth]{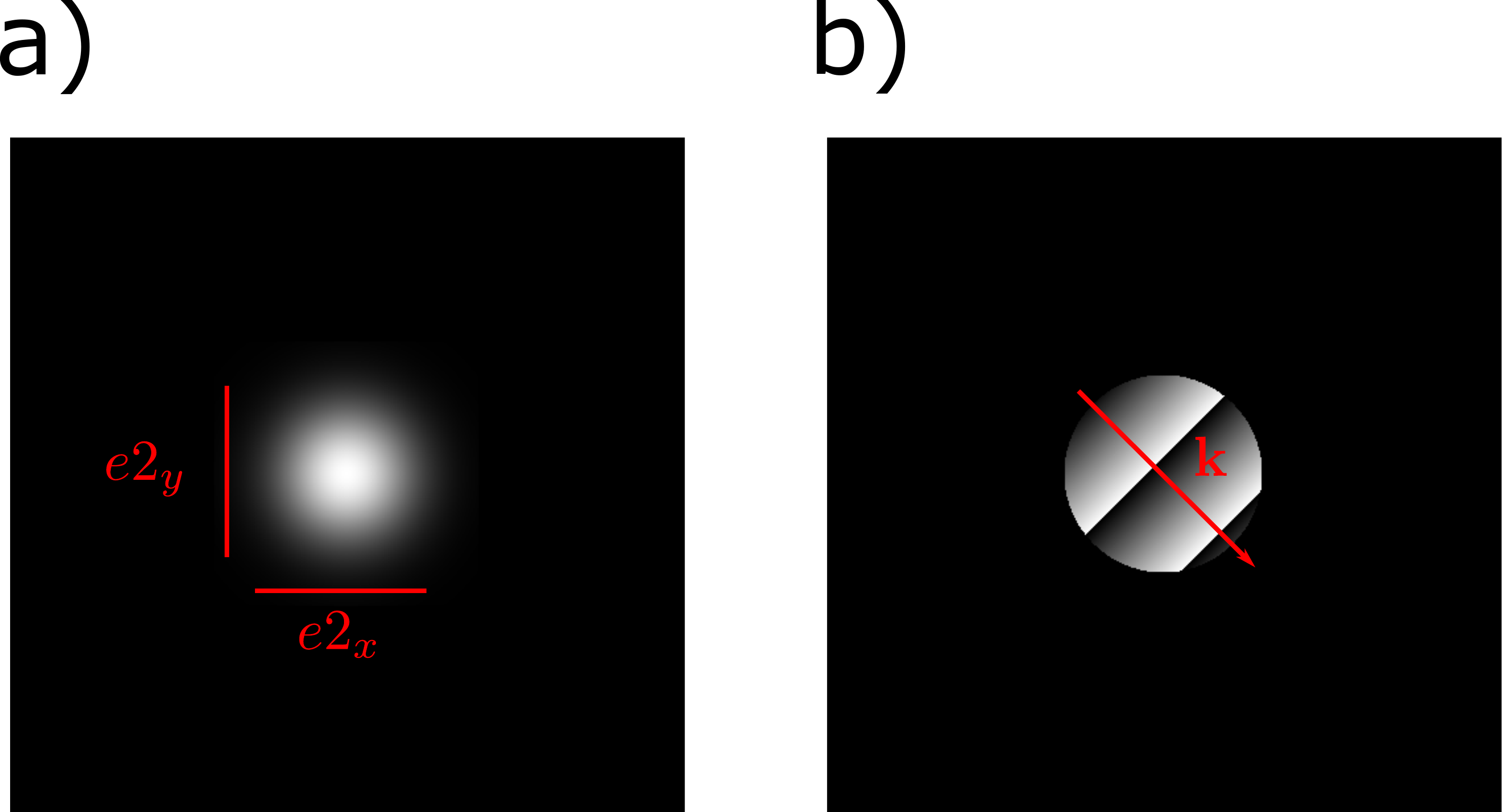} 
	\caption{a) Gaussian intensity profile. b) phase gradient applied in direction $\bf{k}$.}
	\label{fig:GaussGrad}
\end{figure}

In Figure \ref{fig:GaussGrad} an amplitude a) and phase image b) of the desired field in the plane of the SLM are shown.
$e2_{x/y}$ is the $e2$ width in x and y [pixels] of the Gaussian beam on the surface of the SLM. The phase $\phi$ is a constant offset which is added to the phase profile. 
By applying a phase gradient $\textbf{k}$ locally around the position of the beam we adjust the pointing as is shown in \ref{fig:GaussGrad} for a particular $\textbf{k}$. %
In order to verify that we indeed control the phase $\phi$ of our light field we can interfere two separate beams. When inserting two coherent laser beams of sum intensity $I_{inp}$ into a 50:50 beam-splitter, we expect the output intensity to obey
\begin{eqnarray}
I_{out_1}(\Delta\phi)=I_{inp}\sin^2(\Delta\phi/2)\\
I_{out_2}(\Delta\phi)=I_{inp}\cos^2(\Delta\phi/2),
\end{eqnarray} 
where $\Delta\phi$ is the relative phase of the two beams.
\begin{figure}[h!]
	\centering
	\includegraphics[width=0.7\textwidth]{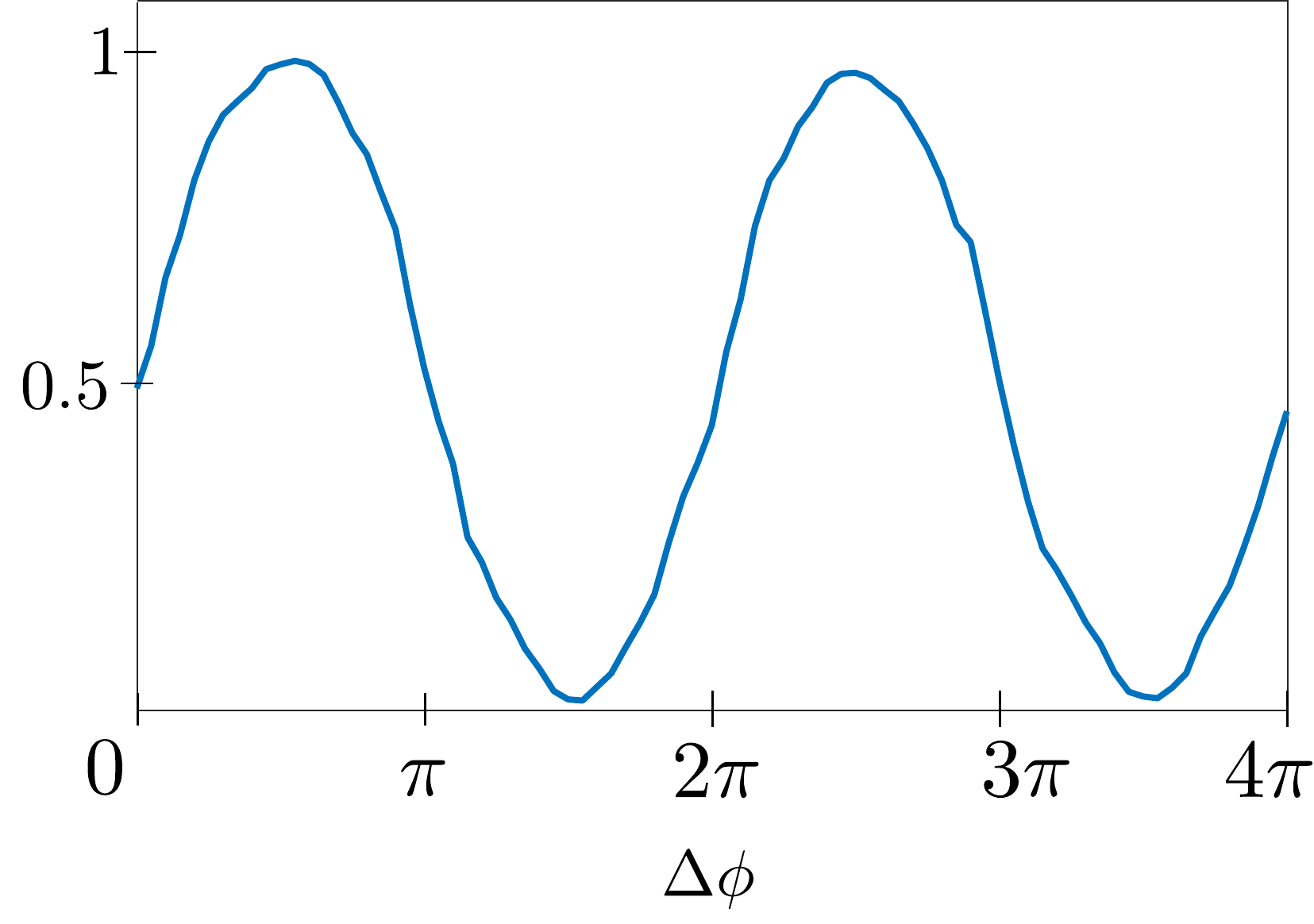} 
	\caption{Output intensity on a single output port of an integrated 50:50 beamsplitter. One input beam is kept at a constant phase while the other phase is scanned from 0 to $4\pi$.}
	\label{fig:PhaseSweep}
\end{figure}
In Figure \ref{fig:PhaseSweep} I plot the output intensity at one of the ouptut ports of a femtosecond written 50:50 beamspitter, as we scan the relative phase of the two beams.  We can clearly see the expected sinusoidal behaviour, confirming that we can control the full phase range for the input beams. %
To verify that we have control over the pointing, I measure a single Gaussian beam in the Fourier plane. 

  \begin{figure}[h!]
  	\centering
`  	\includegraphics[width=0.7\textwidth]{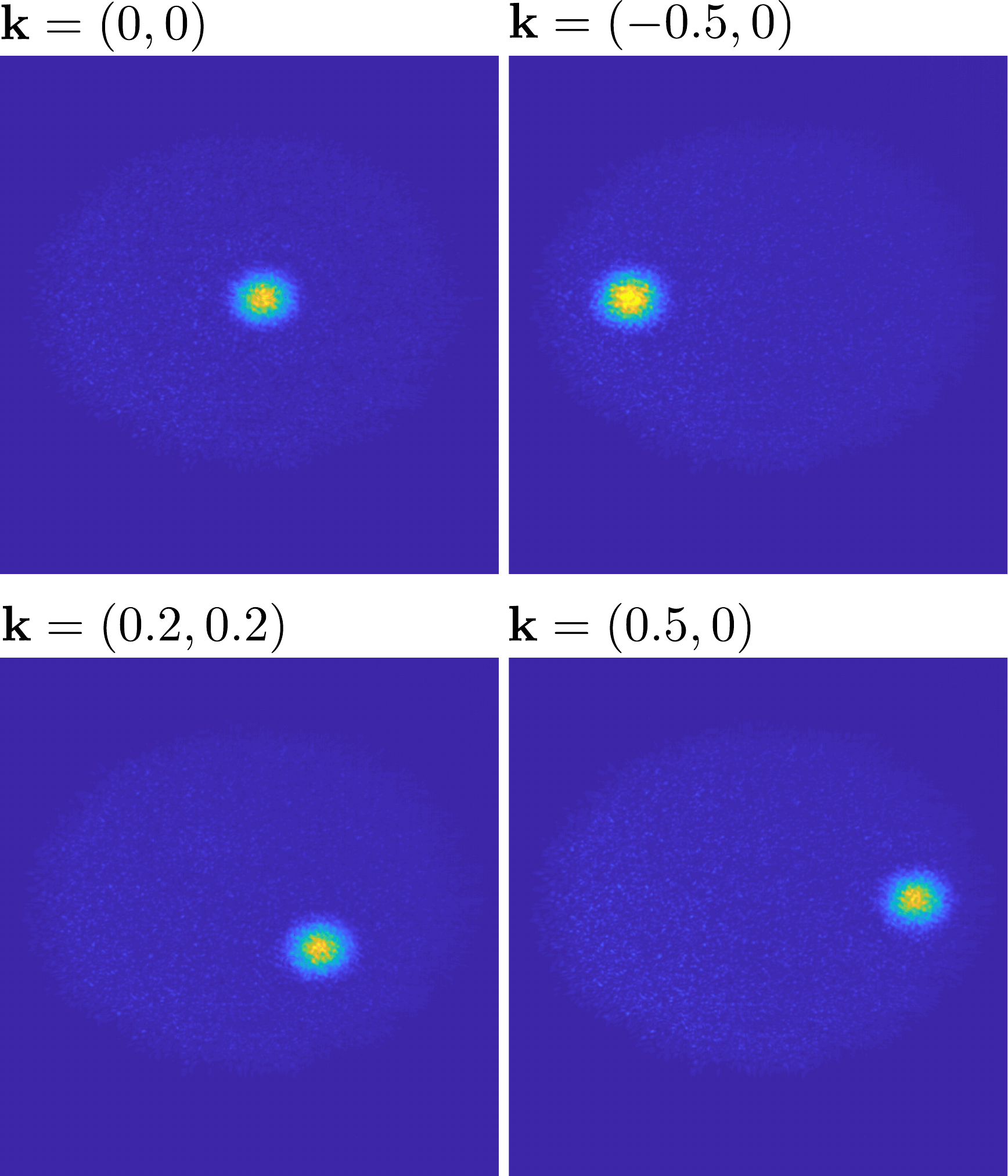} 
  	\caption{Fourier plane camera image of a single beam with different ${{\bf k}=(k_x,k_y)}$ applied. The units of the k vector are $[\text{rad}/\text{pixel}]$. The circular halo marks the edge of the pinhole aperture.}
  	\label{fig:KVec}
  \end{figure}

In Figure \ref{fig:KVec} the camera image of the beam is shown for several different ${\bf k}$-vectors. The circular halo is an image of the pinhole aperture placed in the Fourier plane.   
We notice that given the size of the aperture, the maximum $|{\bf k}|$ is a bit more than 0.5 [rad/pixel], which is significantly smaller than the required $\pi$ $[\text{rad}/\text{pixel}]$.
\subsection{Coupling to waveguides}
\begin{figure}[h!]
	\centering
	\includegraphics[width=0.8\textwidth]{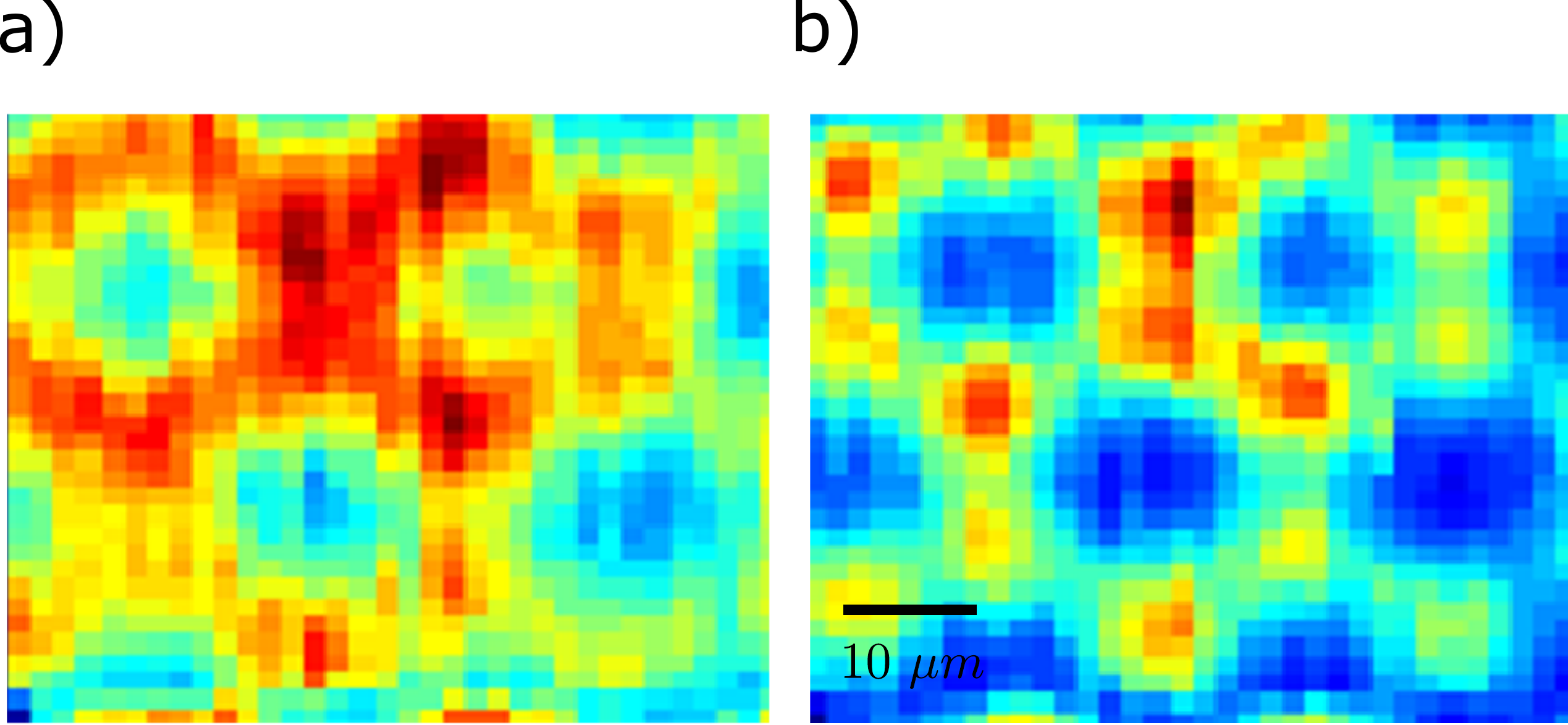} 
	\caption{Scan of the chip surface: a) sample out of focus b) sample in focus.}
	\label{fig:SLMScan}
\end{figure}

\begin{figure}[h!]
	\centering
	\includegraphics[width=1\textwidth]{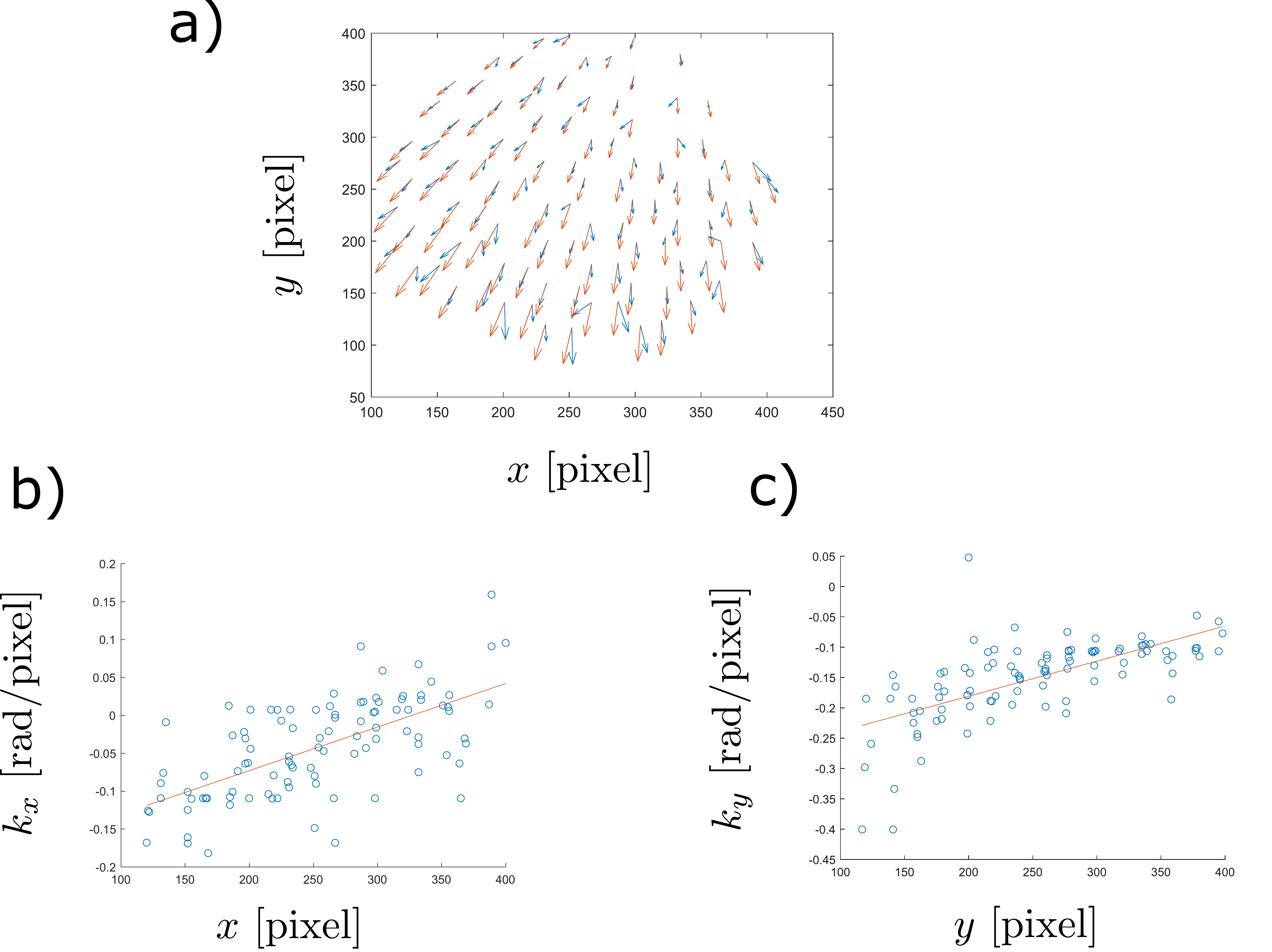} 
	\caption{Result of the optimisation of the ${\bf k}$-vectors. a) Blue vector-field, representing the direction of the applied ${\bf k}$ vector that optimises coupling to each waveguide. On the $x$ and $y$ axes the location of the waveguides is noted (as pixel coordinates on the SLM). The orange vector-field represents a linear fit. b) and c) Linear fits to $k_x$ and $k_y$ components.}
	\label{fig:KVectorfield}
\end{figure}

Having established how we can exert full control over the light field, I want to now describe how we utilise this powerful experimental tool to excite waveguides.
Firstly, we note that the amplitude and phase degrees of freedom of the light field are accessible entirely through the software controlling the SLM. This enables a great deal of automation in procedures which otherwise would have been performed manually. I will show how we can find the position of the waveguides on the surface of the photonic crystal and how to optimise coupling into a waveguide.
We have already stated that we image the surface of the SLM onto the surface of the chip. We need to first verify that the chip is indeed in the Fourier plane of the optics. This is done by scanning a beam across the surface of the chip (by moving a dot across the SLM). Simultaneously, the total light intensity at the output is recorded and plotted in a diagram as a function of the beam position on the chip (dot position on the SLM). The result is shown in Figure \ref{fig:SLMScan}.
We can clearly see the individual waveguides of the chip's hexagonal waveguide  lattice. As the beam crosses a waveguide on the surface of the chip, light is transmitted through the waveguides, the better the coupling, the greater the intensity of the transmitted light. If the chip is out of focus, the resulting image will be blurred, as seen in Figure \ref{fig:SLMScan} a). We can then adjust the focal position to achieve better coupling. Once the focus is adjusted, the waveguides are sharply imaged, as seen in \ref{fig:SLMScan} b). So far we have only scanned the position by moving a dot across the surface of the SLM. To optimise coupling to a waveguide we also need to access the pointing of the beam. The position coordinates are labelled $(x,y)$ and the direction of the beam $(k_x,k_y)$. To optimise coupling to a waveguide, we need to utilise all four of these degrees of freedom.
In Figure \ref{fig:KVectorfield} the result of the optimisation is shown. Both, the position of the individual beams (in SLM pixel coordinates) as well as the pointing $(k_x,k_y)$ have been optimised using a gradient descent method. As initial values for the optimisation, we can obtain the beam positions $x,y$ from scanning the beam across the chip, as shown in Figure \ref{fig:SLMScan} and for the initial values of $(k_x,k_y)$. I start with a constant value across the chip. The values for $(k_x,k_y)$ as a function of position are then fitted linearly. The variation of the values for $(k_x,k_y)$ across the surface of the chip are likely due to an offset from the main optical axis and a small aberration error in the optics. 

\section{Introduction to mode-coupling theory}
\label{sect:modeCoup}
\begin{figure}[h!]
	\centering
	\includegraphics[width=0.7\textwidth]{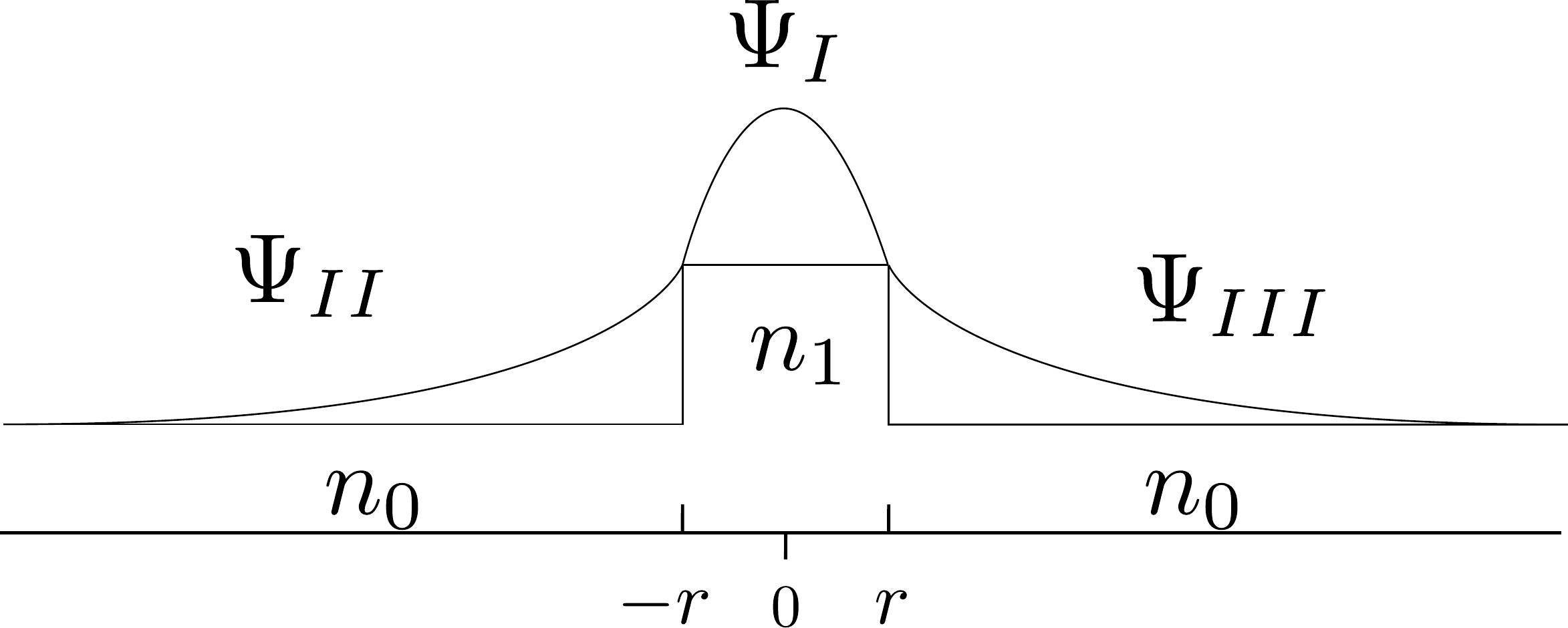}
	
	\caption{Illustration of the different regions of solutions to the Helmholtz equation. The waveguide depicted has a rectangular profile and refractive index $n_1$ while the surrounding material has refractive index $n_0$.}
	
	\label{fig:WGRegions}
\end{figure}

In this section I will discuss the basics of mode-coupling theory. A waveguide guides light because of a refractive index difference between the waveguide and the surrounding material. In our case the refractive index difference is caused by a material change in the silica substrate induced by a strong femtosecond laser.
There are two distinct regions, within the waveguide (refractive index $n_1$) and outside (refractive index ($n_0$)). In Figure \ref{fig:WGRegions} the configuration is illustrated. In the following section we are going to assume a constant refractive index change across the waveguide. 

Solving the Helmholtz equation leads to solutions for the electrical field.
For the transverse component we obtain

\begin{eqnarray}
\Psi_I^{sym}(x)&=&E_0\cdot\cos(k_x x) \label{eqn:symsol}\\
\Psi_I^{asym}(x)&=&E_0\cdot\sin(k_x x)\\\label{eqn:ccwgs}
\Psi_{II}(x)&=&E_0\cdot\exp(\gamma_m(x+r))\\ \label{eqn:ccwgs2}
\Psi_{III}(x)&=&E_0\cdot\exp(-\gamma_m(x-r))\label{eqn:expsol}.
\end{eqnarray}

A derivation of these equations can for example be found in \cite{Saleh2007}.
$\gamma_m$ is called the decay constant of evanescent field.
Within the waveguide the solution is a cosine/sine, while outside of the waveguide the light decays exponentially from the waveguide.
In Figure \ref{fig:WGRegions} the different solutions are illustrated. Within the waveguide the 0th order cosine is sketched (with 0 nodes), one speaks of single mode excitation of the waveguide.
In the longitudinal direction the wave oscillates at a frequency given by the propagation constant $\beta_m$, which is proportional to an effective refractive index that depends on the shape of the waveguide and mode of the light field. Even for a rectangular waveguide of constant refractive index there is no analytic solution, however $\beta_m$ can be determined numerically \cite{Saleh2007}.
When two waveguides are brought close to each other, such that the evanescent light field of one waveguide overlaps with the mode of the other waveguide, light can couple between them. The coupling strength of two waveguides $i,j$ is expressed in terms of the coupling coefficient $k_{ij}$. 
Its value for two waveguides $i,j$ of rectangular refractive index profile can be calculated with the waveguides' refractive indices $n_j$, the  refractive index of the surrounding material $n_0$, the propagation constant $\beta_i$ and the shape of modes $\Psi_{i,j}$ \cite{huang1994coupled,Saleh2007}
\begin{equation}
k_{ij}\sim k_0^2\frac{1}{\beta_i}\int(n^2_{j}-n_0^2)\Psi_i\Psi_j^*dA,
\label{eqn:modecoupl}
\end{equation}
with $k_0=\frac{2\pi}{\lambda}$.
The integration is executed over the waveguide area in the plane transversal to the propagation direction.
The input/output relation of an electrical field entering two coupled waveguides can be written in terms of a matrix equation
\begin{spacing}{1.2}
\begin{equation}
\bm{ E}=\begin{pmatrix}
	E_1\\ 
	E_2
\end{pmatrix}\\
\end{equation}
\begin{equation}
\hat{K}=\begin{pmatrix}
k_{1,1} & k_{1,2}\\ 
k_{2,1}&k_{2,2} 
\end{pmatrix}
\end{equation}
\begin{equation}
\bm{ E}^{out}(l)=\exp(-i\cdot \hat{K}\cdot l)\bm{ E}^{in},
\end{equation}
\end{spacing}
where $l$ is the distance the two waveguides propagate next to each other (coupling region). We can choose $k_{1,2}=k_{2,1}=k$, and $k_{1,1}=k_{2,2}=0$ for identical waveguides, as these only contribute an additional phase that is identical for all waveguides. 
The solution is
\begin{spacing}{1.2}
\begin{equation}
\begin{pmatrix}
E^{out}_1\\ 
E^{out}_2
\end{pmatrix}=\begin{pmatrix}
\cos(kl) &-i\sin(kl) \\ 
-i\sin(kl) &\cos(kl) 
\end{pmatrix}\cdot\begin{pmatrix}
E^{in}_1\\ 
E^{in}_2
\end{pmatrix}.
\end{equation}
\end{spacing}
\hspace{0.5cm}

An interesting special case is that of a single input beam into port 1. The (normalised) intensities at the output are then going to be
\begin{eqnarray}
I^{out}_1=\cos^2(kl)\\
I^{out}_2=\sin^2(kl).
\end{eqnarray}
I define $\frac{2\pi}{k}$ as the beat length.
Another important characteristic is the dependence of the coupling coefficient on the distance between the waveguides. The field is decaying exponentially outside of the waveguide.
Solving equation \ref{eqn:modecoupl} by inserting \ref{eqn:ccwgs} and \ref{eqn:ccwgs2}, we obtain that the coupling coefficient will behave as %
$k_{ij}\sim 1/a\cdot e^{-\gamma d}$, where $d$ is the distance between waveguides.
In the next section we are going to use this relationship to characterise the coupling coefficients between waveguides. We will be measuring the beat length $2\pi/k_{i,j}$ between different sets of waveguides. %
\clearpage
\section{Characterising coupling strengths}
In order for us to be able to design the desired Hamiltonian of the lattice and verify beforehand that the desired mode-structure is indeed present, we need to find an adequate model which closely resembles experimental reality. There are two main effects we need to consider: waveguide mode ellipticity and waveguide variation across different depths. While we expect the waveguides to have a near circular mode, even a slight ellipticity of the evanescent mode field could lead to different coupling strengths. This will depend on the waveguides' relative orientation with respect to each other. Thanks to the aberration corrected waveguide fabrication process, we achieve a great deal of homogeneity across the entire lattice. However, there are still small differences in the waveguides as a function of depth. In order to characterise these differences we measured the coupling strength between waveguides across the entire depth of the lattice.  To characterise mode ellipticity and variation across depth we fabricated a large array of two-mode couplers (beamspitters) with three different orientations of the waveguides in the coupling region, at four different depths in the chip. For each depth and configuration twelve beams spitters at different coupling lengths were manufactured. In order to characterise the exponential mode decay away from the waveguide we additionally fabricated them at five different distances from each other. In total the number of beam splitters is therefore:
\capstartfalse

\begin{table}[h!]
	\begin{tabular}{|l|l|l|l|l|}
		\hline
		Orientations & Depths & Coupling lengths $l$ & Distances $d$& Total Number of couplers \\ \hline
		3            & 4      & 12               & 5         & 720                      \\ \hline
	\end{tabular}
\end{table}
\capstarttrue

Since the lattice we are investigating is a hexagonal one (with small distortions), the coupling strength between wavguides at orientations that appear in this lattice are measured. 

\begin{figure}[h!]
	\centering
	\includegraphics[width=1\textwidth]{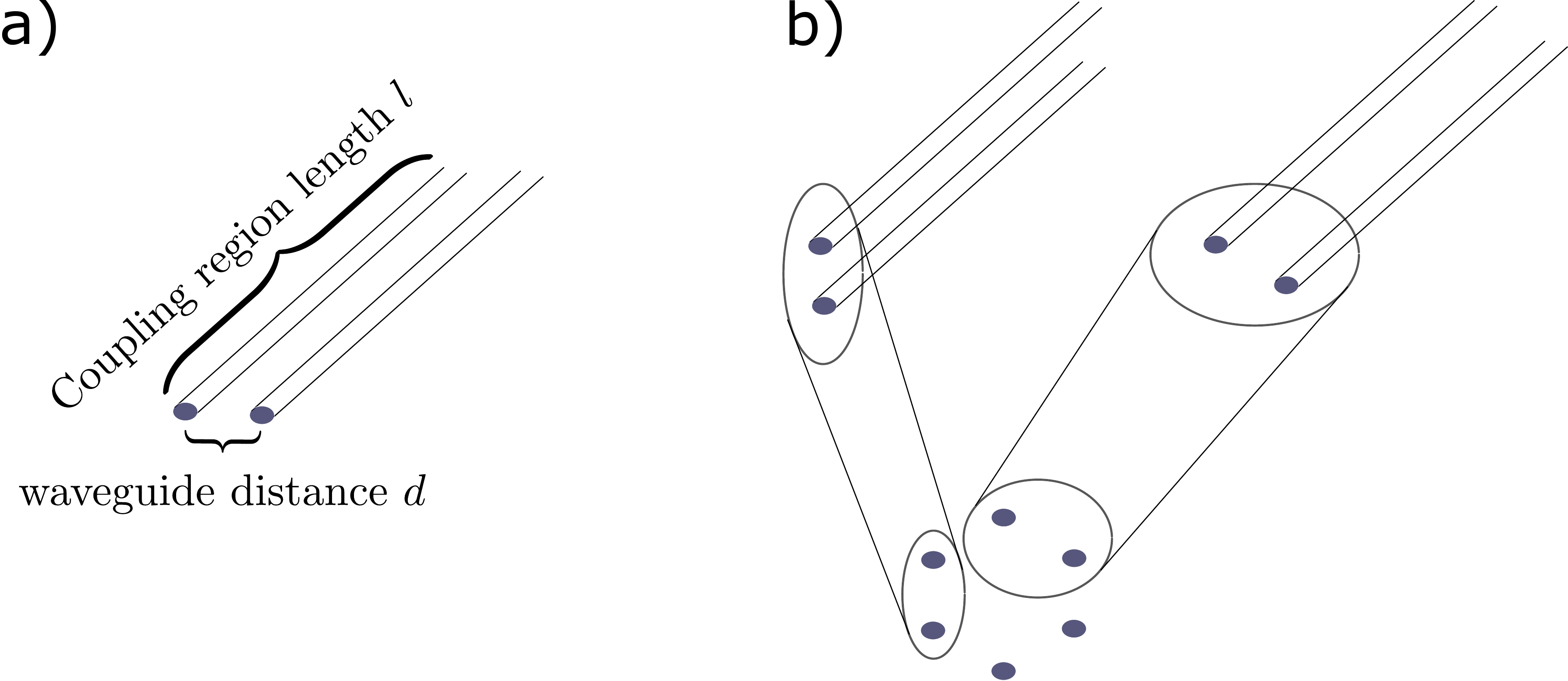}
	
	\caption{a) Waveguides in horizontal orientation with coupling length and waveguide distance illustrated. b) Vertical and 60 Degree orientations with their placement within the hexagonal lattice shown.}
	
	\label{fig:WGConfigs}
\end{figure}

\begin{figure}[h!]
	\centering
	\includegraphics[width=1\textwidth]{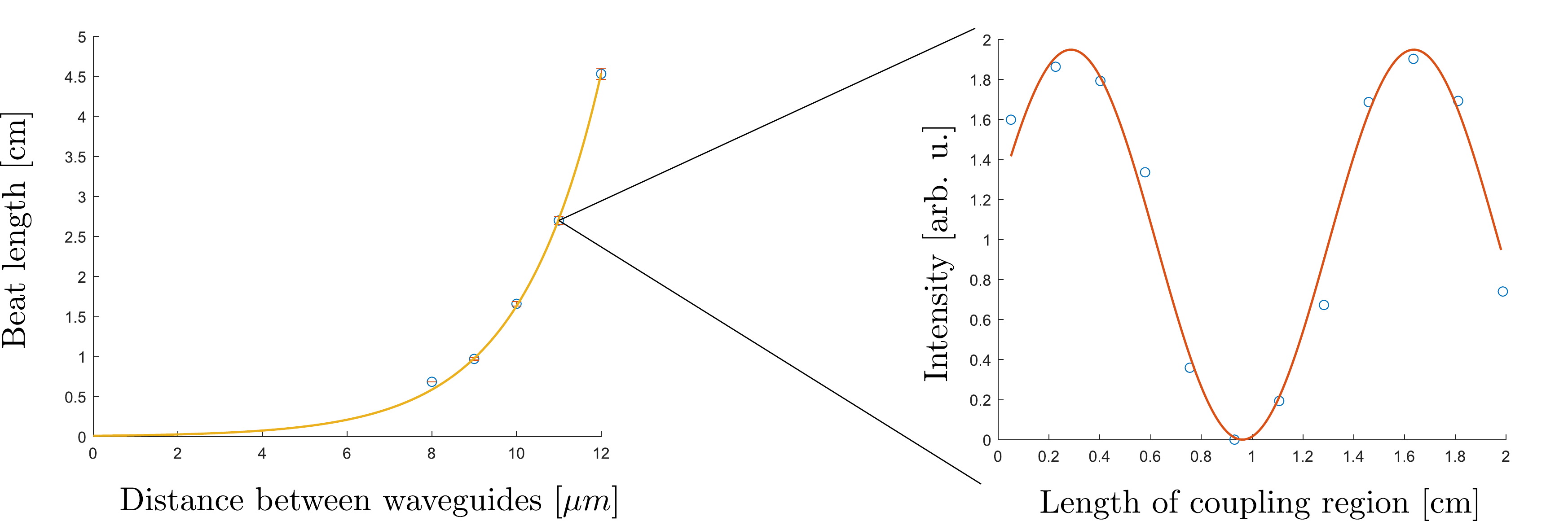}
 
		\caption{Left panel: exponential fit to beat length for different waveguide distances. Right panel: sinusoidal fit to measurement data of beamsplitters at twelve different coupling lengths. The waveguides are in a horizontal configuration. The light polarisation is horizontal.}

	\label{fig:ExpFitFS}
\end{figure}

In Figure \ref{fig:ExpFitFS} example data for a depth of 100 $\mu m$ is shown. The exponential decay of the mode is reflected in an increasing beat-length as a function of the distance between waveguides. 
 
\begin{figure}[h!]
	\centering
	\includegraphics[width=1\textwidth]{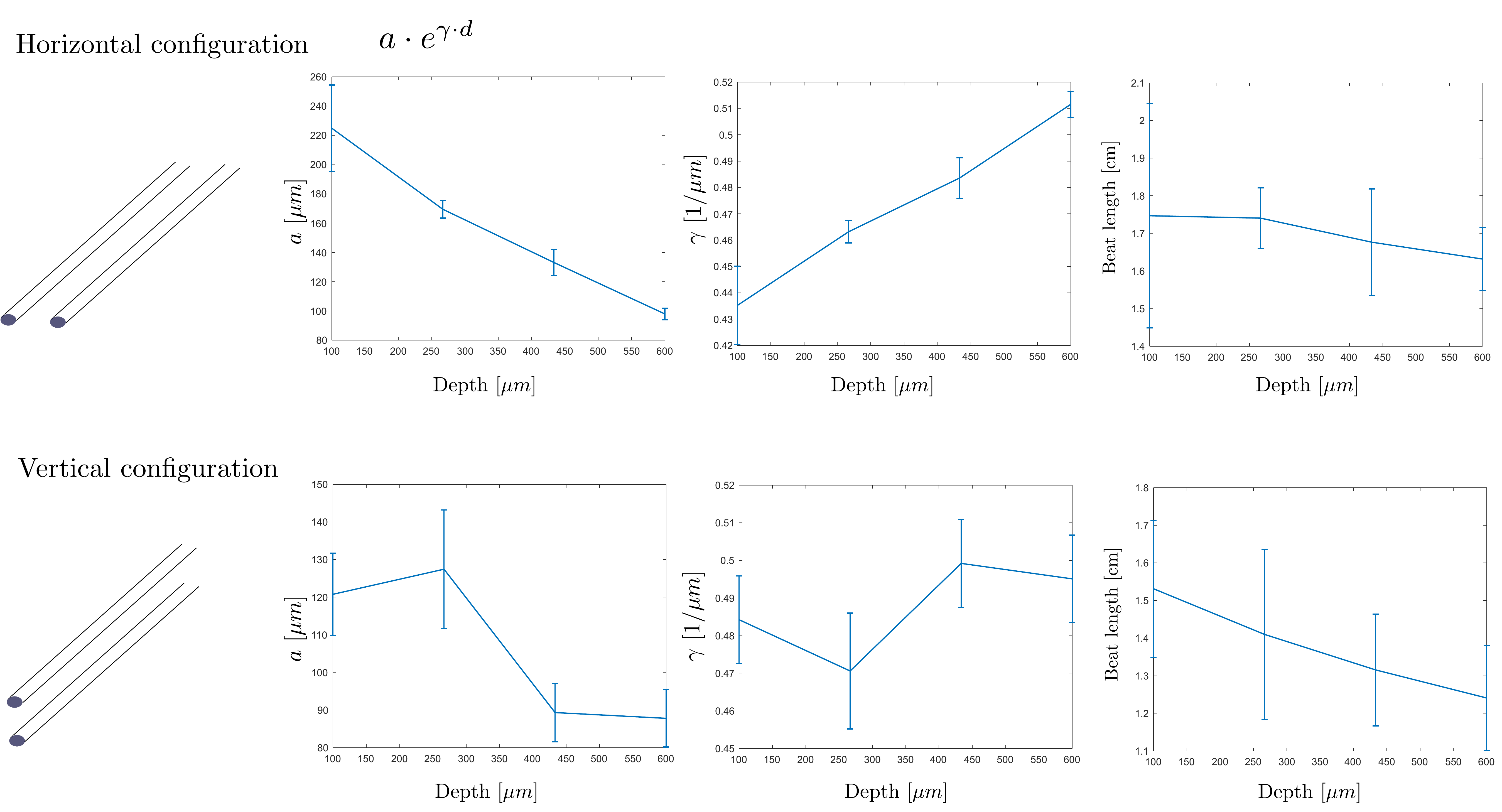}
	
	\caption{Plots of the fitting parameters of the exponential fit of the beat length, shown in \ref{fig:ExpFitFS}, as a function of writing depth. Horizontal and vertical configurations are shown. The last column shows the beat length at a waveguide distance of 10 $\mu m$.}
	
	\label{fig:ExpFitDepthDep}
\end{figure}

In Figure \ref{fig:ExpFitDepthDep} the depth dependence of the exponential fitting parameters is shown. As we can see there is a smooth dependence of the parameters as a function of depth, allowing us to linearly interpolate the respective parameter between adjacent depths. 

\begin{figure}[h!]
	\centering
	\includegraphics[width=0.7\textwidth]{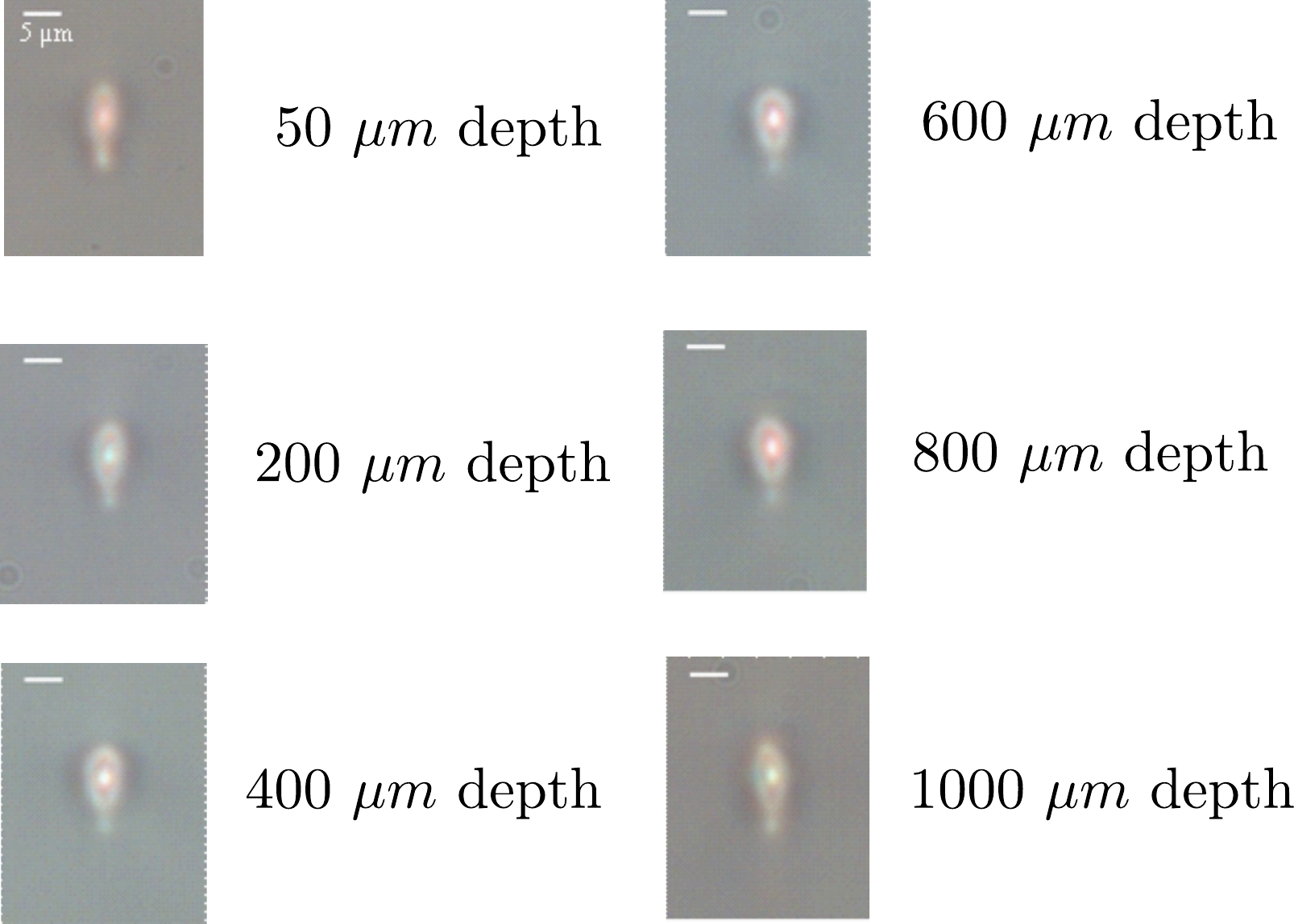}
	
	\caption{Microscope images of the waveguides (waveguide going into the plane).}
	
	\label{fig:WGImDepth}
\end{figure}

In Figure \ref{fig:WGImDepth} microscope images of the waveguides at various depths are depicted. The width of the whole structure is on the order of 5 $\mu m$ with a smaller core region. %
We note that the coupling is asymmetric in horizontal and vertical directions. This is due to the elliptical shape of the waveguide profile. This can be clearly seen in the microscope images shown in Figure \ref{fig:WGImDepth}. This asymmetry leads to a mode profile that is elliptical, with the long axis in the vertical direction. The beat length is measured to be smaller for the vertical configuration (Figure \ref{fig:ExpFitDepthDep}, rightmost column), confirming a stronger coupling of the waveguides in this configuration. We can also see a slight trend of increasing coupling strength as a function of depth. 

\section{Simulating the lattice Hamiltonian}
In the previous section we have discussed the characterisation of coupling strengths between waveguides at different depths.
We can now use these data to reconstruct the coupling matrix for a given spatial arrangement of waveguides. This allows us to predict the mode structure we can expect in the experiment.
\begin{figure}[h!]
	\centering
	\includegraphics[width=1\textwidth]{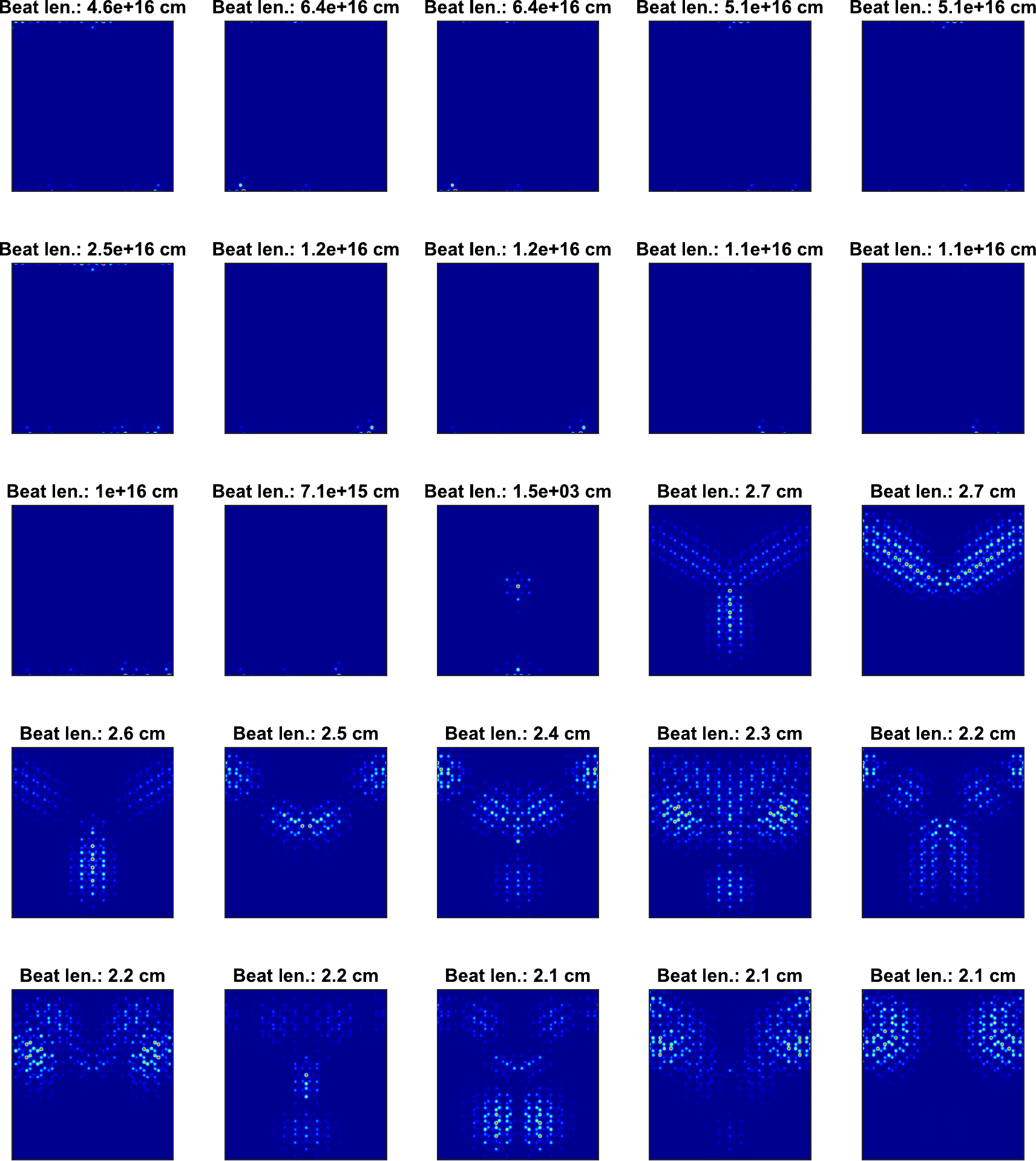}
	
	\caption{Spatial modes and eigenvalues of the 25 lowest lying modes. The desired zero mode is in the centre panel with a beat length of 15 meters. }
	
	\label{fig:Modes}
\end{figure}
The 25 lowest lying eigenmodes of a lattice with a vortex distortion giving rise to a localised zero mode are shown in Figure \ref{fig:Modes} (nearest neighbour coupling only). From this modelling I was able to correct the real lattice geometry such that it exhibits the zero mode. I choose to optimise the value of the zero mode to be as close to zero as possible. I can achieve a sufficiently low energy zero mode by stretching the lattice by around two to five percent in the vertical direction. This compensates for the slight mode-ellipticity in the vertical direction as discussed in the previous section. If we did not apply this correction, the zero mode would additionally be slightly vertically stretched.%
\clearpage
\section{Conclusion}
In this chapter I have presented to the reader the experimental methods employed to fabricate photonic crystals. I then discussed an experimental method I conceived, based on a spatial light modulator, which allows for the excitation of modes in the photonic crystal that span multiple waveguides. This method offers a great flexibility in exciting integrated photonic structures. I have also utilised the experimental setup to characterise large sets of integrated components to determine the coupling behaviour of a photonic crystal lattice. The high degree of automation and flexibility in the setup presented here offers an exciting perspective for further use cases. In Appendix \ref{app:FSwriting} I also discuss how the SLM based method can be used to determine the unitary transformation of large integrated photonic circuits. I anticipate that this experimental method will also enable further studies of topological features in a wide variety of photonic structures.
\clearpage 

%% file: chapter7.tex
\chapter{Experimental demonstration of topological zero modes in photonic graphene} 
\label{chap:ExpTopModes} The work in this chapter is in parts based on a paper in preparation for publication \cite{menssen2019photonic}.
\section{Introduction}
Topological photonics sheds light on some of the surprising effects seen in condensed matter physics that arise with the appearance of topological invariants \cite{teo2010topological}. Optical waveguides provide a well-controlled platform to investigate effects that relate to different topological phases of matter, providing insight into phenomena such as topological insulators and superconductors by direct simulation of the states that are protected by the topology of the system.  Whereas previous work \cite{rechtsman2013topological,plotnik2014observation,stutzer2018photonic,noh2018topological,hafezi2013imaging}  has focused on realising protected edge states in topological insulators, here, we observe a topologically protected mode within the bulk of a 2+1D photonic material by introducing a vortex distortion into a graphene lattice \cite{iadecola2016non,jackiw1981zero,hou2007electron}. The observed modes lie mid-gap at zero energy. This is the first experimental demonstration of a mode that is a solution to the Dirac equation in the presence of a vortex defect, as proposed by Jackiw and Rossi \cite{jackiw1981zero,Jackiw2007}. Further, we show adiabatic transport of the mode as the vortex is moved, as well as  robustness of the mode to imperfections in the waveguide lattice. These vortex modes exhibit non-abelian exchange statistics \cite{lo1993non}, similar to Majorana bound states \cite{teo2010topological}, and could therefore be used to realise a simulation of quantum logic gates in a topological quantum computer \cite{nayak2008non}. Our research promises to open new avenues in the rapidly developing field of topological photonics \cite{ozawa2018topological} and provide new insights in the physics of topological solid state systems. 
\section{The Jackiw-Rossi model in photonic graphen}
By virtue of the analogy between light propagating through a photonic crystal and the tight binding Hamiltonian of an electronic system, topological effects can be observed in their photonic counterpart. A photonic platform allows for a high degree of control over the system parameters and is thus a powerful tool to investigate topological effects. In solid state systems this degree of control is notoriously difficult. \\
We demonstrate a stationary zero mode in the bulk and show full control over the topological mode by adiabatically translating it on the 2D surface of the graphene lattice. We also demonstrate it is topologically protected against random errors.\\
Enabled by major advances \cite{Huang2016} in femtosecond-laser waveguide-fabrication technology and by implementing a new experimental method to excite modes that are spread over many lattice sites, we are able for the first time to study the spatial features of photonic zero modes and benchmark their topological protection in a photonic crystal. 
Our photonic crystal is fabricated by writing waveguides in a glass substrate with a femtosecond laser.

\begin{figure}[h!]
	\centering
	\includegraphics[width=0.55\textwidth]{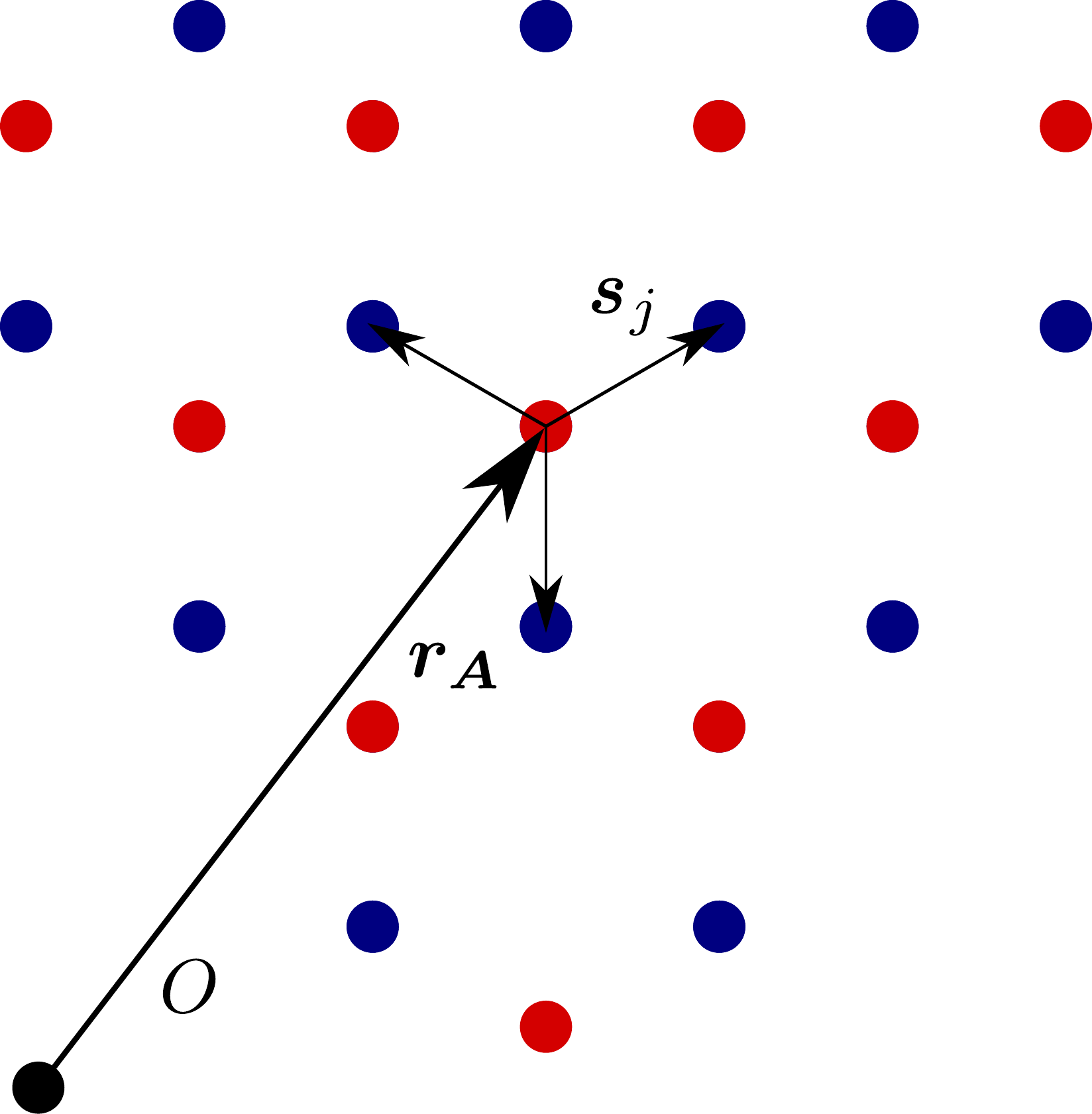}
	\caption{Graphene lattice with two triangular sublattices in red (A) and blue (B), lattice vector $\bm{r}$ and vectors between nearest neighbours $\bm{s}_j$ indicated.
		Hopping of electrons between lattice sites is analogous to photons hopping between the waveguides comprising the lattice. }
	\label{fig:sublat}
\end{figure}

The hexagonal lattice can be decomposed into two different triangular sublattices A and B as shown in Figure \ref{fig:sublat} in red and blue respectively. The $\bm{s}_j$ are vectors connecting a site from sublattice A to its nearest neighbours, which lie on sublattice B.

It has recently been proposed \cite{iadecola2016non,hou2007electron} that that effect of  introducing a distortion in the coupling between lattice sites, $\delta t_{\bm{r},\bm{s}_j}$, which is a function of a spatially varying complex order parameter $\Delta(\bm{r})$ to the tight binding Hamiltonian of the honeycomb lattice, gives rise to localised topological modes. The Hamiltonian reads in terms of lattice coordinates:

\begin{eqnarray} \label{eqn:vorham}
H_{\bm{r},\bm{r}+\bm{s}_j}=-t-\delta t_{\bm{r},\bm{s}_j}\\
\delta t_{\bm{r},\bm{s}_j}=\frac{1}{3}\Delta(\bm{r})e^{i\bm{K}_+\bm{s}_j}e^{2i\bm{K}_+\bm{r}}+\text{\text{c.c.}}.
\end{eqnarray}
$\bm{K}_+=(4\pi/(3\sqrt{3}a),0)$ points to a Dirac point in the reciprocal lattice of the undistorted graphene lattice with lattice constant $a$, $t$ is the nearest neighbour hopping strength.
It was first shown in \cite{chamon2000solitons} that a modulation of this form couples the two Dirac points, leading to the Hamitonian in momentum space (equation \ref{eqn:fullHamilt}), as was discussed in Chapter \ref{chap:TheoryTopModes}.
For a constant $\Delta_0(\bm{r})=\Delta_0$, a band gap at the Dirac points in graphene opens, realising the dispersion relation of a Dirac fermion with mass $\Delta_0$. $\epsilon(\bm{p})=\pm \sqrt{\bm{p}^2+\Delta_0^2}$.
If we introduce a position dependent order parameter which contains a vortex, we obtain a localised mode at the centre of the vortex

\begin{eqnarray}
\Delta(\bm{r})=\Delta_0(\bm{r})e^{i(\alpha+N\text{arg}(\bm{r}-\bm{R}_0))}\\
\Delta_0(\bm{r})=\Delta_0\tanh(|(\bm{r}-\bm{R}_0)|/l_0).
\end{eqnarray}

The $\text{arg}(\bm{r}-\bm{R_0}))$ signifies the polar angle of the connecting vector between the centre of the vortex $R_0$ and the lattice site at position $\bm{r}$. $N$ is called winding number; in this work we set its value to 1. The sign of the winding number determines which sublattice supports the mode ($+1$  for sublattice B, $-1$ for sublattice A). $l_0$ is the width of the vortex. The mode is confined to sublattice B whereas the distortion is applied to sublattice A. 
These topological modes are lattice analogues of Dirac fermions which acquire mass through the coupling to a scalar field \cite{jackiw1981zero} (cf. Higgs mechanism). If that field contains a vortex this additionally leads to charge fractionalisation and non-abelian exchange symmetry \cite{hou2007electron}. $\Delta(\bm{r})$ in the lattice theory takes the role of the scalar field. The solution of the Dirac equation of a single electron in the presence of a vortex $\Delta(\bm{r})$ is, as previously discussed \cite{jackiw1981zero,hou2007electron,Jackiw2007,iadecola2016non}
\begin{equation}
u(r)\sim e^{i(\alpha/2-\pi/4)}e^{-\int_{0}^{r}dr'\Delta_0(r')}.
\label{eqn:zmode}
\end{equation}
For photons in a lattice, the wavefunction is additionally modulated by $e^{i\bm{K}_+ r}$ giving the expression for the electrical field strength as a function of the lattice vector \cite{iadecola2016non}

\begin{equation}
E(\bm{r},z,t)\sim \text{Re}[(u(r)e^{i\bm{K}_+\bm{r}}+ \text{c.c.})e^{ik_\omega z-i\omega t}],
\end{equation}
where $\bm{r}$ points to a lattice site on the supporting sublattice B. 
The shape of a vortex $\Delta(\bm{r})$ and the mode of the is visualised in Figure \ref{fig:LatticeGeomTheoComp} a). The perturbations $\delta t_{\bm{r},\bm{s}_j}$ of the graphene coupling Hamiltonian which realise the vortex are implemented by small shifts in the wave-guides’ positions, assuming exponential decay of the field away from the waveguides. The exponential decay was measured experimentally for different waveguide positions and relative orientations. These data allow us to implement the Hamiltonian faithfully. In \cite{chamon2000solitons,iadecola2016non} an analytic expression for the waveguide displacement implementing the Hamiltonian \ref{eqn:vorham} is given.
For a vector $\bm{r}_A$ on sublattice A of the undistorted graphene lattice and the displacements $\bm{u}_A(\bm{r})$, the displaced vector is given by: $\bm{r}_A'=\bm{r}_A+\bm{u}_A(\bm{r})$,
where
\begin{equation}
\bm{u}_A(\bm{r})=\frac{i}{2}\xi\Delta(\bm{r})e^{-i\bm{K}_+\bm{r}}\binom{1}{i}+\text{c.c.}
\end{equation}
and $\xi$ a constant of units $[\text{length}^2]$.
A plot illustrating relative shifts of the waveguides from the hexagonal lattice configuration is shown in Figure \ref{fig:LatticeGeomTheoComp} b). It is important to note here, that it is the the collective effect of small distortions applied to every lattice site which gives rise to the topologically confined mode at the centre of the vortex defect.

\section{Experiment}
In order to excite the topological mode, we developed a method based on a SLM (Spatial Light Modulator) to simultaneously excite multiple waveguides. We succeed for the first time to excite an optical mode consisting of more than a dozen sites by illuminating multiple waveguides with beams of independently controlled phase, amplitude and mode shape. This approach provides a more direct and complete excitation of the mode than has been previously possible by the expedient of exciting a single lattice site at the centre of the topological mode. 
The photonic crystal consists of 1192 waveguides that make up the hexagonal graphene lattice. The distance between lattice sites is around 10 $\mu m$. This is sufficiently large to ensure suppression of next nearest neighbour coupling. We estimate the ratio of nearest to next nearest neighbour coupling strength to be less than $5\%$. The lattice has a rectangular boundary of approximately 400x400 $\mu m$. 
\begin{figure}[h!]
		\centering
		\includegraphics[width=0.9\textwidth]{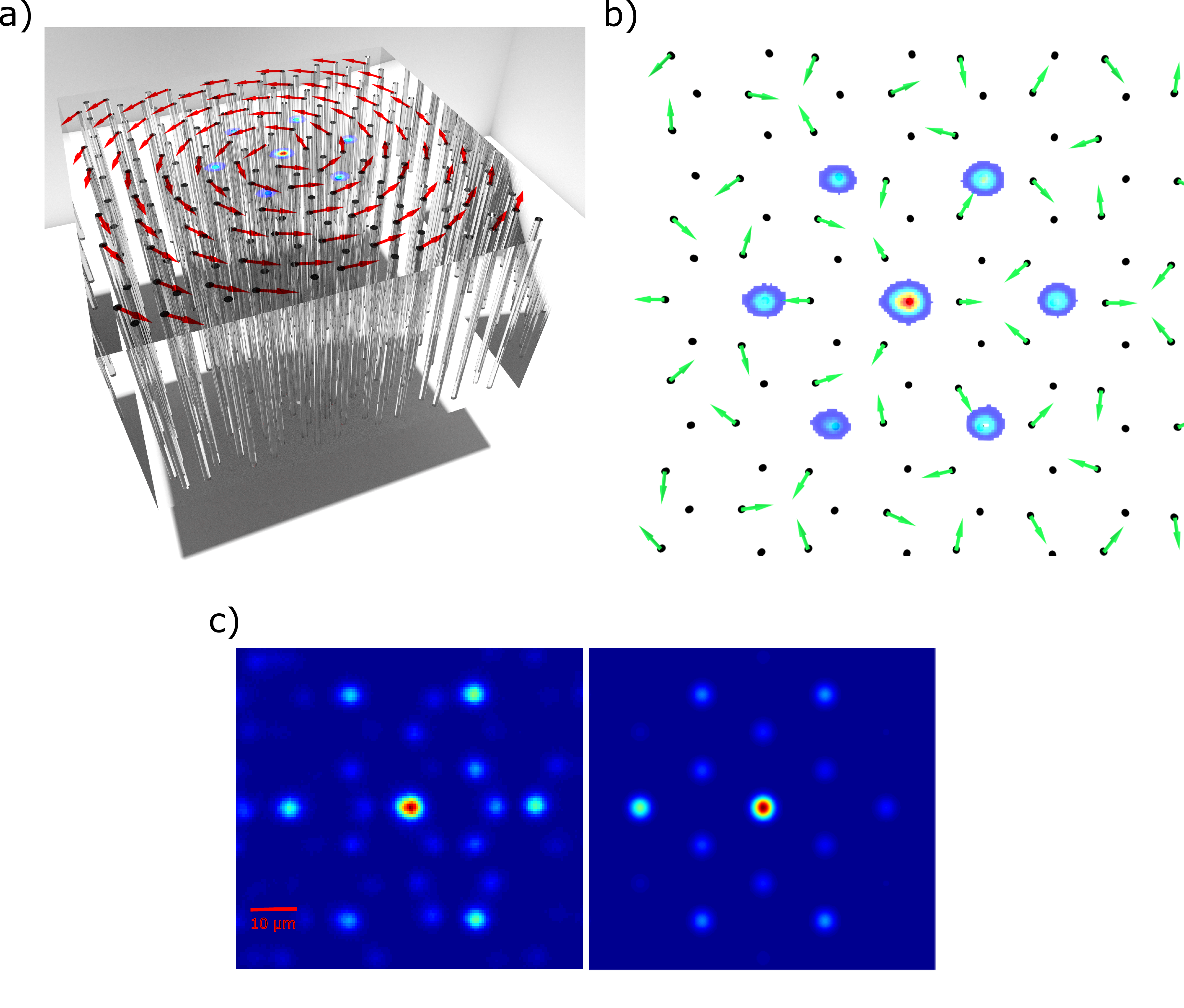}
		
		\caption{a) Photonic crystal lattice with $\Delta(\bm{r})$ represented by red arrows. Zero-mode intensity for dominant modes indicated at the centre of the vortex. The phase offset is $\alpha=\pi/2$. For visualising $\Delta(\bm{r})$, the complex plane has been identified with 2D surface of the crystal. b) Shift of the waveguides from the graphene lattice configuration which implements the topological vortex distortion. The direction of the shift is indicated by green arrows. c) Left: Experimental result for a stationary topological mode Right: theoretical simulation. The mode has been convolved a with a Gaussian to approximate the modes of the waveguides. The ratio of light intensity in the centre waveguide to the averaged intensity in a waveguide lying on the outer bright hexagon is approximately 3:1, consistent with the expected mode decay of the vortex-bound wavefunction (equation \ref{eqn:zmode}).
		}
		
		\label{fig:LatticeGeomTheoComp}
\end{figure}

We excite up to 13 of the waveguides carrying appreciable intensity of the zero mode. Coupling to the zero mode is optimised by varying the input phases and amplitudes of the exciting beams. 
In a first experiment we demonstrate a stationary topological mode. The vortex distortion is located at the centre of the lattice. In Figure \ref{fig:LatticeGeomTheoComp} c), comparing the experimental result (left) to the theoretical calculation (right), we can see that the shape of the simulated mode and the experimental result match very well. The brightest peaks are in each case the central one and six peaks that form a large hexagon around the central mode. Most of the intensity is confined to the sublattice which carries the topological mode, as designed. In order to quantify the degree to which this intensity pattern represents the zero-order mode, we introduce the ratio of light intensity between the two sublattices
$\gamma_{AB}$=(Light intensity in sublattice B)/(Light intensity in sublattice A)
as a measure of fidelity for the excitation of the mode, which should be confined to one sublattice only. For a mode supported by sublattice B, we expect $\gamma_{AB}>1$.  The measured mode displayed in Figure 2 c) has $\gamma_{AB}=5.9$.
The light in the zero mode is tightly confined to the centre of the vortex and decays quickly outside the radius of the vortex $l_0=20$ $\mu m$.

\begin{figure}[h!]
	\centering
	\includegraphics[width=1\textwidth]{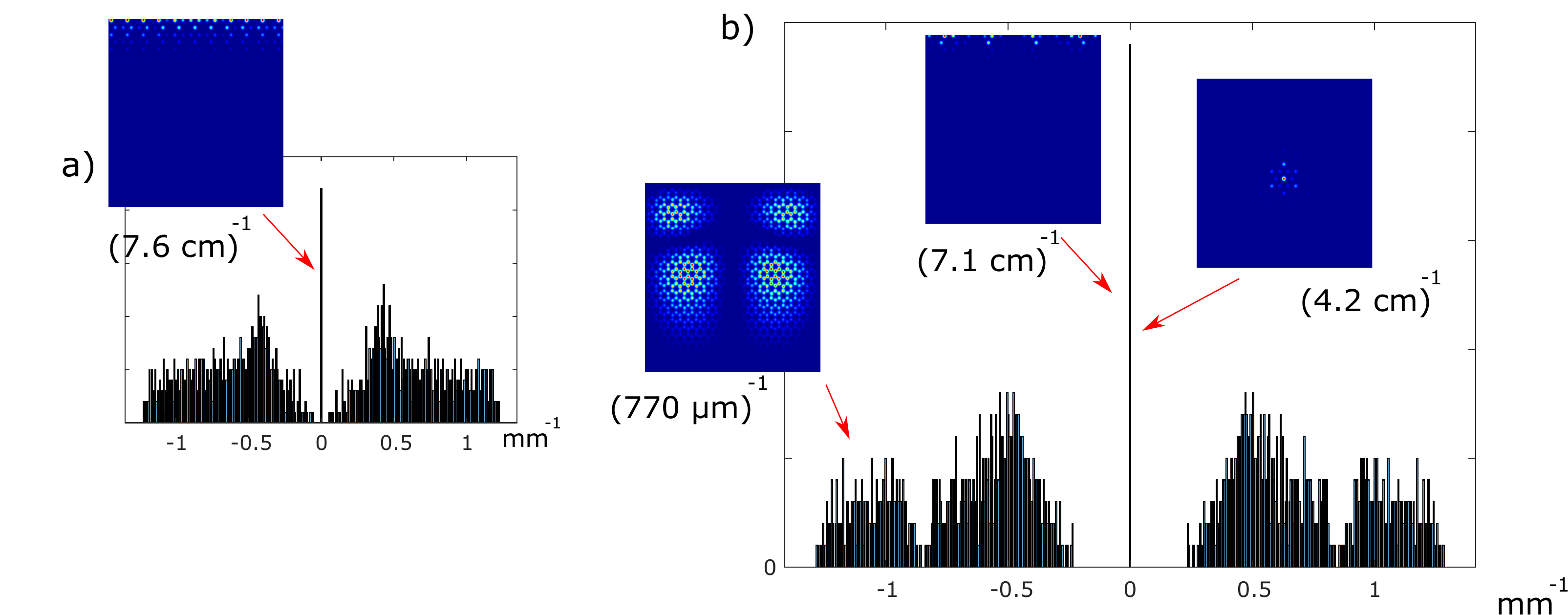}
	\caption{a) Numerically calculated density of states of a graphene lattice with 1192 sites. b) Density of states with central vortex distortion. The energy is given in terms of the beat length of the mode. Inset images of eigenmodes at different energies are shown. The energy of the zero mode corresponds to a beat length of several centimetres. In contrast, high energy bulk modes oscillate in the range of less than a millimetre. 
	}
	\label{fig:BandStructure}
\end{figure}

Next, we numerically calculate the energy spectrum of the real lattice Hamiltonian from the waveguide positions, using experimental data to obtain the coupling strengths. Comparing \ref{fig:BandStructure} a) and b), we can clearly see a band gap opening around zero energy, as the vortex distortion is introduced to the graphene lattice. The topological mode has zero energy and lies together with the edge-states at the centre of the band gap. The energy of the zero-mode is not exactly zero due to residual next nearest neigbour coupling, finite lattice size and other effects. In terms of the oscillation length it is on the order of a few centimetres for a lattice of 1192 sites.

When there is a zero mode present this also gives rise to a corresponding state at the edge of the lattice.%

\begin{figure}[h!]
	\centering
	\includegraphics[width=0.7\textwidth]{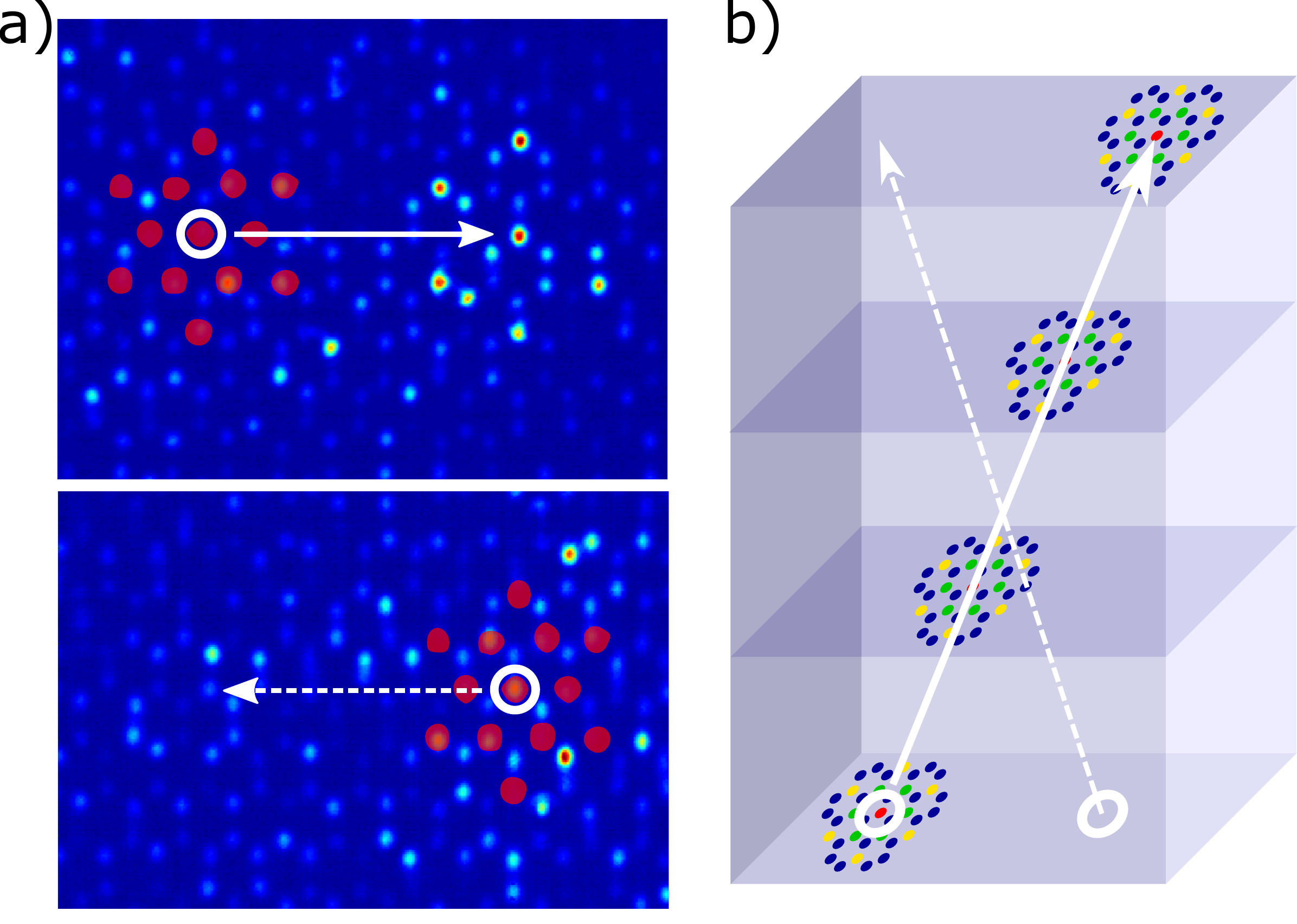}
	\caption{a) Top: mode is translated from left to right on the lattice. The circled waveguide marks the position of the centre of the input light field. We optimise the phases and amplitude of the input light across 13 waveguides around the white circle and optimise on the amount and confinement of light at the expected output (centre of the vortex at the output indicated by arrow). We observe light transport from left to right as the light emerges at the output vortex in the correct sublattice, $\gamma_{AB}=4.7$. Bottom: Same region as above, mode is excited on the other side of the lattice, where no vortex is present. The circled waveguide again marks the centre of the input light field as in the top picture. $\gamma_{AB}=0.67$ in the bottom picture.  b) Illustration of mode shifting by adiabatically translating the centre of the vortex as a function of depth z. Image credit: D. Felce and A. Menssen.
	}
	\label{fig:ComboShifty}
\end{figure}

Next, we translate zero mode across the lattice by adiabatically shifting the vortex distortion from one side of the lattice to the other by around 100 $\mu m$. A chip of 9 cm length is sufficient to ensure adiabaticity as the zero mode is translated. For the mode to move 100 $\mu m$ across the lattice we need at least 4 cm of propagation distance, this was verified by propagating the mode numerically. 
We can observe transport of the zero-mode with most of the transmitted intensity measured confined around the centre of the shifted vortex and in the correct sublattice (with a ratio of $\gamma_{AB}=4.7$) (\ref{fig:ComboShifty} a), top panel). The mode is no longer symmetrically excited - this is due to small variations in fabrication which easily can shift the relative intensities with which waveguides in the supporting sublattice are excited -. In \ref{fig:ComboShifty} a), bottom panel, we illustrate what happens if we try to excite the mode at a position in the lattice where there is no vortex present. We attempt to find a set of input phases and amplitudes that maximises light confinement at an output position indicated by the arrow, however, as expected we fail to excite a mode and see no light transport. The ratio of intensities $\gamma_{AB}=0.67$ is consistent with a random excitation of waveguides.  

\begin{figure}[h!]
	\centering
	\includegraphics[width=0.45\textwidth,height=0.4\textwidth]{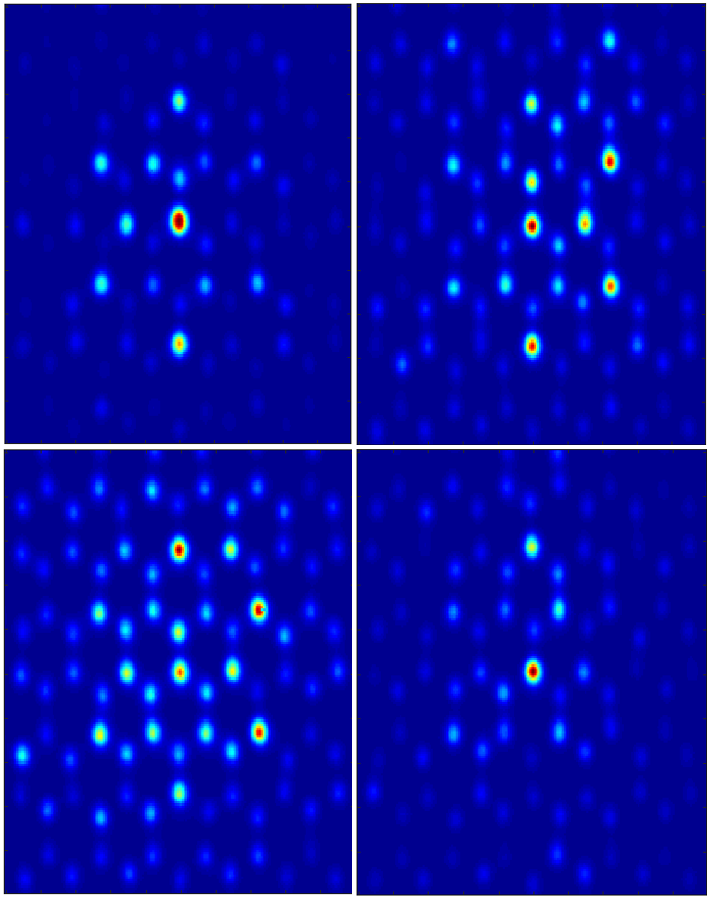}
	\caption{Topological mode with random distortions 
		Top left: no error, Top right: 200 nm, Bottom left: 400 nm, Bottom right: 600 nm. The ratios of light in the correct sublattice to the wrong one are $\gamma_{AB}= 3.8, 2.9, 2.2, 2.1$.
	}
	\label{fig:LatError}
\end{figure}

To demonstrate that the mode is topologically protected against random errors of the lattice, we introduce a distortion to the position of the waveguides by shifting them by a random distance, sampled from a two-dimensional uniform distribution, where the radius $r_d$ of the distribution corresponds to the maximum shift applied. %
In Figure \ref{fig:LatError}, we show four different distortions, $r_d={0,200,400,600}$ nm. The systematic distortion introduced to the hexagonal lattice due to the vortex is of order 800 nm. We can observe that the mode remains visible even for random errors that are on the order of the change introduced by the topological order parameter itself. The larger the distortion the more light leaks into the other sublattice. We measure a steadily decreasing amount of light in the correct sublattice $\gamma_{AB}=\{3.8,2.9,2.2,2.1\}$ as the random distortion increases. 
The protection is a consequence of the symmetries of the Hamiltonian \cite{altland1997nonstandard} and the band gap opening due to the vortex distortion. The zero mode lies far away in energy from other states that might be excited. To illustrate the effect of the distortion we average the output state over several input configurations, all of which have large overlap with the distorted mode. The averaged pictures are shown in Figure \ref{fig:LatError}.

\subsection{Excitation of the topological mode}
\begin{figure}[h!]
	\centering
	\includegraphics[width=0.8\textwidth]{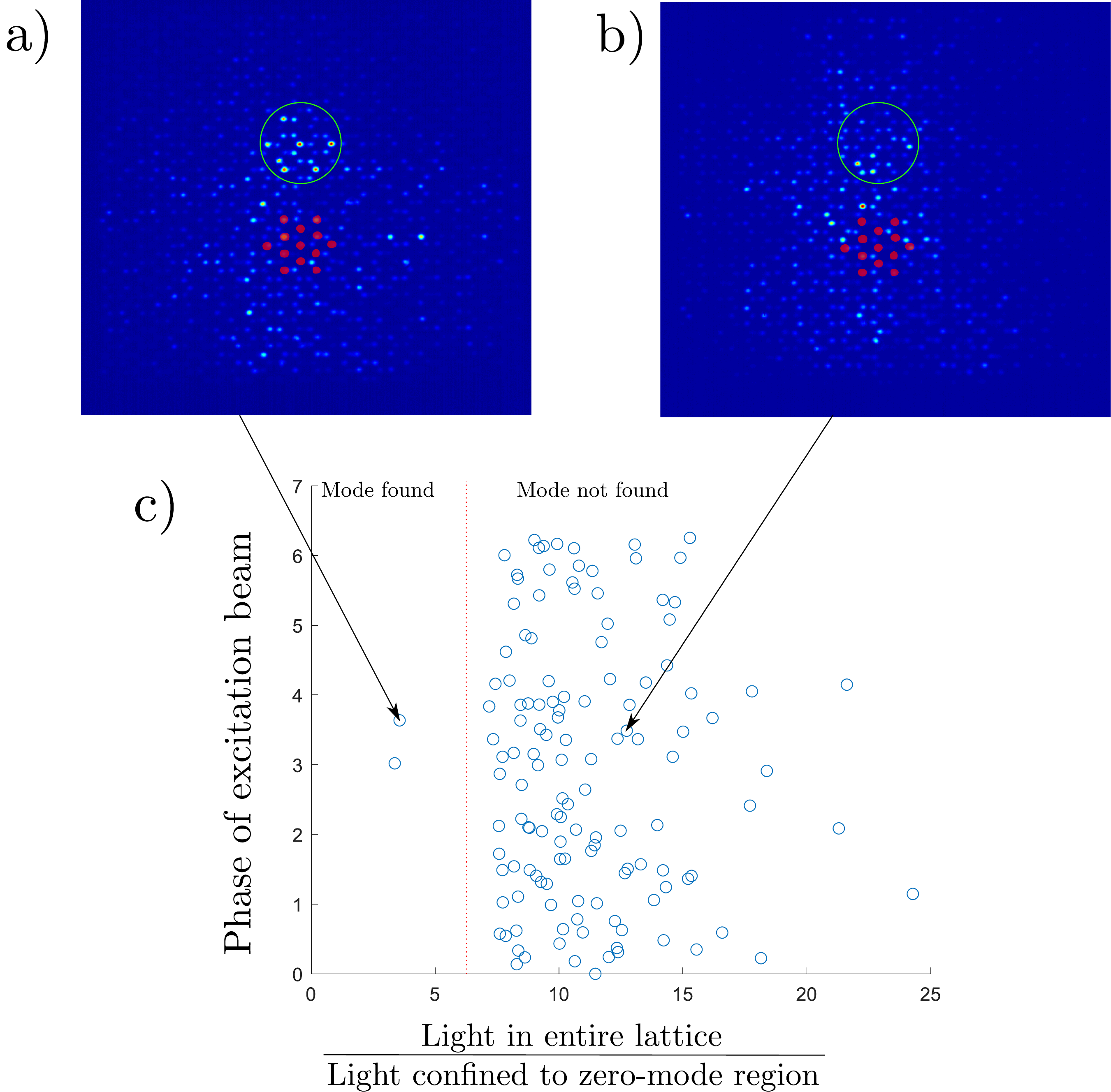}
	\caption{a) Successful excitation of the zero mode. Excitation pattern indicated by 13 red dots. Green circle marks region of confinement for which the light intensity is maximised. The characteristic pattern of the zero mode is visible within the green circle. Value of the optimisation function is below the critical threshold. b) Unsuccessful excitation of the zero mode. Value of the optimisation function is above the critical threshold. No mode structure is visible. c) Phase of one of the exciting beams vs. the ratio of light intensity in the entire lattice and confined within zero-mode region. Points correspond to trials with varying random initial values for the phases and amplitudes passed to the gradient descent algorithm. Threshold indicated by red, dotted line separates values of the optimisation function where the excitation of the zero mode was successful/unsuccessful.}
	\label{fig:ExcitationMode}
\end{figure}

The excite the topological mode, we illuminate up to 13 waveguides around the expected location of the zero mode.
A gradient descent optimisation routine varies the input phases and amplitudes of the beams. The objective of the optimisation is to maximise the light intensity within a set of waveguides at the expected location of the mode at the output of the chip. The waveguides over which the output light is integrated are deliberately chosen on both sublattices, such that we can exclude the possibility of exciting a faux mode simply because the number of available parameters might be large enough to excite a mode that looks like the zero mode.
In Figure \ref{fig:ExcitationMode} an optimisation result is shown. For the adiabatically shifted mode we use 13 input beams, marked in locations in Figure \ref{fig:ExcitationMode} a). The gradient descent algorithm is started with a set of 13 random phases and amplitudes. The optimisation function to be minimised is the ratio of light intensity within the region where the zero mode is expected (green circle in Figure \ref{fig:ExcitationMode} a)/b)), and the light intensity spread over the entire lattice. Each trial either leads to a successful or unsuccessful excitation of the zero mode. Once the mode is found, the procedure is terminated. In Figure \ref{fig:ExcitationMode} c) the phase of a single one of the 13 excitation beams is plotted vs. the value of the optimisation target. Each point corresponds to a trial.  $\sim 150$ trials to excite the zero mode are shown, two are successful. We can clearly see that once the mode is found the value of the optimisation function jumps to a significantly lower value. There is a threshold (marked by the red line in Figure \ref{fig:ExcitationMode}) which marks the light confinement to a specific region, which is achievable without the presence of a localised mode.

\clearpage
\section{Outlook: braiding of two modes}

The authors of \cite{iadecola2016non} have suggested that the multi-particle exchange-statistics of these photonic zero modes is non-abelian. I will be showing first efforts aimed towards demonstrating their non-abelian statistics.
Non-abelian anyons are quasi particles that appear in low dimensional systems, such as planar solid state materials where particles can only travel in 2 (spatial) + 1 (time) dimensions \cite{leinaas1977theory}\footnote{it has been suggested that in some circumstances \cite{teo2010majorana} there exist non-abelian exciations in 3D materials}; they exhibit exotic exchange statistics which is neither fermionic nor bosonic. Non-abelian anyons have been at the centre of considerable attention due to their potential applications in realising a topological quantum computer \cite{nayak2008non}.
So far we have investigated in great detail a single topological zero-mode. 
In this section I will document first attempts at investigating the particle exchange statistics of two of these modes as proposed in \cite{iadecola2016non}. Following their proposal, I attempt an implementation of a braiding operation on two photonic zero modes. We have previously stated that these modes should exhibit non-abelian exchange statistics. They are related to Majorana bound states that occur in superconductors \cite{chamon2010quantizing,teo2010topological,milovanovic2008fractionalization}.
\subsection{Simulating the braiding operation}
I simulate the braiding of two modes using the data measured for the coupling coefficients to reconstruct the coupling matrix as discussed in Chapter \ref{chap:ExpTopModes}. Time propagation is achieved by implementing the unitary evolution operator for a small length step \cite{iadecola2016non} (Appendix)
\begin{equation}
\mathcal{U}(L)=\prod_{n=0}^{L/\delta l}\exp(-iH(n\delta l)\delta l).
\end{equation}
$L$ is the length of the chip and $\delta l$ is the distance between adjacent layers in the depth of the chip. 
To realise the evolution of the mode, at each length $\delta l$ the waveguide configuration is adjusted to accommodate the new position of the two vortices. The waveguides in between are linearly interpolated. In the present set I use 19 time steps over a lattice of 6 cm length.

\begin{figure}[h!]
	\centering
	\includegraphics[width=0.8\textwidth]{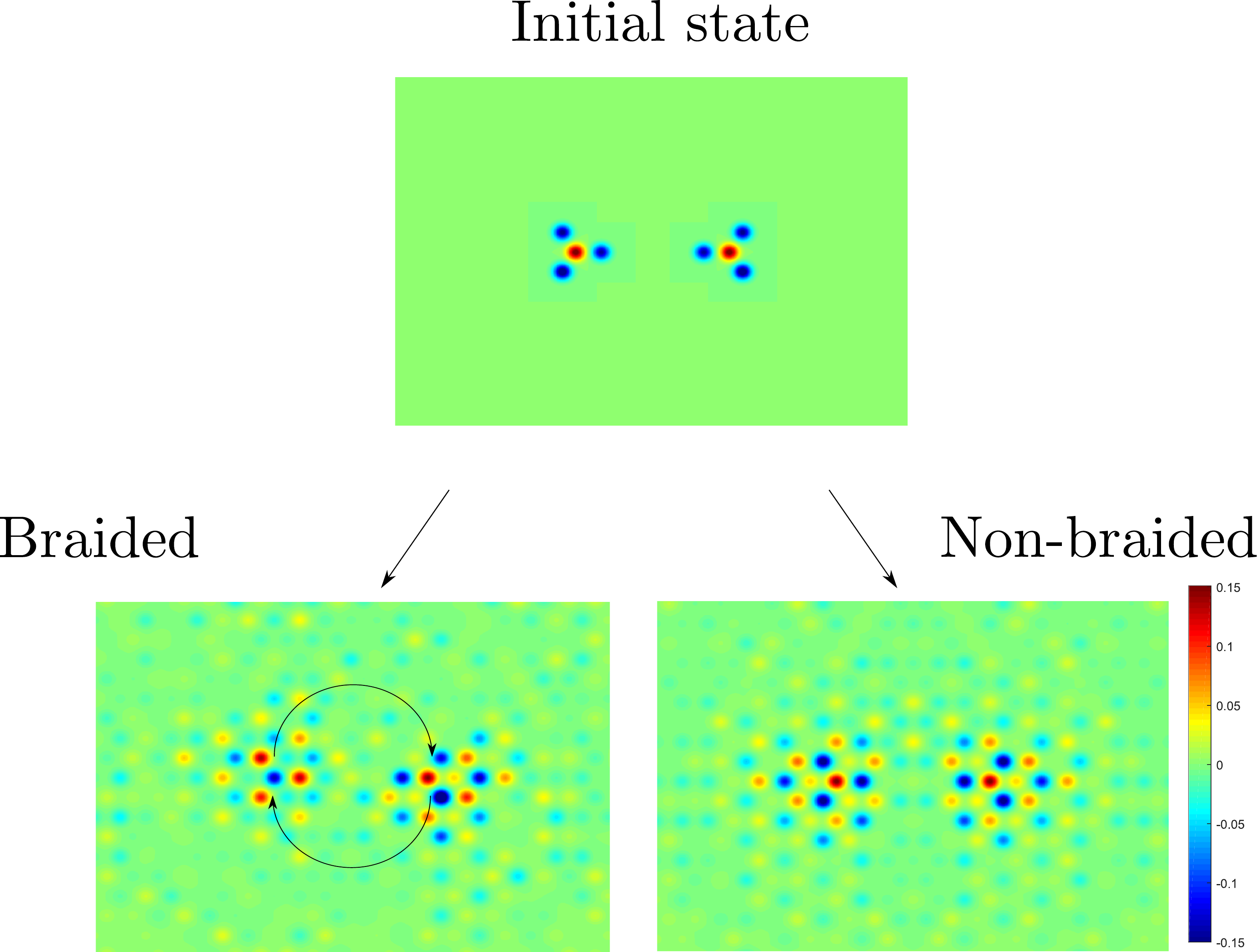}
	\caption{Comparison of two zero-modes as they are either braided (bottom left panel) or propagate along a straight path (bottom right panel). Shown is the real part of the wavefunction. The phase difference between sites at the braided vortex is $\pi$, while there is no phase difference in the non-braided case.}
	\label{fig:BraidedModes}
\end{figure}
The propagated initial state is then
\begin{equation}
{\bf v}_{final}=\mathcal{U}(L){\bf v}_{initial}.
\end{equation}
We notice in Figure \ref{fig:BraidedModes} that in the braided case there is a phase of $\pi$ imparted onto the left mode, while the right mode does not change in phase. It was argued in \cite{stern2004geometric,iadecola2016non} that these kind of phases are a building block for non-abelian statistics. %

We consider a two-particle state with photons occupying either of the two vortex zero-modes (``left'' or ``right'' (L/R)).
The state in one vortex is a superposition of being occupied by a single photon, or vacuum
\begin{equation}
\frac{1}{\sqrt{2}}(\ket{0}_{L/R}+\ket{1}_{L/R}).
\end{equation}
The ket represents the number occupation of the vortex zero mode.
For two vortices the initial state will be%
\begin{equation}
\ket{\psi}=\frac{1}{\sqrt{2}}(\ket{0}_L+\ket{1}_L)\otimes\frac{1}{\sqrt{2}}(\ket{0}_R+\ket{1}_R).
\end{equation}
Upon a braiding operation, where the two vortices are swapped, as in Figure \ref{fig:BraidedModes}, the state transforms as \cite{iadecola2016non}
\begin{equation}
\frac{1}{\sqrt{2}}(\ket{0}_L+\ket{1}_L)\otimes\frac{1}{\sqrt{2}}(\ket{0}_R+\ket{1}_R)\rightarrow\frac{1}{\sqrt{2}}(\ket{0}_L-\ket{1}_L)\otimes\frac{1}{\sqrt{2}}(\ket{0}_R+\ket{1}_R).
\end{equation}
The bosonic creation operators for a zero mode in the right/left vortex transform as
\begin{eqnarray}
b^\dagger_L\rightarrow b^\dagger_R\\ \nonumber
b^\dagger_R\rightarrow -b^\dagger_L.
\end{eqnarray}
After a second swap the state becomes
\begin{equation}
\frac{1}{\sqrt{2}}(\ket{0}_L-\ket{1}_L)\otimes\frac{1}{\sqrt{2}}(\ket{0}_R+\ket{1}_R)\rightarrow\frac{1}{\sqrt{2}}(\ket{0}_L-\ket{1}_L)\otimes\frac{1}{\sqrt{2}}(\ket{0}_R-\ket{1}_R).
\end{equation}
The initial state before swapping and the final state after a complete exchange are orthogonal.
This is quite different from the case of bosons or fermions, where after swapping the particles twice, the original state is restored.
We remember our requirement of a non-abelian exchange
\begin{equation}
\psi_\alpha\rightarrow U_{\alpha}^\beta\psi_\beta.
\end{equation}
I.e. the wavefunction changes state when the particle exchange occurs.
\vspace{0.5cm }
\begin{spacing}{1}
Here, $U=\frac{1}{2}\left(\begin{matrix}-1&0\\0&1\end{matrix}\right)\otimes\left(\begin{matrix}-1&0\\0&1\end{matrix}\right)$ and $\ket{\psi}=\frac{1}{2}(\ket{00}+\ket{01}+\ket{10}+\ket{11})$.
\end{spacing}
\vspace{0.5 cm}
In contrast, an abelian exchange is characterised by the accumulation of only a global phase in the multi-particle wavefunction
\begin{equation}
\psi\rightarrow e^{i\phi}\psi.
\end{equation}

\subsection{Experimental implementation}
\begin{figure}[h!]
	\centering
	\includegraphics[width=0.9\textwidth]{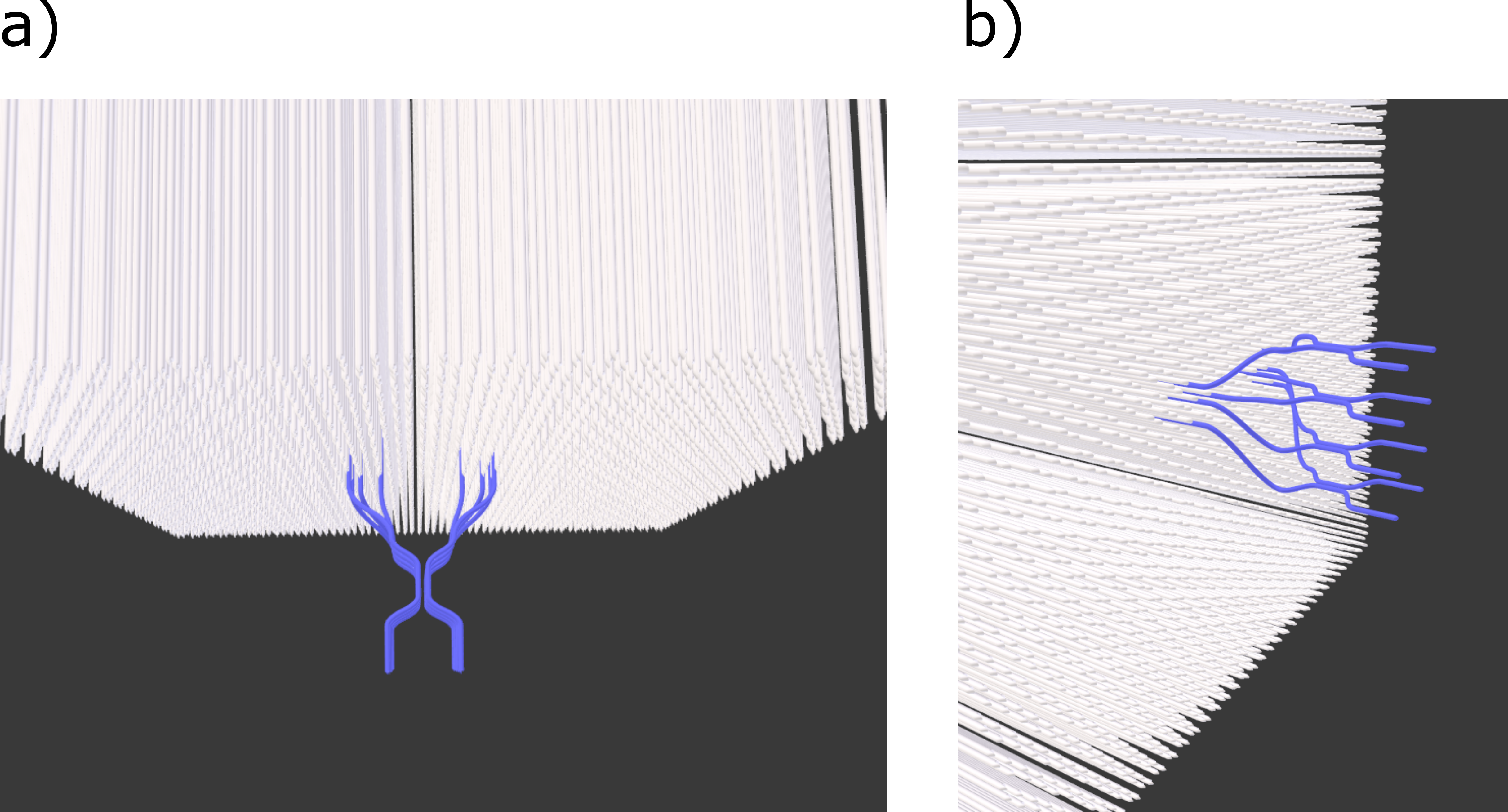}
	\caption{Design for exciting two photonic topological modes using two-mode couplers. lattice waveguides in white, two-mode couplers in blue a) top perspective b) view from the side.}
	\label{fig:LatBraid}
\end{figure}

The experimental challenge lies in coherently exciting the two zero-modes at the input of the lattice and then measuring their relative phase at the output, which in the case that the modes have been braided should be different by $\pi$. The two vortex-modes need to be phase-stably excited at the same intensity. To achieve this I use a set of 50:50 beamsplitters which are inscribed into the chip before the lattice, while the lattice itself is shifted deeper into the chip.
In Figure \ref{fig:LatBraid}, this design is illustrated. A series of four beam-spitters (blue) is used to couple to the zero mode at each site. For each beamsplitter the output arm on the left goes to the vortex mode on the left and the right arm to the vortex on the right hand side. An identical set of beamsplitters is positioned at the output to measure the phase difference. In essence this is a series of Mach-Zender interferometers surrounded by a lattice, where the task is to measure a phase offset on one of the arms.  Using the SLM setup we couple light into one input arm at each beam-splitter and maximise coupling to the mode by maximising throughput to the other side of the lattice. When comparing a lattice where the two modes braid to lattices where the vortices are stationary, we should observe a phase-shift of $\pi$ at the output, the signature of which would be that the light intensity is on the opposite arm of the output beamsplitter, compared to where the modes have not been braided. This is experimentally quite challenging, as this requires a consistent phase across the 9 cm propagation length (6 cm of lattice and 1.5 cm for the couplers on each side of the lattice) of the chip. Even small variations of the waveguides are detrimental.

\section{Conclusion and outlook}
I have demonstrated that we can simulate a bulk topological zero mode in photonic graphene, which binds to a vortex. Through coherent excitation of multiple waveguides we were able to investigate the spatial features of this zero-mode. We could demonstrate that it is topologically protected and showed adiabatic transport. The high degree of control we have demonstrated over these localised, protected topological modes may in the future enable applications such as the protection of information encoded in quantum states \cite{blanco2018topological,rechtsman2016topological} against inevitable fabrication errors in linear optical circuits by injecting multiple entangled particles into different protected vortex modes.  This is also the first experimental realisation of the Jackiw-Rossi model, opening the possibility for further studying models employed in high energy physics, using this platform. Furthermore, a promising use for applications could lie in systems that make use of protected topological modes such as topological lasers \cite{bandres2018topological,harari2018topological}, which have already been demonstrated in edge-states. While experimentally challenging, demonstrating a braiding operation by showing the accumulation of a relative phase, as suggested in \cite{iadecola2016non}, may be feasible. I have presented a possible experimental realisation for such an experiment.

%% file: Conclusion.tex
\chapter{Conclusion and outlook}

\section{Multi-photon interference}
I have studied three- and four-photon interference effects. Are there any new physical effects that can be explored in the investigation of even greater number of photons? 
Generally, the question of computational complexity of partially distinguishable scattering arises and was studied in a recent publication \cite{renema2018efficient}. This work investigates the effects of partial distinguishability in ``boson sampling'' \cite{aaronson2011computational}. It is noted in \cite{renema2018efficient} that as distinguishability is increased, the higher order interference terms do not contribute to the particle statistics. $N$-photon interference terms scale as $r^N$, where $r$ is the modulus overlap between photons (assumed identical for simplification) and for large $N$ and small overlaps $r$ these contributions vanish exponentially in the number of photons. However, in \cite{renema2018efficient} the new degrees of freedom of multi-photon interference that arise from multi-photon phases are not exploited, as all overlap integrals are assumed real. It may be worth studying the implications of these phase degrees of freedom for complexity theory.\\  Further studies could investigate the possibility of solving graph-theoretical problems through multi-photon interference, such as the ``Hamilton Path'' problem, which asks if there is a path that traverses every vertex for a given graph exactly once. Some instances of this problem have been shown to be NP-complete \cite{itai1982hamilton}. Such a graph could be prepared using partially distinguishable photons. However, a potential issue could be the diminishing contribution from higher order interference terms as necessary degrees of distinguishability are introduced.\\
When investigating multi-particle indistinguishability an important task is to verify for a given ensemble if the participating photons are distinguishable or indistinguishable. This question was investigated recently by Viggianiello et. al. \cite{viggianiello2018optimal}. Notably, ``Sylvester'' type interferometers \cite{crespi2015suppression} are identified as an optimal device for discriminating fully distinguishable and indistinguishable states for two photons in four modes. We have previously noted that for four-photon interference the fringe visibility is optimal for a certain phase of the quitter. The visibility of four-photon interference is maximised for the Fourier configuration of $\chi=\pi/2$. In the case of $\chi=0$ the interferometer is represented by a Sylvester matrix and the interference visibility is minimal. An interesting question to investigate could be what optimal indistinguishability tests are for an ensemble of partially distinguishable photons. In \cite{brod2019witnessing} a more general criterion for multi-particle indistinguishability is developed. A state exhibiting ``genuine n-particle indistinguishability'' is defined as a convex combination of an n-photon state with only indistinguishable photons and a state with at least two orthogonal photons, where the contribution from the set of fully indistinguishable photons cannot vanish.
The question arises as to what the relationship between ``genuine n-particle indistinguishability'' as defined in \cite{brod2019witnessing} and multi-photon interference is. Interestingly, as we have shown in Chapter \ref{chap:FourPhot}, photons in orthogonal states can exhibit genuine multi-photon interference. The absence of ``genuine n-particle indistinguishability'' as defined in \cite{brod2019witnessing} does not seem preclude genuine n-particle interference. However, the converse argument holds: when ``genuine n-particle indistinguishability'' has been demonstrated, there will be n-particle interference.  %

\section{Topological modes and light-control using a spatial light modulator}
The experimental SLM based method for exciting complex modes in photonic structures, I conceived, promises a wide range of further studies to be conducted. There are several interesting characteristics of the Jackiw-Rossi model which merit further investigation. It has for example been noted\footnote{Internal communication with Claudio Chamon} that for higher winding numbers, sets of orthogonal, localised zero modes exist \cite{Jackiw1976}. In other words a multimode-optical structure, where all modes have the same dispersion. This is unlike in a multicore fibre, where every mode has a different propagation constant. Further uses of this experimental technique include the characterisation of large unitary interferometers. In Appendix \ref{app:FSwriting} I demonstrate how the control of the light field provided by our experimental method allows for the implementation of an algorithm for the characterisation of unitary interferometers introduced in \cite{rahimi2013direct}.
Currently the light efficiency is $<10\%$. The reason for this is that most of the light is filtered out by the spatial filter. Due to the high losses a single photon use of this setup would be problematic. However, one could conceive to make use of phase SLMs which, in contrast to the Ferroelectric Liquid Crystal (FLC) SLM used here, are better suited to control the phase of the light field for one polarisation. If we use two separate phase SLMs that are placed in the real and Fourier plane respectively, one could achieve low-loss control of position, direction and phase of individual beams. This could enable studying quantum interference effects in large photonic lattices.\\ Currently, most experiments in the field have been limited to classical excitation of photonic topological structures. An exciting prospect would be to take topological photonics into the quantum realm \cite{tambasco2018quantum,blanco2018topological} and investigate non-classical multi-photon interference effects in photonic topological structures. 
For example one could demonstrate protection of quantum information in photonic systems, where qubits are transmitted through multiple topologically protected modes.
\chapter*{Acknowledgements}
I would like to first and foremost thank all the colleagues I have had the pleasure of working with together. Alex Jones, especially, with whom I spent countless hours in the lab working hard and at the blackboard thinking hard. Nothing beats playing hangman in ancient Greek at 4 a.m. while running an experiment. I want to thank David Felce who I had the pleasure of working with on the topological photonics project. I will miss dearly discussing controversial political topics next to an optical table! William Clements, we had the best time discussing physics and trying out all the restaurants in Oxford. Gil for being a great house- and lab mate. Jun Guan, for working with me on the topological modes project and working tirelessly on the waveguide fabrication. I want to thank the post-docs who guided me during my DPhil.: Stefanie Barz, Steve Kolthammer, Helen Chrzanowski, Bryn Bell and Andreas Eckstein. I would like to thank Ian Walmsley for his supervision and the trust he placed in me, giving me the opportunity to pursue research in the tremendous group that is the Ultrafast Group.

Finally, I thank Merton College for providing a home and community and Dr. James Buckee for his generous support of my work.
\afterpage{\blankpageC}

%% file: Appendix.tex
\appendix

\chapter{Work on single photon sources}
To enable the type of multi-photon interference experiments which are discussed in the first part of this thesis, we require a source of single photons. The type of source that I have used in multi-photon interference experiments relies on a non-linear process in silica: Spontaneous Four Wave Mixing (FWM). Two photons from a bright, coherent ``pump'' beam are converted into a signal and an idler photon. The signal photons are detected immediately after their creation and herald the presence of an idler photon in the other spectral mode. The idler photon is then sent into the experiment, a multi-mode interferometer in our case, and measured thereafter. In the following, I will first give a brief introduction to non-linear light generation and four-wave mixing. I do not execute a full derivation but rather motivate and present the final result: the bi-photon state generated in a four wave mixing process. A complete derivation can be found in \cite{eckstein2014high,spring2013chip,posner2018high}. I then present the source setup used for measuring three-photon interference which has been described in \cite{spring2013chip,spring2017chip} as well as a setup I built for the generation of single photons in the telecom range together with measurements of the spectral characteristics of that source.
\section{Four wave mixing}
\label{app:FWM}

The polarisation density $P(t)$ in a medium can be written in terms of the electrical field $E(t)$ and the non-linear coefficients $\chi$ of the material
\begin{equation}
P(t)=\chi^{(1)}E(t)+\chi^{(2)}E(t)^2+\chi^{(3)}E(t)^3+...+.
\end{equation}
We choose $E(t)=E_1e^{i\omega_1t}+E_2e^{i\omega_2t}+\text{c.c.}$, a plane wave electrical field with two frequencies, and examine the second order term in the polarisation $\chi^{(2)}E(t)^2$. Substituting field results in:
\begin{equation}
\begin{split}
\chi^{(2)}E(t)^2=\chi^{(2)}(|E_1|^2+|E_2|^2+E_1^2e^{i2\omega_1t}\\+E_2^2e^{i2\omega_2t}+2E_1E_2e^{i(\omega_1+\omega_2)t}+E_1E_2^*e^{i(\omega_1-\omega_2)t})+\text{c.c.}.
\end{split}
\end{equation}
We can see that the dipole density $P(t)$ oscillates at frequencies of the higher harmonics $2\omega_1$, $2\omega_2$, as well as the sum and the difference of the two input fields $\omega_1+\omega_2$, $\omega_1-\omega_2$. Since the polarisation density acts as a source of the electrical field, light at these frequencies is generated.
\clearpage
In a similar fashion the third order non-linearity $\chi^{(3)}$ gives rise to four wave mixing, where the frequencies of four fields (two pump field as well as signal and idler) $E(t)=E_1e^{i\omega_{p1}t}+E_2e^{i\omega_{p2}t}+E_3e^{i\omega_st}+E_4e^{i\omega_it}+\text{c.c.}$ are combined. The number of different frequencies is much larger than in the previous case. However, due to the constraints imposed by a phase matching condition -the fields generated at every point in the medium need to constructively interfere to contribute an output signal- only a small range of frequencies is allowed. Also, energy conservation holds between the two input pump photons and the two output signal and idler photons: $\omega_{p1}+\omega_{p2}=\omega_s+\omega_i$. We consider here a waveguide source with a single spatial mode. The propagation direction of the light is along the $z$-axis.

The phase matching function $\phi(\omega_s,\omega_i)$ at each point in the waveguide is:
\begin{equation}
\phi(\omega_s,\omega_i)=\int_{0}^{L}dze^{i(k_{s}+k_{i}-k_{p1}-k_{p2})z}=Le^{i\Delta k L/2}\text{sinc}(\Delta k L/2),
\end{equation}
with $\Delta k=k_{s}+k_{i}-k_{p1}-k_{p2}$.

The output state of the signal and idler photons is given by:
\begin{equation}
\ket{\psi_{s,i}}=c\int\int d\omega_s d\omega_i F(\omega_s,\omega_i)a^\dagger(\omega_s)a^\dagger(\omega_i)\ket{0},
\end{equation}
where $F(\omega_s,\omega_i)$ is call the ``joint spectral amplitude'' (JSA). Assuming identical pump fields $\omega_{p1}=\omega_{p2}$ the JSA is given by:
\begin{equation}
F(\omega_s,\omega_i)=\int d\omega_p\phi(\omega_s,\omega_i)\alpha(\omega_p)\alpha(\omega_s+\omega_i-\omega_p),
\end{equation}
where $\alpha(\omega_p)$ is the spectral envelope of the pump beam.
The factorability of the JSA determines the purity of the heralded single photon. If $F(\omega_s,\omega_i)=f_s(\omega_s)f_i(\omega_i)$, the bi-photon state is completely separable and the heralded state is pure. If the state is not factorable, the state has spectral entanglement between the signal and idler.
\section{Spectral data NIR}
I measured signal and idler spectra using a single photon spectrometer. We notice that the idler has a larger background than the signal. 
\begin{figure}[h]
	\centering
	\includegraphics[width=0.6\textwidth]{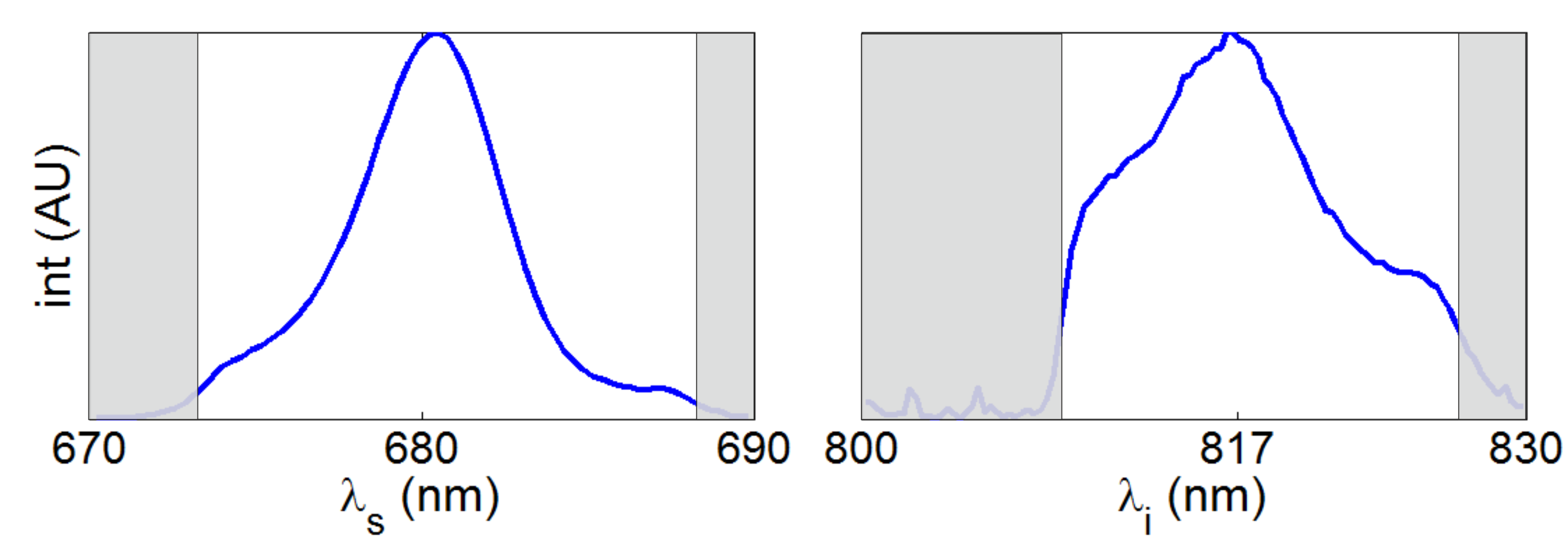}
	\caption{Spectrum of signal and idler photons. Spectral filters indicated by
		shaded area. The pump is at 740 nm. Signal and idler photons are at 680 nm and 817 nm.
	}
\end{figure}
\begin{figure}[h!]
	\centering
	\includegraphics[width=0.6\textwidth]{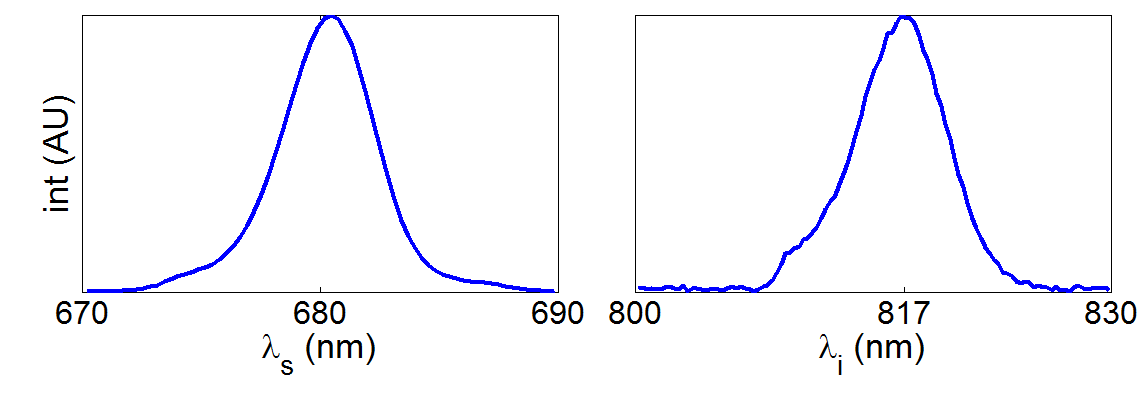}
	\caption{Spectrum of signal and idler photons with background subtraction.
	}
\end{figure}
The heralded $g_H^{(2)}\sim 0.1$ and the unheralded $g_U^{(2)}\sim1.8$ \cite{spring2017chip}.
\section{Experimental setup telecom}
\begin{figure}[h!]
	\centering
	\includegraphics[width=1\textwidth]{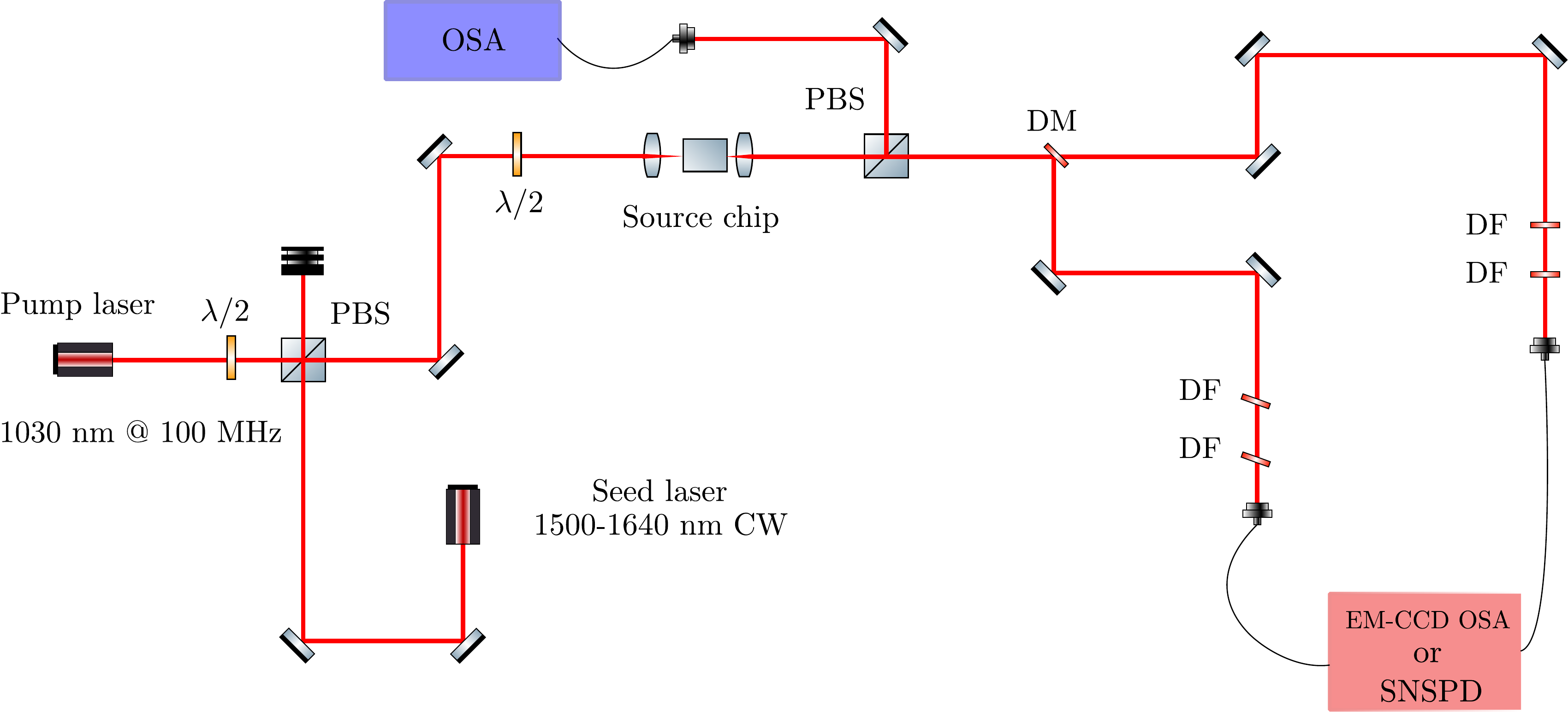}
	\label{fig:Hipster}
	\caption{Experimental setup. DM: dichroic mirror, DF: dielectric filter, EM-CCD: electron multiplying CCD camera, OSA: optical spectrum analyser, SNSPD: superconducting single photon detector.}
\end{figure}

In Figure \ref{fig:Hipster} The experimental setup, realising FWM in the telecom range is depicted.

\section{Data telecom}
\begin{figure}[h]
	\centering
	\includegraphics[width=0.7\textwidth]{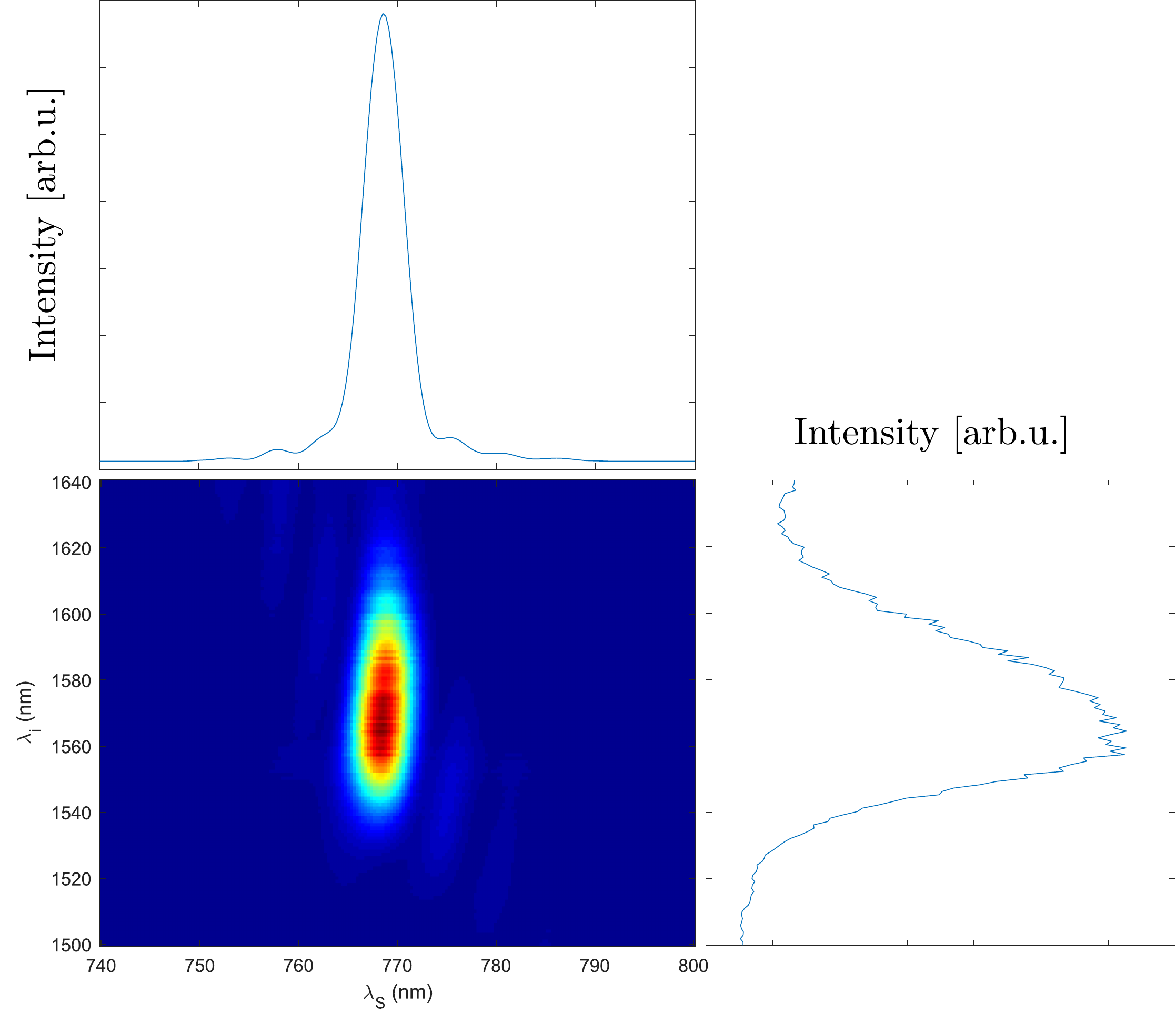}
	\label{fig:SourceSpectr1}
	\caption{Measured JSI with projection of signal and idler photons.}
\end{figure}

I measured the Joint Spectral Intensity ($|F(\omega_s,\omega_i)|^2$) of the FWM process using a method introduced in \cite{eckstein2014high}. Along with the classical pump beam at 1030 nm a second classical ``seed beam'' around the idler wavelength of 1550 nm is inserted into the waveguide. This seed causes in a difference frequency generation process (DFG) between the pump and the seed to enhance emission into the signal mode. Since the JSI is the same for the FWM and the DFG processes, this allows us to probe spectral correlations in FWM. The seed beam is narrow-band and tunable. We scan the frequency of the seed beam across the spectral width of the idler photon and simultaneously record the spectrum at the signal wavelength. In this way we can build up, slice by slice, the JSI. The measurement is shown in figure $\ref{fig:SourceSpectr1}$. We can see in this measurement, with remarkable resolution, the detailed features of the JSI. As anticipated we observe lobes which are a result of the sinc dependence of the phase-matching function.

\begin{figure}[h]
	\centering
	\includegraphics[width=0.7\textwidth]{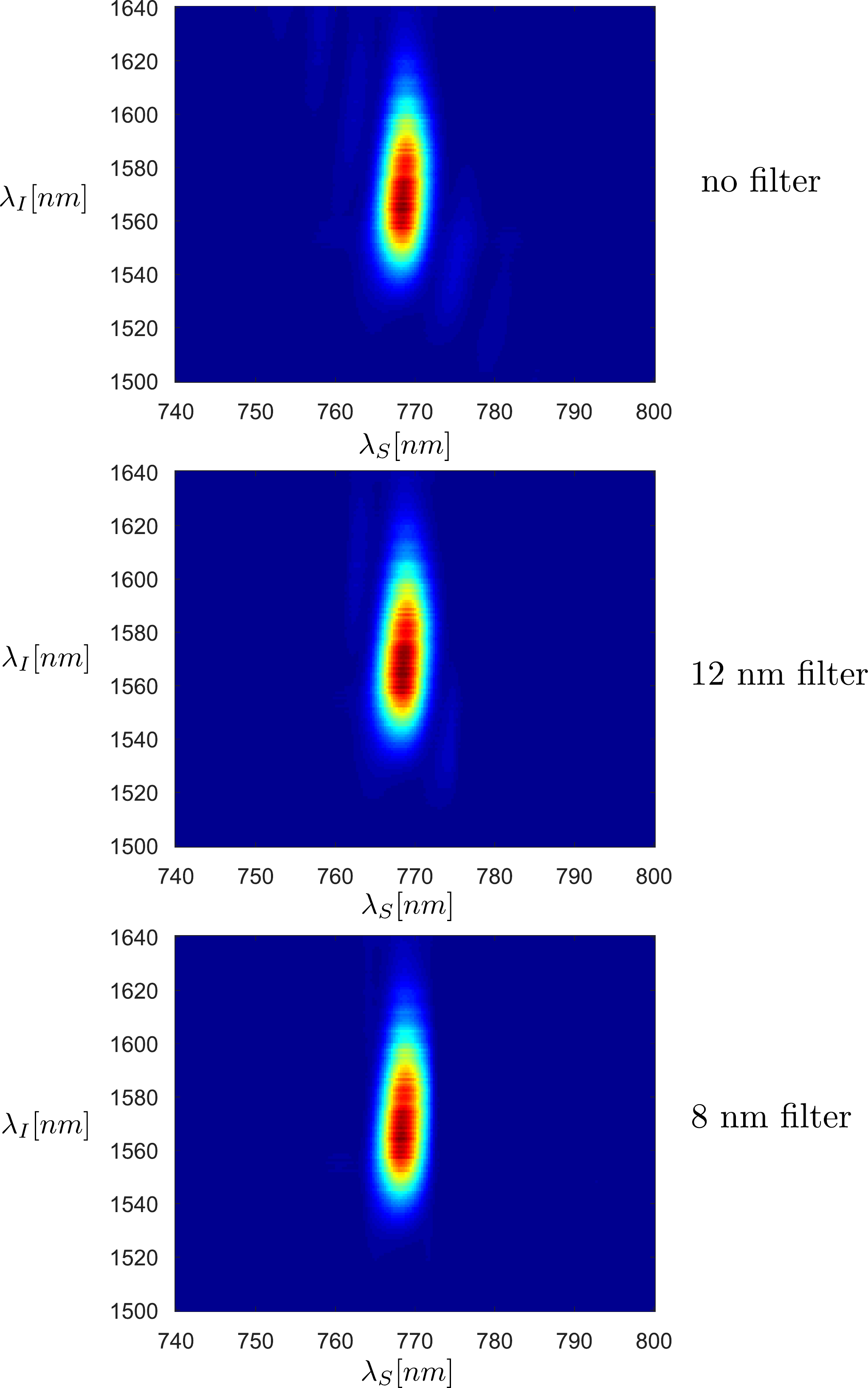}
	\label{fig:JSITelco}
	\caption{Joint spectral amplitudes with different filter-widths.}
\end{figure}

\begin{figure}[h]
	\centering
	\includegraphics[width=0.7\textwidth]{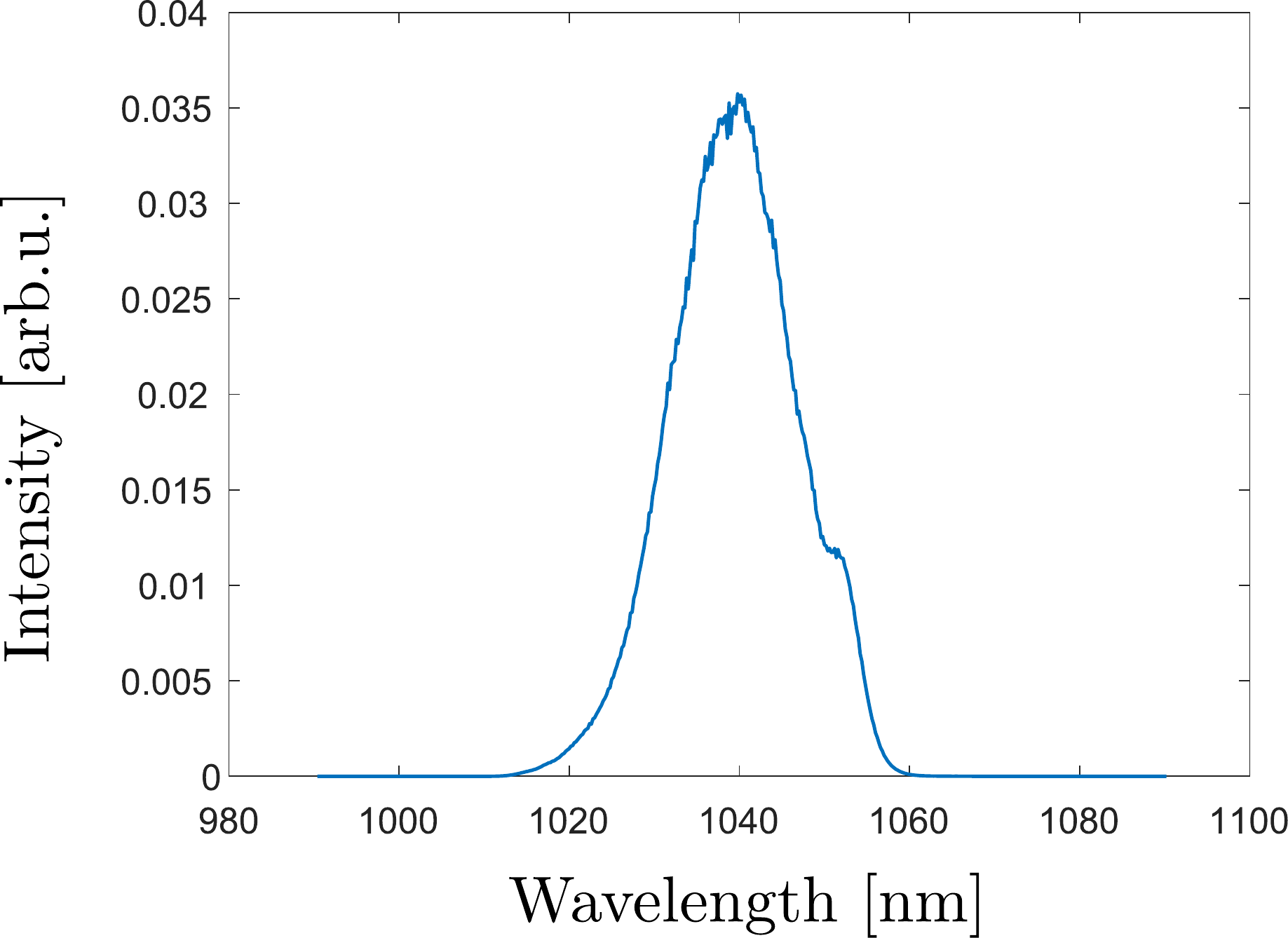}
	\label{fig:SourceSpectr}
	\caption{Pump spectrum.}
\end{figure}

\chapter{Three-photon interference: Raw experimental data and simulation of the experiment}

Taken from \cite{menssen2017distinguishability} Appendix.
\section{Raw experimental data}
\subsection*{Polarisations set for $\varphi=0$}
\subsubsection*{HOM dips for temporal alignment of photons}
In order to align the generated photons temporally and verify their indistinguishability, we perform heralded HOM measurements for the three possible pairs injected into the tritter. We also use these to verify our polarisation state preparations. The results are shown in the Figures below.

\begin{figure}[h!]
	\centering
	\includegraphics[width=0.6\linewidth]{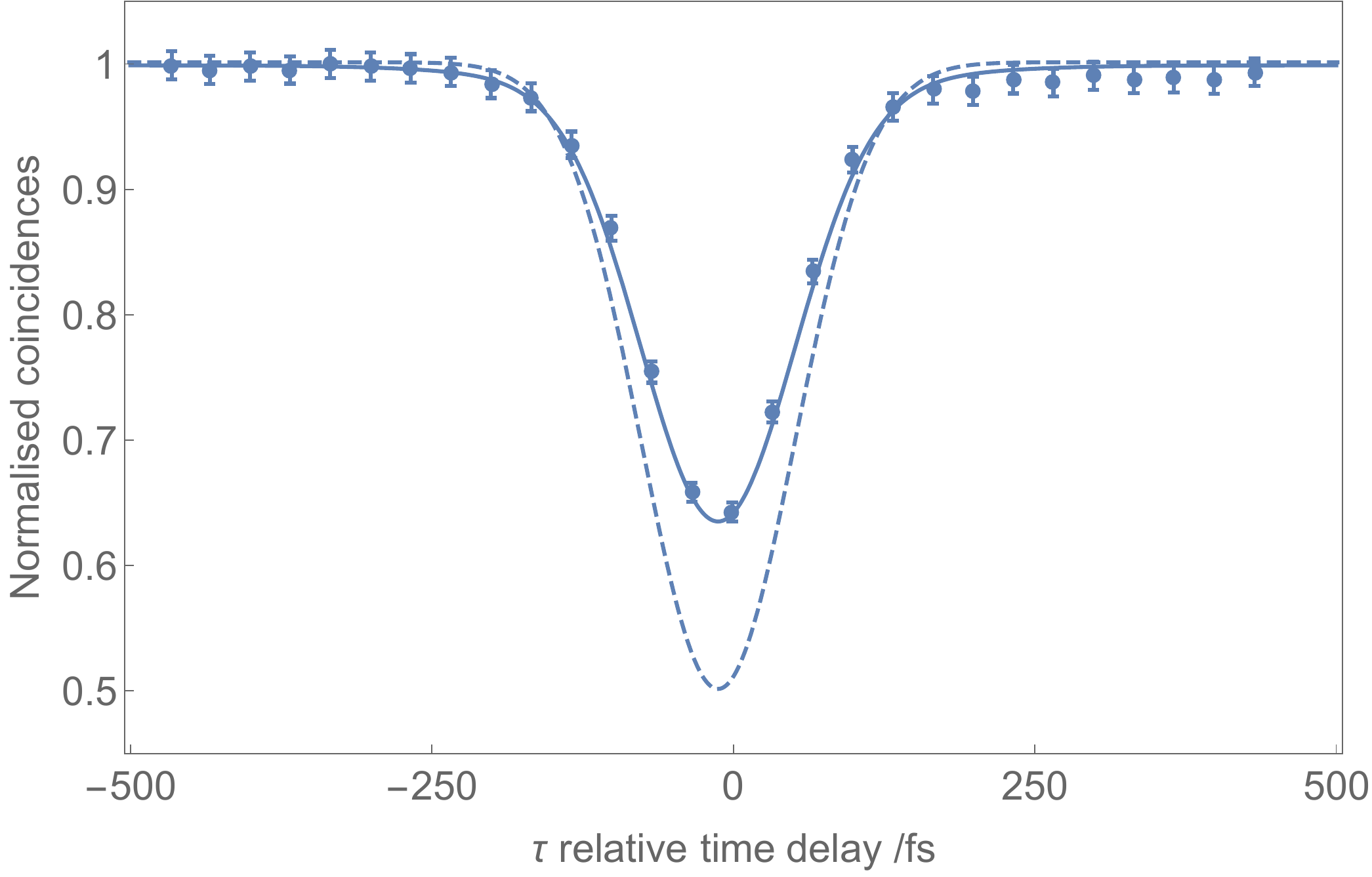}
	\caption{Plot of normalised heralded two-photon coincidences through the tritter when the injected photons have identical polarisations. In this case we inject photons into the first and second tritter inputs and monitor the first and second output ports. The solid line is the model curve, whilst the dashed line is an ideal theory curve. The model yields a visibility of 36$\%$, while for the ideal curve the visibility is 50\%. The FWHM of the model dip is $\sim$ 150 fs.}
	\label{fig:HOM1212aligned}
\end{figure}

\begin{figure}[h!]
	\centering
	\includegraphics[width=0.6\linewidth]{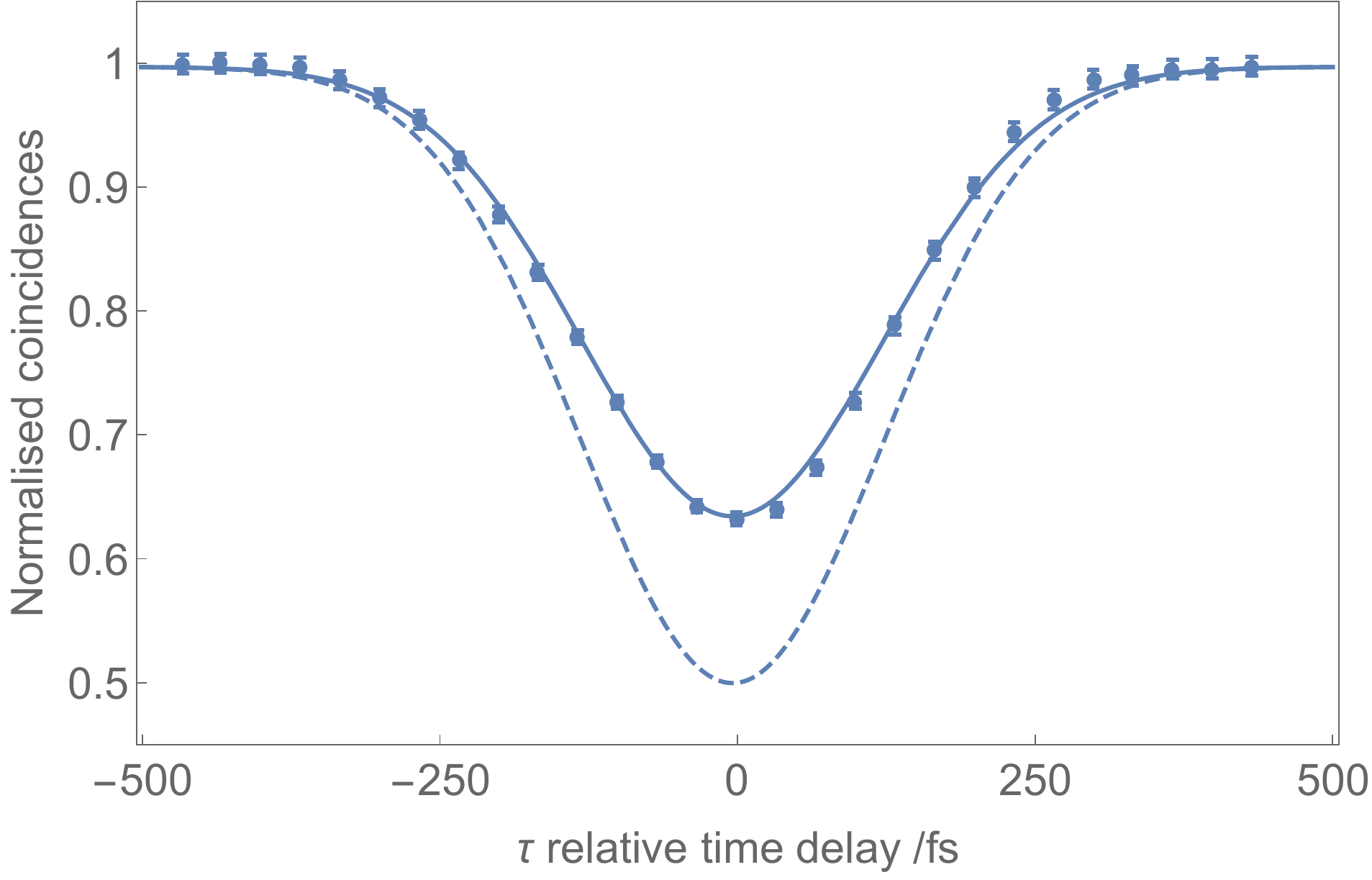}
	\caption{Plot of normalised heralded two-photon coincidences through the tritter when the injected photons have identical polarisations. In this case we inject photons into the first and third tritter inputs and monitor the first and third output ports. The solid line is the model curve, whilst the dashed line is an ideal theory curve. The model yields a visibility of 36$\%$, while for the ideal curve the visibility is 50\%. The FWHM of the model dip is $\sim$ 300 fs.}
	\label{fig:HOM1313aligned}
\end{figure}
\begin{figure}[h!]
	\centering
	\includegraphics[width=0.6\linewidth]{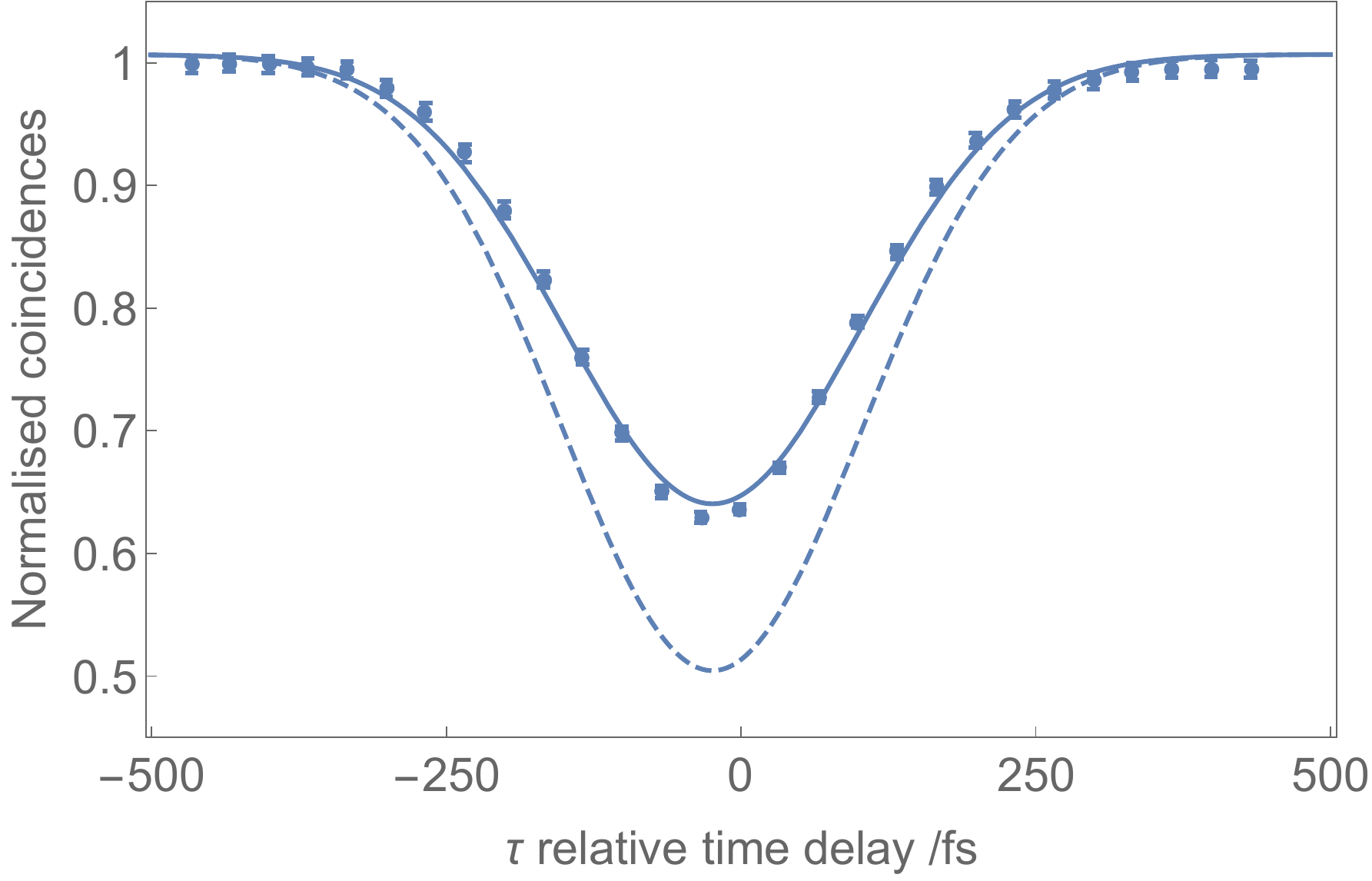}
	\caption{Plot of normalised heralded two-photon coincidences through the tritter when the injected photons have identical polarisations. In this case we inject photons into the second and third tritter inputs and monitor the second and third output ports. The solid line is the model curve, whilst the dashed line is an ideal theory curve. The model yields a visibility of 36$\%$, while for the ideal curve the visibility is 50\%. The FWHM of the model dip is $\sim$ 300 fs.}
	\label{fig:HOM2323aligned}
\end{figure}
\clearpage
\subsubsection*{Additional output event plots}
Here we present plots for count rates corresponding to $P_{210},P_{201},P_{300}$ in the case where all photons are injected into the tritter with the same polarisation. In the ideal case when all photons are completely indistinguishable in time and polarisation, $P_{210}=P_{201}=0$ and these outputs are completely suppressed~\cite{tichy2010zero}. Our simulations demonstrate this is not the case when taking into account experimental imperfections, and the visibility is reduced from 100\% to around 57\%. The theory and simulation curves have been rescaled for comparison with experimental count rates.
\begin{figure}[h!]
	\centering
	\includegraphics[width=0.6\linewidth]{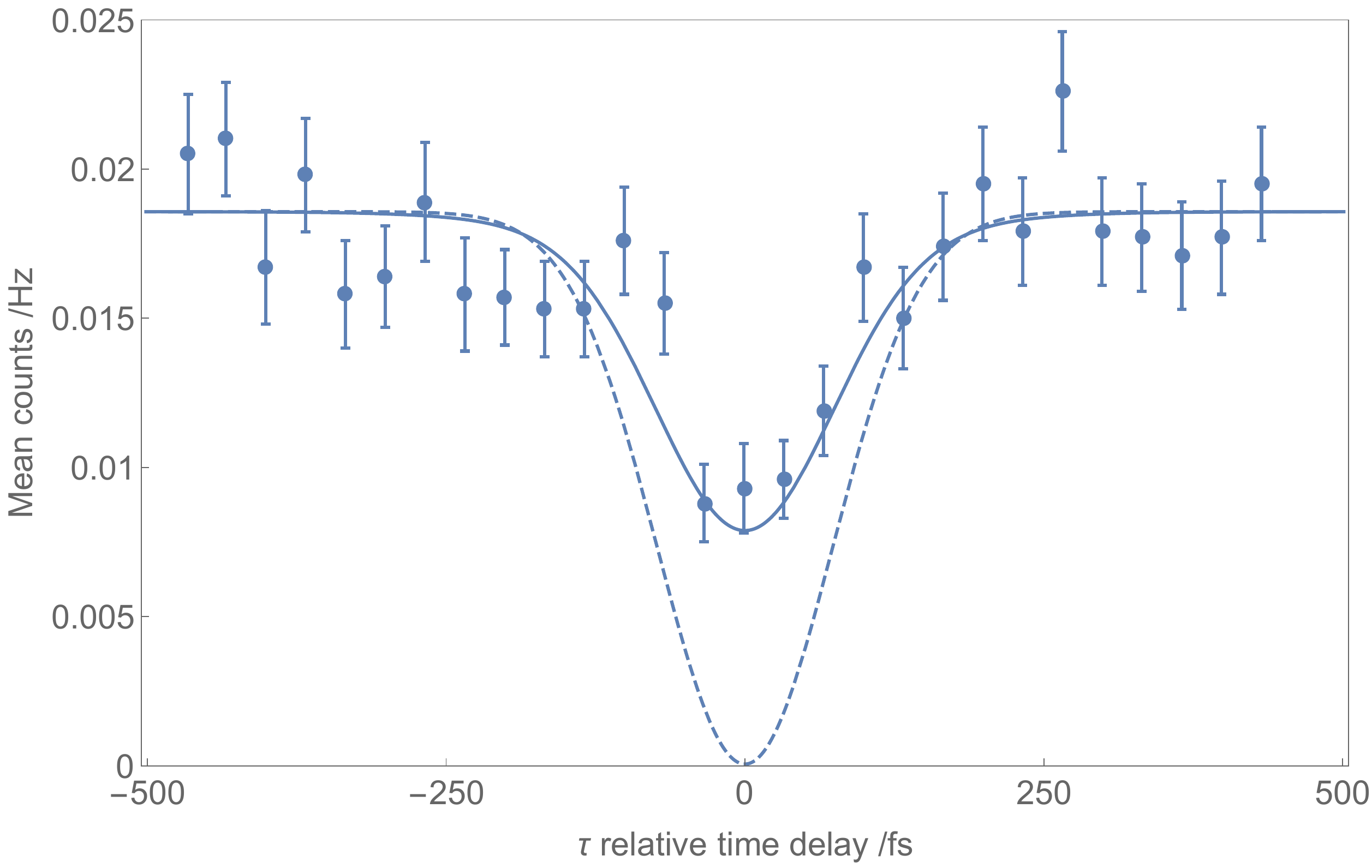}
	\caption{Plot of mean counts for the outputs corresponding to $P_{210}$ as illustrated in Figure~\ref{fig:DetConf} b). The model yields a visibility of 57$\%$, while for the ideal curve the visibility is 100\%. The FWHM of the model dip is $\sim$ 180 fs.}
	\label{fig:210aligned}
\end{figure}
\begin{figure}
	\centering
	\includegraphics[width=0.6\linewidth]{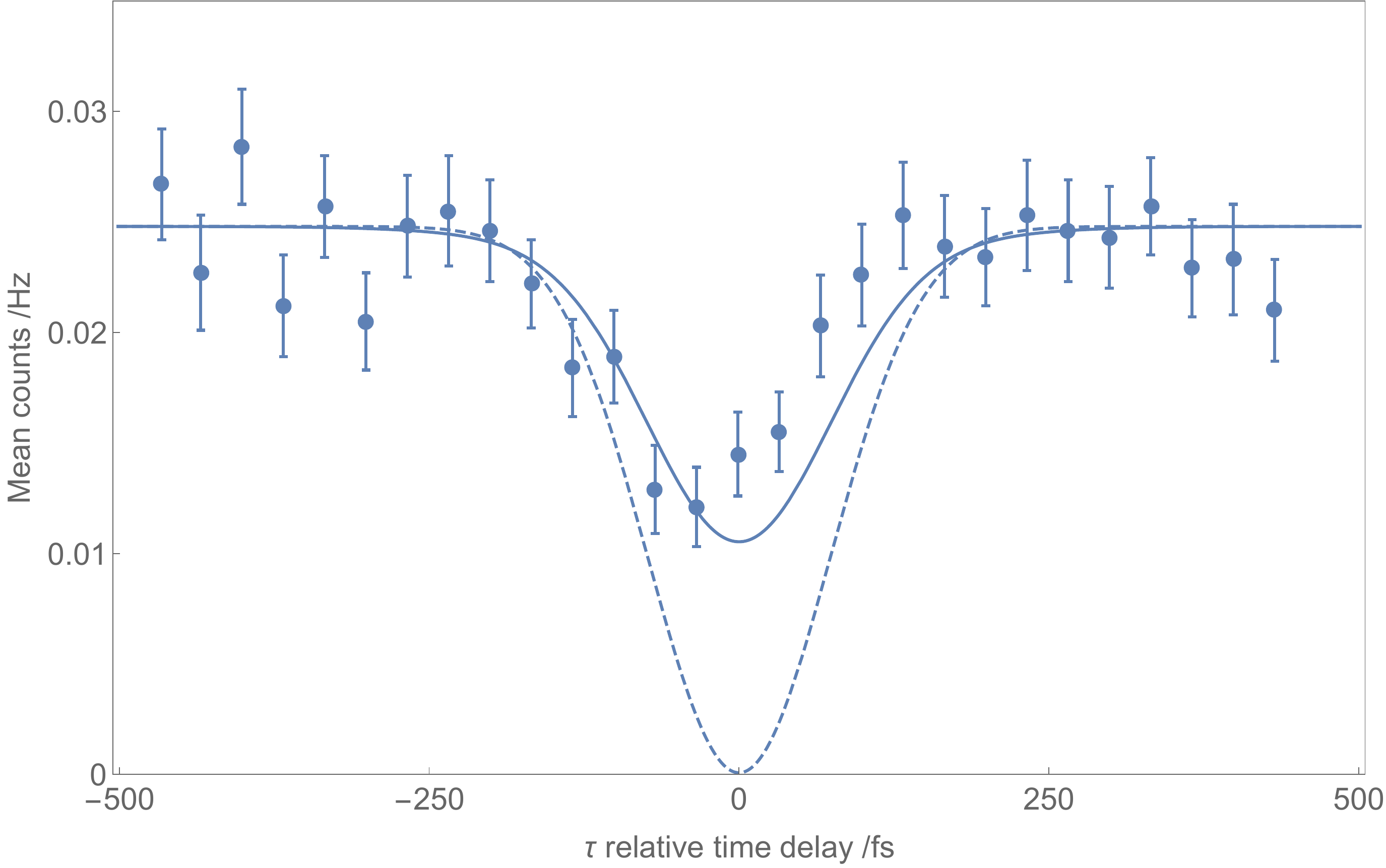}
	\caption{Plot of mean counts for the outputs corresponding to $P_{201}$ as illustrated in Figure~\ref{fig:DetConf} b). The solid line is the model curve, whilst the dashed line is an ideal theory curve. The model yields a visibility of 57$\%$, while for the ideal curve the visibility is 100\%. The FWHM of the model dip is $\sim$ 180 fs.}
	\label{fig:201aligned}
\end{figure}
\begin{figure}[h!]
	\centering
	\includegraphics[width=0.8\textwidth]{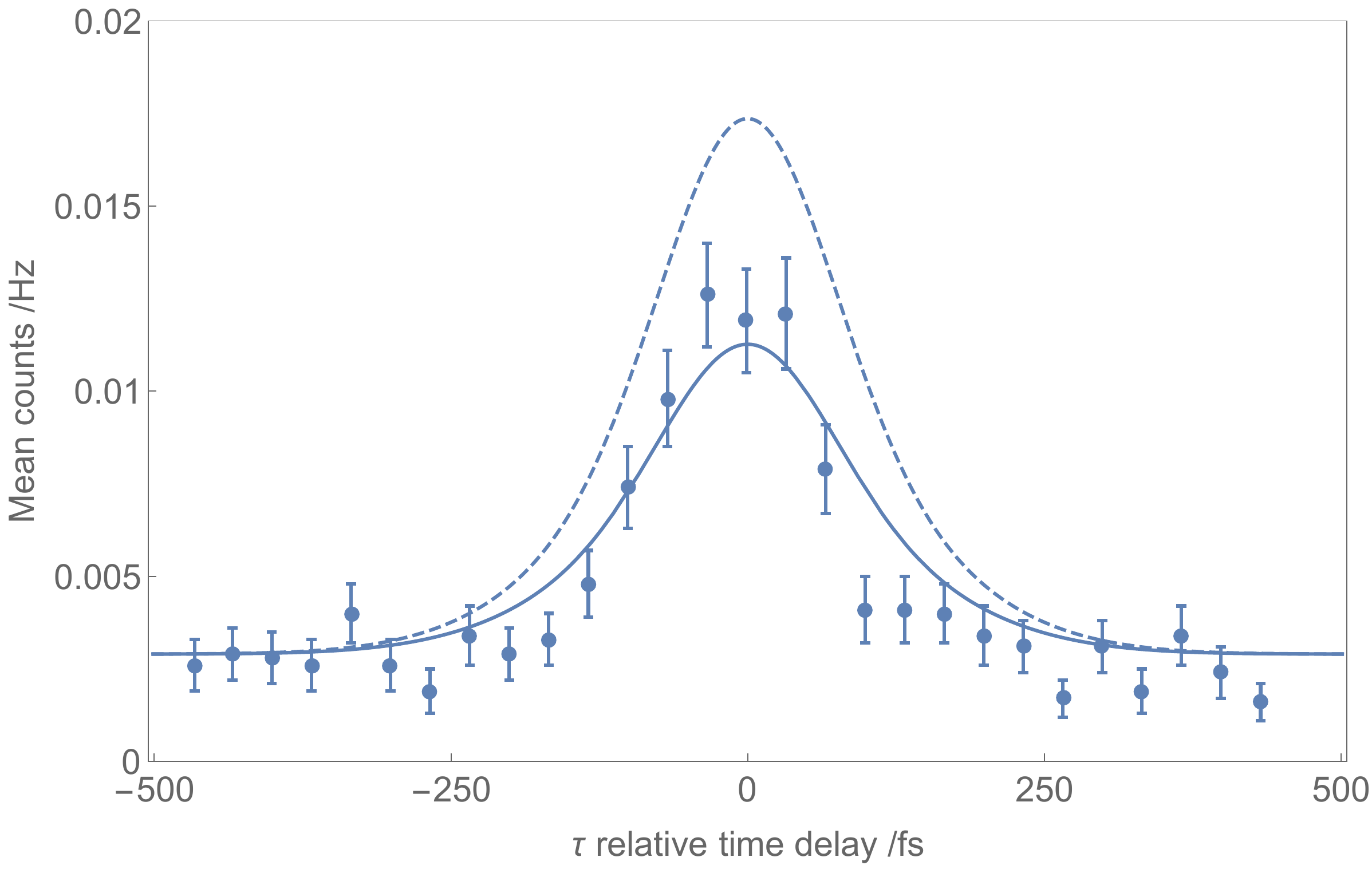}
	\caption{Plot of count rates corresponding to $P_{300}$ as measured using the setup in Figure~\ref{fig:DetConf} a), when all photons have the same polarisation.}
\end{figure}
\clearpage

\subsection*{Polarisations set for $\varphi=\pi$}
\subsubsection*{HOM dips for temporal alignment of photons}
\begin{figure}[h!]
	\centering
	\includegraphics[width=0.63\linewidth]{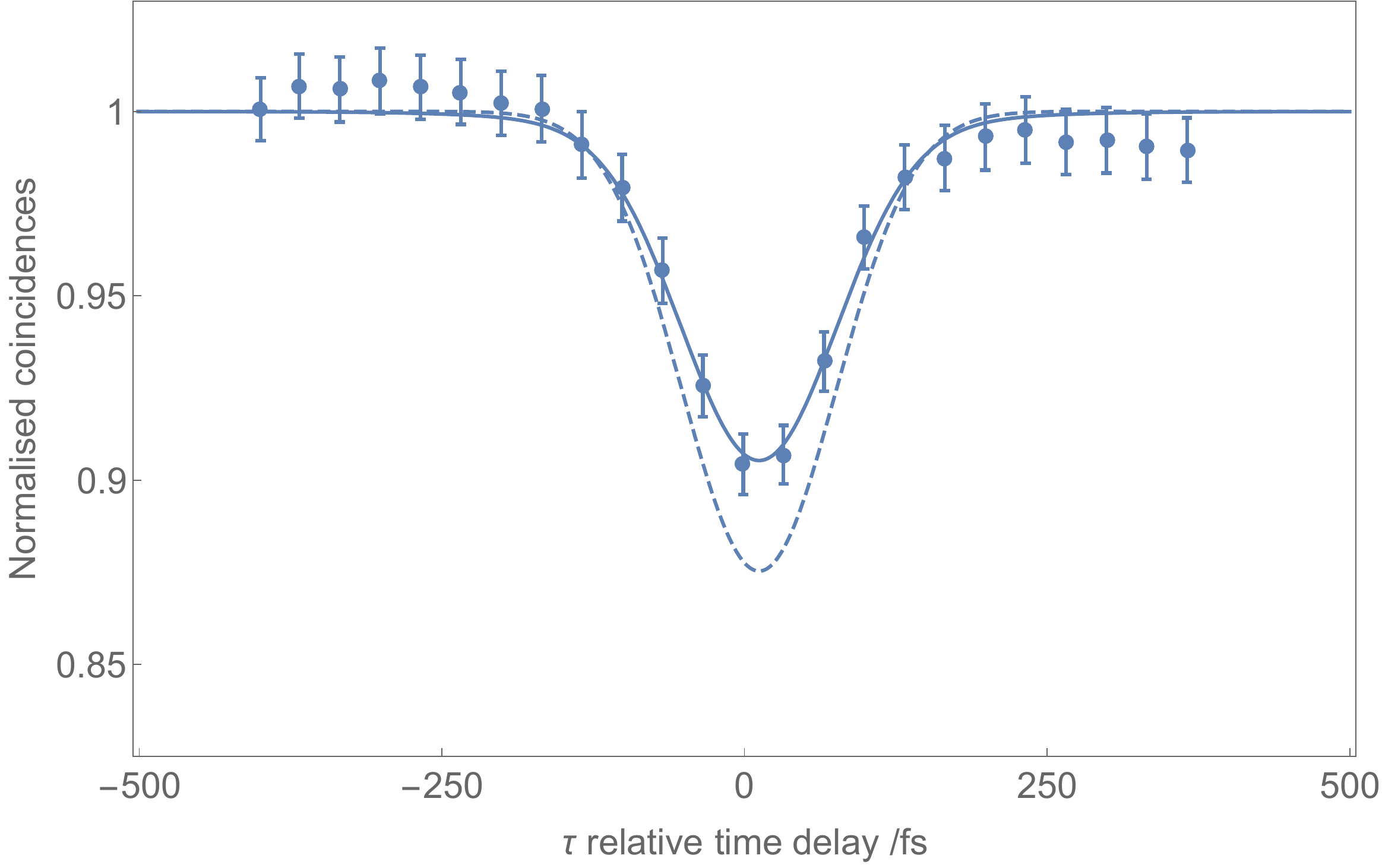}
	\caption{Plot of normalised heralded two-photon coincidences through the tritter when the injected photons have polarisations as in equation \ref{eqn:staticpol2}. In this case we inject photons into the first and second tritter inputs and monitor the first and second output ports.  The solid line is the model curve, whilst the dashed line is an ideal theory curve. The model yields a visibility of 10$\%$, while for the ideal curve the visibility is 12.5\%. The FWHM of the model dip is $\sim$ 150 fs. The slight gradient in the wings of the distribution is due to a change in coupling as a function of the translation stage position.}
	\label{fig:HOM1212merc}
\end{figure}

Again, to align the three photons temporally before injection into the tritter, we perform HOM measurements for the three pairs of photons. We expect 12.5\% visibility but record closer to 10\%, again due to the effects mentioned previously. The dip in Figure~\ref{fig:HOM1212merc} is twice as narrow as the others, corresponding to the dip between two photons which are both being translated in time on injection. The other two dips are from when only one of the photons injected into the tritter is translated in time. The dips are all centred such that the three photons overlap in time when the stages are at their zero positions.

\begin{figure}[h!]
	\centering
	\includegraphics[width=0.63\linewidth]{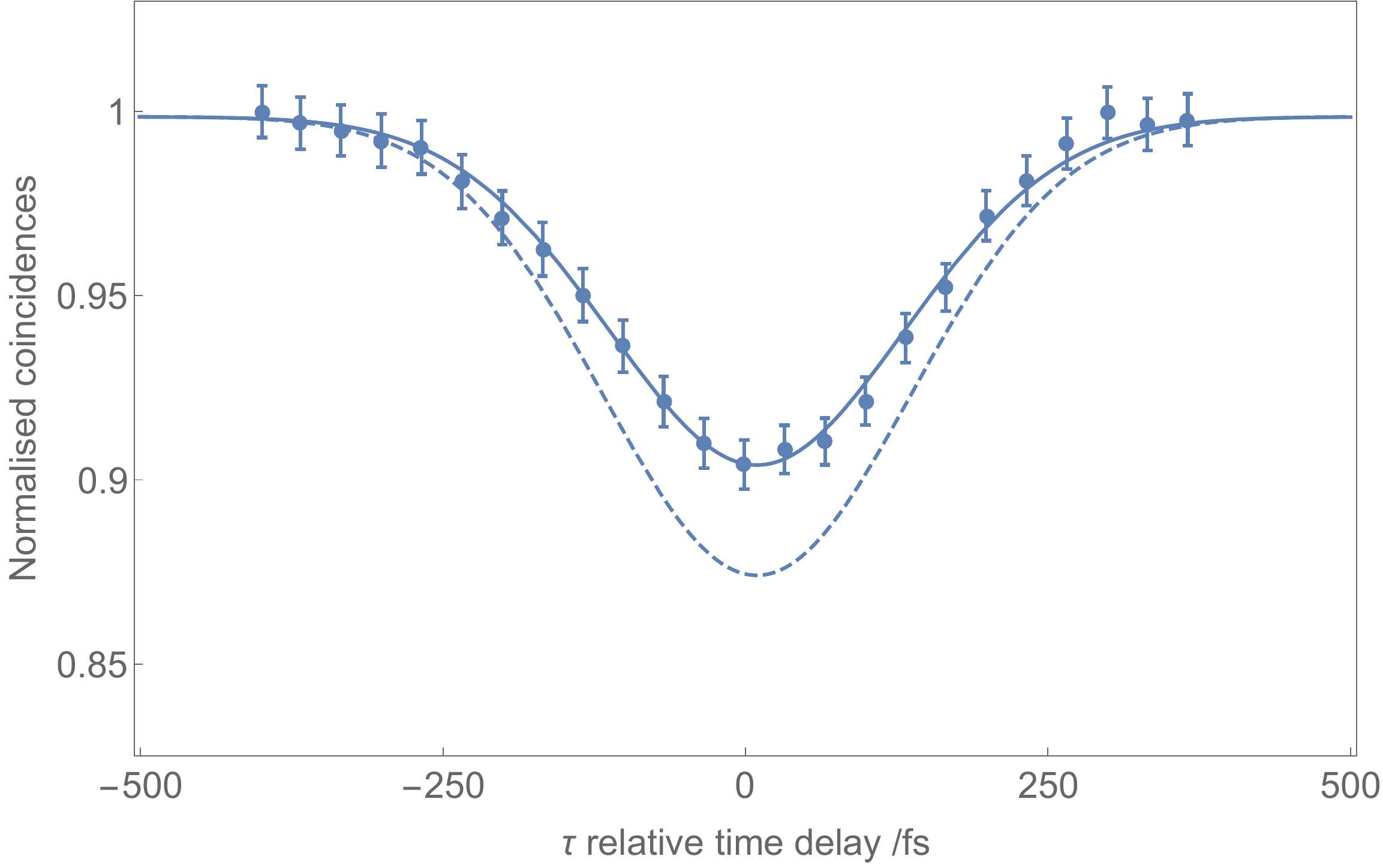}
	\caption{Plot of normalised heralded two-photon coincidences through the tritter when the injected photons have polarisations as in equation \ref{eqn:staticpol2}. In this case we inject photons into the first and third tritter inputs and monitor the first and third output ports. The solid line is the model curve, whilst the dashed line is an ideal theory curve. The model yields a visibility of 10$\%$, while for the ideal curve the visibility is 12.5\%. The FWHM of the model dip is $\sim$ 300 fs.}
	\label{fig:HOM1313merc}
\end{figure}
\begin{figure}[h!]
	\centering
	\includegraphics[width=0.63\linewidth]{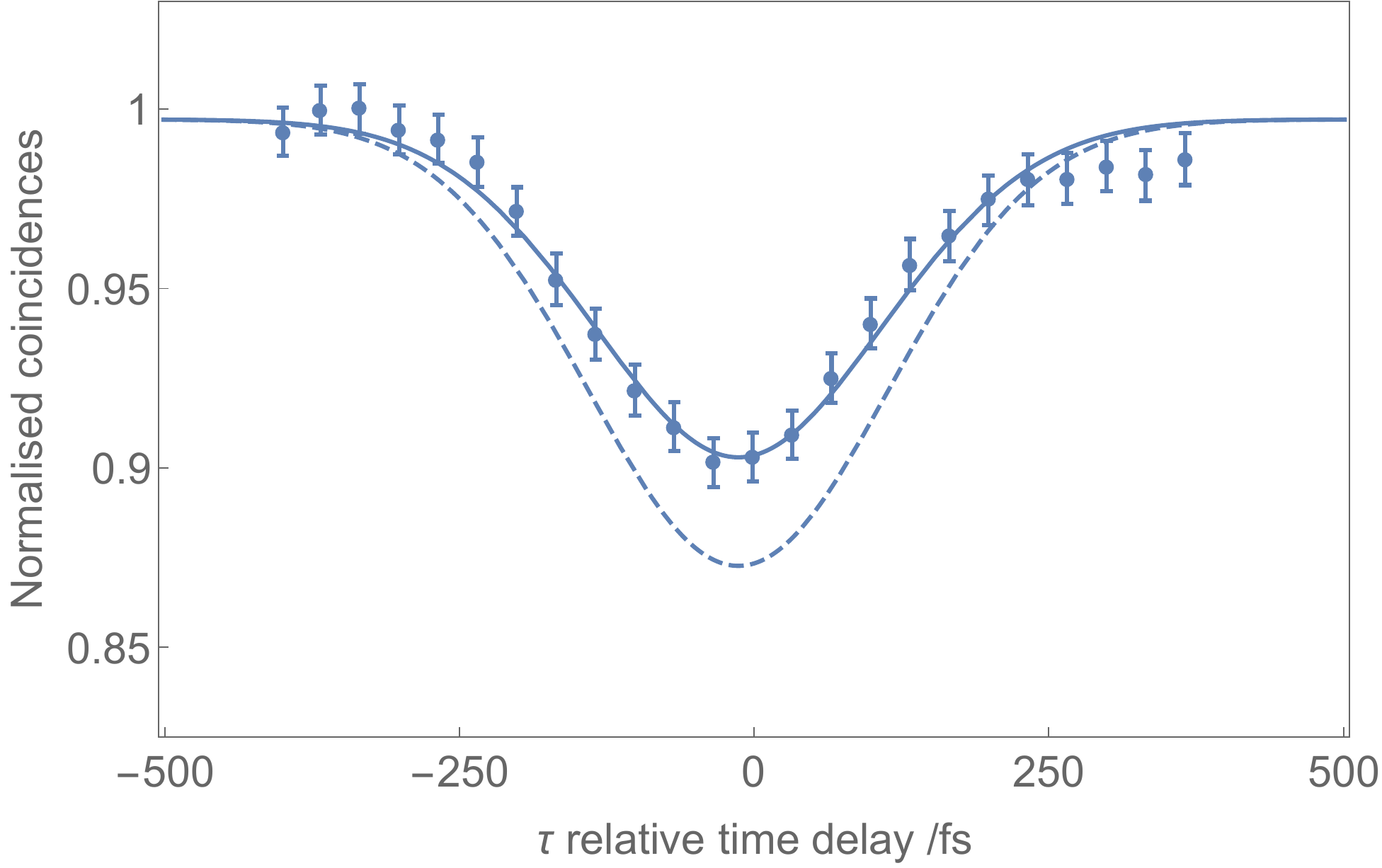}
	\caption{Plot of normalised heralded two-photon coincidences through the tritter when the injected photons have polarisation as in equation \ref{eqn:staticpol2}. In this case we inject photons into the second and third tritter inputs and monitor the second and third output ports.  The solid line is the model curve, whilst the dashed line is an ideal theory curve. The model yields a visibility of 10$\%$, while for the ideal curve the visibility is 12.5\%. The FWHM of the model dip is $\sim$ 300 fs. The slight difference in counts in the wings of the distribution is due to a change in coupling as a function of translation stage position.}
	\label{fig:HOM2323merc}
\end{figure}
We also recorded coincidences for the outputs corresponding to $P_{210},P_{201},P_{300}$ but these statistics are all predicted to have lower visibilities for this case of $\varphi=\pi$ compared to $\varphi=0$. Our recorded statistics are not sufficient to resolve these features.

\subsection*{Probing the triad phase}
\subsubsection*{Polarisation dependence of the tritter}
\begin{figure}[h!]
	\centering
	\includegraphics[width=0.75\textwidth]{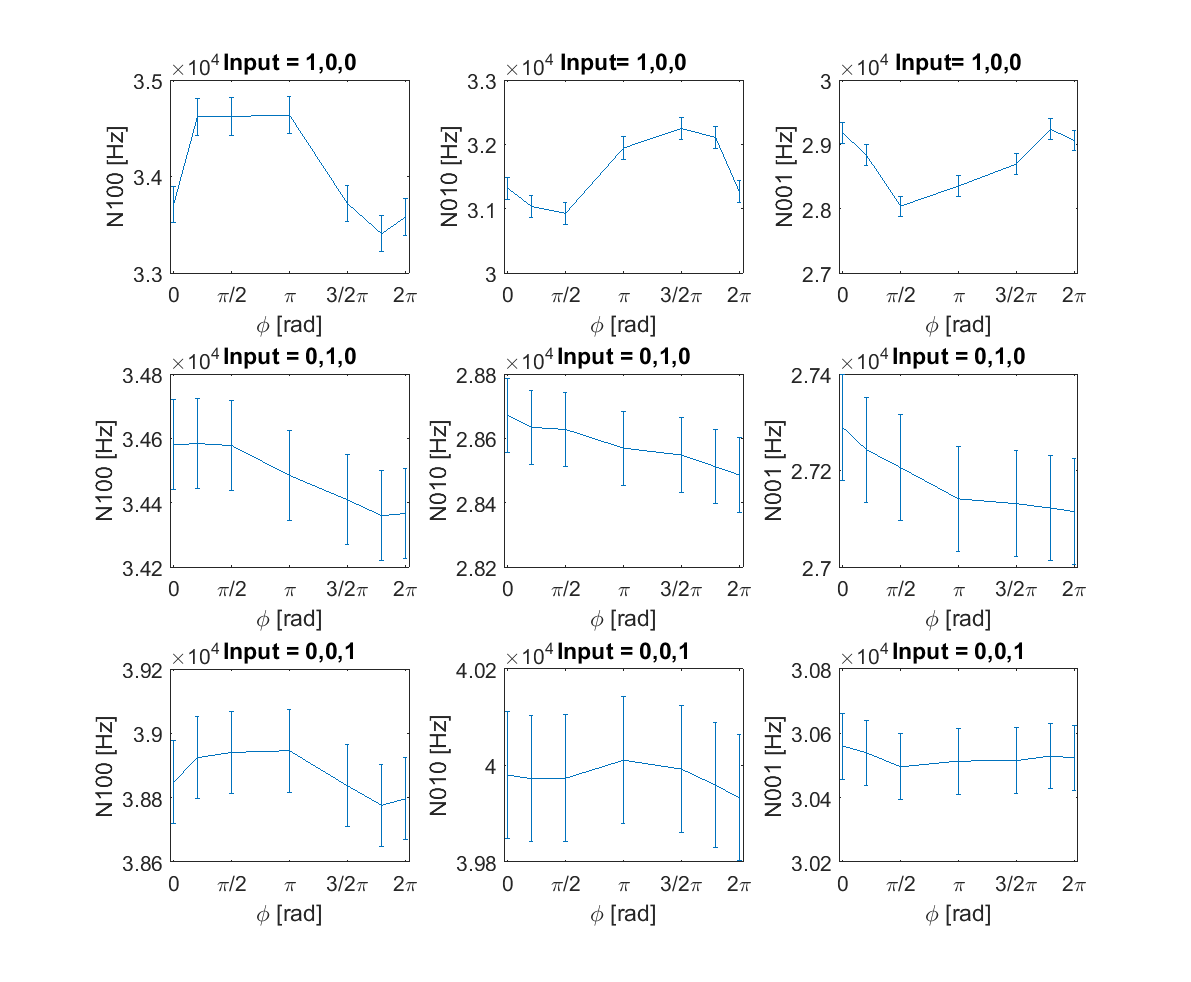}
	\caption{\label{fig:SinglesSweep} All input and output combinations for heralded single-photon events. The y-axis labels the count rates for a particular output configuration, and the x-axis is the triad phase we scan. The input port for the injected photon is labelled above each plot. The variation of the counts for the case where the polarisation of the photon is varied before injection (first row) shows that the tritter is slightly polarisation dependent: the coupling between spatial modes varies as a function of the triad phase. The slight drop of counts shown in the second row (where a single photon is injected into the second tritter input) is due to imperfect fibre coupling.}
\end{figure}

\begin{figure}
	\centering
	\includegraphics[width=0.50\textwidth]{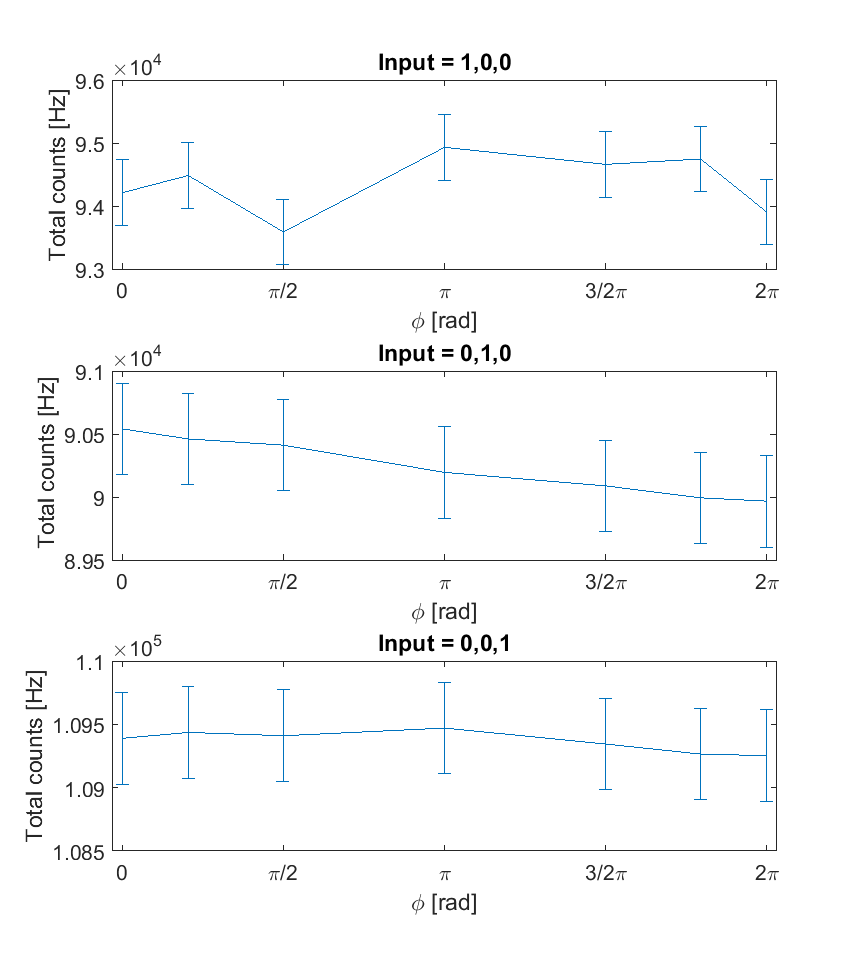}
	\caption{\label{fig:SumSinglesSweep} We plot the sum of all heralded single counts for different inputs into the tritter (total counts=N100+N010+N001, corresponding to summing the counts in the rows appearing in Figure~\ref{fig:SinglesSweep}).}
\end{figure}

For isolating three-photon interference, we scan the triad phase by varying the polarisation of one of the photons. In order to study the polarisation-dependence of the tritter, we send heralded single photons into different tritter inputs and record the output counts (see Figures \ref{fig:SinglesSweep} and \ref{fig:SumSinglesSweep}).

The total number of counts is relatively constant (see Fig.~\ref{fig:SumSinglesSweep}), whilst some individual heralded singles events in the top row of Figure \ref{fig:SinglesSweep} vary as the triad phase (and thus polarisation of the photon injected into the first input) changes. This suggests that the couplings of the  tritter have a slight polarisation dependence.

\subsubsection*{Heralded two-photon coincidences}
\begin{figure}[h!]
	\centering
	\includegraphics[width=0.75\textwidth]{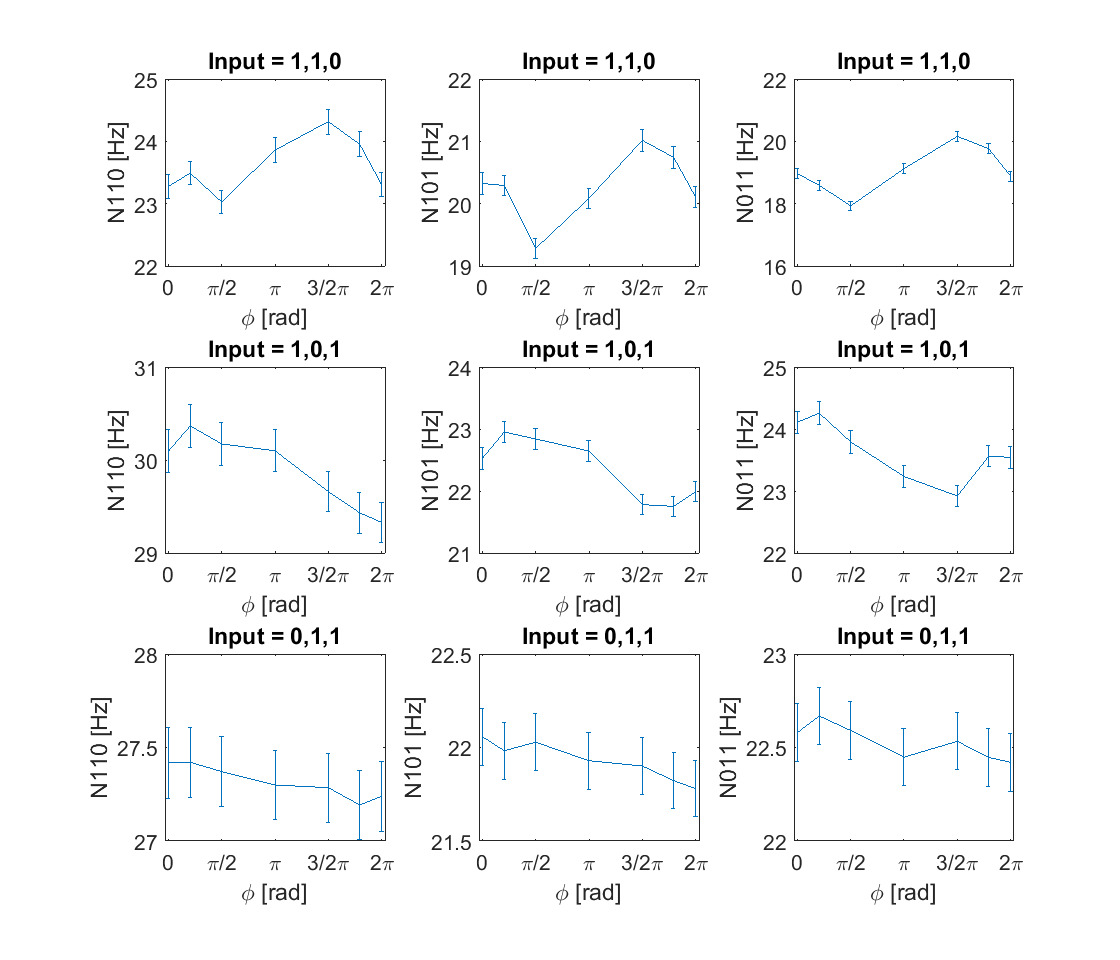}
	\caption{\label{fig:HOMDipsSweep} All input and output combinations for heralded two-photon events. We plot the number of heralded two-fold coincidences in the first and second (N110), first and third (N101), and second and third (N011) spatial output modes when changing the triad phase (and thus polarisation of the photon injected into the third input). The channels with the highest variation are those involving the first input channel, and this suggests it is due to the tritter's polarisation dependence.}
\end{figure}

We monitored the heralded two-fold coincidences to verify that we have as little variation as possible as a function of the triad phase. In Figure~\ref{fig:HOMDipsSweep} all possible combinations of heralded two-photon events are displayed. The largest variation in counts is observed for channels containing the first input channel, arising, as discussed in the previous section, from the tritter's polarisation dependence. 

\clearpage
\subsubsection*{Additional output event plots}
\label{app:additionalEv}
\begin{figure}[h!]
	\centering
	\includegraphics[width=0.9\textwidth]{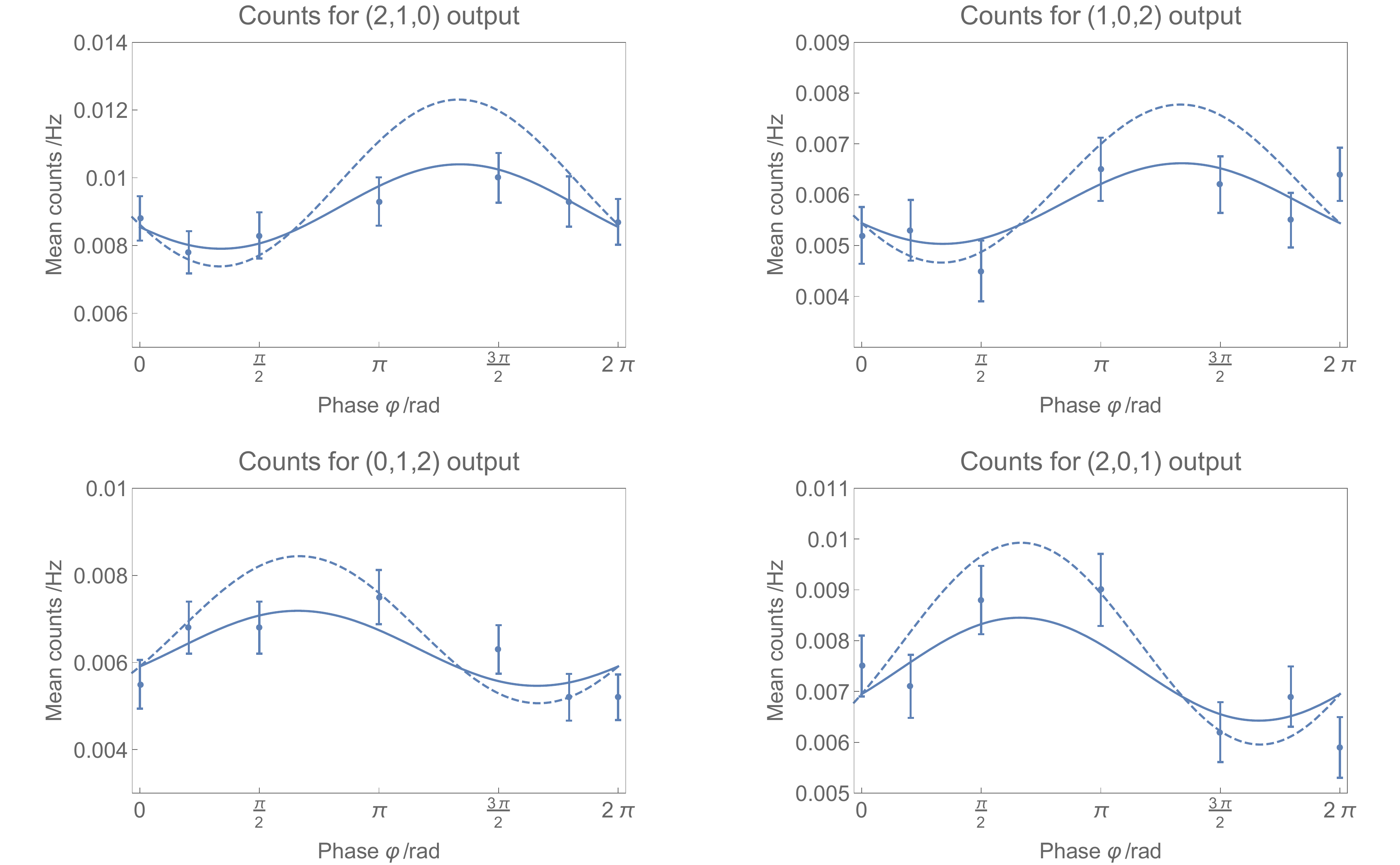}
	\caption{Plots for count rates corresponding to cases where two photons exit the same output port, whilst the third exits in a different port. From Eqns.~\ref{eqn:probs6} and \ref{eqn:probs7} we expect cosine curves shifted by $-\pi/3$ for (2,1,0) and (1,0,2), and by $+\pi/3$ for (0,1,2) and (2,0,1). The solid lines are simulation curves and the dashed lines are ideal theory, and both have been normalised to fit the data at $\varphi=0,2\pi$ via a multiplicative scaling factor for comparison.} 
\end{figure}

\section{Simulation of the experiment}
\label{sect:simExp}
Our experimental data show the expected behaviour, but there are some deviations from the probabilities given by theory.
These are primarily due to an imperfect tritter operation, imperfections in the photon preparation (polarisation, purity ($P>90\%$), distinguishability), and higher-order photon emission. Furthermore, along with photons that are produced by the SFWM-process, uncorrelated photons are created in other processes such as Raman scattering and fluorescence~\cite{spring2017chip}. To understand the influence of all these effects on the measured visibilities, we performed a simulation of our experiment. We used the formalism developed in \cite{Tichy2015,shchesnovich2015} to simulate general mixed, squeezed states, contaminated with distinguishable noise photons, that are input into a lossy unitary.
Our model includes terms corresponding to up to $N=8$ photons in total (signals and heralds) and up to 3 uncorrelated noise photons. This provides sufficient accuracy as terms corresponding to higher photon numbers are negligibly small.

\subsection*{Impure input states}

It was noted previously in~\cite{Tichy2015, shchesnovich2015} that the counting statistics for a mixed state input can be expressed as a function of the density matrices $\rho_i$ for each photon in input mode $i$. 
For three photons input to an interferometer described by the unitary $U$ this leads to the following expression for the coincidence probability $P_{111}$:

\begin{equation}
\begin{split}
\label{eqn:P111MixMalte}
P_{111}=\text{perm}(U*U^\star)
+\text{Tr}(\rho_1\rho_2)\text{perm}(U*U_{2,1,3}^\star)\\
+\text{Tr}(\rho_1\rho_3)\text{perm}(U*U_{3,2,1}^\star)\\
+\text{Tr}(\rho_2\rho_3)\text{perm}(U*U_{1,3,2}^\star)\\
+2\text{Re}( Tr(\rho_1\rho_2\rho_3))\text{Re}(\text{perm}(U*U_{2,3,1}^\star))\\
-2\text{Im}( \text{Tr}(\rho_1\rho_2\rho_3))\text{Im}(\text{perm}(U*U_{2,3,1}^\star)).
\end{split}
\end{equation}

For simplicity in the simulation we make the assumption that we can decompose the density matrix into a mixed and a pure subspace, where the full density matrix for each photon is given by their tensor product:
\begin{equation}
\label{eqn:model}
\rho_i=\rho_{pure,i}\otimes\rho_{mixed,i}.
\end{equation}

$\rho_{pure}$ may be represented as the tensor product of a density matrix which contains the temporal modes and another containing the polarisation degree of freedom 
\begin{equation}
\label{eqn:model2}
\rho_{pure,i}=\rho_{temp,i}\otimes\rho_{pol,i}.
\end{equation}

For general temporal modes $\ket{t_1},\ket{t_2},\ket{t_3}$, we find a representation of the states in terms of orthonormal modes $\ket{\tau_1},\ket{\tau_2},\ket{\tau_3}$ using the Gram-Schmidt decomposition: 
\begin{eqnarray}
\ket{t_1}&=&\ket{\tau_1}\\
\ket{t_2}&=&\braket{t_1}{t_2}\ket{\tau_1}+\sqrt{1-|\braket{t_1}{t_2}|^2}\ket{\tau_2}\\
\ket{t_3}&=&\braket{t_1}{t_3}\ket{\tau_1}+\alpha\ket{\tau_2}+\sqrt{1-|\alpha|^2-|\braket{t_1}{t_3}|^2}\ket{\tau_3},
\end{eqnarray}
where $\alpha=\frac{\braket{t_2}{t_3}-\braket{t_2}{t_1}\braket{t_1}{t_3}}{\sqrt{1-|\braket{t_1}{t_2}|^2}}$
and  $\ket{t_1} ,\ket{\tau_2},\ket{\tau_3}$ are a set of orthonormal vectors. We can then construct the density matrices in mode basis:
\begin{equation}
\rho_{temp,i}=\ket{t_i}\bra{t_i}.
\end{equation}

The polarisation density matrix is constructed from basis states $\ket{H}$ and $\ket{V}$. Mixedness is modelled on a two dimensional Hilbert-space which is chosen to be orthogonal to time-frequency and polarisation modes.

\subsection*{Higher order photon contributions}
\label{app:higherord}
The state of a single ideal two-mode-squeezer is given by:
\begin{equation}
\ket{\Psi}=\sqrt{1-\lambda^2}\sum^\infty_{n=0}\lambda^n\ket{n_{s}n_{i}}.
\label{eqn_squeeze1}
\end{equation}
Furthermore, we assume each source generates uncorrelated photons which are created with probabilities $P_I$ for the idlers and $P_S$ for the signals. In particular $(1-P_I)(1-P_S)$ is the probability of producing no uncorrelated noise photons.  $(1-P_I)P_I(1-P_S)P_S$ is the probability of creating exactly one uncorrelated photon pair.
We can then construct the density matrix for one source's emission:
\begin{equation}
\hat{\rho}=(1-\lambda^2)\cdot(1-P_I)\cdot(1-P_S)\sum^\infty_{n,k,l=0}\lambda^{2n}P_I^{k}P_S^{l}\ket{n_{s}n_{i},k_{s}l_{i}}\bra{n_{s}n_{i},k_{s}l_{i}},
\label{eqn_squeeze3}
\end{equation}
where for each total number of photons $2n+k+l$, we include cases where they come from four-wave mixing or noise processes. The indices $k$ and $l$ label the number of signal and idler noise photons which are assumed to be completely distinguishable from all other photons.

\subsection*{Parameter values}

In the following table we give the parameter values that were used for the simulation:

\begin{center}
	\begin{tabular}{ |l |l|l| }
		\hline
		Name & Symbol & Value \\
		\hline
		Squeezing-parameter & $\lambda$ & 0.16 \\
		\hline
		Purity & $\mathcal{P}$ & 0.9 \\
		\hline
		Fluorescence probability idler & $P_I$ & 0.035   \\
		\hline
		Fluorescence probability signal & $P_S$ & 0.009   \\
		\hline   
	\end{tabular}
\end{center}

The squeezing parameter was taken to be the same as in~\cite{spring2017chip}; the experiment reported in~\cite{spring2017chip} was performed with the same power of the pump beam). The purity is a lower bound estimate and primarily affected by our ability to filter out non-factorable components in the (signal/idler) joint spectral distribution. We were limited in the signal/idler filtering bandwidth as we used a single pair of angle tuned bandpass filters in the beam path of signal and idler photons, immediately after a dichroic mirror. Since the three beams pass through the filters at slightly different angles the filters' spectral edges are slightly shifted with respect to each other, effectively limiting our tuning range. We calculate the degree of spectral purity for the given filter bandwidth of $10-15$ nm and obtain a value of approximately $\approx 90\%$ purity. The uncorrelated noise probability is obtained from a measurement of the heralded $g^{(2)}(0)$ in \cite{spring2017chip} (supplementary). We perform a fit of the $g^{(2)}(0)$ to our model and use $P_I$ as a free parameter. $P_S$ is chosen to be $1/4$ of $P_I$ as the background noise for the signals is significantly smaller. The ratio of $\frac{P_S}{P_I}\approx 0.25$ was obtained by comparing background noise levels of signal and idler photons with a single photon spectrometer. When the pump polarisation is rotated by 90 degree we lose phase-matching, allowing us to observe the background noise only at the given input power.

\chapter{Proof integer Chern number}
\label{app:proofChern}
Equation \ref{eqn:chern} due to the Stokes theorem is zero for continuously defined function $\mathcal{A}_n$:
\begin{equation}
\mathcal{C}_n=\frac{1}{2\pi i}\int_{\partial BZ}\mathcal{A}_n.
\label{eqn:chern2}
\end{equation}

This is due to the Brillouin zone being equipped with torus topology $T^d$. Since the Torus has no boundary the integral vanishes as it is executed over an empty set $\partial BZ=\partial T^d=\emptyset$. The requirement for a non-zero Chern-number is therefore that the Berry connection $\mathcal{A}_n$ cannot be continuously defined over the Brillouin zone. The authors of \cite{ozawa2018topological,nakahara2003geometry} partition the integral into two regions,
\begin{equation}
\mathcal{C}_n=\frac{1}{2\pi i}(\int_{\partial S}\mathcal{A}_n +\int_{\partial S'}\mathcal{A'}_n ),
\label{eqn:chern4}
\end{equation}
where different gauges are chosen $\mathcal{A}_n$, $\mathcal{A}_n'$ in the regions $S$ and $S'$, which remove the singularity over the integration region.  $\partial S=-\partial S'$ is a boundary between the regions. This can be expressed in terms of the Berry phase as
\begin{equation}
\mathcal{C}_n=\frac{1}{2\pi i}(\gamma-\gamma'),
\end{equation}
where the difference $(\gamma-\gamma')$ will be an integer multiple of $2 \pi i$, since the Berry phases are integrated along the same path $\partial S$ and the Berry phase along a closed path is defined up to an integer multiple of $2\pi$.
To illustrate this, I will give a brief example from \cite{nakahara2003geometry}.
They consider a vector on the Bloch-sphere, the underlying manifold in this case is $S^2$. This could for example describe the polarisation state of a single photon.
\begin{equation}
\ket{\psi}_A=\binom{\cos(\theta/2)}{e^{i\phi}\sin(\theta/2)}
\end{equation}
The state is singular for $\theta=\pi$, since the $\phi$ becomes undefined as $\theta$ passes through the pole.
We are free to choose a phase for every point on the Bloch-sphere $e^{i\Phi(\theta,\phi)}$. This corresponds to a $U(1)$ gauge freedom. The vector expressed in another possible gauge is then:
\begin{equation}
\ket{\psi}_{A'}=\binom{e^{-i\phi}\cos(\theta/2)}{\sin(\theta/2)}.
\end{equation}
Here, the singularity appears for $\theta=0$. We note that the Berry connection takes the function of a $U(1)$ gauge field, which can be seen in the way it behaves under a gauge transformation.
The Berry gauge potential $\mathcal{A}$ and $\mathcal{A}'$ become:
\begin{eqnarray}
\mathcal{A}&=&{}_A\bra{\psi} \text{d}\ket{\psi}_A=\frac{1}{2}i(1-\cos(\theta))\text{d}\phi\\
\mathcal{A}'&=&{}_{A'}\bra{\psi} \text{d}\ket{\psi}_{A'}=-\frac{1}{2}i(1+\cos(\theta))\text{d}\phi. \nonumber
\end{eqnarray} Here, d is the Cartan exterior derivative.
It is also noted in \cite{nakahara2003geometry} that the transformation from one $U(1)$ gauge to another takes the familiar form of the original potential to which the gradient of a scalar field is added:
\begin{eqnarray}
\mathcal{A}'&=&\mathcal{A}-i\text{d}\phi.
\end{eqnarray}
We execute the integration over the Berry curvature on the two hemispheres of $S^2$. We choose for each integration a gauge that removes the singularity in each integration region. This means we select $\mathcal{A}$ for $\theta\in[0,\pi/2]$ and $\mathcal{A}'$  for $\theta\in[\pi/2,\pi]$. This is a valid approach, because of the gauge invariance of the Berry curvature. Using Gauss's theorem we can replace the integration of the Berry curvature over the hemisphere by an integration of the Berry phase over the edge of the hemisphere (cf. equations \ref{eqn:chern},\ref{eqn:chern2}), which is the equator at $\theta=\pi/2$. 
We now integrate the Berry phase in both gauges $\mathcal{A}$ and $\mathcal{A}'$ along the equator $\theta=\pi/2$ and calculate the Chern number \ref{eqn:chern4}.
\begin{eqnarray}
\mathcal{C}=\frac{1}{4\pi }(\int_{0}^{2\pi}\text{d}\phi-\int_{2\pi}^{0}\text{d}\phi)=1.
\end{eqnarray}
We indeed obtain an integer value for the Chern number.
\chapter{Developing femtosecond written multi-mode interferometers for multi-photon interferometry}
\label{app:FSwriting}
\section{Multi-mode interferometers}
\begin{figure}[h!]
	\centering
	\includegraphics[width=0.9\textwidth]{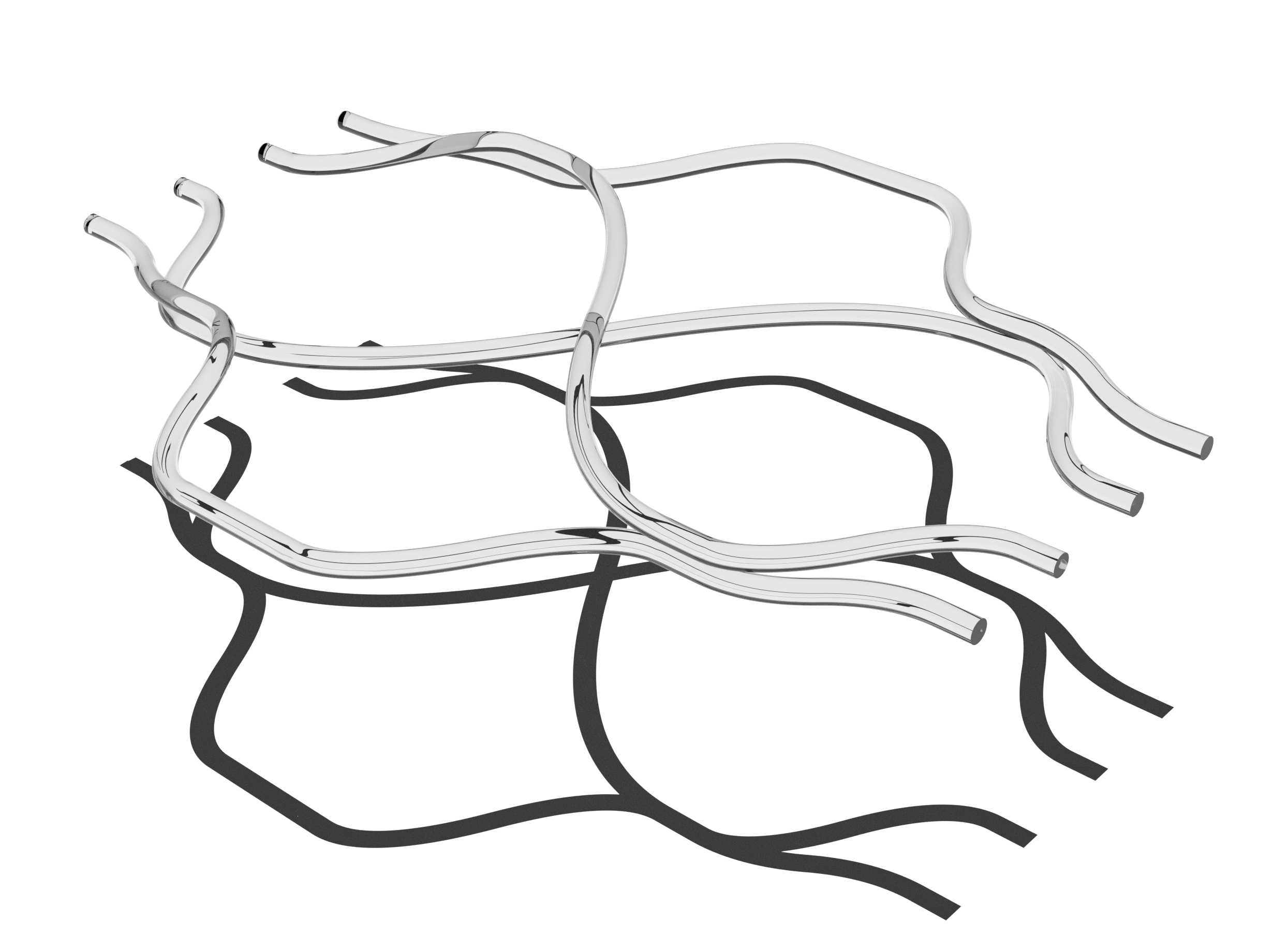}
	\caption{4x4 integrated interferometer ``Quitter''.
	}
	\label{fig:QuitterInt}
\end{figure}

\begin{figure}[h!]
	\centering
	\includegraphics[width=0.9\textwidth]{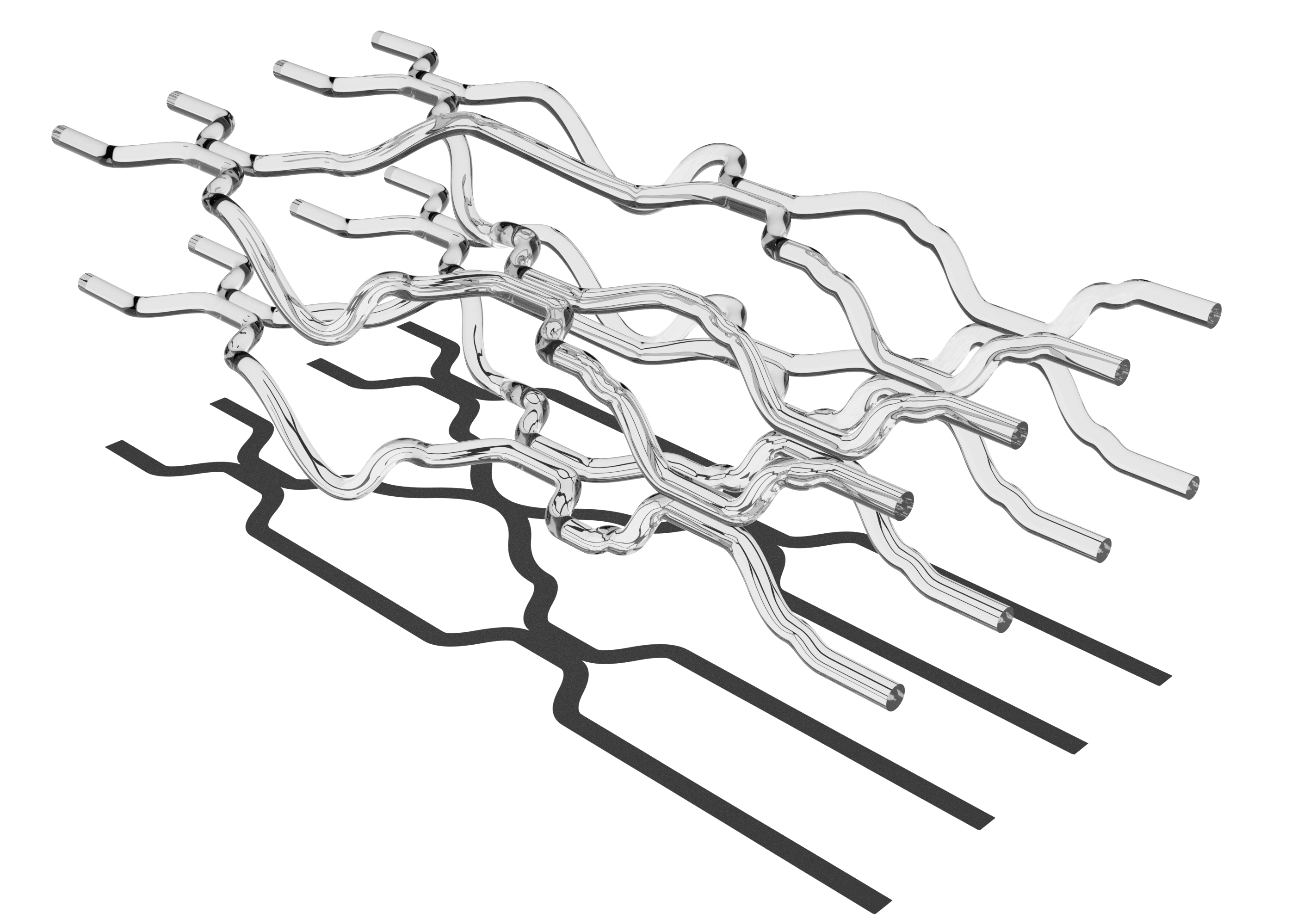}
	\caption{8x8 integrated interferometer ``Octopus''.
	}
	\label{fig:Octopus}
\end{figure}

We can also use femtosecond writing to fabricate integrated interferometers \cite{Marshall:09,Sansoni2010,crespi2013integrated,Spagnolo2013,meany2015} for use in multi-photon interference experiments. To this end I developed a Matlab based function library enabling the design of integrated circuits. In Figures \ref{fig:QuitterInt} and \ref{fig:Octopus} examples of a 4x4 ``Quitter'' and an 8x8 ``Octopus'' are shown.  

This 4x4 interferometer consists of a series of four evanescently coupled beam-splitters connected by path-length matched waveguides.

\section{The SLM as a characterisation tool for large multi-mode interferometers}
\begin{figure}[h!]
	\centering
	\includegraphics[width=0.7\textwidth]{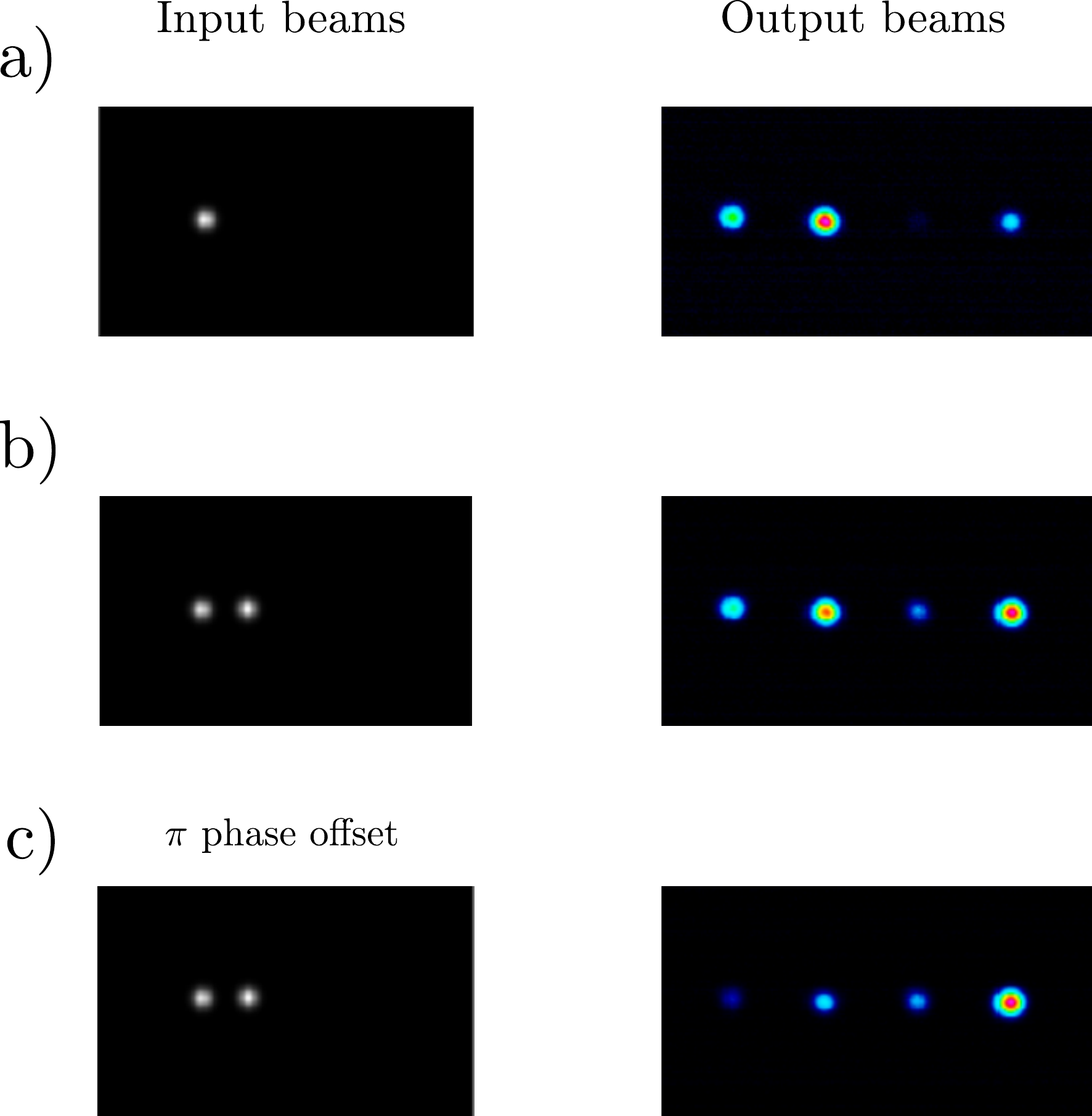}
	\caption{4x4 four coupler. Left panel: SLM image of the input beams. Right panel: Camera image of the output beams. a) Single input beam. b) Reference and probe beam. c) Probe beam with a $\pi$ phase offset.}
	\label{fig:SLMGUI}
\end{figure}

Since we can easily adjust the phase, position and intensity of the beams projected by the SLM we can implement a characterisation method suggested by Rahimi et. al. \cite{rahimi2013direct}. In a first step individual beams are inserted into each arm of the on-chip interferometer to obtain the splitting ratios. Then a reference beam is coupled to one port of the interferometer, while a moving second probe beam couples to every other port, simultaneously sweeping the phase offset between the reference and the probe beam. At the output of the device the variation in intensity is recorded for each input configuration and relative phase angle. 
If we insert coherent states $\ket{\alpha_1}$ and $\ket{\alpha_2} = \ket{e^{i\phi}\alpha_2}$ into port 1 and j of the interferometer,
the intensity $I_k$ at the kth output port is given by \cite{rahimi2013direct}:
\begin{equation}
I_k=I(r^2_{1k}+r^2_{jk}+2r_{1k}r_{jk}\cos(\phi+\theta_{jk})),
\end{equation} 
where $r_{jk}$ and $\theta_{jk}$ are amplitude and phase of the unitary interferometer respectively.
$U_{jk}=r_{jk}e^{i\theta_{jk}}$.
In Figure \ref{fig:SLMGUI} several input configurations and measured output beams  for a random 4x4 coupler are illustrated.

\begin{figure}[h!]
	\centering
	\includegraphics[width=0.8\textwidth]{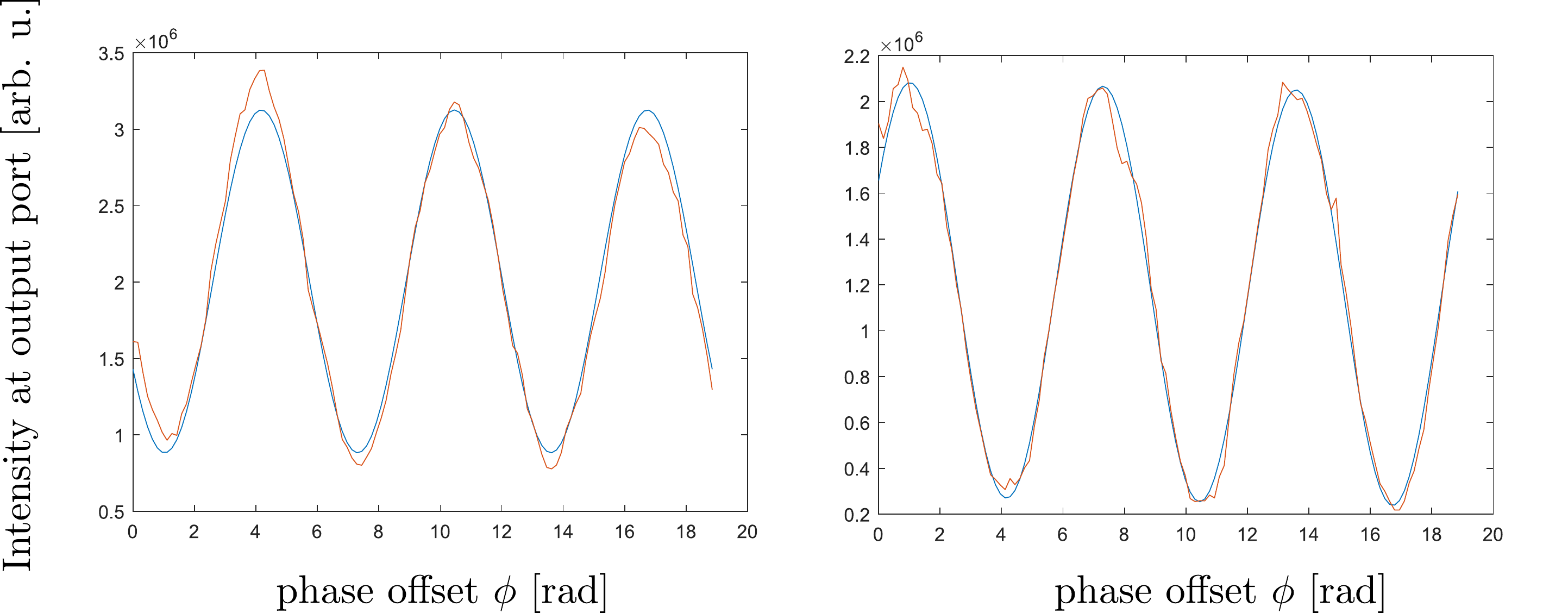}
	\caption{Recorded output intensity at two ports, as the relative phase between reference and probe beam is scanned.}
	\label{fig:UnitaryCharac}
\end{figure}
\begin{figure}[h!]
	\centering
	\includegraphics[width=0.5\textwidth]{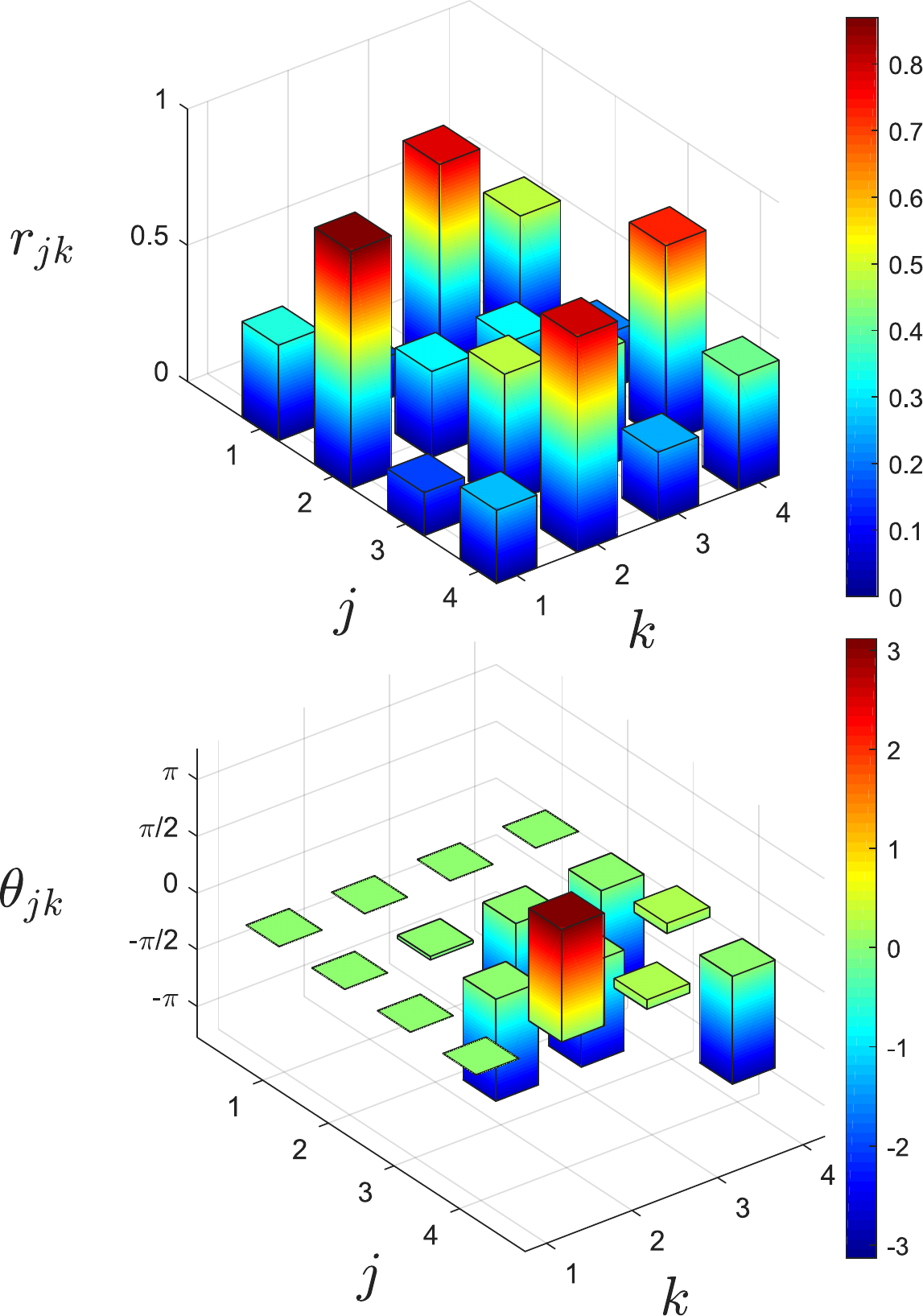}
	\caption{Characterisation data of a 4x4 unitary interferometer.}
	\label{fig:UnitaryCharac2}
\end{figure}
\begin{figure}[h!]
	\centering
	\includegraphics[width=0.3\textwidth]{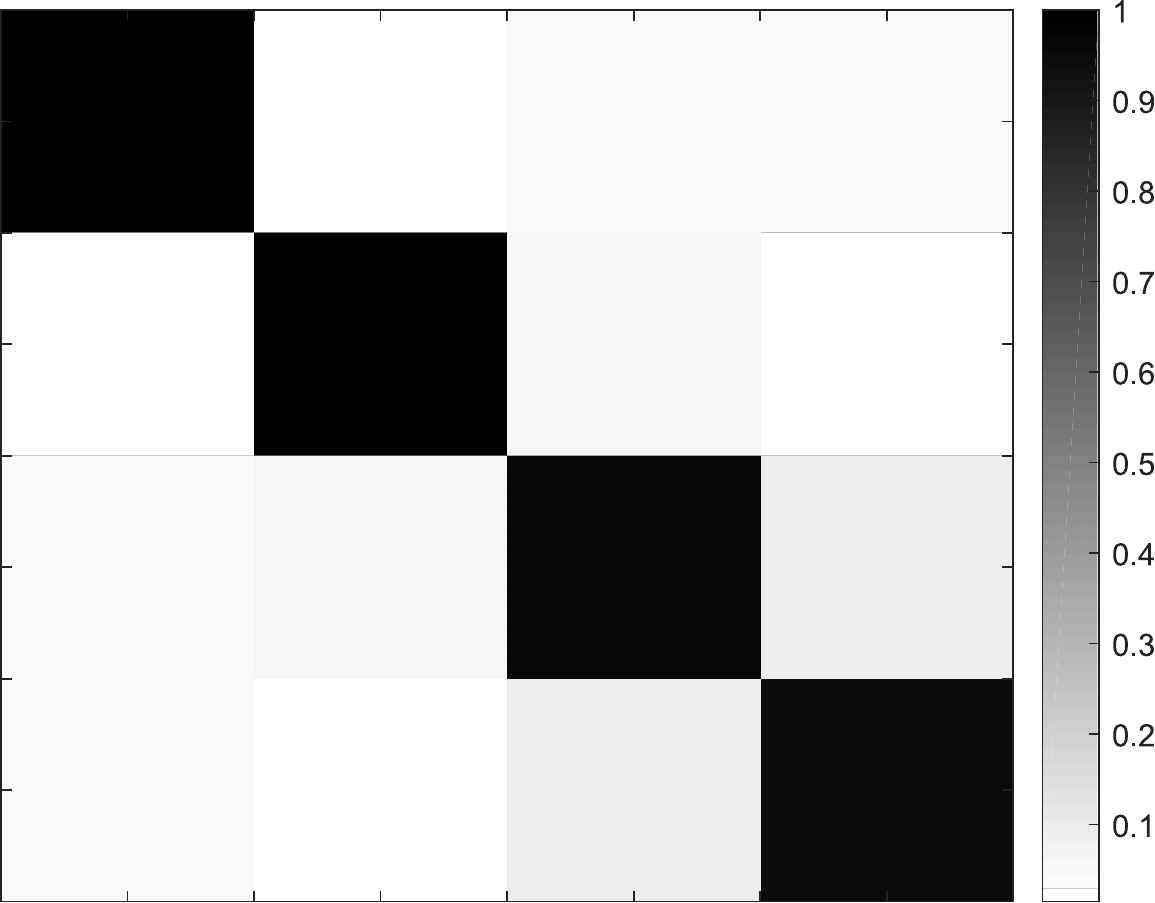}
	\caption{$|UU^\dagger|$}
	\label{fig:Unitarity}
\end{figure}

In Figure \ref{fig:UnitaryCharac2} the characterisation data for a 4x4 unitary interferometer is shown.
To verify the unitarity of the matrix obtained, I calculate $UU^\dagger$. The result is shown in Figure \ref{fig:Unitarity}. As expected the resulting matrix is close to unity.

\clearpage
\chapter{Derivation of waveguide distortion}
\label{app:WGDist}
Following and elaborating \cite{chamon2000solitons}:
our starting point are displacements of the hexagonal lattice sites of the form: $u_A(\bm{r})=\mathcal{A}e^{-i\bm{G\cdot r}}$, $u_B(\bm{r})=\mathcal{B}e^{i\bm{G\cdot r}}$, where we have written
the vectors $\bm{u}_A(\bm{r})$, $\bm{u}_B(\bm{r})$ as complex numbers. $\bm{G}=\bm{K}_+-\bm{K}_-=2\bm{K}_+$. The associated displacement vector is simply: $\bm{u}_{AB}(\bm{r})=\binom{\text{Re}[u_{AB}]}{\text{Im}[u_{AB}]}$.
\begin{figure}[h!]
	\centering
	\includegraphics[width=0.5\textwidth]{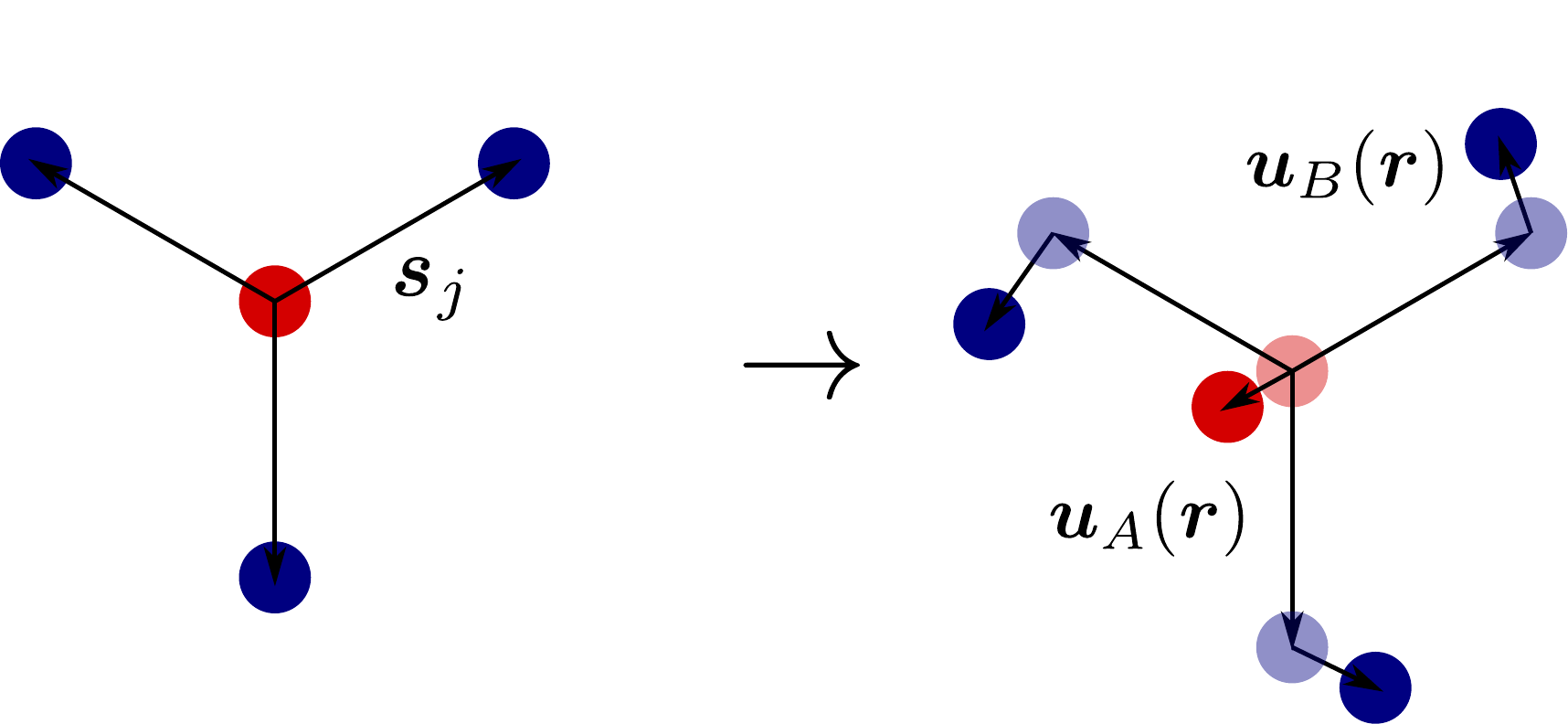}
	\caption{Graphene lattice displacement with two triangular sub-lattices in blue and red, vectors between nearest neighbours $\bm{s_j}$ and displacement vectors $\bm{u}_A(\bm{r})$ and $\bm{u}_B(\bm{r})$ indicated.}
	\label{fig:sublatDisp2}
\end{figure}
In Figure \ref{fig:sublatDisp2} the site displacements are illustrated.
For a lattice site in sublattice $A$ at position $\bm{r}_{AB}$ the site is displaced to $\bm{r}_{AB}\rightarrow\bm{r}_{AB}+\bm{u}_{AB}(\bm{r})$.
We now introduced the cubic roots of unity:
$z_j=e^{i2\pi/3(j-1)}$, which satisfy:
\begin{eqnarray}
\sum_{j=1}^3 z_j=0\\
\sum_{j=1}^3 z_j^2=0\\
\sum_{j=1}^3 z_j^3=3.
\end{eqnarray}
The same set of relations applies to their complex conjugate $z^*_j$.
We can also write them in the form $z_j=e^{i\bm{K_}+\cdot \bm{s}_j}$ and $z^*_j=e^{i\bm{K_-}\cdot \bm{s}_j}$. Another useful relation is: $z_j^2=z^*_j$.
Furthermore, we can write the complex nearest neighbour ``vectors'' as: $s_j=-ia_0z_j$. Using $z_j^2=z^*_j$ we obtain: $u_B(\bm{r}+\bm{s}_j)=u_B(\bm{r})z^*_j$.
A change in bond length along the $j$th bond can then be written as:

\begin{eqnarray}
\delta s_j/a_0&=&\frac{1}{a_0}|s_j-u_A(\bm{r})+u_B(\bm{r}+\bm{s}_j)|-1\\
&=&\frac{1}{a_0}\{(s_j-u_A(\bm{r})+u_B(\bm{r}+\bm{s}_j))\\\nonumber &\cdot&(s^*_j-u^*_A(\bm{r})+u^*_B(\bm{r}+\bm{s}_j))\}^{1/2}-1\\
&\approx&\{1-\frac{1}{a_0^2}s^*_j( u_A(\bm{r})-u_B(\bm{r}+\bm{s}_j))+\text{c.c.}\}^{1/2}-1\\
&\approx&-\frac{1}{2}\frac{s^*_j}{a_0^2}(u_A(\bm{r})-u_B(\bm{r}+\bm{s}_j))+\text{c.c.} .
\end{eqnarray}
Here we expanded the square root to first order in the displacements. We then substitute the explicit expressions for the displacements and using $s_j=-ia_0z_j$ as well as $z_j^2=z^*_j$, we write:
\begin{eqnarray}
\delta \bm{s}_j&=&-\frac{i}{2}\{z_j^*(\mathcal{A}e^{-i\bm{G\cdot r}}-\mathcal{B}e^{i\bm{G\cdot r}}z^*_j)\}+\text{c.c.}\\
&=&-\frac{i}{2}z^*_je^{-i\bm{G\cdot r}}(\mathcal{A}+\mathcal{B}^*)+\text{c.c.} .
\end{eqnarray}
Let us now assume that we have an exponential coupling between sites at $\bm{r}$ and $\bm{r}+\bm{s}_j$, which is of the form:
$t_{\bm{r},\bm{s}_j}=te^{-\gamma\delta \bm{s}_j}$
\begin{eqnarray}
t_{\bm{r},\bm{s}_j}&\approx& t-\gamma t \delta\bm{s}_j\\
&=&t+\delta t_{\bm{r},\bm{s}_j}
\end{eqnarray}
\begin{eqnarray}
\delta t_{\bm{r},\bm{s}_j}=\alpha z_j e^{i\bm{G\cdot r}}+\alpha^* z^*_je^{-i\bm{G\cdot r}},
\end{eqnarray}
where $\alpha=-\frac{i}{2}\gamma t(\mathcal{A}^*+\mathcal{B}).$
The Hamiltonian $H_c$ then reads:

\begin{eqnarray}
H_c=-\sum_{\bm{r}\in A}\sum_{j=1}^{3}(\alpha z_je^{i\bm{G\cdot r}}+\alpha^* z^*_je^{-i\bm{G\cdot r}})a^\dagger(\bm{r})b(\bm{r}+\bm{s}_j)+\text{h.c.} .
\end{eqnarray}
We now insert the Fourier transform of the ladder operators:
\begin{spacing}{1.2}
	\begin{equation}
	\Bigg\{ \begin{matrix}
	\displaystyle a(\bm{r}) \\[0.0cm]
	\displaystyle b(\bm{r})
	\end{matrix} \Bigg\}=\frac{1}{\sqrt{N}}\sum_{\bm {k\in BZ}}e^{i\bm {k r}}
	\Bigg\{ \begin{matrix}
	\displaystyle a(\bm{k}) \\[0.0cm]
	\displaystyle b(\bm{k})
	\end{matrix} \Bigg\}
	\end{equation}
\end{spacing}
\vspace{0.5cm}
and obtain:
\begin{eqnarray}
H_c=-1/N\sum_{\bm{r}\in A}\sum_{\bm{k},\bm{k}'}\sum_{j=1}^{3}(e^{i(\bm{k}'-(\bm{k}-\bm{G}))\bm{r}}e^{i\bm{k}'\bm{s}_j}\alpha z_j+...)a^\dagger(\bm{k})b(\bm{k'})+\text{h.c.}.
\end{eqnarray}
With the definition of the delta function we obtain:
\begin{eqnarray}
H_c=-\sum_{\bm{k}}\sum_{j=1}^{3}(e^{i(\bm{k-G})\bm{s}_j}\alpha z_j)a^\dagger(\bm{k})b(\bm{k-G})+...+\text{h.c.}.
\end{eqnarray}
Next, we separate the sum over $\bm{k}$ into two parts around the two Dirac points. $\bm{k}=\bm{p} +\bm{K}^+$ and $\bm{k}=\bm{p} +\bm{K}^-$.

\begin{eqnarray}
H_c&=&-\sum_{\bm{p}}\sum_{j=1}^{3}(e^{i(\bm{p}+\bm{K}_-)\bm{s}_j}\alpha z_j)a^\dagger(\bm{p} +\bm{K}_+)b(\bm{p} +\bm{K}_-)+...+\text{h.c.}\\
&\approx&-\sum_{\bm{p}}3\alpha a^\dagger(\bm{p} +\bm{K}_+)b(\bm{p} +\bm{K}_-)+...+\text{h.c.}\\
&=&-\sum_{\bm{p}}3\alpha a_+^\dagger(\bm{p})b_-(\bm{p})+... +\text{h.c.}
\end{eqnarray}
In the second step we have expanded the exponential to first order in $\bm{p}$ and used that $\sum_{j=1}^3 \bm{ps_j}=0$ and $ z_j^*z_j=1$. Next, we identify $3\alpha=-\frac{3i}{2}\gamma t(\mathcal{A}^*+\mathcal{B})=\Delta(\bm{r})$
and read off $\mathcal{A}^*=\frac{1i}{3\gamma t}\Delta(\bm{r})$, $\mathcal{B}=\mathcal{A}^*$.
The complex displacements vectors are then:
\begin{eqnarray}
u_A(\bm{r})&=&-\frac{1i}{3\gamma t}\Delta^*(\bm{r})e^{-i\bm{G\cdot r}}\\
u_B(\bm{r})&=&\frac{1i}{3\gamma t}\Delta(\bm{r})e^{i\bm{G\cdot r}},
\end{eqnarray}
and the vector valued displacements are:
\begin{eqnarray}
\bm{u}_A(\bm{r})&=&\frac{1i}{6\gamma t}\Delta(\bm{r})e^{i\bm{G\cdot r}}\binom{1}{+i}+\text{c.c.}\\
\bm{u}_B(\bm{r})&=&\frac{1i}{6\gamma t}\Delta(\bm{r})e^{i\bm{G\cdot r}}\binom{1}{-i}+\text{c.c.} .
\end{eqnarray}
Note that can make the replacement $\bm{G}=2\bm{K_+}\rightarrow -\bm{K_+}$ because of the periodicity of the Brillouin zone, and identify $3\gamma t=\frac{1}{\xi}$.

%% file: MenssenThesis.bbl
\begin{thebibliography}{100}

\bibitem{einstein1905erzeugung}
A.~Einstein, ``{{\"{U}}ber einen die Erzeugung und Verwandlung des Lichtes
  betreffenden heuristischen Gesichtspunkt},'' {\em Annalen der physik},
  vol.~322, no.~6, pp.~132--148, 1905.

\bibitem{hong1987measurement}
C.-K. Hong, Z.-Y. Ou, and L.~Mandel, ``{Measurement of subpicosecond time
  intervals between two photons by interference},'' {\em Physical review
  letters}, vol.~59, no.~18, p.~2044, 1987.

\bibitem{pittman1996can}
T.~B. Pittman, D.~V. Strekalov, A.~Migdall, M.~H. Rubin, A.~V. Sergienko, and
  Y.~H. Shih, ``{Can two-photon interference be considered the interference of
  two photons?},'' {\em Physical Review Letters}, vol.~77, no.~10, p.~1917,
  1996.

\bibitem{rarity1990two}
J.~G. Rarity, P.~R. Tapster, E.~Jakeman, T.~Larchuk, R.~A. Campos, M.~C. Teich,
  and B.~E.~A. Saleh, ``{Two-photon interference in a Mach-Zehnder
  interferometer},'' {\em Physical review letters}, vol.~65, no.~11, p.~1348,
  1990.

\bibitem{kwiat1990correlated}
P.~G. Kwiat, W.~A. Vareka, C.~K. Hong, H.~Nathel, and R.~Y. Chiao,
  ``{Correlated two-photon interference in a dual-beam Michelson
  interferometer},'' {\em Physical Review A}, vol.~41, no.~5, p.~2910, 1990.

\bibitem{bennett2009interference}
A.~J. Bennett, R.~B. Patel, C.~A. Nicoll, D.~A. Ritchie, and A.~J. Shields,
  ``{Interference of dissimilar photon sources},'' {\em Nature Physics},
  vol.~5, no.~10, p.~715, 2009.

\bibitem{patel2010two}
R.~B. Patel, A.~J. Bennett, I.~Farrer, C.~A. Nicoll, D.~A. Ritchie, and A.~J.
  Shields, ``{Two-photon interference of the emission from electrically tunable
  remote quantum dots},'' {\em Nature photonics}, vol.~4, no.~9, p.~632, 2010.

\bibitem{matthews2013observing}
J.~C.~F. Matthews, K.~Poulios, J.~D.~A. Meinecke, A.~Politi, A.~Peruzzo,
  N.~Ismail, K.~W{\"{o}}rhoff, M.~G. Thompson, and J.~L. O'brien, ``{Observing
  fermionic statistics with photons in arbitrary processes},'' {\em Scientific
  reports}, vol.~3, p.~1539, 2013.

\bibitem{mandel1991coherence}
L.~Mandel, ``{Coherence and indistinguishability},'' {\em Optics letters},
  vol.~16, no.~23, pp.~1882--1883, 1991.

\bibitem{brod2019witnessing}
D.~J. Brod, E.~F. Galv{\~{a}}o, N.~Viggianiello, F.~Flamini, N.~Spagnolo, and
  F.~Sciarrino, ``{Witnessing Genuine Multiphoton Indistinguishability},'' {\em
  Physical review letters}, vol.~122, no.~6, p.~63602, 2019.

\bibitem{kocsis2011observing}
S.~Kocsis, B.~Braverman, S.~Ravets, M.~J. Stevens, R.~P. Mirin, L.~K. Shalm,
  and A.~M. Steinberg, ``{Observing the average trajectories of single photons
  in a two-slit interferometer},'' {\em Science}, vol.~332, no.~6034,
  pp.~1170--1173, 2011.

\bibitem{schmidt2013momentum}
L.~P.~H. Schmidt, J.~Lower, T.~Jahnke, S.~Sch{\"{o}}{\ss}ler, M.~S.
  Sch{\"{o}}ffler, A.~Menssen, C.~L{\'{e}}v{\^{e}}que, N.~Sisourat,
  R.~Ta{\"{i}}eb, H.~Schmidt-B{\"{o}}cking, and Others, ``{Momentum transfer to
  a free floating double slit: realization of a thought experiment from the
  Einstein-Bohr debates},'' {\em Physical review letters}, vol.~111, no.~10,
  p.~103201, 2013.

\bibitem{scully1982quantum}
M.~O. Scully and K.~Dr{\"{u}}hl, ``{Quantum eraser: A proposed photon
  correlation experiment concerning observation and ``delayed choice'' in
  quantum mechanics},'' {\em Physical Review A}, vol.~25, no.~4, p.~2208, 1982.

\bibitem{menssen2017distinguishability}
A.~J. Menssen, A.~E. Jones, B.~J. Metcalf, M.~C. Tichy, S.~Barz, W.~S.
  Kolthammer, and I.~A. Walmsley, ``{Distinguishability and many-particle
  interference},'' {\em Physical review letters}, vol.~118, no.~15, p.~153603,
  2017.

\bibitem{Tichy2015}
M.~C. Tichy, ``{Sampling of partially distinguishable bosons and the relation
  to the multidimensional permanent},'' {\em Physical Review A}, vol.~91,
  no.~2, p.~022316, 2015.

\bibitem{shchesnovich2015}
V.~S. Shchesnovich, ``{Partial indistinguishability theory for multiphoton
  experiments in multiport devices},'' {\em Phys. Rev. A}, vol.~91, p.~13844,
  jan 2015.

\bibitem{ra2013nonmonotonic}
Y.-S. Ra, M.~C. Tichy, H.-T. Lim, O.~Kwon, F.~Mintert, A.~Buchleitner, and
  Y.-H. Kim, ``{Nonmonotonic quantum-to-classical transition in multiparticle
  interference},'' {\em Proceedings of the National Academy of Sciences},
  vol.~110, no.~4, pp.~1227--1231, 2013.

\bibitem{tichy2011four}
M.~C. Tichy, H.-T. Lim, Y.-S. Ra, F.~Mintert, Y.-H. Kim, and A.~Buchleitner,
  ``{Four-photon indistinguishability transition},'' {\em Physical Review A},
  vol.~83, no.~6, p.~62111, 2011.

\bibitem{shchesnovich2017interference}
V.~S. Shchesnovich and M.~E.~O. Bezerra, ``{Interference of identical
  particles: collective phases, ``circle dancing'' and graph theory},'' {\em
  arXiv preprint arXiv:1707.03893}, 2017.

\bibitem{jones2018interfering}
A.~E. Jones, A.~J. Menssen, H.~M. Chrzanowski, V.~S. Shchesnovich, and I.~A.
  Walmsley, ``{Interfering photons in orthogonal states},'' in {\em CLEO:
  QELS{\_}Fundamental Science}, pp.~FTh1H----4, Optical Society of America,
  2018.

\bibitem{campos2000three}
R.~A. Campos, ``{Three-photon Hong-Ou-Mandel interference at a multiport
  mixer},'' {\em Physical Review A}, vol.~62, no.~1, p.~13809, 2000.

\bibitem{lim2005generalized}
Y.~L. Lim and A.~Beige, ``{Generalized Hong--Ou--Mandel experiments with bosons
  and fermions},'' {\em New Journal of Physics}, vol.~7, no.~1, p.~155, 2005.

\bibitem{mahrlein2017hong}
S.~M{\"{a}}hrlein, S.~Oppel, R.~Wiegner, and J.~von Zanthier,
  ``{Hong--Ou--Mandel interference without beam splitters},'' {\em Journal of
  Modern Optics}, vol.~64, no.~9, pp.~921--929, 2017.

\bibitem{agne2017observation}
S.~Agne, T.~Kauten, J.~Jin, E.~Meyer-Scott, J.~Z. Salvail, D.~R. Hamel, K.~J.
  Resch, G.~Weihs, and T.~Jennewein, ``{Observation of genuine three-photon
  interference},'' {\em Physical review letters}, vol.~118, no.~15, p.~153602,
  2017.

\bibitem{tillmann2015generalized}
M.~Tillmann, S.-H. Tan, S.~E. Stoeckl, B.~C. Sanders, H.~{De Guise},
  R.~Heilmann, S.~Nolte, A.~Szameit, and P.~Walther, ``{Generalized multiphoton
  quantum interference},'' {\em Phys. Rev. X}, vol.~5, p.~41015, oct 2015.

\bibitem{stanisic2018discriminating}
S.~Stanisic and P.~S. Turner, ``{Discriminating distinguishability},'' {\em
  Physical Review A}, vol.~98, no.~4, p.~43839, 2018.

\bibitem{tichy2017extending}
M.~C. Tichy and K.~M{\o}lmer, ``{Extending exchange symmetry beyond bosons and
  fermions},'' {\em Physical Review A}, vol.~96, no.~2, p.~22119, 2017.

\bibitem{green1953generalized}
H.~S. Green, ``{A generalized method of field quantization},'' {\em Physical
  Review}, vol.~90, no.~2, p.~270, 1953.

\bibitem{branning2000interferometric}
D.~Branning, W.~Grice, R.~Erdmann, and I.~A. Walmsley, ``{Interferometric
  technique for engineering indistinguishability and entanglement of photon
  pairs},'' {\em Physical Review A}, vol.~62, no.~1, p.~13814, 2000.

\bibitem{tichy2012many}
M.~C. Tichy, M.~Tiersch, F.~Mintert, and A.~Buchleitner, ``{Many-particle
  interference beyond many-boson and many-fermion statistics},'' {\em New
  Journal of Physics}, vol.~14, no.~9, p.~93015, 2012.

\bibitem{stern2004geometric}
A.~Stern, F.~von Oppen, and E.~Mariani, ``{Geometric phases and quantum
  entanglement as building blocks for non-Abelian quasiparticle statistics},''
  {\em Physical Review B}, vol.~70, no.~20, p.~205338, 2004.

\bibitem{teo2010topological}
J.~C.~Y. Teo and C.~L. Kane, ``{Topological defects and gapless modes in
  insulators and superconductors},'' {\em Physical Review B}, vol.~82, no.~11,
  p.~115120, 2010.

\bibitem{asboth2016short}
J.~K. Asb{\'{o}}th, L.~Oroszl{\'{a}}ny, and A.~P{\'{a}}lyi, ``{A short course
  on topological insulators},'' {\em Lecture notes in physics}, vol.~919, 2016.

\bibitem{ozawa2018topological}
T.~Ozawa, H.~M. Price, A.~Amo, N.~Goldman, M.~Hafezi, L.~Lu, M.~Rechtsman,
  D.~Schuster, J.~Simon, O.~Zilberberg, and Others, ``{Topological
  photonics},'' {\em arXiv preprint arXiv:1802.04173}, 2018.

\bibitem{rechtsman2013photonic}
M.~C. Rechtsman, J.~M. Zeuner, Y.~Plotnik, Y.~Lumer, D.~Podolsky, F.~Dreisow,
  S.~Nolte, M.~Segev, and A.~Szameit, ``{Photonic Floquet topological
  insulators},'' {\em Nature}, vol.~496, no.~7444, p.~196, 2013.

\bibitem{rechtsman2013strain}
M.~C. Rechtsman, J.~M. Zeuner, A.~T{\"{u}}nnermann, S.~Nolte, M.~Segev, and
  A.~Szameit, ``{Strain-induced pseudomagnetic field and photonic Landau levels
  in dielectric structures},'' {\em Nature Photonics}, vol.~7, no.~2, p.~153,
  2013.

\bibitem{rechtsman2016topological}
M.~C. Rechtsman, Y.~Lumer, Y.~Plotnik, A.~Perez-Leija, A.~Szameit, and
  M.~Segev, ``{Topological protection of photonic path entanglement},'' {\em
  Optica}, vol.~3, no.~9, pp.~925--930, 2016.

\bibitem{noh2018topological}
J.~Noh, W.~A. Benalcazar, S.~Huang, M.~J. Collins, K.~P. Chen, T.~L. Hughes,
  and M.~C. Rechtsman, ``{Topological protection of photonic mid-gap defect
  modes},'' {\em Nature Photonics}, p.~1, 2018.

\bibitem{blanco2018topological}
A.~Blanco-Redondo, B.~Bell, D.~Oren, B.~J. Eggleton, and M.~Segev,
  ``{Topological protection of biphoton states},'' {\em Science}, vol.~362,
  no.~6414, pp.~568--571, 2018.

\bibitem{iadecola2016non}
T.~Iadecola, T.~Schuster, and C.~Chamon, ``{Non-abelian braiding of light},''
  {\em Physical review letters}, vol.~117, no.~7, p.~73901, 2016.

\bibitem{jackiw1981zero}
R.~Jackiw and P.~Rossi, ``{Zero modes of the vortex-fermion system},'' {\em
  Nuclear Physics B}, vol.~190, no.~4, pp.~681--691, 1981.

\bibitem{szameit2010discrete}
A.~Szameit and S.~Nolte, ``{Discrete optics in femtosecond-laser-written
  photonic structures},'' {\em Journal of Physics B: Atomic, Molecular and
  Optical Physics}, vol.~43, no.~16, p.~163001, 2010.

\bibitem{Huang2016}
L.~Huang, P.~S. Salter, F.~Payne, and M.~J. Booth, ``{Aberration correction for
  direct laser written waveguides in a transverse geometry},'' {\em Optics
  Express}, vol.~24, no.~10, p.~10565, 2016.

\bibitem{chamon2010quantizing}
C.~Chamon, R.~Jackiw, Y.~Nishida, S.-Y. Pi, and L.~Santos, ``{Quantizing
  Majorana fermions in a superconductor},'' {\em Physical Review B}, vol.~81,
  no.~22, p.~224515, 2010.

\bibitem{milovanovic2008fractionalization}
M.~V. Milovanovi{\'{c}}, ``{Fractionalization in dimerized graphene and
  graphene bilayer},'' {\em Physical Review B}, vol.~78, no.~24, p.~245424,
  2008.

\bibitem{shchesnovich2018collective}
V.~S. Shchesnovich and M.~E.~O. Bezerra, ``{Collective phases of identical
  particles interfering on linear multiports},'' {\em Physical Review A},
  vol.~98, no.~3, p.~33805, 2018.

\bibitem{Jackiw1976}
R.~Jackiw and C.~Rebbi, ``{Solitons with fermion number},'' {\em Physical
  Review D}, vol.~13, no.~12, pp.~3398--3409, 1976.

\bibitem{Jackiw2007}
R.~Jackiw and S.~Y. Pi, ``{Chiral gauge theory for graphene},'' {\em Physical
  Review Letters}, vol.~98, no.~26, pp.~1--4, 2007.

\bibitem{sansoni2012two}
L.~Sansoni, F.~Sciarrino, G.~Vallone, P.~Mataloni, A.~Crespi, R.~Ramponi, and
  R.~Osellame, ``{Two-particle bosonic-fermionic quantum walk via integrated
  photonics},'' {\em Physical review letters}, vol.~108, no.~1, p.~10502, 2012.

\bibitem{Barcy1977}
H.~Bacry, ``{Lectures on group theory and particle theory},'' 1977.

\bibitem{leinaas1977theory}
J.~M. Leinaas and J.~Myrheim, ``{On the theory of identical particles},'' {\em
  Il Nuovo Cimento B (1971-1996)}, vol.~37, no.~1, pp.~1--23, 1977.

\bibitem{roos2017revealing}
C.~F. Roos, A.~Alberti, D.~Meschede, P.~Hauke, and H.~H{\"{a}}ffner,
  ``{Revealing quantum statistics with a pair of distant atoms},'' {\em
  Physical review letters}, vol.~119, no.~16, p.~160401, 2017.

\bibitem{carpenter1970green}
K.~M. Carpenter, ``{Green's paraparticles and the unified theory of identical
  particles},'' {\em Annals of Physics}, vol.~60, no.~1, pp.~1--26, 1970.

\bibitem{artin1947theory}
E.~Artin, ``{Theory of braids},'' {\em Ann. of Math}, vol.~48, no.~2,
  pp.~101--126, 1947.

\bibitem{nayak2008non}
C.~Nayak, S.~H. Simon, A.~Stern, M.~Freedman, and S.~D. Sarma, ``{Non-Abelian
  anyons and topological quantum computation},'' {\em Reviews of Modern
  Physics}, vol.~80, no.~3, p.~1083, 2008.

\bibitem{smith2007photon}
C.~H. Monken, B.~J. Smith, and M.~G. Raymer, ``{Photon wave functions,
  wave-packet quantization of light, and coherence theory},'' {\em New Journal
  of Physics}, vol.~9, no.~11, p.~414, 2007.

\bibitem{Fedorov2005}
M.~V. Fedorov, M.~A. Efremov, A.~E. Kazakov, K.~W. Chan, C.~K. Law, and J.~H.
  Eberly, ``{Spontaneous emission of a photon: Wave-packet structures and
  atom-photon entanglement},'' {\em Phys. Rev. A}, vol.~72, p.~32110, sep 2005.

\bibitem{schwinger1960unitary}
J.~Schwinger, ``{Unitary transformations and the action principle},'' {\em
  Proceedings of the national academy of sciences of the United States of
  America}, vol.~46, no.~6, p.~883, 1960.

\bibitem{gu2019quantum}
X.~Gu, M.~Erhard, A.~Zeilinger, and M.~Krenn, ``{Quantum experiments and graphs
  II: Quantum interference, computation, and state generation},'' {\em
  Proceedings of the National Academy of Sciences}, vol.~116, no.~10,
  pp.~4147--4155, 2019.

\bibitem{bradler2018gaussian}
K.~Br{\'{a}}dler, P.-L. Dallaire-Demers, P.~Rebentrost, D.~Su, and
  C.~Weedbrook, ``{Gaussian boson sampling for perfect matchings of arbitrary
  graphs},'' {\em Physical Review A}, vol.~98, no.~3, p.~32310, 2018.

\bibitem{benjamin2006brokered}
S.~C. Benjamin, D.~E. Browne, J.~Fitzsimons, and J.~J.~L. Morton, ``{Brokered
  graph-state quantum computation},'' {\em New Journal of Physics}, vol.~8,
  no.~8, p.~141, 2006.

\bibitem{lu2007experimental}
C.-Y. Lu, X.-Q. Zhou, O.~G{\"{u}}hne, W.-B. Gao, J.~Zhang, Z.-S. Yuan,
  A.~Goebel, T.~Yang, and J.-W. Pan, ``{Experimental entanglement of six
  photons in graph states},'' {\em Nature physics}, vol.~3, no.~2, p.~91, 2007.

\bibitem{Spagnolo2013}
N.~Spagnolo, C.~Vitelli, L.~Aparo, P.~Mataloni, F.~Sciarrino, A.~Crespi,
  R.~Ramponi, and R.~Osellame, ``{Three-photon bosonic coalescence in an
  integrated tritter.},'' {\em Nature communications}, vol.~4, p.~1606, 2013.

\bibitem{spring2017chip}
J.~B. Spring, P.~L. Mennea, B.~J. Metcalf, P.~C. Humphreys, J.~C. Gates, H.~L.
  Rogers, C.~S{\"{o}}ller, B.~J. Smith, W.~S. Kolthammer, P.~G.~R. Smith, and
  Others, ``{Chip-based array of near-identical, pure, heralded single-photon
  sources},'' {\em Optica}, vol.~4, no.~1, pp.~90--96, 2017.

\bibitem{tan2013}
S.-H. Tan, Y.~Y. Gao, H.~de~Guise, and B.~C. Sanders, ``{SU (3) quantum
  interferometry with single-photon input pulses},'' {\em Physical review
  letters}, vol.~110, no.~11, p.~113603, 2013.

\bibitem{deGuise2014}
H.~de~Guise, S.-H. Tan, I.~P. Poulin, and B.~C. Sanders, ``{Coincidence
  landscapes for three-channel linear optical networks},'' {\em Physical Review
  A}, vol.~89, no.~6, p.~63819, 2014.

\bibitem{shchesnovich2014}
V.~S. Shchesnovich, ``{Sufficient condition for the mode mismatch of single
  photons for scalability of the boson-sampling computer},'' {\em Physical
  Review A}, vol.~89, no.~2, p.~22333, 2014.

\bibitem{tamma2016multi}
V.~Tamma and S.~Laibacher, ``{Multi-boson correlation sampling},'' {\em Quantum
  Information Processing}, vol.~15, no.~3, pp.~1241--1262, 2016.

\bibitem{tichy2014interference}
M.~C. Tichy, ``{Interference of identical particles from entanglement to
  boson-sampling},'' {\em Journal of Physics B: Atomic, Molecular and Optical
  Physics}, vol.~47, no.~10, p.~103001, 2014.

\bibitem{shchesnovich2015tight}
V.~S. Shchesnovich, ``{Tight bound on the trace distance between a realistic
  device with partially indistinguishable bosons and the ideal
  BosonSampling},'' {\em Physical Review A}, vol.~91, no.~6, p.~63842, 2015.

\bibitem{zeilinger1993einstein}
A.~Zeilinger, M.~Zukowski, M.~A. Horne, H.~J. Bernstein, and D.~M. Greenberger,
  ``{Einstein-Podolsky-Rosen correlations in higher dimensions},'' {\em
  Fundamental Aspects of Quantum Theory, Singapore: World Scientific}, 1993.

\bibitem{spring2013chip}
J.~B. Spring, P.~S. Salter, B.~J. Metcalf, P.~C. Humphreys, M.~Moore,
  N.~Thomas-Peter, M.~Barbieri, X.-M. Jin, N.~K. Langford, W.~S. Kolthammer,
  M.~J. Booth, and I.~A. Walmsley, ``spr,'' {\em Optics Express}, vol.~21,
  no.~11, pp.~13522--13532, 2013.

\bibitem{lepert2011demonstration}
G.~Lepert, M.~Trupke, E.~A. Hinds, H.~Rogers, J.~C. Gates, and P.~G.~R. Smith,
  ``{Demonstration of UV-written waveguides, Bragg gratings and cavities at 780
  nm, and an original experimental measurement of group delay},'' {\em Optics
  express}, vol.~19, no.~25, pp.~24933--24943, 2011.

\bibitem{knill2001scheme}
E.~Knill, R.~Laflamme, and G.~J. Milburn, ``{A scheme for efficient quantum
  computation with linear optics},'' {\em nature}, vol.~409, no.~6816,
  pp.~46--52, 2001.

\bibitem{hofmann2012heralded}
J.~Hofmann, M.~Krug, N.~Ortegel, L.~G{\'{e}}rard, M.~Weber, W.~Rosenfeld, and
  H.~Weinfurter, ``{Heralded entanglement between widely separated atoms},''
  {\em Science}, vol.~337, no.~6090, pp.~72--75, 2012.

\bibitem{I4phot}
A.~E. Jones, A.~J. Menssen, H.~M. Chrzanowski, V.~S. Shchesnovich, and I.~A.
  Walmsley, ``{Interfering photons in orthogonal states},'' {\em In
  preparation}, 2019.

\bibitem{shih1993four}
Y.~H. Shih and M.~H. Rubin, ``{Four photon interference experiment for the
  testing of the Greenberger-Horne-Zeilinger theorem},'' {\em Physics Letters
  A}, vol.~182, no.~1, pp.~16--22, 1993.

\bibitem{ou1999observation}
Z.~Y. Ou, J.-K. Rhee, and L.~J. Wang, ``{Observation of four-photon
  interference with a beam splitter by pulsed parametric down-conversion},''
  {\em Physical review letters}, vol.~83, no.~5, p.~959, 1999.

\bibitem{de2003quantum}
H.~{De Riedmatten}, I.~Marcikic, W.~Tittel, H.~Zbinden, and N.~Gisin,
  ``{Quantum interference with photon pairs created in spatially separated
  sources},'' {\em Physical Review A}, vol.~67, no.~2, p.~22301, 2003.

\bibitem{liu2007four}
B.~H. Liu, F.~W. Sun, Y.~X. Gong, Y.~F. Huang, G.~C. Guo, and Z.~Y. Ou,
  ``{Four-photon interference with asymmetric beam splitters},'' {\em Optics
  letters}, vol.~32, no.~10, pp.~1320--1322, 2007.

\bibitem{wong2005quantum}
H.~M. Wong, K.~M. Cheng, and M.-C. Chu, ``{Quantum geometric phase between
  orthogonal states},'' {\em Physical review letters}, vol.~94, no.~7,
  p.~70406, 2005.

\bibitem{bernstein1974must}
H.~J. Bernstein, ``{Must quantum theory assume unrestricted superposition?},''
  {\em Journal of Mathematical Physics}, vol.~15, no.~10, pp.~1677--1679, 1974.

\bibitem{mosley2008heralded}
P.~J. Mosley, J.~S. Lundeen, B.~J. Smith, P.~Wasylczyk, A.~B. U'Ren,
  C.~Silberhorn, and I.~A. Walmsley, ``{Heralded generation of ultrafast single
  photons in pure quantum states},'' {\em Physical Review Letters}, vol.~100,
  no.~13, p.~133601, 2008.

\bibitem{zou1991induced}
X.~Y. Zou, L.~J. Wang, and L.~Mandel, ``{Induced coherence and
  indistinguishability in optical interference},'' {\em Physical review
  letters}, vol.~67, no.~3, p.~318, 1991.

\bibitem{rice1994multiparticle}
D.~A. Rice, C.~F. Osborne, and P.~Lloyd, ``{Multiparticle interference},'' {\em
  Physics Letters A}, vol.~186, no.~1-2, pp.~21--28, 1994.

\bibitem{greenberger1990bell}
D.~M. Greenberger, M.~A. Horne, A.~Shimony, and A.~Zeilinger, ``{Bell's theorem
  without inequalities},'' {\em American Journal of Physics}, vol.~58, no.~12,
  pp.~1131--1143, 1990.

\bibitem{menssen2019photonic}
A.~J. Menssen, J.~Guan, D.~Felce, M.~J. Booth, and I.~A. Walmsley, ``{A
  Photonic Topological Mode Bound to a Vortex},'' {\em arXiv:1901.04439}, 2019.

\bibitem{WITTEN1985557}
E.~Witten, ``{Superconducting strings},'' {\em Nuclear Physics B}, vol.~249,
  no.~4, pp.~557--592, 1985.

\bibitem{thouless1982}
D.~J. Thouless, M.~Kohmoto, M.~P. Nightingale, and M.~den Nijs, ``{Quantized
  Hall Conductance in a Two-Dimensional Periodic Potential},'' {\em Phys. Rev.
  Lett.}, vol.~49, pp.~405--408, aug 1982.

\bibitem{haldane2017nobel}
F.~D.~M. Haldane, ``{Nobel lecture: Topological quantum matter},'' {\em Reviews
  of Modern Physics}, vol.~89, no.~4, p.~40502, 2017.

\bibitem{rechtsman2013topological}
M.~C. Rechtsman, Y.~Plotnik, J.~M. Zeuner, D.~Song, Z.~Chen, A.~Szameit, and
  M.~Segev, ``{Topological creation and destruction of edge states in photonic
  graphene},'' {\em Physical review letters}, vol.~111, no.~10, p.~103901,
  2013.

\bibitem{plotnik2014observation}
Y.~Plotnik, M.~C. Rechtsman, D.~Song, M.~Heinrich, J.~M. Zeuner, S.~Nolte,
  Y.~Lumer, N.~Malkova, J.~Xu, A.~Szameit, and Others, ``{Observation of
  unconventional edge states in ‘photonic graphene'},'' {\em Nature
  materials}, vol.~13, no.~1, p.~57, 2014.

\bibitem{stutzer2018photonic}
S.~St{\"{u}}tzer, Y.~Plotnik, Y.~Lumer, P.~Titum, N.~H. Lindner, M.~Segev,
  M.~C. Rechtsman, and A.~Szameit, ``{Photonic topological Anderson
  insulators},'' {\em Nature}, vol.~560, no.~7719, p.~461, 2018.

\bibitem{hafezi2013imaging}
M.~Hafezi, S.~Mittal, J.~Fan, A.~Migdall, and J.~M. Taylor, ``{Imaging
  topological edge states in silicon photonics},'' {\em Nature Photonics},
  vol.~7, no.~12, p.~1001, 2013.

\bibitem{bandres2018topological}
M.~A. Bandres, S.~Wittek, G.~Harari, M.~Parto, J.~Ren, M.~Segev, D.~N.
  Christodoulides, and M.~Khajavikhan, ``{Topological insulator laser:
  Experiments},'' {\em Science}, vol.~359, no.~6381, p.~eaar4005, 2018.

\bibitem{su1979solitons}
W.~Su, J.~R. Schrieffer, and A.~J. Heeger, ``{Solitons in polyacetylene},''
  {\em Physical review letters}, vol.~42, no.~25, p.~1698, 1979.

\bibitem{nakahara2003geometry}
M.~Nakahara, {\em {Geometry, topology and physics}}.
\newblock CRC Press, 2003.

\bibitem{lu2014topological}
L.~Lu, J.~D. Joannopoulos, and M.~Solja{\v{c}}i{\'{c}}, ``{Topological
  photonics},'' {\em Nature Photonics}, vol.~8, no.~11, p.~821, 2014.

\bibitem{raussendorf2007topological}
R.~Raussendorf, J.~Harrington, and K.~Goyal, ``{Topological fault-tolerance in
  cluster state quantum computation},'' {\em New Journal of Physics}, vol.~9,
  no.~6, p.~199, 2007.

\bibitem{bargmann1964note}
V.~Bargmann, ``{Note on Wigner's theorem on symmetry operations},'' {\em
  Journal of Mathematical Physics}, vol.~5, no.~7, pp.~862--868, 1964.

\bibitem{rabei1999bargmann}
E.~M. Rabei, N.~Mukunda, R.~Simon, and Others, ``{Bargmann invariants and
  geometric phases: A generalized connection},'' {\em Physical Review A},
  vol.~60, no.~5, p.~3397, 1999.

\bibitem{hatsugai1993edge}
Y.~Hatsugai, ``{Edge states in the integer quantum Hall effect and the Riemann
  surface of the Bloch function},'' {\em Physical Review B}, vol.~48, no.~16,
  p.~11851, 1993.

\bibitem{altland1997nonstandard}
A.~Altland and M.~R. Zirnbauer, ``{Nonstandard symmetry classes in mesoscopic
  normal-superconducting hybrid structures},'' {\em Physical Review B},
  vol.~55, no.~2, p.~1142, 1997.

\bibitem{angelakis2014probing}
D.~G. Angelakis, P.~Das, and C.~Noh, ``{Probing the topological properties of
  the Jackiw-Rebbi model with light},'' {\em Scientific reports}, vol.~4,
  p.~6110, 2014.

\bibitem{hou2007electron}
C.-Y.~Y. Hou, C.~Chamon, and C.~Mudry, ``{Electron fractionalization in
  two-dimensional graphenelike structures},'' {\em Physical review letters},
  vol.~98, no.~18, p.~186809, 2007.

\bibitem{chamon2000solitons}
C.~Chamon, ``{Solitons in carbon nanotubes},'' {\em Physical Review B},
  vol.~62, no.~4, p.~2806, 2000.

\bibitem{bhardwaj2005femtosecond}
V.~R. Bhardwaj, E.~Simova, P.~B. Corkum, D.~M. Rayner, C.~Hnatovsky, R.~S.
  Taylor, B.~Schreder, M.~Kluge, and J.~Zimmer, ``{Femtosecond laser-induced
  refractive index modification in multicomponent glasses},'' {\em Journal of
  applied physics}, vol.~97, no.~8, p.~83102, 2005.

\bibitem{eaton2005heat}
S.~M. Eaton, H.~Zhang, P.~R. Herman, F.~Yoshino, L.~Shah, J.~Bovatsek, and
  A.~Y. Arai, ``{Heat accumulation effects in femtosecond laser-written
  waveguides with variable repetition rate},'' {\em Optics Express}, vol.~13,
  no.~12, pp.~4708--4716, 2005.

\bibitem{Saleh2007}
B.~E.~A. Saleh and M.~C. Teich, {\em {Fundamentals of photonics}}, vol.~10.

\bibitem{Goorden2014}
S.~A. Goorden, J.~Bertolotti, and A.~P. Mosk, ``{Superpixel-based spatial
  amplitude and phase modulation using a digital micromirror device},'' {\em
  Optics Express}, vol.~22, no.~15, p.~17999, 2014.

\bibitem{huang1994coupled}
W.-P. Huang, ``{Coupled-mode theory for optical waveguides: an overview},''
  {\em JOSA A}, vol.~11, no.~3, pp.~963--983, 1994.

\bibitem{lo1993non}
H.-K. Lo and J.~Preskill, ``{Non-Abelian vortices and non-Abelian
  statistics},'' {\em Physical Review D}, vol.~48, no.~10, p.~4821, 1993.

\bibitem{teo2010majorana}
J.~C.~Y. Teo and C.~L. Kane, ``{Majorana fermions and non-Abelian statistics in
  three dimensions},'' {\em Physical review letters}, vol.~104, no.~4,
  p.~46401, 2010.

\bibitem{harari2018topological}
G.~Harari, M.~A. Bandres, Y.~Lumer, M.~C. Rechtsman, Y.~D. Chong,
  M.~Khajavikhan, D.~N. Christodoulides, and M.~Segev, ``{Topological insulator
  laser: Theory},'' {\em Science}, vol.~359, no.~6381, p.~eaar4003, 2018.

\bibitem{renema2018efficient}
J.~J. Renema, A.~Menssen, W.~R. Clements, G.~Triginer, W.~S. Kolthammer, and
  I.~A. Walmsley, ``{Efficient classical algorithm for boson sampling with
  partially distinguishable photons},'' {\em Physical review letters},
  vol.~120, no.~22, p.~220502, 2018.

\bibitem{aaronson2011computational}
S.~Aaronson and A.~Arkhipov, ``{The computational complexity of linear
  optics},'' in {\em Proceedings of the forty-third annual ACM symposium on
  Theory of computing}, pp.~333--342, ACM, 2011.

\bibitem{itai1982hamilton}
A.~Itai, C.~H. Papadimitriou, and J.~L. Szwarcfiter, ``{Hamilton paths in grid
  graphs},'' {\em SIAM Journal on Computing}, vol.~11, no.~4, pp.~676--686,
  1982.

\bibitem{viggianiello2018optimal}
N.~Viggianiello, F.~Flamini, M.~Bentivegna, N.~Spagnolo, A.~Crespi, D.~J. Brod,
  E.~F. Galv{\~{a}}o, R.~Osellame, and F.~Sciarrino, ``{Optimal photonic
  indistinguishability tests in multimode networks},'' {\em Science Bulletin},
  vol.~63, no.~22, pp.~1470--1478, 2018.

\bibitem{crespi2015suppression}
A.~Crespi, ``{Suppression laws for multiparticle interference in Sylvester
  interferometers},'' {\em Physical Review A}, vol.~91, no.~1, p.~13811, 2015.

\bibitem{rahimi2013direct}
S.~Rahimi-Keshari, M.~A. Broome, R.~Fickler, A.~Fedrizzi, T.~C. Ralph, and
  A.~G. White, ``{Direct characterization of linear-optical networks},'' {\em
  Optics express}, vol.~21, no.~11, pp.~13450--13458, 2013.

\bibitem{tambasco2018quantum}
J.-L. Tambasco, G.~Corrielli, R.~J. Chapman, A.~Crespi, O.~Zilberberg,
  R.~Osellame, and A.~Peruzzo, ``{Quantum interference of topological states of
  light},'' {\em Science advances}, vol.~4, no.~9, p.~eaat3187, 2018.

\bibitem{eckstein2014high}
A.~Eckstein, G.~Boucher, A.~Lema{\^{i}}tre, P.~Filloux, I.~Favero, G.~Leo,
  J.~E. Sipe, M.~Liscidini, and S.~Ducci, ``{High-resolution spectral
  characterization of two photon states via classical measurements},'' {\em
  Laser {\&} Photonics Reviews}, vol.~8, no.~5, pp.~L76----L80, 2014.

\bibitem{posner2018high}
M.~T. Posner, T.~Hiemstra, P.~L. Mennea, R.~H.~S. Bannerman, U.~B. Hoff,
  A.~Eckstein, W.~S. Kolthammer, I.~A. Walmsley, D.~H. Smith, J.~C. Gates, and
  Others, ``{High-birefringence direct UV-written waveguides for heralded
  single-photon sources at telecommunication wavelengths},'' {\em arXiv
  preprint arXiv:1805.12058}, 2018.

\bibitem{tichy2010zero}
M.~C. Tichy, M.~Tiersch, F.~de~Melo, F.~Mintert, and A.~Buchleitner,
  ``{Zero-transmission law for multiport beam splitters},'' {\em Physical
  review letters}, vol.~104, no.~22, p.~220405, 2010.

\bibitem{Marshall:09}
G.~D. Marshall, A.~Politi, J.~C.~F. Matthews, P.~Dekker, M.~Ams, M.~J.
  Withford, and J.~L. O'Brien, ``{Laser written waveguide photonic quantum
  circuits},'' {\em Opt. Express}, vol.~17, pp.~12546--12554, jul 2009.

\bibitem{Sansoni2010}
L.~Sansoni, F.~Sciarrino, G.~Vallone, P.~Mataloni, A.~Crespi, R.~Ramponi, and
  R.~Osellame, ``{Polarization entangled state measurement on a chip},'' {\em
  Physical review letters}, vol.~105, no.~20, p.~200503, 2010.

\bibitem{crespi2013integrated}
A.~Crespi, R.~Osellame, R.~Ramponi, D.~J. Brod, E.~F. Galvao, N.~Spagnolo,
  C.~Vitelli, E.~Maiorino, P.~Mataloni, F.~Sciarrino, E.~F. Galva, C.~Vitelli,
  E.~Maiorino, P.~Mataloni, F.~Sciarrino, E.~F. Galvao, N.~Spagnolo,
  C.~Vitelli, E.~Maiorino, P.~Mataloni, and F.~Sciarrino, ``{Integrated
  multimode interferometers with arbitrary designs for photonic boson
  sampling},'' {\em Nature Photonics}, vol.~7, no.~7, p.~545, 2013.

\bibitem{meany2015}
T.~Meany, M.~Gr{\"{a}}fe, R.~Heilmann, A.~Perez-Leija, S.~Gross, M.~J. Steel,
  M.~J. Withford, and A.~Szameit, ``{Laser written circuits for quantum
  photonics},'' {\em Laser {\&} Photonics Reviews}, vol.~9, no.~4,
  pp.~363--384, 2015.

\end{thebibliography}
